\documentclass[10pt,french,twoside]{book}

\usepackage[french]{babel}
\usepackage[latin1]{inputenc} 
\usepackage[T1]{fontenc} 

\usepackage{graphicx}
\usepackage{epsfig,psfig}
\usepackage{latexsym}
\usepackage{amsfonts}
\usepackage{amsmath,amssymb}
\usepackage{comment}
\usepackage{float}

\usepackage{fabdoc1,fabmacromath}
\usepackage{subfig}

\newcommand{\RR}{{\mathbb R}}
\newcommand{\NN}{{\mathbb N}}

\newtheorem{theorem}{Theorem}[section]
\newtheorem{lemma}{Lemma}[section]
\newtheorem{corollary}{Corollary}[section]

\newtheorem{proposition}[theorem]{Proposition}
\newtheorem{definition}{Definition}[section]

\newcommand{\be}{\begin{equation}}
\newcommand{\ee}{\end{equation}}
\newcommand{\bea}{\begin{eqnarray}}
\newcommand{\eea}{\end{eqnarray}}
\newcommand{\sh}{\sinh}
\newcommand{\ch}{\cosh}
\newcommand{\prf}{\noindent{\bf Proof}\ }
\newcommand{\qed}{{\hfill $\Box$}}
\newcommand{\proof}{{\bfseries Proof.}}

\newcommand{\beq}{\begin{equation}}
\newcommand{\eeq}{\end{equation}}
\newcommand{\beqa}{\begin{eqnarray}}
\newcommand{\eeqa}{\end{eqnarray}}
\newcommand{\noi}{\noindent}
\newcommand{\om}{\Omega}
\newcommand{\tom}{\tilde \Omega}
\newcommand{\e}{\varepsilon}

\begin{document}
\begin{titlepage}

\includegraphics[width=40mm]{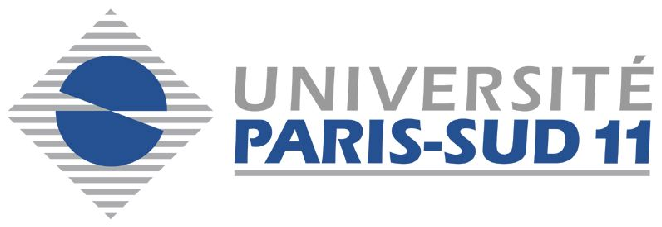}
\hfill
\includegraphics[width=20mm]{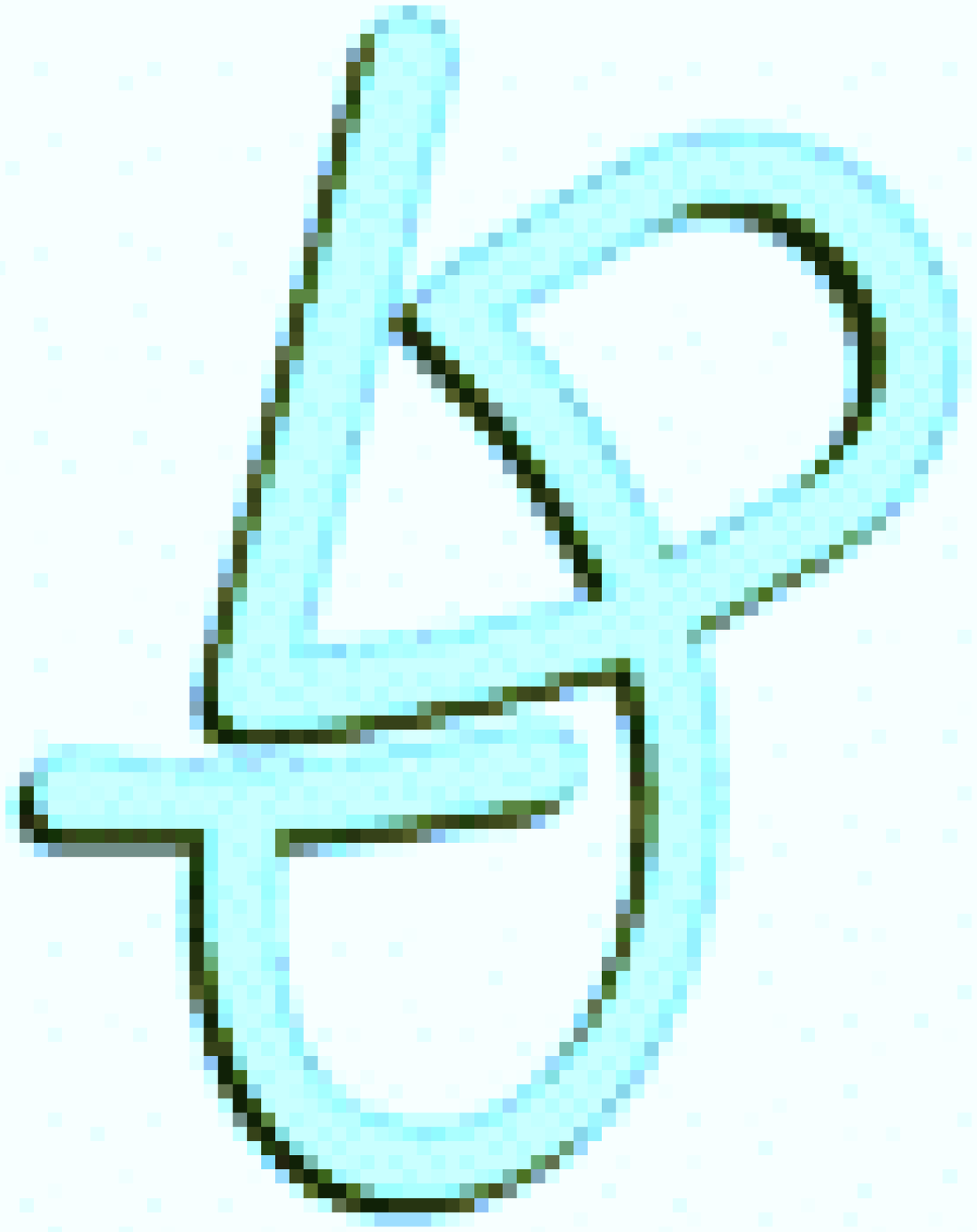}

\begin{center}
\fbox{{\bf UNIVERSIT\'E PARIS 11}}

\vskip1cm

{\bf TH\`ESE}
\vskip1cm

Sp\'ecialit\'e: {\bf  PHYSIQUE TH\'EORIQUE}\\

Pr\'esent\'e\\
 pour obtenir le grade de
\vskip0.5cm
\large {\bf Docteur de l'Universit\'e Paris 11}

\vskip0.75cm

par

\vskip0.75cm

{\sc \bf R\u azvan-Gheorghe Gur\u au}

\vskip0.5cm

Sujet: \\

\Large{\bf La renormalisation dans la th\'eorie non commutative des champs}

\vskip.5cm

Soutenue le 19 d\'ecembre 2007 devant le jury compos\'e de: 
\vskip.25cm
$\begin{array}{lll}
  & \mbox{Jean Bernard Zuber} & \mbox{pr\'esident} \\
  & \mbox{Vincent Rivasseau} & \mbox{directeur de th\`ese}\\
  & \mbox{Harald Grosse} & \mbox{rapporteur}\\
  & \mbox{Edwin Langmann} & \mbox{rapporteur}\\
  & \mbox{Jacques Magnen} & \mbox{examinateur}
   \end{array}$ 
\end{center}

\end{titlepage}

\newpage 

\thispagestyle{empty}
$ $
\newpage

\begin{center}
{\bf Remerciements}
\end{center}

Avant tout, je voudrais exprimer ma reconaissance \'a Vincent. Sans lui 
cette th\`ese n'aurait pas \'et\'e possible. Il a \'et\'e le meilleur directeur 
de th\`ese qu'on puisse avoir, non seulement pour ses calit\'es scientifiques 
mais aussi pour ses qualit\'es humaines. Au-del\`a de me former 
scientifiquement, il m'a appris l'ouverture, la rigueur et la critique 
(mod\'er\'ee par la bienveillance) et pour cela je lui serai endett\'e \`a jamais.
Je ne pense pas avoir \'et\'e le plus commode th\'esard, mais j'ai eu la chance 
et le privil\`ege d'avoir \'et\'e son \'el\`eve. Il a toujours trouv\'e le temps 
de r\'epondre \`a mes questions, m\^eme les plus b\^etes, m\^eme r\'ep\'et\'ees la 
dixi\`eme fois. Il a toujours pr\'ef\'er\'e m'aider plut\^ot que me laisser 
me d\'ebrouiller tout seul. J'esp\`ere que notre collaboration se poursuivra 
par la suite, car il est pour moi un mod\'ele humain et scientifique.

Je remercie \`a Jacques, de facto mon directeur de th\`ese adjoint. Nos 
collaborations m'ont appris surtout \`a ne jamais l\^acher un probl\`eme. Il 
n'y a aucun probl\`eme trop difficile pour lui, et j'esp\`ere 
apprendre \`a lui ressembler. Sa fa\,con de toujours r\'eduire un probl\`eme
\`a l'essentiel, quoi qu'il devienne des fois difficile \`a suivre, est 
enviable. Apr\`es ces ann\'ees pass\'ees ensemble j'esp\`ere avoir apris \`a
utiliser mon intelect de la m\^eme fa\,con. 

Je remercie \`a Jean Bernard Zuber, qui a eu la gentillesse d'accepter d'\^etre le 
pr\'esident de mon jury de th\`ese. Dans les trois ann\'ees que nous 
avons enseign\'e ensemble le LP206 il m'a fait apprendre qu'il vaut mieux 
accepter les autres avant d'essayer \`a les am\'eliorer. Je vous promets 
d'essayer de vous tutoyer!

A mes deux rapporteurs Harald Grosse et Edwin Langmann je voudrais
les remercier d'avoir tout de suite accept\'e de faire partie de mon jury 
de th\`ese malgr\'e les longues et peu commodes voyages qu'ils ont d\^u faire.

J'admire enorm\'ement les travaux de Harald Grosse, {\it les} travaux fondateurs pour les 
th\'eories des champs non commutatives renormalisables. De ses travaux 
j'ai appris que la rigueur peut nous guider dans un labyrinthe de 
techniques et qu'il faut parfois faire un long calcul avant d'arriver \`a 
la compr\'ehension. J'esp\`ere qu'en suivant son exemple je ne vais 
jamais me laisser intimid\'e par les dificult\'ees techniques.

Les travaux d'Edwin Langmann sur la dualit\'e Langmann-Szabo m'ont appris 
qu'il faut poursuivre des id\'ees non standard et audacieuses 
pour leur beaut\'e. De plus, les discussions avec lui ont ouvert 
mon int\'er\^et pour les mod\`eles matriciels que je voudrais
approfondir dans le futur.

Je remercie Nikita Nekrasov qui a \'ecrit un rapport sur ma th\`ese. Ses observations pertinentes m'ont amen\'e
a revoir et am\'eliorer une partie de celle-ci. Il m'a aussi sugg\'er\'e la r\'ef\'erence \cite{NekIns} qui introduit un terme harmonique dans les th\'eories noncommutatives sur le plan de Moyal. Ses  tr\`es vastes travaux dans notre domaine
montrent un esprit encyclop\'edique que j'esp\`ere acqu\'erir un jour.

Je voudrais ensuite exprimer ma reconaissance \`a Monsier Gheorghe 
Nenciu, mon premier mentor. Les lectures qu'il m'a conseill\'ees
sont \`a la base de ma formation, et si jamais je trouve qu'une 
id\'ee m'est famili\`ere c'est surtout grace \`a ses premiers conseils. Il est 
pour moi un example et un symbole d'honn\^etet\'e scientifique dans un 
contexte adverse. Ses encouragements ont beaucoup contribu\'e \`a ma 
d\'ecision de continuer mes \'etudes en France et c'est pour cela aussi 
que je lui suis reconnaissant.

Je remercie \`a Madame Florina Stan qui m'a b\'en\'evolement enseign\'e 
les fondements de la physique. Pour avoir accept\'e de m'apprendre 
la physique \`a un moment tr\`es important de ma formation sans jamais 
demander rien en retour, elle reste pour moi une des grandes chances
que j'ai eu.

Je voudrais aussi remercier \`a ma famille. Ma m\`ere, mon p\`ere et ma soeur
m'ont toujours soutenu dans ma decision de faire de la science. 
Ils ont toujours su m'aider et me conseiller, ainsi qu'\`a m'encourager de poursuivre 
mes id\'ees. Ma d\'ecision d'apprendre la physique est d\^ue en grande partie
\`a ma m\`ere, qui a su me donner le bon livre \`a lire au bon moment de mon 
enface. 

Je voudrais aussi exprimer mes remerciements \`a Marie-France Rivasseau pour avoir 
corrig\'e minutieusement le fran\,cais dans ma th\`ese. Si mon
travail n'est pas pr\'esent\'e dans un fran\,cais parfait c'est uniquement d\^u 
aux corrections ajout\'ees apr\`es sa derni\`ere lecture.
Je voudrais aussi lui remercier de m'avoir si souvent 
re\,cu si chalereusement chez elle au cours de ces ann\'ees.

Et surtout, je remercie \`a Delia. Elle a enti\'erement relu 
mon manuscrit et m'a \'enorm\'erment aid\'e \`a mieux transmettre 
mes idees. Elle m'a toujours encourag\'e \`a me perfectionner, 
scientifiquement et personellement. Elle sait toujours
me soutenir quand je passe des moments difficiles et je suis 
privil\'egi\'e de l'avoir \`a mes c\^ot\'es. 
Elle m'a appris \`a \^etre plus fort et plus confiant en moi. Avoir 
arriv\'e aujord'hui \`a pr\'esenter ce travail est tout autant d\^u \`a 
la confiance qu'elle a eu en moi qu'\`a son effort et 
sacrifice chaque jour. 
C'est pour cela que cette th\`ese lui est d\'edi\'ee.
\newpage

\begin{center}
{\bf Resum\'e}
\end{center}

	La th\'eorie non commutative des champs est un candidat possible pour la 
quantification de la gravit\'e. Dans notre th\`ese nous nous int\'eressons plus pr\'ecis\'ement au mod\'ele $\phi^4$ sur le plan de Moyal dans un potentiel harmonique, introduit par Grosse et Wulkenhaar.  Dans ce mod\`ele la dualit\'e de Langmann-Szabo pr\'esente au niveau du vertex est \'etendue au propagateur.
	Le long de nos \'etudes plusieurs r\'esultats ont \'et\'e obtenus concernant 
ce mod\`ele. Nous avons ainsi prouv\'e la renormalisabilite \`a touts les ordres directement dans l'espace des positions. Nous avons introduit la repr\'esentation param\'etrique du mod\`ele, ainsi que la repr\'esentation de Mellin Compl\`ete. Nous avons prouv\'e que le flot de la constante de couplage est born\'e \`a tous les ordres de la th\'eorie des perturbations. De plus nous avons introduit la r\'egularisation et la renormalisation dimmensionelle du mod\`ele. 
	Les directions futures de recherche comprennent l'\'etude des th\'eories de 
jauge sur le plan de Moyal et leur possible pertinence pour la quantification de la gravitation. Les liens avec la th\'eorie des cordes et la gravitation quantique \`a boucles pourront aussi \^etre d\'etaill\'es.
\begin{center}
{\bf Abstract}
\end{center}
	Non commutative quantum field theory is a possible candidate for the 
quantization of gravity.
	In our thesis we study in detail the $\phi 4$ model on the Moyal plane with 
an harmonic potential. Introduced by Grosse and Wulkenhaar, this model exhibits the Langmann-Szabo duality not only for  the vertex but also for the propagator. 
	We have obtained several results concerning this model. We have proved the 
renormalisability of this theory at all orders in the position space. We have introduced the parametric and Complete Mellin representation for the model. Furthermore we have proved that the coupling constant has a bounded flow at all orders in perturbation theory. Finally we have achieved the dimensional regularization and renormalization of the model.
	Further possible studies include the study of gauge theory on the Moyal plane 
and there possible applications for the quantization of gravity. The connections with string theory and loop quantum gravity should also be investigated.

\newpage
\thispagestyle{empty}

\tableofcontents
\newpage
\thispagestyle{empty}

\chapter{Introduction}

\medskip

Aujourd'hui la physique est une des sciences les plus accomplies. Elle explique avec une pr\'ecision remarquable les ph\'enom\`enes les plus divers, sur une plage d'\'echelles comprise entre l'\'echelle du syst\`eme solaire et l'\'echelle nucl\'eaire. Entre ces deux limites les lois physiques fondamentales (la m\'ecanique quantique, l'\'electromagn\'etisme, la gravitation, etc.) sont aussi bien pr\'edictives que simples. Elles comportent un nombre r\'eduit de param\`etres et de postulats. Les ph\'enom\`enes \'emergents, comme la chimie, la biologie, etc., sont tr\`es complexes \`a cause du grand nombre des constituants et non pas de la complexit\'e des principes fondateurs.

La situation d\'eg\'en\`ere au fur et \`a mesure qu'on s'\'ecarte de cette fen\^etre d'\'energies: on manque de mod\`eles non seulement suffisamment pr\'edictifs mais surtout math\'ematiquement satisfaisants.
Le mod\`ele standard, qui  s'est impos\'e pour la description de la physique \`a petite \'echelle, a plus de $30$ param\`etres ind\'ependants et son Lagrangien est compos\'e de plus de $100$ termes.

Toutes les th\'eories physiques comprennent deux aspects: les lois valides \`a une certaine \'echelle d'\'energie et la fa\c con dont elles influencent les lois "effectives" \`a d'autres \'echelles d'\'energie. Par exemple la m\'ecanique quantique fournit une description presque parfaite des ph\'enom\`enes \`a l'\'echelle atomique, mais il n'est pas raisonnable de l'appliquer directement au calcul de la pression d'un gaz dans une bo\^ite. Pour ce calcul on doit utiliser la m\'ecanique statistique classique.

Les ph\'enom\`enes \`a l'\'echelle nucl\'eaire sont d\'ecrits par la th\'eorie des champs. Elle combine la m\'ecanique quantique et la relativit\'e restreinte (ce qui constitue un pas vers l'unification) et elle a une puissance pr\'edictive impressionnante. L'histoire de la th\'eorie des champs est remarquable en elle-m\^eme. Propos\'ee dans les ann\'ees $40$ elle a failli \^etre abandonn\'ee dans les ann\'ees $60$ pour revenir en force dans les ann\'ees $70$.

L'outil conceptuel le plus neuf de cette th\'eorie est la renormalisation. Au d\'ebut simple "recette de cuisine" utile pour cacher des quantit\'es infinies dans des param\`etres non physiques, la renormalisation est devenue un des concepts de base de notre compr\'ehension de la physique \`a tr\`es courte \'echelle gr\^ace \`a l'interpr\'etation moderne \`a la Wilson.

Etant donn\'e une th\'eorie \`a une \'echelle d'\'energie \'elev\'ee, la renormalisation nous indique comment cette th\'eorie se transforme \`a toutes les \'echelles  d'\'energie plus basses (donc \`a plus grande distance). Ainsi, il se peut que la structure de la th\'eorie se reproduise, mais les valeurs de ses param\`etres changent (et dans ce cas la th\'eorie est dite renormalisable), ou il se peut que la th\'eorie initiale engendre une autre th\'eorie effective compl\`etement diff\'erente.

Comme nous ne connaissons pas la th\'eorie ultime, nous pouvons supposer qu'\`a une \'energie tr\`es \'elev\'ee on a un r\'eservoir avec toutes les th\'eories possibles.
Si l'une de ces th\'eories ne survit pas sur plusieurs \'echelles alors, plut\^ot qu'une v\'eritable th\'eorie universelle, elle est un accident du point de vue des autres \'echelles. Par contre, si la th\'eorie se reproduit sur plusieurs \'echelles nous allons la rencontrer g\'en\'eriquement.  Cette remarque nous m\`ene \`a consid\'erer de pr\'ef\'erence des th\'eories renormalisables pour d\'ecrire la physique.

Parmi les th\'eories des champs possibles nous nous int\'eressons aux th\'eories des champs sur des espaces non commutatifs (TCNC).
Ces th\'eories pourraient \^etre adapt\'ees \`a la description de la physique au-del\`a du mod\`ele standard. De plus, elles sont certainement pertinentes pour la description de la physique en champ fort. La question de leur renormalisabilit\'e est donc fondamentale pour savoir si ces th\'eories sont g\'en\'eriques et universelles au sens d\'ecrit ci-dessus. C'est l'\'etude de cette question qui est le th\`eme central de notre th\`ese.

Comme la renormalisation est une technique assez compliqu\'ee nous devons nous poser la question du niveau de rigueur des r\'esultats que nous souhaitons avoir. Il y a trois niveaux de rigueur que nous pourrions nous proposer dans notre \'etude.

Un r\'esultat peut \^etre v\'erifi\'e jusqu'\`a un certain ordre de la th\'eorie de 
perturbations. C'est le niveau auquel s'arr\^ete le physicien exp\'erimentateur. Le mod\`ele standard a \'et\'e v\'erifi\'e dans les exp\'eriences de haute \'energie jusqu'\`a l'ordre trois au moins dans certains effets fins comme le calcul du facteur $g-2$ de l'\'electron pr\'edit avec dix chiffres en \'electrodynamique quantique \footnote{Ce facteur est la quantit\'e physique pr\'edit avec la plus grande pr\'ecision par une th\'eorie.}.
Du point de vue du groupe de renormalisation la plus spectaculaire v\'erification est la variation de la constante de structure fine par deux pour cent sur la plage d'\'echelle accessibles aux exp\'eriences au LEP.

Le deuxi\`eme niveau de rigueur que nous pouvons nous proposer est de prouver un r\'esultat \`a tous les ordres dans la th\'eorie des perturbations. C'est le niveau du physicien th\'eoricien, et c'est \`a ce niveau-ci que nous pr\'esentons tous les travaux de notre th\`ese.

Le troisi\`eme niveau est constructif. C'est le niveau ultime du physicien math\'ematicien. Il s'agit par exemple de prouver la ressommation au sens de Borel des s\'eries de perturbations en une fonction, et ensuite d'\'etendre les propri\'et\'es prouv\'ees pour la s\'erie \`a sa somme de Borel. Des d\'eveloppements r\'ecents prometteurs existent dans cette direction qui nous permettent d'esp\'erer que la construction des mod\`eles de th\'eories des champs sur des espaces non commutatifs ne va pas tarder.

Dans le chapitre suivant nous pr\'esentons une introduction g\'en\'erale aux th\'eories des champs sur des espaces non commutatifs. Le chapitre trois comprend une introduction g\'en\'erale au mod\`ele $\Phi^{\star 4}_4$ sur l'espace de Moyal et pr\'esente d'autres mod\`eles trait\'es par nous. Les chapitres quatre \`a huit pr\'esentent diff\'erents r\'esultats obtenus au cours de notre th\`ese. Le chapitre neuf pr\'esente les conclusions de cette th\`ese. Le chapitre dix pr\'esente quelques aspects techniques concernant le plan de Moyal. Les appendices contiennent les diff\'erents travaux qui sont \`a la base de nos r\'esultats.

\newpage
\thispagestyle{empty}

\chapter{Historique et probl\'ematique des TCNC}

L'id\'ee de chercher une repr\'esentation g\'eom\'etrique nouvelle du monde \`a tr\`es petite \'echelle est bien plus ancienne qu'on ne pourrait le croire. Des id\'ees dans ce sens ont \'et\'e avanc\'ees par Schr\"odinger \cite{Sch} d\`es l'ann\'ee 1934. Reprises par Heisenberg \cite{Heis}, Pauli et Oppenheimer, ces id\'ees sont pour la premi\`ere fois formul\'ees math\'ematiquement par Snyder \cite{Sny}. Les \'etudes initiales ont \'et\'e motiv\'ees d'une part par l'\'etude des particules dans des forts champs magn\'etiques et d'autre part par l'espoir que le comportement \`a petite \'echelle (UV) de telles th\'eories sera am\'elior\'e. 

Apr\`es ces travaux initiaux, l'id\'ee de combiner une non-commutativit\'e des coordonn\'ees avec la physique a \'et\'e longtemps 
n\'eglig\'ee \`a cause principalement des difficult\'es math\'ematiques (notamment li\'ees \`a l'invariance de Lorentz) et du d\'eveloppement spectaculaire des th\'eories des champs commutatives.

Nous avons aujourd'hui plusieurs raisons de nous int\'eresser de nouveau aux TCNC. Nous les rencontrons comme candidates pour la physique des hautes \'energies mais aussi comme une description naturelle de la physique des particules non relativistes dans un champ magn\'etique fort. 
L'effet Hall quantique d\'ecouvert en 1980 par von Klitzing \cite{Klitz} a \'et\'e l'une des plus grandes surprises de la physique moderne: les \'electrons dans un champ fort ont un comportement collectif qui conduit \`a la quantification de la r\'esistance. 
L'effet Hall fractionnaire, mis en \'evidence par Tsui et St\"ormer \cite{Tsui} n'a pas \'et\'e pr\'evu par les th\'eoriciens et il est encore plus surprenant. Les travaux th\'eoriques sur ce sujet ont culmin\'e avec l'introduction par Laughlin \cite{Laugh} d'une fonction d'onde non locale d\'ecrivant le liquide de Hall. Cette fonction d'onde peut \^etre interpr\'et\'ee en termes d'un gaz d'anyons, des objets ayant des charges fractionnaires. Ces objets ont \'et\'e mis en \'evidence exp\'erimentalement sous la forme de bruit de gr\`enaille (voir \cite{Glat} pour une pr\'esentation d\'etaill\'ee). Dans ces exp\'eriences nous "voyons" en quelque sorte concr\`etement les charges fractionaires.

En plus de la description de Laughlin nous disposons d'une description d'un syst\`eme de Hall en terms d'une th\'eorie de champ de Chern Simons. Une interpr\'etation de cette th\'eorie en terme de particules compos\'ees form\'ees d'\'electrons habill\'es avec des quanta du champ magn\'etique \`a \'et\'e propos\'ee par Read \cite{Read}.

Les deux descriptions de l'effet Hall font intervenir des objets non locaux. Leur analyse devrait faire intervenir en cons\'equence des th\'eories de champs non locales. La non-localit\'e peut, \`a son tour, \^etre traduite comme provenant d'une description non commutative de l'espace. Ces consid\'erations ont men\'e Susskind \cite{Susk} \`a conjecturer que l'effet Hall devrait \^etre bien d\'ecrit 
par une th\'eorie de Chern Simons non commutative. Ses travaux formels ont \'et\'e affin\'es par
Polychronakos \cite{Polych} qui a propos\'e une version matricielle avec des coupures raides (sharp cutoffs) du mod\`ele de Chern Simons non commutatif comme th\'eorie finitaire utilisable dans la description de l'effet Hall. Van Raamsdonk et al. \cite{Raam} ont mis en correspondance les \'etats propres de la th\'eorie de Polychronakos avec les fonctions d'onde de Laughlin. Ces id\'ees sont \`a la base de l'\'etude des liquides non commutatives (voir \cite{Bour} pour une pr\'esentation du sujet). 

Nous esp\'erons que le d\'eveloppement dans cette th\`ese du groupe de renormalisation adapt\'ee aux TCNC s'appliquera \`a ces probl\`emes. Il n'existe pas encore une \'etude de la renormalisation des th\'eories de champs qui s'appliquent \`a l'effet Hall. Les comportements \`a longue distance caract\'eristiques de l'effet Hall devraient pouvoir \^etre compris en termes d'un flot du groupe de renormalisation. Une telle \'etude apporterait l'explication {\it ab initio} de l'effet Hall fractionnaire ainsi qu'une description pr\'ecise et unifi\'ee des transitions entre tous les plateaux de Hall.

La deuxi\`eme raison en faveur de l'\'etude des TCNC est la physique des hautes \'energies. Nous allons maintenant nous concentrer sur les arguments apport\'es par cette physique car la bonne th\'eorie de la gravitation quantique reste \`a ce jour le "Graal" de la physique th\'eorique moderne.

Une fois le mod\`ele standard formul\'e dans les ann\'ees $70$ les physiciens ont commenc\'e \`a se poser s\'erieusement le probl\`eme de la quantification de la gravitation. Apr\`es de nombreux efforts dans cette direction nous avons aujourd'hui plusieurs candidats pour cette th\'eorie quantique de la gravitation. Parmi ces propositions la th\'eorie des cordes et la gravit\'e quantique \`a boucles sont les plus \'elabor\'ees.

La th\'eorie des cordes \`a \'et\'e introduite dans les ann\'ees $70$ par Veneziano comme th\'eorie effective des interactions fortes.
A partir des travaux de $1984$  de Green et Schwarz, qui prouvent l'absence d'anomalies, elle s'est impos\'ee comme la proposition dominante dans la physique au del\`a du mod\`ele standard.
Les cordes supersym\'etriques semblent promettre une bonne th\'eorie de la gravitation quantique. Le principal succ\`es de cette th\'eorie reste \`a ce jour la pr\'esence du graviton (particule de masse nulle et spin $2$) dans 
 le spectre de la corde ferm\'ee. Ainsi elle nous fournit un formalisme qui unifie 
naturellement les champs de jauge avec les gravitons.

De plus l'entropie des trous noirs \`a \'et\'e calcul\'ee par Strominger et Vafa \cite{StrVafa} et ensuite Maldacena et al. \cite{Malda}. Ce calcul, un grand succ\`es de la th\'eorie des cordes, est cependant limit\'e \`a des 
 trous noirs tr\`es particuliers ($Q\approx M$, appell\'ee aussi quasi-extr\'emaux).

D'autre part, malgr\'e de nombreux efforts dans ce sense, la seconde quantification de la th\'eorie des cordes et l'identification du vide restent non compl\`etement elucid\'es. La possibilit\'e de vides multiples compatibles avec la phenom\'enologie actuelle (le probl\`eme du "paysage") rendent les pr\'edictions physiques difficiles.

La gravit\'e quantique \`a boucles renonce totalement aux notions g\'eom\'etriques habituelles et essaie de les retrouver comme engendr\'ees dynamiquement. L'objet fondamental dans cette th\'eorie sont les r\'eseaux de spins. Nous essayons d'associer \`a chaque configuration d'un r\'eseau de spins un espace-temps. Cette hypoth\`ese, certes s\'eduisante est tr\`es difficile \`a manipuler et donne difficilement des pr\'edictions.

Parmi les avanc\'ees r\'ecentes dans la domaine de la gravit\'e quantique on doit aussi mentionner les th\'eories de la relativit\'e modifi\'ees comme la relativit\'e doublement restreinte. Cette derni\`ere rend compte d'une longueur minimale \'egale pour tous les observateurs, fix\'ee \`a l'\'echelle de Planck. On ne dispose pas pour l'instant d'un sch\'ema de quantification satisfaisant de telles th\'eories.

Pour tous ces raisons une th\'eorie mathematiquement plus simple qui rende compte d'une partie des caract\'eristiques des th\'eories pr\'esent\'ees auparavant serait fort souhaitable.

Les TCNC \'etudi\'ees dans cette th\`ese sont un autre candidat \`a la quantification de la gravit\'e. A cause de leur \'enorme succ\`es exp\'erimental les th\'eories des champs commutatives sont notre point de d\'epart. Nous voulons trouver une description des ph\'enom\`enes reconstituant dans une certaine limite la th\'eorie des champs commutative mais qui, en plus, prenne en compte des effets de la relativit\'e g\'en\'erale.

La th\'eorie math\'ematique de la g\'eom\'etrie non commutative a \'et\'e d\'evelopp\'ee dans les ann\'ees 80-90 par des math\'ematiciens comme Alain Connes, Michel Dubois-Violette, John Madore, etc.. Leurs travaux math\'ematiques ont sembl\'e au d\'ebut peu li\'es \`a la physique.

Tout a chang\'e dans les ann\'ees 90. Plusieurs r\'esultats de la th\'eorie des cordes ont point\'e vers un r\'egime limite de la th\'eorie de type IIA d\'ecrit par la th\'eorie non commutative sur le plan de Moyal. Cela, et l'espoir de construire des th\'eories automatiquement r\'egularis\'ees dans l'UV, a relanc\'e la recherche sur les TCNC. Assez t\^ot Filk \cite{Filk:1996dm} s'est rendu compte que dans les nouvelles TCNC le secteur planaire a le m\^eme comportement que les th\'eories commutatives. La renormalisation du secteur planaire de ces th\'eories, qui contient une infinit\'e de graphes, est donc n\'ecessaire.

Les th\'eories de jauge non commutatives sur le plan de Moyal ont attir\'e l'int\'er\^et \`a partir des travaux de Martin et Sanchez-Ruiz en 1999 \cite{Sanchez}. Des \'etudes ont \'et\'e men\'ees sur le tore non commutatif par Sheikh-Jabbari \cite{Sheikh}. Les auteurs ont v\'erifi\'e la renormalisabilit\'e \`a une boucle de ces th\'eories.

Ces travaux ont culmin\'e avec les travaux de Seiberg et Witten en 1999 \cite{SeiWitt}. Ces auteurs ont donn\'e une relation (strictement formelle) entre les th\'eories de Yang-Mills non commutatives et les th\'eories ordinaires. Cette relation a engendr\'e une activit\'e f\'ebrile et de milliers des papiers sur le sujet. Dans la m\^eme p\'eriode, les travaux de Maldacena \cite{Malda2} ont mis en relation la th\'eorie des cordes et les th\'eories de jauge maximalement supersym\'etriques. Nous voyons ainsi un triangle de relation se d\'evelopper entre les th\'eories des champs commutatives, les TCNC et la th\'eorie des cordes.

D'autre part, dans les ann\'ees 97 Chamseddine et Connes \cite{Connes} ont trouv\'e une reformulation du mod\`ele standard (au niveau classique) en terme d'un principe d'action spectrale sur une g\'eom\'etrie non commutative tr\`es simple. Leurs travaux ont inspir\'e un vif int\'er\^et (sp\'ecialement du c\^ot\'e des math\'ematiciens) et les actions spectrales ont aujourd'hui une vie scientifique propre. Malgr\'e l'incontestable r\'eussite de condenser le mod\`ele standard sous une forme tr\`es compacte ces travaux suivent encore les sch\'emas de quantification habituelle. 
En particulier ils ne fournissent pas un nouveau groupe de renormalisation pour le traitement du mod\`ele standard ni une compr\'ehension nouvelle de la gravit\'e quantique. En ce sens les action spectrales semblent \^etre seulement le sommet d'un iceberg non commutatif encore largement \`a d\'ecouvrir.

Les TCNC ont \'et\'e ensuite mis en relation avec la M-th\'eorie compactifi\'ee sur des tores, et une \'etude approfondie des instantons et solitons dans les TCNC a \'et\'e men\'ee par Nekrasov \cite{NekIns}. En plus la g\'eom\'etrie non commutative s'est av\'er\'ee un outil tr\`es puissant pour la quantification par d\'eformation dans la th\'eorie des cordes. L'\'etude des groupes et groupo\"ides quantiques a \'et\'e entreprise \`a l'aide de la g\'eom\'etrie non commutative.

Entre les ann\'ees 1995 et 2000 les progr\`es rapides dans les TCNC amen\`erent l'espoir de faire des calculs plus s\^urs et plus pouss\'es que dans la th\'eorie des cordes. Les TCNC avaient \'et\'e prouv\'ees renormalisables \`a une et deux boucles et tout le monde supposait qu'elles le seraient \`a tous les ordres. Un papier de 2000 par Chepelev et Roiban \cite{CheRoi} (r\'ev\'el\'e aujourd'hui faux) semblait dire que c'\'etait vraiment le cas. On avait alors une reformulation \'el\'egante du mod\`ele standard, et on commen\c cait \`a comprendre tr\`es bien les th\'eories de jauge non commutatives. Un compte-rendu de l'\'etat des TCNC dans l'ann\'ee 2001 se trouve dans \cite{DN}.

Les premiers troubles sont apparus peu apr\`es le papier de Chepelev et Roiban. Minwalla, Van Raamsdonk et Seiberg \cite{MvRS} ont trouv\'e un probl\`eme qui avait \'echapp\'e \`a ces auteurs: il existe des graphes non-planaires qui sont convergents UV gr\^ace \`a certains facteurs oscillants mais qui, \`a cause de leur comportement IR, ins\'er\'es dans des graphes plus grands engendrent des divergences (ce ph\'enom\`ene est expliqu\'e en d\'etail dans le chapitre suivant). Ces divergences ne sont pas renormalisables et portent le nom de m\'elange UV/IR. Malgr\'e ce probl\`eme, le comptage de puissance de Chepelev et Roiban semblait \^etre correct. On sait aujourd'hui que ce n'est pas le cas. M\^eme leur th\'eor\`eme de comptage des puissances (sans preuve publi\'ee) s'est r\'ev\'el\'e en fait erron\'e.

Le m\'elange UV/IR est un probl\`eme fondamental des TCNC. Toute th\'eorie des champs r\'eels va souffrir de ces divergences. Malgr\'e de nombreux efforts afin de surmonter ce probl\`eme on n'a pas trouv\'e de solution satisfaisante pendant quelques ann\'ees. Entre temps les TCNC ont \'et\'e abandonn\'ees par une bonne partie de la communaut\'e des cordes.

Pourquoi revenir sur des th\'eories qui semblent inconsistantes? Nous allons pr\'esenter un "Gedanken Experiment" en faveur des TCNC.

Commen\c cons par un argument classique de m\'ecanique quantique, pr\'esent\'e par Feynman. Quand nous mesurons la position d'une particule (supposons un \'electron) nous l'illuminons avec un photon. Nous observons le processus d'absorbtion du photon par l'\'electron.

La pr\'ecision de la mesure de la position de l'\'electron 
est donn\'ee par la longueur d'onde du photon. Cette limitation est due au fait que la probabilit\'e de pr\'esence du photon est \'etendue sur une largeur spatiale donn\'ee par sa longueur d'onde.
Pour l'augmenter on doit utiliser des photons d'\'energie de plus en plus haute. Les photons vont subir un choc avec les \'electrons, et vont leur communiquer une certaine impulsion. Comme le choc est \'elastique on ne peut pas affirmer avec pr\'ecision quelle impulsion va \^etre transf\'er\'ee entre les photons et les \'electrons. L'impulsion de transfert maximal nous donne l'incertitude sur l'impulsion de l'\'electron. Les photons avec de hautes \'energies vont perturber l'impulsion de l'\'electron de plus en plus. Les deux mesures sont concurrentes: accro\^itre la pr\'ecision de la mesure de la position d'un \'electron va augmenter l'incertitude de la mesure de son impulsion. Nous sommes ainsi forc\'es d'introduire le principe d'incertitude de Heisenberg.

Toutes les th\'eories qui essaient de d\'ecrire l'au-del\`a du mod\`ele standard ont converg\'e sur l'id\'ee qu'il existe une \'echelle fondamentale dans la nature o\`u les effets de la quantification de la gravitation seront ressentis: l'\'echelle de Planck $\ell_P$. Le rapport entre cette \'echelle et l'\'echelle de la corde joue un r\^ole essentiel dans la th\'eorie des cordes. En m\^eme temps c'est l'\'echelle unit\'e du r\'eseau de spin dans la gravitation quantique, et l'\'echelle invariante dans la relativit\'e doublement restreinte.

Il est assez naturel de conjecturer qu'\`a cette \'echelle la nature m\^eme de l'espace-temps pr\'esente des changements profonds. Il devrait \^etre impossible de localiser un objet avec une pr\'ecision plus petite que la longueur de Planck. Ce qui est plus profond et moins \'evident au sujet de l'\'echelle de Planck c'est qu'elle est plus profond\'ement une \'echelle d'aire qu'une \'echelle de longueur. 

Faisons une exp\'erience de pens\'ee inspir\'ee par l'exp\'erience classique de Feynman.
Nous voudrions mesurer la position d'un \'electron dans l'espace-temps. Supposons qu'on est loin de toute mati\`ere donc que l'espace temps est plat. Nous envoyons un photon (ou toute autre particule) dans la direction de l'\'electron. Nous essayons de mesurer la position de l'\'electron \`a l'aide de ce photon. La longueur d'onde associ\'ee \`a un photon libre d'\'energie $E$ est $\lambda\approx hcE^{-1}$. C'est la pr\'ecision maximale avec laquelle on peut localiser le photon dans sa direction de mouvement (et dans le temps). C'est ainsi la pr\'ecision de la mesure de la position du point dans ces deux directions. Jusqu'ici on n'a pas innov\'e par rapport \`a l'exp\'erience de Feynman. Mais supposons maintenant qu'on veuille mesurer en m\^eme temps la position du point sur les directions transversales au mouvement du photon. A tout photon d'\'energie $E$ on associe un rayon de Schwarzschild $r_S\approx GEc^{-4}$. Ce rayon repr\'esente l'horizon du photon. Aucune information (comme la position d'un \'electron \`a l'int\'erieur de cet horizon) ne peut le franchir. Ce que nous pouvons affirmer avec certitude c'est que le point est soit \`a l'int\'erieur soit \`a l'ext\'erieur de l'horizon. Le rayon de Schwarzschild nous donne une borne sur la pr\'ecision de la localisation du point sur les deux directions orthogonales au mouvement du photon. 
Comme $r_S$ augmente avec l'\'energie, les mesures de la position d'un point sur les directions parall\`eles et orthogonales \`a son mouvement sont incompatibles!

Le produit de ces deux incertitudes est $\lambda r_S=\ell_P ^2$. On ne peut pas sonder la g\'eom\'etrie sur des aires plus petites que cette aire fondamentale, le concept d'espace-temps continu doit \^etre repens\'e!

Ce argument est classique et a \'et\'e largement d\'ebattu au cours du temps. La principale critique qu'on lui a oppos\'e c'est qu'il repose sur le pr\'ejug\'e que la relativit\'e g\'en\'erale s'applique \`a des \'echelles o\`u elle n'a jamais \'et\'e test\'ee. 
Cette critique n'est pas totalement fond\'ee. En effet le seul ingr\'edient de relativit\'e g\'en\'erale dont on s'est servi est la notion d'horizon. Cette notion est beaucoup plus large que la relativit\'e g\'en\'erale. La thermodynamique des trous noirs, dont la th\'eorie microscopique est gouvern\'ee par une th\'eorie quantique de la gravit\'e encore \`a d\'ecouvrir, est fond\'ee sur cette notion\footnote{En effet, comme tout objet ne peut \^etre observ\'e qu'\`a travers son horizon, nous disposons aujourd'hui d'une enti\`ere th\'eorie physique bas\`ee sur le principe holographique, disant que tout ph\'enom\`ene observable doit \^etre d\'ecrit comme une th\'eorie sur la fronti\`ere du ph\'nom\`ene.}.

La fa\c con la plus simple d'engendrer un principe d'incertitude entre les positions est d'introduire la non-commutation des coordonn\'ees (en tant qu'op\'erateurs quantiques):
\bea
   [x^i,x^j]=\imath \theta^{ij} \; ,
\eea
o\`u $\theta$ est une nouvelle constante analogue \`a la constante de Planck.
Un espace de ce type porte le nom d'espace de Moyal. Nous pouvons \'evidemment essayer de traiter des commutateurs plus compliqu\'es $\theta^{ij}(x)$.

Cet argument simple nous dit que les TCNC sont fondamentales pour la compr\'ehension de la physique \`a l'\'echelle de Planck! Nous devons donc chercher des solutions afin de contourner les probl\`emes de ces th\'eories comme le m\'elange UV/IR!

La solution du m\'elange a d\^u attendre les ann\'ees 2003-2004 et les travaux fondateurs de Grosse et Wulkenhaar \cite{GrWu03-1, GWR2x2, GrWu04-3} portant sur le mod\`ele $\phi^4_4$ sur l'espace de Moyal $\RR^4$. Le vertex de cette th\'eorie poss\`ede une sym\'etrie non triviale\footnote{Introduit par Langmann et Szabo \cite{LaSz}.}: il a la m\^eme forme dans l'espace des positions et dans l'espace des impulsions. L'id\'ee de Grosse et Wulkenhaar a \'et\'e de modifier aussi le propagateur du mod\`ele pour ob\'eir \`a cette sym\'etrie\footnote{Un t\`erme du m\^eme type a \'et\'e deja introduit dans \cite{NekraHarmonique} mais dans un contexte tr\`es different.}! Cette modification relie les r\'egimes UV et IR au niveau de l'action libre et constitue la bonne solution du m\'elange. 
Grosse et Wulkenhaar ont pu prouver que le mod\`ele $\Phi_4^{\star 4}$ ainsi modifi\'e est renormalisable \`a tous les ordres de perturbation. Nous allons dor\'enavant appeler ce mod\`ele vulcanis\'e.

Leurs travaux ont \'et\'e ensuite repris et d\'evelopp\'es et la grande majorit\'e des r\'esultats de notre th\`ese porte sur ce sujet.

Nos \'etudes nous poussent \`a compl\`etement changer notre intuition habituelle sur les petites et les grandes \'echelles. Il s'av\'ere que penser aux \'echelles en termes de distance n'a plus de sens dans les TCNC: il y a une sym\'etrie entre les comportements \`a longue et courte distance. Seul a un sens de parler de zones de haute et basse \'energie. Comme l'\'energie est born\'ee inf\'erieurement on a une seule direction du groupe de renormalisation! Ce point de vue peut \^etre qualifi\'e de spectral, car on base notre analyse sur le spectre du hamiltonien. Il rejoint ainsi les id\'ees de l'action spectrale et nous fournit probablement le bon cadre pour leur quantification future.

Nous pouvons adopter deux attitudes envers ce probl\`eme: soit essayer de trouver le plus vite possible le maximum de mod\`eles ayant la sym\'etrie de Langmann-Szabo et qui ne pr\'esentent pas de m\'elange, soit essayer de comprendre en plus grand d\'etail la th\'eorie $\Phi_4^{\star 4}$ et le nouveau groupe de renormalisation qui lui est associ\'e. Le long de cette th\`ese nous allons prendre le deuxi\`eme point de vue. Nous esp\'erons ainsi d\'egager le plus grand nombre de traits sp\'ecifiques des TCNC sur l'espace de Moyal, et d\'evelopper une panoplie d'outils plus compl\`ete.

Parmi les mod\`eles r\'ecents vulcanises de TCNC nous avons le mod\`ele de Gross-Neveu \'etudi\'e par Vignes-Tourneret \cite{RenNCGN05} et des propositions de th\'eories de jauge d\'evelopp\'ees ind\'ependemment par Grosse et al. \cite{GrosseYM}, de Goursac et al. \cite{WalletYM}. Tandis que le mod\`ele de Vignes-Tourneret a \'et\'e prouv\'e renormalisable nous ne disposons pas encore de preuve que les th\'eories de jauge r\'ecemment propos\'e le soient.
Le mod\`ele $\Phi_6^{\star 3}$ a \'et\'e prouv\'e renormalisable par Gross et Steinacker \cite{GrossePhi3}, et le mod\`ele $\Phi_3^{\star 6}$ par Zhi Tuo Wang.

Le premier trait sp\'ecifique des TCNC est la non-localit\'e. Comme la notion de point perd son sens il est in\'evitable de perdre la notion de localit\'e. Cette perte et tous les probl\`emes correspondants d'unitarit\'e et perte d'invariance sous le groupe de Poincar\'e font que les TCNC ont des difficult\'es \`a s'imposer 
dans la communaut\'e des physiciens.

Cependant une non-commutation de type Moyal est covariante sous le groupe de Lorentz restreint. Nous consid\'erons les TCNC comme une premi\`ere \'etape en direction de la physique de l'\'echelle de Planck. Nous conjecturons qu'au fur et \`a mesure qu'on monte dans les \'energies on va rencontrer des comportements de TCNC sur l'espace de Moyal qui seront ensuite remplac\'es par des comportements plus compliqu\'es avec un param\`etre $\theta$ qui deviendra sans doute dynamique, jusqu'\`a l'\'echelle de Planck o\`u nous esp\'erons atteindre une formulation r\'eellement ind\'ependante du fond.

\newpage
\thispagestyle{empty}

\chapter{Le mod\`ele $\Phi^{\star 4}_4$}

Nous voulons entreprendre une \'etude qui nous r\'ev\`ele avec pr\'ecision le plus grand nombre des traits sp\'ecifiques des TCNC. Pour cela nous sommes forc\'es de choisir un espace non commutatif tr\`es simple. Par la suite nous allons nous placer sur l'espace de Moyal. Sa description pr\'ecise se trouve dans l'appendix \ref{sec:planMoyal}. En quelques mots, il s'agit de l'espace $\RR^D$ muni d'un produit non commutatif de fa\c con que le commutateur des deux  coordonn\'ees soit une constante.

\section{D\'efinition du mod\`ele}

Nous voulons g\'en\'eraliser le mod\`ele $\Phi^4_4$ commutatif d\'efini par l'action:
\bea
\label{eq:phi44com}
 S=\int d^Dx [\frac{1}{2} \partial_{\mu} \bar{\phi}(x) \partial_{\mu} \phi(x) + 
 \frac{1}{2} \mu^2 \bar{\phi}(x)\phi(x)+
 \frac{\lambda}{4!}\bar{\phi}(x)\phi(x)\bar{\phi}(x)\phi(x)]\; ,
\eea
qui d\'ecrit un boson charg\'e muni d'une interaction \`a quatre corps.

Dans un premier temps on va simplement remplacer les produits par des produits de Moyal:
\bea
\label{eq:nonvulcphi44}
S=\int d^Dx [\frac{1}{2}\partial_{\mu}\bar{\phi}(x)\star\partial^{\mu}\phi(x)+
\frac{1}{2}\mu^2 \bar{\phi}\star\phi(x)+
 \frac{\lambda}{2}\bar{\phi}\star\phi\star\bar{\phi}\star\phi(x)]\; ,
\eea
o\`u $\star$ est le produit de Moyal d\'efini par (\ref{eq:prodMoyal}).

Une propri\'et\'e (\ref{eq:integralMoyal}) du produit de Moyal est que dans toute int\'egrale quadratique on peut le remplacer par le produit habituel. Par cons\'equent la partie libre de l'action est celle habituelle (commutative), la diff\'erence entre ce mod\`ele et le mod\`ele correspondant commutatif provenant exclusivement du terme d'interaction (explicit\'e en eq. (\ref{eq:prod4Moyal})).

\begin{figure}
  \centerline{\epsfig{figure=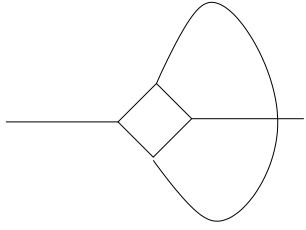,width=4cm}}
\caption{Le tadpole nonplanaire}
\label{fig:nonplTadpole}
\end{figure}

Pour un champ scalaire r\'eel, le mod\`ele na\"if souffre d'un nouveau type de divergence: le m\'elange ultraviolet infrarouge.
Il existe un graphe (appel\'e tadpole non planaire pour des raisons qui seront plus claires par la suite), repr\'esent\'e dans la figure \ref{fig:nonplTadpole}, qui est convergent ultraviolet, mais qui, encha\^in\'e un nombre arbitraire de fois dans une boucle, engendre une divergence infrarouge aussi forte que l'on veut. 

Son amplitude \`a impulsion de transfert $p$ est:
\bea
{\cal A}\propto \int d^4k \frac{e^{ip_{\mu}k_{\nu}\theta^{\mu \nu}}}{k^2+\mu^2}
\simeq_{p \to 0} p^{-2}\; ,
\eea
et l'on peut engendrer une puissance n\'egative de $p$ toute aussi grande que l'on d\'esire!
Cette divergence ne peut pas \^etre r\'eabsorb\'ee dans la red\'efinition des param\`etres initiaux de la th\'eorie et rend la th\'eorie non renormalisable.

Ce type de divergence s'av\`ere \^etre tr\`es g\'en\'erique dans la TCNC. Non seulement elle est pr\'esente dans toutes les th\'eories r\'eelles d\'efinies sur l'espace de Moyal, mais elle est aussi pr\'esente pour des th\'eories d\'efinies sur des vari\'et\'es non commutatives plus g\'en\'erales. 

Il est vrai que ce m\'elange n'est pas pr\'esent dans une th\'eorie complexe, car les graphes comme le tadpole non planaire ne sont pas permis dans un tel mod\`ele, mais cette solution n'est pas satisfaisante. C'est plut\^ot un accident d\^u aux r\`egles de contraction qu'une solution de principe.

Malgr\'e de nombreux efforts sur ce sujet, le m\'elange n'a pu \^etre r\'esolu qu'avec l'introduction par Grosse et Wulkenhaar d'une modification de l'action initiale du mod\`ele. 

Le vertex de ce mod\`ele poss\`ede une sym\'etrie non triviale: son expression dans l'espace des impulsions et des positions est identique! On appelle cette sym\'etrie la dualit\'e de Langmann-Szabo.
La perc\'ee de Grosse et Wulkenaar consiste dans l'addition d'un nouveau terme quadratique dans l'action pour rendre covariante sous cette transformation aussi la partie libre du mod\`ele. Le mod\`ele de $\Phi^{\star 4}_4$ "vulcanis\'e" qui sera analys\'e en d\'etail tout le long de cette th\`ese est donc d\'ecrit par l'action:
\bea
\label{eq:vulcphi44}
S=\int d^Dx [\frac{1}{2}\partial_{\mu}\bar{\phi}\star\partial^{\mu}\phi+
\frac{1}{2}\Omega^2 \tilde{x}^2 \bar \phi \star \phi
+\frac{1}{2}\mu^2 \bar{\phi}\star\phi+
 \frac{\lambda}{2}\bar{\phi}\star\phi\star\bar{\phi}\star\phi]\; ,
\eea
avec $\tilde{x}=2\theta^{-1}x$.

Les \'etudes initiales de ce mod\`ele ont \'et\'e men\'ees dans la base matricielle (BM) introduite dans l'appendice \ref{sec:planMoyal}. Nous pr\'esentons par la suite un bref survol des principaux r\'esultats obtenus par ces m\'ethodes.

\section{L'action dans la BM}

L'action  (\ref{eq:vulcphi44}) s'\'ecrit dans la base matricielle comme:
\bea
\label{eq:ActionMatrix}
S=2\pi\theta \sum_{m,n,k,l}\Big{(}\bar\phi_{mn}H_{mn,kl}\phi_{kl}+
\frac{\lambda}{2} \bar \phi_{mn}\phi_{nk} \bar \phi_{kl}\phi_{lm}\Big{)} \, ,
\eea
ou par $\bar \phi$ on a not\'e la matrice hermitienne conjugu\'ee de $\phi$.
Les indices $m$, $n$, etc. sont des paires de nombres entiers positifs
$m=(m^1,m^2) \in \NN^2$.

La partie quadratique de l'action est donn\'ee par (voir \cite{GWR2x2}):
\bea
\label{eq:PropaVulcMatrix}
 H_{mn,kl}&=&\Big{[}\mu^2+\frac{2(1+\Omega^2)}{\theta}(m+n+1)\Big{]}
      \delta_{nk}\delta_{ml} - \frac{2(1-\Omega^2)}{\theta}
      \nonumber\\
      &\times&\Big{[}\sqrt{(n+1)(m+1)}\delta_{n+1k}\delta_{m+1l}+\sqrt{nm}
\delta_{n-1k}\delta_{m-1l}\Big{]}\; .
\eea
L'inverse de la partie quadratique est le propagateur. 
La grande r\'eussite technique de Grosse et Wulkenhaar consiste dans le calcul explicite de ce propagathor, faisant intervenir des familles de polyn\^omes orthogonaux. 

Une preuve alternative et directe de leur r\'esultat se trouve en appendice \ref{sec:Propagators}. Le calcul r\`epose sur l'astuce de Schwinger. Nous repr\'esentons la matrice inverse par l'int\'egrale:
\bea
   H^{-1}=\frac{\theta}{8\Omega}\int_{0}^{1}\frac{d\alpha}
   {1-\alpha}(1-\alpha)^{\frac{\theta}{8\Omega}H} \, .
\eea 
Les seuls \'el\'ements matriciels non nuls sont donn\'es par:
\bea
  &&[(1-\alpha)^{\frac{\theta}{8\Omega}H}]_{mm+h,l+hl}=
  \nonumber\\
  &&\sum_{u=max(0,-h)}^{min(m,l)}A(m,l,h,u)
  \Big{(}C\frac{1+\Omega}{1-\Omega}\Big{)}^{m+l-2u}E(m,l,h,u)\; ,
\eea 
  avec $C=\frac{(1-\Omega)^2}{4\Omega}$, et
\bea 
&&A(m,l,h,u)= \frac{\sqrt{m!(m+h)!l!(l+h)!}}{(m-u)!(l-u)!(h+u)!u!}\nonumber\\  &&E(m,l,h,u)=\frac{(1-\alpha)^{\frac{h+2u+1}{2}}\alpha^{m+l-2u}}
  {(1+C\alpha)^{m+l+h+1}} \; .
\eea

\subsection{Les graphes de Feynman}

\begin{figure}[ht]
\centering{
\includegraphics[width=80mm]{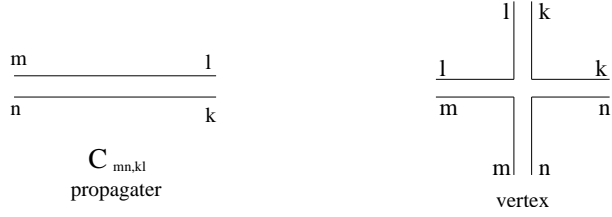}}
\caption{Le propagateur et le vertex dans l'espace des matrices}
\label{fig:propamatrix}
\end{figure}

Le mod\`ele (\ref{eq:ActionMatrix}) est un mod\`ele dit 
"pseudomatriciel\footnote{Pseudomatriciel car le propagateur n'est pas une matrice diagonale: il ne conserve pas parfaitement les indices matriciels}". Les graphes de Feynman d'un tel mod\`ele sont beaucoup plus riches en information topologique que ceux du mod\`ele commutatif. Comme le propagateur lie deux indices \`a deux autres, nous le repr\'esentons par un ruban (voir fig. (\ref{fig:propamatrix})). 

De m\^eme, le vertex est repr\'esent\'e comme un carrefour de quatre rubans (voir fig. \ref{fig:propamatrix}).
Les graphes de Feynman de ce mod\`ele ressemblent \`a ceux d'un mod\`ele matriciel \`a la diff\'erence que les indices ne sont pas parfaitement conserv\'es le long des rubans.

\begin{figure}[ht]
\centering{
\includegraphics[width=60mm]{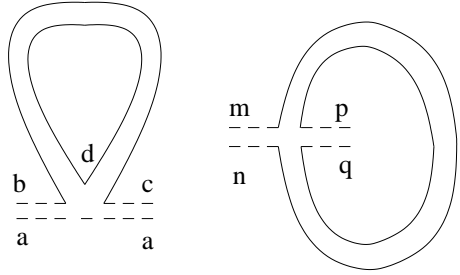}}
\caption{Le tadpole planaire et non planaire}
\label{fig:2tadpoles}
\end{figure}

Prenons l'exemple des deux tadpoles, pr\'esent\'e dans la figure \ref{fig:2tadpoles}.

Pour le premier tadpole l'indice $d$ est somm\'e sur la boucle interne et les indices $a$, $b$ et $c$ sont des indices externes. Pour le deuxi\`eme tous les indices sont des indices externes. Le tadpole dans la partie gauche de la figure a une face interne (avec indice de face $d$) et une face bris\'ee par des pattes externes. Nous l'appelons "planaire".
Le tadpole de droite a deux faces bris\'ees et nous l'appelons (de mani\`ere impropre) non-planaire.

Tout graphe avec $N$ vertex, $L$ propagateurs et $F$ faces dont $B$ bris\'ees par les $N_e$ pattes externes peut \^etre dessin\'e sur une surface de Riemann de genre $g$, donn\'e par:
\bea
2-2g=N-L+F \; .
\eea
et avec $B$ fronti\`eres.

A chaque ligne on associe une orientation de $\bar \phi$ \`a $\phi$. Autour d'un vertex les fl\`eches entrantes et sortantes se succ\`edent. A cause de cette propri\'et\'e le mod\`ele est appel\'e orientable. S\'eparons les deux bords d'un ruban en bord droit et bord gauche par rapport \`a cette orientation.

Pour un graphe donn\'e, son complexe conjugu\'e est donn\'e par le graphe ayant
toutes les fl\`eches renvers\'ees. Dans le graphe complexe conjugu\'e tous les bords droits sont devenus gauches et inversement.

Le propagateur de ce mod\`ele est r\'eel, et le vertex aussi. Par cons\'equent nous devons toujours extraire la partie r\'eelle des fonctions de corr\'elation. Nous sym\'etrisons toujours un graphe et son complexe conjugu\'e. Par cons\'equent notre mod\`ele a une sym\'etrie parfaite entre la gauche et 
la droite\footnote{Ceci n'est pas le cas pour un mod\`ele g\'en\'eral. Nous verrons dans la sous-section suivante que les mod\`eles qui ne poss\`edent pas cette sym\'etrie sont consid\'erablement plus compliqu\'es.}.

Une notion tr\`es importante est celle de "graphe dual". A tout "graphe direct" on associe le graphe ayant toutes les faces remplac\'ees par des vertex et tous les vertex par des faces. Ainsi nous le construisons de la mani\`ere suivante:
\begin{itemize}
\item A toute face on associe son vertex dual. 
\item A chaque indice que l'on rencontre en tournant autour de la face on associe un coin sur le vertex dual.
\item Tout propagateur de $m n$ \`a $k l$ est remplac\'e par son dual allant de $m k$ \`a $n l$.
\end{itemize}

\begin{figure}[ht]
\centering{
\includegraphics[width=75mm]{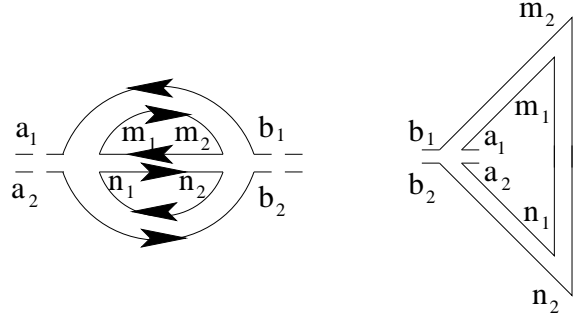}}
\caption{Le sunshine et son dual}
\label{fig:dualsunshine}
\end{figure}

L'exemple du graphe "sunshine" et de son dual est pr\'esent\'e dans la figure \ref{fig:dualsunshine}.

Les caract\'eristiques topologiques du graphe ${\cal G}$ dual de $G$ sont:
\bea
&&{\cal N}({\cal G})=F(G), \quad {\cal F}({\cal G})=N(G), \quad {\cal L}({\cal G})=L(G) ,\nonumber\\
&& {\cal N}_e({\cal G})=B(G),\quad {\cal B}({\cal G})=N_e(G), \quad \gamma({\cal G})=g(G) \;,
\eea
o\`u nous avons not\'e en calligraphique le nombre de vertex, faces et lignes du graphe dual et avec $\gamma$ son genre.

\subsection{Comptage de puissances}

On s'int\'eresse \`a \'etablir le comptage de puissances de ce mod\`ele. D'une mani\`ere opportune on se place \`a $\Omega=1$, o\`u le mod\`ele devient purement matriciel. Cela signifie que dans la figure \ref{fig:dualsunshine} on a $m_1=m_2$ et $n_1=n_2$. Nous devons par cons\'equent sommer un indice par face. Les modifications \`a $\Omega\neq 1$ seront indiqu\'ees au long du texte.

Nous d\'ecomposons les propagateurs en tranches:
\bea
\frac{1}{m+n+A}=\int_0^{\infty} dte^{-t(m+n+A)}=\sum_{i=0}^{\infty}C^i \; ,
\eea
avec
\bea
C^i=\int_{M^{-(i+1)}}^{M^{-i}} dt e^{-t(m+n+A)}\le M^{-i}e^{-KM^{-i}(m+n+A)} \; .
\eea

Nous voulons sommer tous les indices de faces, \`a attribution des indices de tranches donn\'ee (la somme sur les attributions est facile).

Pour tout graphe, consid\'erons son dual. Il sera tr\`es important par la suite. Les indices des faces du graphe direct sont maintenant associ\'es aux vertex du graphe dual.

A attribution d'\'echelle donn\'ee, nous notons \`a l'\'etage $i$ les composantes connexes
$G_k^i$ du graphe form\'e par toutes les lignes avec des indices sup\'erieurs ou \'egaux \`a $i$. Il est facile de prouver que si un graphe n'est pas connexe alors son dual ne l'est pas non plus.

Supposons que le dual de $G_k^i$, not\'e ${\cal G}_k^i$, n'est pas connexe. Alors $G_k^i$ ne sera pas connexe non plus, ce qui est faux. Nous concluons que les composantes connexes \`a l'\'echelle $i$ du graphe dual ${\cal G}_k^i$ sont les duaux des composantes connexes $G_k^i$ du graphe direct \`a l'\'echelle $i$.

Les pr\'efacteurs des propagateurs donnent:
\bea
\label{eq:gain}
\prod_{\ell}M^{-i_{\ell}} =\prod_{i;k}M^{-{\cal L}({\cal G}_k^i)} \; .
\eea

Tout indice somm\'e avec la d\'ecroissance d'un propagateur d'\'echelle $i$ va nous co\^uter $M^{2i}$. Nous avons donc int\'er\^et \`a sommer les indices avec les propagateurs les plus infrarouges possible.

Appellons ${\cal J}$ l'ensemble des propagateurs qu'on utilise pour sommer. Pour le construire commen\c cons par la plus basse tranche. Parmi les lignes qui ont l'\'echelle exactement \'egale \`a $i$, on choisit toutes les lignes de tadpole et l'on efface les vertex sur lesquels elles se branchent. Ensuite on choisit le maximum des lignes qui relient des vertex, \`a condition qu'elles ne forment pas des boucles. 
Une fois qu'un maximum de lignes de l'\'echelle $i$ a \'et\'e choisi nous passons \`a l'\'echelle $i+1$.

On obtient ainsi l'ensemble ${\cal J}$ des lignes les plus infrarouges tel que tout vertex ait une ligne de tadpole, ou bien soit reli\'e au reste des vertex par une ligne d'arbre.

Ensuite nous analysons notre graphe des hautes \'echelles vers les basses. Pour autant qu'un vertex porte encore des demi-lignes non contract\'ees nous ne sommons pas son index. A l'\'echelle o\`u toutes les demi-lignes du vertex se branchent sur des propagateurs nous sommons son index avec la ligne de l'ensemble ${\cal J}$ qui le touche. 
Les sommes nous co\^utent:
\bea
\label{eq:loss}
  \prod_{\ell\in {\cal J}} M^{2i_{\ell}}=\prod_{i,k}M^{2[{\cal N}({\cal G}_k^i)- 
{\cal N}_e({\cal G}_k^i)]} \; .
\eea
En mettant ensemble (\ref{eq:gain}) et (\ref{eq:loss}) et en passant aux nombres topologiques du graphe direct on obtient:
\bea
\prod_{i,k}M^{-4g(G_k^i)-2[B(G_k^i)-1]-\frac{N_e(G_k^i)-4}{2}}\; .
\eea

Nous pouvons maintenant comprendre pourquoi cette th\'eorie ne poss\`ede pas le m\'elange UV/IR.

Dans la th\'eorie non vulcanis\'ee on ne peut pas sommer les indices avec des lignes de tadpole du graphe dual. Nous ne devons choisir dans ${\cal J}$ que des lignes d'arbre. Supposons qu'\`a l'\'echelle basse $i$ les demi lignes d'un vertex se branchent dans un tadpole. Nous sommes oblig\'es de sommer son index avec une ligne d'arbre qui est d'\'echelle plus \'elev\'ee $j$. La contribution totale est donc:
\bea
M^{-i}M^{-j}M^{2j}=M^{j-i}\; ,
\eea
qui n'est pas sommable sur $j$ par rapport \`a $i$. Une telle contribution divergente, ayant plus d'une face bris\'ee n'a pas la forme du lagrangien et ne peut pas \^etre renormalis\'ee!

Dans notre th\'eorie, par contre, c'est possible de sommer avec les lignes de tadpole du graphe dual. On somme toujours un vertex avec la ligne la plus infrarouge qui le touche! Par cons\'equent, pour l'exemple ci-dessus on a:
\bea
M^{-i}M^{-j}M^{2i}=M^{-(j-i)}\; ,
\eea 
qui est sommable sur $j$ par rapport \`a $i$.

\subsection{Renormalisation dans la BM}

Le comptage de puissances \'etabli dans la section pr\'ec\'edente nous dit que les contributions divergentes viennent seulement des sous fonctions planaires avec une seule face bris\'ee avec deux ou quatre pattes externes
\footnote{Pour le mod\`ele \`a $\Omega\neq 1$ une \'etude plus d\'etaill\'ee aboutit \`a s\'electionner parmi ces derni\`eres seules les fonctions \`a quatre points avec des index conserv\'es et les fonctions \`a deux points avec index conserv\'e, ou avec un saut accumul\'e d'index de $2$ le long de la face externe.}:
\bea
 g(G_k^i)=0, \quad B(G_k^i)=1, \quad N_e(G_k^i)=2 \text{ ou } 4 \; .
\eea

Nous notons $\Gamma^4(m,n,k,l)$ la fonction \`a quatre points, planaire avec une seule face bris\'ee, une particule irr\'eductible amput\'ee avec indices externes $m$, $n$, $k$ et $l$.

Nous notons $\Sigma(m,n)$ la fonction \`a deux points, planaire avec une seule face bris\'ee, une particule irr\'eductible amput\'ee avec des indices externes $m$ et $n$.

Nous effectuons un d\'eveloppement de Taylor de la fonction \`a quatre points:
\bea
\Gamma^4(m,n,k,l)=\Gamma^4(0,0,0,0)+m\partial_m\Gamma^4+\dots  \; .
\eea
Le premier terme de la relation ci-dessus est logarithmiquement divergent. Pour le soustraire il suffit de rajouter dans l'action un contre-terme de la forme:
\bea
\Gamma^4(0,0,0,0)\bar\phi_{mn}\phi_{nk}\bar\phi_{kl}\phi_{lm}\; .
\eea

La fonction \`a deux points a un d\'eveloppement de Taylor de la forme:
\bea
\Sigma(m,n)=\Sigma(0,0)+(m+n)\partial\Sigma(0,0)+\dots \; ,
\eea
o\`u l'on a not\'e $\partial\Sigma$ la d\'eriv\'ee de $\Sigma$ par rapport \`a un de ses indices externes (\`a cause de la sym\'etrie gauche droite on peut choisir librement
l'index par rapport auquel on d\'erive). Cette d\'eriv\'ee fait intervenir les fonctions \`a deux points avec un saut d'index sur les pattes externes.

La fonction \`a deux points s'\'ecrit comme:
\bea
G^2(m,n)=H_{mn,nm}-\Sigma(0,0)-(m+n)\partial\Sigma(0,0) \; ,
\eea
et on reconna\^it dans le premier terme une renormalisation quadratique de la masse et dans le deuxi\`eme une renormalisation logarithmique de la fonction d'onde
\footnote{\`A $\Omega\neq 1$ on analyse aussi les composantes $G^2(m\pm 1,n\pm 1,n,m)$ qui seront renormalis\'ees par les fonctions \`a deux points avec saut accumul\'e de $2$ au long du cycle ext\'erieur entre les pattes externes $\Sigma(m+1,n+1,n,m)=\sqrt{(m+1)(n+1)}\Sigma(11,00)+\dots$}.

\section{Autres mod\`eles dans la BM}

Nous voulons \'etendre les m\'ethodes pr\'esent\'ees dans ce chapitre
\`a des mod\`eles plus g\'en\'eraux.
Nous voulons traiter un syst\`eme de particules dans un champ magn\'etique constant. Cette \'etude devrait \^etre pertinente pour l'effet Hall. De plus, l'invariance par translation n'est pas bris\'ee dans ces th\'eories, car les translations sont remplac\'ees par des translations magn\'etiques. Suite \`a cette \'etude nous allons pouvoir d\'eterminer si la vulcanisation peut \^etre r\'einterpr\'et\'ee comme la pr\'esence d'un champ externe constant ou non.

Nous ne connaissons pas la forme de l'interaction entre les particules adapt\'ee \`a l'effet Hall. En effet la litt\'erature est tr\`es peu pr\'ecise \`a ce sujet. Le consensus est que les interactions ne devraient pas jouer de r\^ole pour l'effet Hall entier mais qu'elles sont essentielles pour l'effet Hall fractionnaire. Cela dit, l'effet Hall fractionnaire ne repose pas sur la forme pr\'ecise de l'interaction, il suffit d'avoir une interaction d\'ecroissante avec la distance pour l'obtenir.

Nous pourrions \^etre tent\'es de prendre comme interaction celle de Coulomb, traduite par l'\'echange de photons. Certes, cette interaction est l'interaction fondamentale entre les \'electrons. Cependant, comme il s'agit de particules non relativistes il est raisonnable de supposer que l'interaction effective est diff\'erente. Nous supposons que cette interaction est \`a quatre corps. Le point de vue darwinien que nous adoptons dans cette th\`ese nous m\`ene \`a consid\'erer toutes les interactions qui sont juste renormalisables dans l'ultraviolet, \'etant ainsi des points fixes du groupe de renormalisatoin. Une fois achev\'ee l'\'etude d\'etaill\'ee du propagateur, nous allons tester si les interactions de type Moyal entrent dans cette cat\'egorie.

La partie quadratique de l'action est simple.
Dans la jauge sym\'etrique, la d\'eriv\'ee covariante est:
\bea
D_{\mu}=\partial_{\mu}+\imath \Omega \tilde x_\mu \; ,
\eea
et le hamiltonien libre devient:
\bea
H_{\Omega}=-\frac{1}{2}D_{\mu}D_{\mu}=
\frac{1}{2}\left( -\partial^2+4\frac{\Omega^2}{\theta^2}x^2+2\imath \frac{\Omega}{\theta}x\wedge \partial \right) \; .
\eea

Nous pouvons en effet analyser un mod\`ele encore plus g\'en\'eral, avec le Hamiltonien libre donn\'e par:
\bea
H_B=\frac{1}{2}\left( -\partial^2+4\frac{\Omega^2}{\theta^2}x^2+2\imath \frac{B}{\theta}x\wedge \partial \right) \; .
\eea

L'\'etude d\'etaill\'ee du propagateur d'un tel mod\`ele fait l'objet du travail pr\'esent\'e en annexe \ref{sec:Propagators}.

Pour mener l'analyse en tranches de ce propagateur nous obtenons les repr\'esentations suivantes dans l'espace direct et dans la BM:
\begin{lemma}
  Si H est:
  \begin{equation}
    H_B=\frac{1}{2}\big{[}-\partial_0^2-\partial_1^2+
    \Omega^2x^2-2\imath B(x_0\partial_1-x_1\partial_0)\Big{]}.
  \end{equation}
  Le noyau int\'egral de l'op\'erateur $e^{-tH}$ est:
  \begin{equation}
    e^{-tH_B}(x,x')=\frac{\Omega}{2\pi\sh\Omega t}e^{-A},
  \end{equation}
  \begin{equation}
    A=\frac{\Omega\ch\Omega t}{2\sh\Omega t}(x^2+x'^2)-
    \frac{\Omega\ch Bt}{\sh\Omega t}x\cdot x'-\imath
    \frac{\Omega\sh Bt}{\sh\Omega t}x\wedge x'.
  \end{equation}
\end{lemma}

Et dans la base matricielle on a:
\begin{lemma}
\bea
(1-\alpha)^{\frac{\theta}{8\Omega}H_B}_{mm+h,ll+h}=(1-\alpha)^{\frac{\theta}{8\Omega}H}_{mm+h,ll+h}
(1-\alpha)^{-\frac{B}{2\Omega}h} \; .
\eea
\end{lemma}

Gr\^ace \`a la simplicit\'e du vertex de Moyal dans la BM, on peut esp\'erer prouver la renormalisabilit\'e de tels mod\`eles de TCNC une fois que le comportement  d'\'echelle de ce propagateur a \'et\'e d\'etaill\'e dans la BM.

Les principaux r\'esultats de l'annexe \ref{sec:Propagators} sont les suivants:
\begin{itemize}
\item Nous trouvons toujours une d\'ecroissance d'ordre $1$ dans l'indice $l-m$. Les indices sont tr\`es bien conserv\'es le long des rubans. Les mod\`eles avec un tel propagateur sont pseudo-matriciels.
\item A $B\neq \Omega$ le comportement est semblable \`a celui \`a $B=0$. Nous avons une d\'ecroissance d'ordre $M^{i}$ dans l'indice $h=n-m$ (l'\'ecart entre les bords des rubans) ce qui nous permet de reproduire le comptage de puissance de $\phi_4^{\star 4}$. Dans cette cat\'egorie entre le mod\`ele $LSZ$ g\'en\'eral.
\item Pour $\Omega=B$ la situation change drastiquement. Un mod\`ele de ce type est le mod\`ele de Gross-Neveu de \cite{RenNCGN05}. Le comportement d'\'echelle est gouvern\'e par un point col non-trivial dans l'espace d'indices de matrice.
Dans la r\'egion d'indices pr\`es de ce col on perd les d\'ecroissances d'\'echelle en $h$. Il s'av\`ere que le seul cas pour lequel cette perte est dangereuse est le "sunshine". 
\end{itemize}

Le traitement de \cite{RenNCGN05} est men\'e dans l'espace direct. Le propagateur n'a pas de d\'ecroissance dans $v^2$ mais seulement une oscillation $e^{\imath\Omega u\wedge v}$. Un tel facteur oscillant nous permet en principe d'affirmer que $u\wedge v\approx 1$ (et nous donne le bon ordre de grandeur pour $v$) \`a l'aide des int\'egrations par parties. Les int\'egrations par partie requises pour chaque graphe sont d\'ecrites en d\'etail dans \cite{RenNCGN05}. Ces oscillations du propagateur doivent \^etre combin\'ees avec les oscillations des vertex. Il faut choisir un ensemble de variables ind\'ependantes par rapport auxquelles on effectue les int\'egrations par partie.

Nous rencontrons le "probl\`eme du sunshine" \`a $\Omega=1$. Les oscillations des vertex compensent exactement celles des propagateurs! Cette compensation fait que nous ne pouvons pas obtenir le bon comptage des puissances.
Dans l'espace des matrices le statut sp\'ecial de ce point peut \^etre vu comme la rencontre du premier niveau de Landau. Dans la base matricielle le propagateur devient:
\bea
  C_{mn}=\frac{1}{m} \; ,
\eea
et on n'a aucune d\'ecroissance en $n$!

Nous pouvons conclure que le point $\Omega=1$ a un statut tr\`es sp\'ecial pour ces mod\`eles: la renormalisabilit\'e de ces mod\`eles \`a $\Omega=1$ est remise en question.

Nous n'avons pas trouv\'e pour l'instant une fa\c con de d\'ecouper ce propagateur en tranches dans l'espace direct ou dans la base matricielle adapt\'ee \`a l'\'etude du point $\Omega=1$. Des progr\`es r\'ecents dans cette direction ont \'et\'e faits par Disertori et Rivasseau, mais on manque encore de preuves que ces mod\`eles \`a $\Omega=1$ soient renormalisables.

En effet il s'av\`ere que ces probl\`emes sont g\'en\'eriques. Nous les rencontrons chaque fois que nous essayons de traiter la vulcanisatoin comme une pure d\'eriv\'ee covariante dans un champ externe constant.

Si les termes de vulcanisation ne sont pas engendr\'es par un champ de fond, quelle est donc leur origine? De plus, si l'on ne peut pas absorber la brisure de l'invariance par translation dans un changement de jauge, est-ce que ces th\'eories peuvent \^etre vraiment physiques?

Pour r\'epondre \`a ces questions nous pensons que la cl\'e est de trouver un mod\`ele fermionique du type mod\`ele de Gross-Neveu, vulcanis\'e sans singularit\'e \`a $\Omega=1$. Une proposition de th\'eorie fermionique est d\'etaill\'ee dans les conclusions. Malheureusement, pour l'instant ces th\'eories brisent l'invariance par translation.

\newpage
\thispagestyle{empty}

\chapter{Renormalisation dans l'espace direct}

Dans ce chapitre nous pr\'esentons une preuve alternative de la renormalisabilit\'e de la th\'eorie $\Phi^{\star 4}_4$. 

Pourquoi a-t-on besoin d'une nouvelle preuve de la renormalisabilit\'e? Il y a plusieurs arguments en faveur d'une telle d\'emarche.

\`A cause de la complexit\'e du propagateur dans la base matricielle, la preuve dans cette derni\`ere est tr\`es technique. Nous souhaiterions avoir une preuve plus simple.

Plus important, la renormalisabilit\'e d'une th\'eorie repose sur deux caract\'eristiques: le comptage de puissances et la forme des contre-termes. Le comptage de puissances est assez transparent dans la base matricielle, mais la forme des contre-termes n'est pas du tout transparente. Qu'est-ce qui remplace le principe de localit\'e pour notre th\'eorie? L'\'etude qui suit nous m\`ene \`a introduire un principe de Moyalit\'e qui est la bonne g\'en\'eralisation de cette notion.

En $D=4$ le propagateur dans l'espace direct est \cite{Propaga}:
\be
\label{eq:Xpropa}
C(x,x')=\int_{0}^{\infty}dt \frac{\Omega ^2}{[2\pi\sh\Omega t ]^2}
e^{-\frac{\Omega}{4}\coth\frac{\Omega t}{2}(x-x')^2-
\frac{\Omega}{4}\sh\frac{\Omega t}{2}(x+x')^2 - \mu_0^2 t } \; ,
\ee
et le vertex \cite{Filk:1996dm} peut se repr\'esenter comme:
\be
\label{eq:Xvertex}
V(x_1, x_2, x_3, x_4) = \delta(x_1 -x_2+x_3-x_4 )e^{i
\sum_{1 \les i<j \les 4}(-1)^{i+j+1}x_i \theta^{-1}  x_j} \;,
\ee
avec
$x \theta^{-1}  y  \equiv  \frac{2}{\theta} (x_1  y_2 -  x_2  y_1 +  x_3  y_4 - x_4  y_3 )$.

La partie interaction de l'action est r\'eelle. On repr\'esente le vertex comme dans l'\'equation (\ref{eq:Xvertex}), mais on devrait se rappeler que la somme des parties imaginaires est nulle. De plus, le vertex est invariant sous la transformation $1\leftrightarrow 3$ et $2\leftrightarrow 4$. 

La fonction $\delta$ du vertex nous impose que les quatre champs soient positionn\'es dans les coins d'un parall\'elogramme. Le facteur d'oscillation fait intervenir l'aire du parall\'elogramme. Il la contraint \`a \^etre d'ordre $\theta$. 

La valeur de $\theta$, le commutateur des deux coordonn\'ees, nous donne en terme de m\'ecanique quantique la plus petite aire qu'on peut mesurer. Il est donc naturel de trouver que l'interaction de quatre champs est maximale quand ils se distribuent sur une surface d'aire minimale admise.

Pour tout propagateur nous introduisons une variable "courte" $u$ (et une "longue" $v$) \'egale \`a la diff\'erence (somme) entre les positions des deux extr\'emit\'es du propagateur. Dans la zone ultraviolette ($t\approx M^{-2i}$) on a $u\approx M^{-i}$ et $v\approx M^{i}$. Ces variables sont centrales pour la suite.

Nous avons d\'ej\`a vu que dans l'espace direct les vertex sont repr\'esent\'es par des parall\'elogrammes. Chaque parall\'elogramme est muni d'un sens de rotation correspondant \`a l'ordre cyclique des quatre coins. Pour repr\'esenter un graphe nous repr\'esentons tous ses vertex avec le m\^eme sens de rotation (trigonom\'etrique) sur un plan et ensuite nous branchons les propagateurs repr\'esent\'es par des lignes.

\section{Le comptage de puissances}

Le propagateur de la th\'eorie $\Phi^4_4$ commutative est born\'e dans une tranche par:
\bea
C^i\le M^{2i}e^{-M^i |u|} \; .
\eea
Le comptage de puissance est neutre car tout vertex a en moyenne deux propagateurs, donc un pr\'efacteur $M^{4i}$ et un point \`a int\'egrer, ce qui nous rapporte $M^{-4i}$. 

Pour la th\'eorie $\Phi^{\star 4}_4$, nous d\'ecompossons le propagateur (\`a masse nulle) comme:
\bea
C=\sum_{i}C^i\quad C^i=\int_{M^{-2(i+1)}}^{M^{-2i}}dt
\frac{\Omega ^2}{[2\pi\sh\Omega t ]^2}
e^{-\frac{\Omega}{4}\coth\frac{\Omega t}{2}u^2-
\frac{\Omega}{4}\sh\frac{\Omega t}{2}v^2}
\; ,
\eea
et le propagateur dans une tranche est born\'e par:
\bea
C^i\le M^{2i}e^{-M^i |u|-M^{-i} |v|}\; .
\eea
Tout vertex a en moyenne deux propagateurs comme avant, donc un facteur $M^{4i}$, et nous devons int\'egrer deux variables courtes et deux variables longues. Nous utilisons la fonction $\delta$ pour int\'egrer l'une des variables longues. Nous obtenons $M^{-8i}$ pour les int\'egrales courtes et $M^{4i}$ pour l'int\'egrale longue: le comptage de puissances est de nouveau neutre.

Cette argument doit \^etre affin\'e pour une attribution d'\'echelle g\'en\'erale (de nouveau la somme sur les attributions est simple). Choisissons un arbre qui est compatible avec l'attribution d'\'echelle \footnote{Cela signifie qu'il est sous arbre dans tout les composantes connexes a chaque \'etage.}. En int\'egrant les variables longues de l'arbre avec les fonctions $\delta$ et en bornant les facteurs oscillants par $1$, l'amplitude d'un graphe ${\cal A}$ est born\'ee par:
\bea
{\cal A}_G&&\le \prod_{\ell\in G}M^{2i_{\ell}}\prod_{\ell\in T}M^{-4i_{\ell}}=
\prod_{i,k}M^{2L(G_k^i)}\prod_{i,k}M^{-4[N(G_k^i)-1]}
\nonumber\\
&&=\prod_{i,k}M^{-[N_e(G_k^i)-4]} \; ,
\eea
car pour toutes les lignes de boucle le facteur obtenu par l'int\'egration de la variable longue compense exactement celui de la variable courte.

Il faut am\'eliorer ce comptage de puissance pour obtenir les modifications dues \`a la topologie du graphe. Pour cela l'on doit exploiter les facteurs oscillants des vertex. Ils s'expriment \`a l'aide d'une op\'eration topologique, appel\'ee premier mouvement de Filk.

\begin{figure}[ht]
\label{fig:Filk1sunshine}
  \centerline{\epsfig{figure=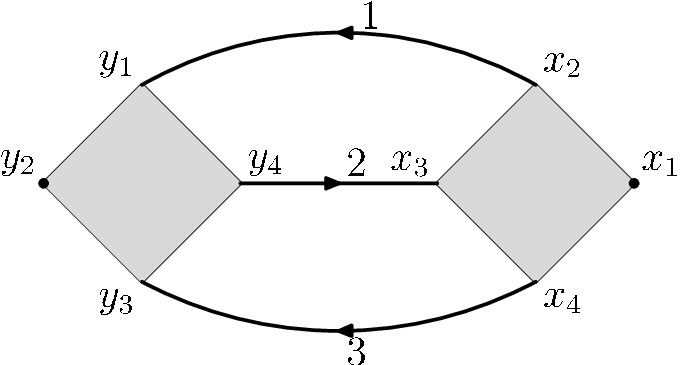,width=6cm}
    \epsfig{figure=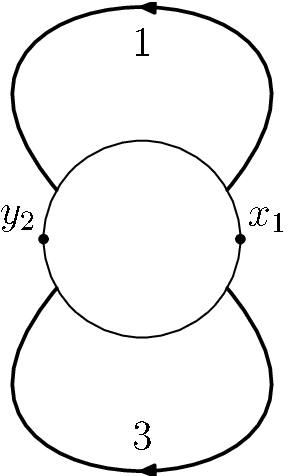,width=2cm}}
\caption{Premier mouvement de Filk: la ligne $2$ est r\'eduite et les vertex se collent.}
\end{figure}

Soit une ligne du graphe qui joint deux vertex. Nous r\'eduisons cette ligne et nous 
collons les deux vertex adjacents dans un seul grand vertex. Cette op\'eration conserve le sens de rotation des vertex.
It\'erons cette op\'eration pour toutes les lignes d'un arbre dans le graphe pour aboutir \`a un graphe r\'eduit, appel\'e rosette, form\'e d'un seul vertex ayant uniquement des lignes de tadpole (un exemple est pr\'esent\'e dans la figure \ref{fig:Filk1sunshine}).

\begin{figure}[ht]
\centerline{\epsfig{figure=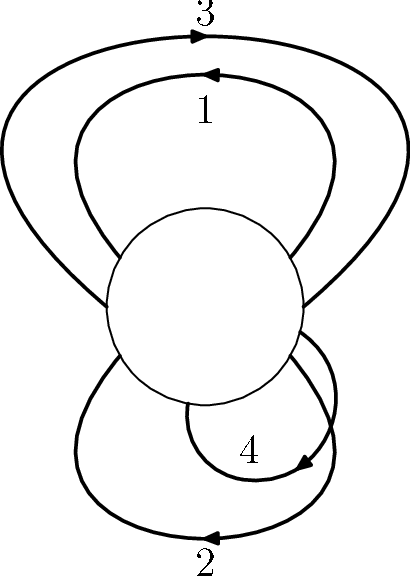,width=3cm}}
\caption{Une rosette nonplanaire}\label{fig:Rosetenonpl}
\end{figure}

Cette op\'eration ne modifie ni le genre $g$, ni le nombre de faces $F$, ni le nombre de faces bris\'ees $B$ du graphe.
Si le graphe est non-planaire il y aura un croisement des deux lignes de la rosette (voir par exemple \ref{fig:Rosetenonpl}).

A un tel croisement correspond un facteur oscillant $e^{\imath v_a\wedge v_b}$ entre les variables longues des deux lignes de boucle ($e^{\imath v_2\wedge v_4}$ dans Fig: \ref{fig:Rosetenonpl}). Si, avant de borner l'oscillation par $1$, on utilise ce facteur pour int\'egrer la variable $v_2$ on arrive \`a gagner un facteur $M^{-4i_a}$ ($M^{-4i_2}$ dans Fig. \ref{fig:Rosetenonpl}) au lieu de payer un facteur $M^{4i_b}$ ($M^{4i_4}$ dans Fig. \ref{fig:Rosetenonpl}), ce qui rend toute fonction non-planaire convergente.
Un argument similaire nous donne l'am\'elioration due aux faces bris\'ees. Nous obtenons ainsi, directement dans l'espace direct, un comptage de puissance partiel de la th\'eorie, suffisant pour prouver la renormalisabilit\'e.

\vspace*{\stretch{2}}

\section{Les soustractions}

Nous venons de comprendre dans le nouveau contexte pourquoi seules les fonctions planaires \`a une seule face bris\'ee avec deux ou quatre pattes externes sont divergentes. Il faut maintenant comprendre pourquoi ces divergences ont la forme du Lagrangien initial. Qu'est-ce qui remplace la notion de localit\'e dans notre th\'eorie? 

\begin{figure}
  \centerline{\epsfig{figure=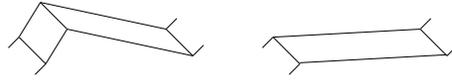,width=6cm}}
\caption{Un pavage divergent}
\label{fig:PavageDiv}
\end{figure}

Soit une fonction divergente \`a quatre points. Si tous les propagateurs internes sont tr\`es ultraviolets leurs extr\'emit\'es sont presque confondues. Dans ces conditions la fonction est un pavage planaire de parall\'elogrammes. Ils vont forc\'ement avoir la forme d'un parall\'elogramme entre les points externes! De plus l'aire orient\'ee du parall\'elogramme r\'esultant est la somme des aires orient\'ees des parall\'elogrammes. La fonction r\'eproduit pr\'ecis\'ement la forme du produit de Moyal! Un exemple est pr\'esent\'e dans la figure \ref{fig:PavageDiv}.

En termes math\'ematiques cela se traduit par le lemme suivant:
\begin{lemma}
La contribution de vertex d'un graphe planaire \`a une face bris\'ee peut s'exprimer sous la forme:
\begin{eqnarray}
&&\delta(\sum_{i}(-1)^{i+1}x_{i}+\sum_{l\in T\cup {\cal L}} u_l)
e^{\imath\sum_{i,j}(-1)^{i+j+1}x_{i}\theta^{-1} x_{j}} 
\nonumber \\ 
&&e^{\imath\sum_{l\in T \cup {\cal L},\;   l \prec j}u_l\theta^{-1} (-1)^{j}x_j
+\imath\sum_{l\in T \cup {\cal L},\; l \succ j }(-1)^j x_{j}\theta^{-1} u_l}
\nonumber \\
&&e^{-\imath\sum_{l,l'\in T \cup {\cal L},\; l \prec l' }u_l\theta^{-1} u_{l'}
-\imath\sum_{l\in  T}\frac{u_l\theta^{-1} v_l}{2}\epsilon(l)
-\imath\sum_{l\in {\cal L}}\frac{u_l\theta^{-1} w_l}{2}\epsilon(l)}
\nonumber \\
&&e^{-\imath\sum_{l\in{\cal L},\, l' \in {\cal L} \cup T;\; l'\subset  l}u_{l'}\theta^{-1} w_l \epsilon(l)} \ ,
\end{eqnarray}  
o\`u $T$ est l'arbre r\'eduit et ${\cal L}$ est l'ensemble des lignes de boucle.
\end{lemma}

Dans la premi\`ere ligne du lemme ci-dessus on reconna\^it la forme du noyau du produit de Moyal. Nous effectuons un d\'eveloppement de Taylor dans la deuxi\`eme ligne ci-dessus. 
Tous les termes correctifs font intervenir au moins une insertion d'une variable courte $u$, ce qui les rend sous-dominants. Comme la fonction \`a quatre points est logarithmiquement divergente l'on peut n\'egliger les contributions sous-dominantes. Nous concluons que la partie divergente de la fonction a quatre points reproduit bien le produit de Moyal.

La fonction \`a deux points est plus compliqu\'ee. Elle se divise dans la contribution des tadpoles qui sont de la forme:
\bea
{\cal A}=\int dudv \,\delta(x_1-x_2+\sum u)e^{\imath x_1\theta^{-1} x_2} F(u,v)\; ,
\eea
et les autres contributions qui peuvent \^etre mises sous la forme:
\bea
{\cal A}=\int dudv \, \delta(x_1-x_2+\sum u)\delta(-x_2+\sum u+\sum v) F(u,v) \; .
\eea
Nous allons effectuer un d\'eveloppement de Taylor des contributions du haut. Commen\c cons par les tadpoles:
\bea
{\cal A}&=&\delta(x_1-x_2)e^{\imath x_1 \theta^{-1} x_2} \int dudv\, F(u,v)
\nonumber\\
&+&\delta'(x_1-x_2)e^{\imath x_1 \theta^{-1} x_2} \int dudv\, (\sum u) F(u,v)
\nonumber\\
&+&\delta''(x_1-x_2)e^{\imath x_1 \theta^{-1} x_2} \int dudv\, (\sum u) (\sum u)F(u,v) \; .
\eea
Toutes les int\'egrales du haut sont gaussiennes. Par cons\'equent, le second terme est nul.

Le premier terme est une contribution locale, de masse, quadratiquement divergente. Le dernier terme est une contribution logarithmiquement divergente. Pour voir \`a quel op\'erateur ce dernier correspond nous allons lisser l'amplitude avec des fonctions test:
\bea
&&\int dx_1 dx_2 \, \delta''(x_1-x_2)\, e^{\imath x_1\theta^{-1} x_2} \bar \phi (x_1)\phi (x_2)\nonumber\\
&&=\int dx_1 dx_2 \, \delta(x_1-x_2) \left( \partial_1 +\imath \theta^{-1} x_2 \right)^2 \bar \phi (x_1) \phi(x_2)\; .
\eea
Cet op\'erateur n'est pas de la forme du lagrangien initial. La subtilit\'e consiste \`a comprendre qu'il faut combiner ce tadpole avec son complexe conjugu\'e. En effet, comme notre mod\`ele est r\'eel, il est naturel de toujours combiner les graphes avec leurs complexes conjugu\'es. D'une mani\`ere \'equivalente, l'on peut brancher un m\^eme graphe sur la combination r\'eelle des patte externes comme:
\bea
\int dx_1 dx_2 \, \delta(x_1-x_2) \left( \partial_1 +\imath \theta^{-1} x_2 \right)^2 [\bar \phi (x_1) \phi(x_2)+\phi(x_1)\bar \phi(x_2)] \; .
\eea
En d\'ev\'eloppant l'expression ci-dessus, nous trouvons que la somme des deux tadpoles nous fournit un contre-terme pour l'op\'erateur:
\bea
\partial^2+\frac{1}{\theta^2}x^2\; .
\eea

Nous devons maintenant effectuer un d\'eveloppement de Taylor des autres contributions \`a la fonction \`a deux points:
\bea
{\cal A}&=&\int dudv \, \delta(x_1-x_2+\sum u)\delta(-x_2+\sum u+\sum v) F(u,v) \nonumber\\
&=&\delta(x_1-x_2)\int dudv \, \delta(\sum u+\sum v) F(u,v)\\
&+&\delta(x_1-x_2)(-x_2)\int du dv \, \delta'(\sum u+\sum v) F(u,v)\nonumber\\
&+&\delta(x_1-x_2) x_2^2 \int dudv \, \delta''(\sum u+\sum v)F(u,v)\nonumber\\
&+&\delta'(x_1-x_2)\int dudv \, (\sum u) \delta(\sum u+\sum v) F(u,v)\nonumber\\
&+&\delta'(x_1-x_2)(-x_2)\int dudv \, (\sum u) \delta'(\sum u+\sum v) F(u,v)\nonumber\\
&+&\delta''(x_1-x_2)\int dudv (\sum u)(\sum u)\delta(\sum u+\sum v)F(u,v) \; . \nonumber
\eea
Le premier terme ci-dessus est une renormalisation de masse. Le deuxi\`eme est nul \`a cause du fait que $F(u,v)$ est gaussienne. Le troisi\`eme est une renormalisation de l'op\'erateur $x^2$. Le quatri\`eme est nul d\^u de nouveau au fait que $F(u,v)$ est gaussienne. Le dernier est une renormalisation du $p^2$.

Nous analysons maintenant le cinqui\`eme terme. De nouveau, nous combinons un graphe avec son complexe conjugu\'e. L'op\'erateur correspondant prend la forme:
\bea
&&\int \delta'(x_1-x_2)(-x_2)[\bar\phi(x_1)\phi (x_2)+\phi(x_1)\bar\phi (x_2)]
\nonumber\\
&=&\int (-x)[\partial \bar \phi \phi+\bar \phi \partial\phi]=\int \bar\phi \phi\; ,
\eea 
et nous concluons qu'il est une sous-divergence de masse.

Nous avons ainsi prouv\'e que la fonction \`a deux points nous donne uniquement des contre-termes de la forme du Lagrangien initial.

A l'aide des m\^emes techniques nous pouvons prouver que le mod\`ele LSZ est renormalisable. La seule diff\'erence vient du fait que les noyaux des propagateurs ne sont pas r\'eels, ce qui fait que les graphes ne se combinent pas exactement avec leurs complexes conjugu\'es et nous avons par cons\'equent aussi une renormalisation du terme $\partial \wedge x$.

\chapter{La fonction b\^eta du mod\`ele $\Phi^{\star 4}_4$}

Pour toute th\'eorie des champs, nous nous int\'eressons uniquement aux fonctions de corr\'elation. Pour les exprimer, nous commencons en imposant un cutoff qui rend toute quantit\'e finie. En suite, nous essayons de prendre la limite du cutoff tendant vers l'infini.

En effectuant le d\'eveloppement perturbatif, nous exprimons les fonctions de corr\'elation en fonction des param\`etres nus de notre th\'eorie. Nous rencontrons des contributions qui divergent avec le cutoff. La renormalisatoin absorbe ces divergences dans une r\'ed\'efinition des param\`etres nus. Ainsi nous r\'eexprimons les fonctions de corr\'elation en fonction des contributions renormalis\'ees (finies) et des param\`etres effectifs (\'egalement finis). Une fois la renormalisabilit\'e d'une th\'eorie \'etablie, la question naturelle qui se pose est d'\'etabir la variation des param\`etres effectifs avec le cutoff. Les flots des diff\'erents param\`etres sont caract\'eris\'es par leurs fonctions b\^eta.

Pour notre mod\`ele, le couplage effectif est donn\'e par la fonction \`a quatre points, une particule irr\'eductible, amput\'ee $\Gamma^4$. A tout vertex correspondent deux propagateurs. Le couplage effectif est donn\'e par:
\bea
\lambda_{i-1}=\frac{\Gamma^4_i}{Z^2_i} \; ,
\eea
o\`u $Z$ est la renormalisation de la fonction d'onde.
Pour une th\'eorie avec un cutoff ultraviolet $\Lambda$, on peut d\'ecrire l'\'evolution du couplage effectif avec l'\'echelle \`a l'aide de la fonction $\beta(\lambda)$ d\'efinie comme:
\bea
\beta(\lambda)=\frac{d\lambda^{nu}}{d\ln\Lambda}\Big{|}_{\lambda^{ren}=ct.} \; .
\eea
Alternativement, dans le cadre de l'analyse multi-\'echelle nous pr\'ef\'erons la d\'efinition:
\bea
\beta(\lambda)_i=\lambda_{i-1}-\lambda_i \; .
\eea

Pour la th\'eorie $\Phi^4_4$ commutative le couplage effectif \'evolue avec l'\'echelle. Cette \'evolution est assez facile \`a comprendre au premier ordre. Nous avons une contribution non nulle pour la fonction $\Gamma^4$ provenant des boules. Le tadpole, \'etant local, est un contre-terme de masse, et $Z=1$ \`a une boucle.

Une \'etude d\'etaill\'ee nous montre que, si l'on veut une constante renormalis\'ee non nulle nous devons commencer avec une constante nue grande. En effet le couplage nu devient infini pour un cutoff UV fini. Inversement, toute constante nue finie donne naissance \`a une th\'eorie libre triviale dans l'infrarouge. Ceci est un grave probl\`eme dans la th\'eorie des champs commutative et a re\c cu le nom "fant\^ome de Landau". Les infinis qu'on \'elimine par la renormalisation ressurgissent dans cette nouvelle divergence.

Ce probl\`eme a failli tuer la th\'eorie des champs. Gr\^ace \`a l'universalit\'e de ce ph\'enom\`ene on l'a presque abandonn\'ee comme description des interactions fondamentales. L'issue de cette situation n'appara\^it qu'avec la d\'ecouverte de la libert\'e asymptotique dans les th\'eories de jauge non ab\'eliennes. Mais m\^eme cette solution n'est pas enti\`erement satisfaisante. La libert\'e asymptotique est le fant\^ome renvers\'e. Cette fois-ci c'est la th\'eorie ultraviolette qui devient triviale! 

Il est vrai qu'une th\'eorie libre dans l'ultraviolet est beaucoup plus physique qu'une th\'eorie \`a couplage fort dans l'ultraviolet, et cela a permis notamment l'introduction du mod\`ele standard pour les interactions fondamentales, n\'eanmoins des probl\`emes subsistent.
La convergence des flots du mod\`ele standard vers une seule constante de couplage demande l'alt\'eration des flots. Cela peut \^etre r\'ealis\'e \`a l'aide de la supersym\'etrie, mais il n'y a eu aucune d\'etection des partenaires supersym\'etriques jusqu'\`a maintenant.

De plus, ce m\^eme fant\^ome a emp\^ech\'e la construction du mod\`ele $\Phi^4_4$ commutatif. Bien qu'on puisse penser que la th\'eorie constructive des champs soit plut\^ot un jeu math\'ematique qu'une question physique, il n'emp\^eche que sans elle les calculs perturbatifs n'ont pas de sens. Si l'on ne peut pas construire une th\'eorie des champs, la s\'erie perturbative risque d'\^etre fausse \`a partir du premier ordre!

Nous concluons donc que les flots des mod\`eles de th\'eories des champs commutatives posent un probl\`eme profond et les rendent math\'ematiquement malades.

La situation change compl\`etement dans les TCNC. Pour le mod\`ele $\Phi^{\star 4}_4$ le tadpole n'est plus local! Il nous donne une renormalisation non triviale de la fonction d'onde \`a partir du premier ordre.
Les calculs \`a une boucle effectu\'es par Grosse et Wulkenhaar \cite{GrWu04-2} nous donnent un flot du param\`etre $\Omega$ vers $1$ et \`a cette valeur le flot du couplage s'arr\^ete. Ce r\'esultat a \'et\'e g\'en\'eralis\'e jusqu'\`a trois boucles par Disertori et Rivasseau en \cite{DR}.

Ces r\'esultats ouvrent la porte vers une possibilit\'e tentante. Est-il possible que le flot de cette th\'eorie soit born\'e sans que la th\'eorie soit triviale? Dans ce cas on aurait une th\'eorie bien d\'efinie et non triviale tout au long de sa trajectoire sous le (semi-)groupe de renormalisation!
Pour le reste du chapitre nous nous pla\c cons \`a $\Omega=1$.

Nous avons le th\'eor\`eme suivant:
\begin{theorem}\label{th:MainBeta}
L'\'equation:
\bea
\Gamma^{4}(0,0,0,0)=\lambda (1-\partial\Sigma(0,0))^2
\eea
est vraie aux corrections irrelevantes pr\`es
\footnote{Elles incluent les termes non- planaires ou ayant plus d'une face bris\'ee.}
 \`a {\bf tous} les ordres de perturbation, 
soit comme \'equation entre les quantit\'es nues avec cutoff UV fix\'e ou comme \'equation entre les quantit\'es renormalis\'ees. Dans ce dernier cas $\lambda$ est la constante nue exprim\'ee comme s\'erie en puissances de $\lambda_{ren}$.
\end{theorem}

Prouver un tel r\'esultat est dans un certain sens tr\`es difficile. Il faut prouver qu'il est vrai \`a tous les ordres de perturbation, et ensuite constructivement!
Comme nous allons le voir par la suite, la preuve de la validit\'e \`a tous les ordres de perturbation est en effet beaucoup plus simple que pr\'evu.

Elle repose sur deux id\'ees: les identit\'es de Ward et l'\'equation de Dyson.
Nous allons travailler dans la base matricielle. Comme $\Omega=1$, nous avons:
\be
C_{m n;k l} = C_{m n} \delta_{m l}\delta_{n k} \ ; \ 
C_{m n}= \frac{1}{A+m+n}\  ,
\ee
avec $A= 2+ \mu^2 /4$, $m,n\in \mathbb{N}^2$ et
\be
\delta_{ml} = \delta_{m_1l_1} \delta_{m_2l_2}\ , \qquad m+n = m_1 + m_2 + n_1 + n_2 
\ .
\ee

La fonctionnelle g\'en\'eratrice est 
\bea\label{eq:actom1matrix}
&&Z(\eta,\bar{\eta})=\int d\phi d\bar{\phi}~e^{-S(\bar{\phi},\phi)+F(\bar{\eta},\eta,;\bar{\phi},\phi)}\nonumber\\
&&F(\bar{\eta},\eta;\bar{\phi},\phi)=  \bar{\phi}\eta+\bar{\eta}\phi \nonumber\\
&&S(\bar{\phi},\phi)=\bar{\phi}X\phi+\phi X\bar\phi+A\bar{\phi}\phi+
\frac{\lambda}{2}\phi\bar{\phi}\phi\bar{\phi}
\eea
o\`u les traces sont sous-entendues et $X_{m n}=m\delta_{m n}$. $S$ est l'action et $F$ sont les sources externes.

\section{Les identit\'es de Ward}

Nous rappelons l'orientation des propagateurs de $\bar{\phi}$ \`a $\phi$.
Pour un champ $\bar{\phi}_{a b}$ l'indice $a$ est un indice gauche
et $b$ un indice droit. Le premier (second) indice de $\bar{\phi}$ se contracte {\it toujours} avec le second (premier) indice de $\phi$.  Par cons\'equent pour $\phi_{c d}$, $c$ est un indice {\it droit} et $d$ est un indice {\it gauche}.

Soit $U=e^{\imath B}$ avec $B$ une matrice petite hermitienne.
Nous faisons un changement de variable ``gauche" (car il agit seulement sur les indices gauches)
\bea
\phi^U=\phi U;\bar{\phi}^U=U^{\dagger}\bar{\phi} \ .
\eea
Comme il s'agit d'un simple changement de variable, nous avons:
\bea
\partial_{\eta}\partial_{\bar \eta}\frac{\delta \ln Z}{\delta B_{ba}}=0 \; ,
\eea
avec $\eta$ des terms de source.
En explicitant cette relation on trouve pour les fonctions planaires \`a deux points l'identit\'e de Ward:
\bea
\label{eq:MainWard}
(a-b)<[ \bar{\phi}\phi]_{a b} \phi_{\nu a} 
\bar{\phi}_{b \nu }>_c=
<\phi_{\nu b} \bar{\phi}_{b \nu}>_c
-<\bar{\phi}_{a \nu} \phi_{\nu a}>_c \; .
\eea
(les indices r\'ep\'et\'es ne sont pas somm\'es).

\begin{figure}[hbt]
\centerline{
\includegraphics[width=100mm]{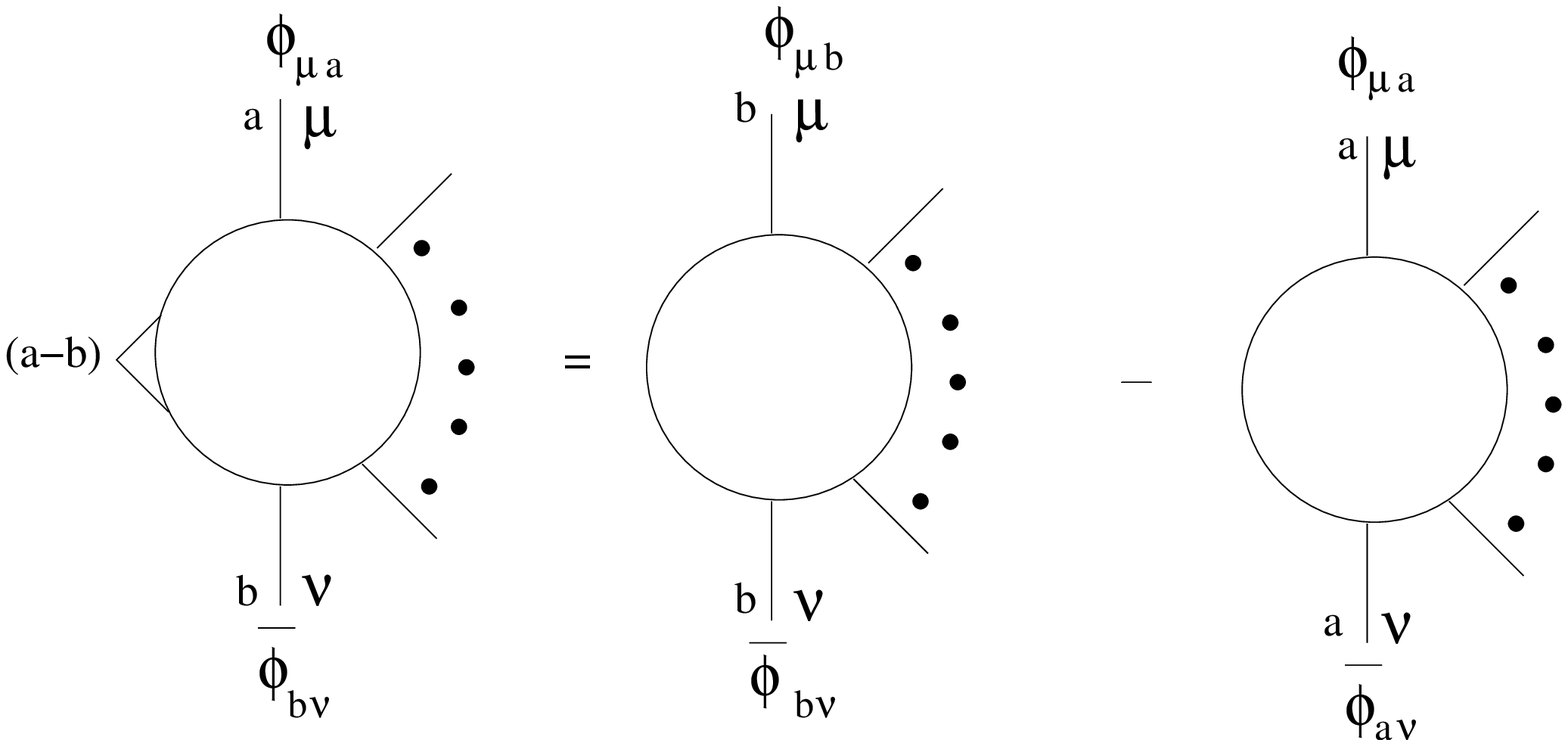}}
\caption{L'identit\'e de Ward pour les fonctions \`a 2p points avec insertion sur une face gauche.}\label{fig:WardMain}
\end{figure}

Les indices $a$ et $b$ sont des indices gauches, et notre identit\'e de Ward d\'ecrit des fonctions avec une insertion sur la face gauche, comme dans la figure \ref{fig:WardMain}. Il est possible d'obtenir une identit\'e similaire avec l'insertion sur la face droite. De plus, en d\'erivant davantage par rapport aux sources externes nous pouvons obtenir de telles identit\'es avec un nombre arbitraire de pattes externes.

Cette \'equation a \'et\'e d\'eduite de l'int\'egrale fonctionnelle. Elle est donc vraie ordre par ordre dans la th\'eorie des perturbations, car celle-ci se d\'eduit \`a partir de l'int\'egrale fonctionnelle. 

On a l'habitude de regarder les identit\'es de Ward comme cons\'equences de l'invariance de jauge. Dans notre traitement on obtient une identit\'e de Ward comme  cons\'equence d'une invariance du vertex sous un changement de variable qui ne laisse pas invariante la partie quadratique. Peut-on interpr\'eter ce changement de variable comme une invariance de jauge? 

Le premier pas dans cette direction est d'introduire des op\'erateurs covariants de multiplication $X$ qui se transforment sous le changement de variable comme:
\bea
X^U=U^{\dagger}X U \; .
\eea
Il faut ensuite construire une action invariante sous cette transformation, telle que sa forme fix\'ee de jauge soit (\ref{eq:actom1matrix}). Cette invariance sera  l'invariance habituelle d'un mod\`ele de matrice sous les transformations unitaires.

Nous devrons de plus utiliser ces "transformations de jauge" pour construire une th\'eorie de Yang-Mills associ\'ee aux th\'eories non commutatives vulcanis\'ees. Elle sera tr\`es certainement diff\'erente des th\'eories Yang-Mills vulcanis\'ees propos\'ees par \cite{GrosseYM} et \cite{WalletYM}, car dans leurs travaux seulement les champs $\phi$ et $\bar \phi$ changent sous les transformations de jauge.

\section{L'\'equation de Schwinger Dyson}

Nous introduisons maintenant le deuxi\`eme outil conceptuel dont nous avons besoin pour prouver le th\'eor\`eme (\ref{th:MainBeta}).

Soit $G^{4}(m,n,k,l)$ la fonction \`a quatre points connexe planaire avec une face bris\'ee avec $m$, $n$, $k$ et $l$ les index de la face externe dans l'ordre cyclique.
Le premier index $m$ est toujours choisi comme index gauche.

Soit $G^{2}(m,n)$ la fonction \`a deux points connexe avec indices externes
$m,n$ (appel\'e aussi propagateur habill\'e).  Comme d'habitude, $\Sigma(m,n)$ est la fonction \`a deux points, une particule irr\'eductible, amput\'ee. $G^{2}(m,n)$ et $\Sigma(m,n)$ sont li\'ees par:
\bea
\label{eq:MainG2Sigmarelation}
  G^{2}(m,n)=\frac{C_{m n}}{1-C_{m n}\Sigma(m,n)}=
  \frac{1}{C_{m n}^{-1}-\Sigma(m,n)} \, .
\eea 

Soit $G_{ins}(a,b;...)$ la fonction connexe \`a deux points planaire \`a une face bris\'ee avec une insertion sur la face gauche avec saut d'index de $a$ \`a $b$.
L'identit\'e de Ward (\ref{eq:MainWard}) s'\'ecrit comme:
\bea
(a-b) ~ G^{2}_{ins}(a,b;\nu)=G^{2}(b,\nu)-G^{2}(a,\nu)\, .
\eea 

\begin{figure}[hbt]
\centerline{
\includegraphics[width=120mm]{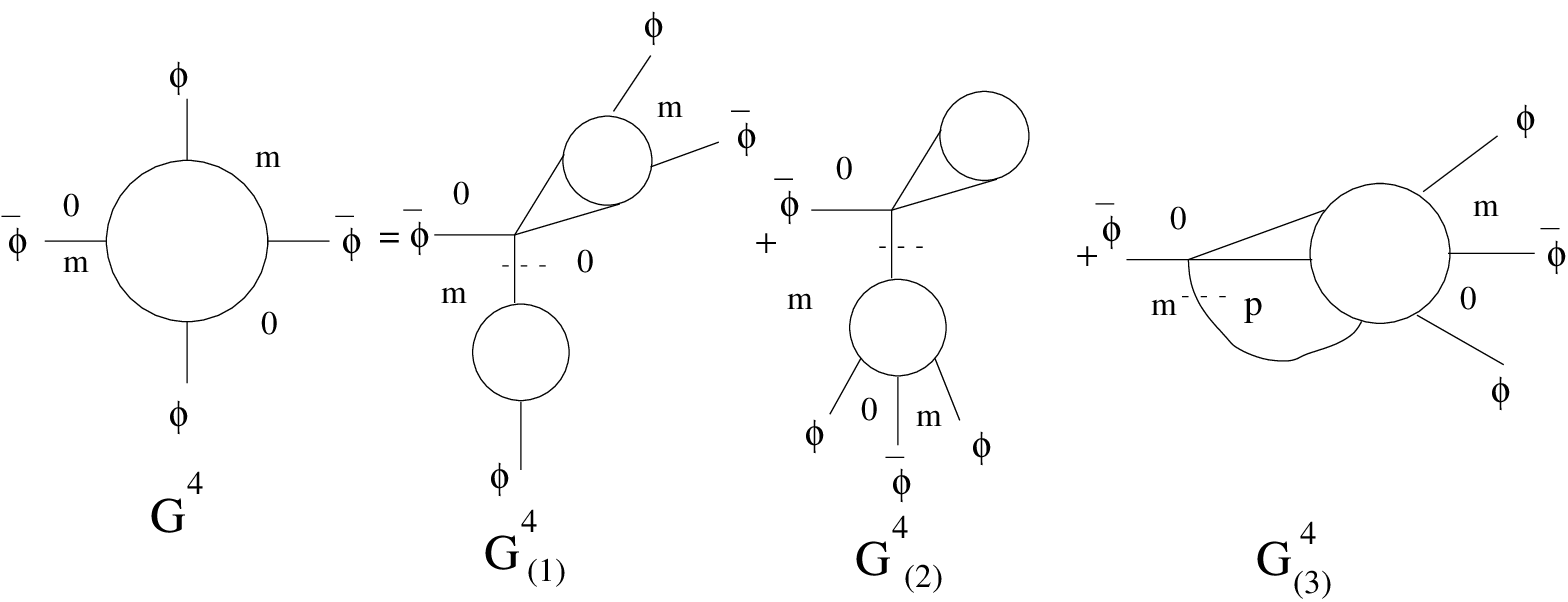}
}
\caption{L'\'equation de Dyson}\label{fig:Maindyson}
\end{figure}

Analysons $G^4(0,m,0,m)$. 
Soit une patte externe $\bar{\phi}$ et le premier vertex qui lui est attach\'e. 
En tournant vers la droite sur le bord $m$ \`a ce vertex nous rencontrons une nouvelle ligne (marqu\'ee dans la figure \ref{fig:Maindyson}). Cette ligne peut soit s\'eparer le graphe en deux composantes d\'econnect\'ees ($G^{4}_{(1)}$ et $G^{4}_{(2)}$ dans \ref{fig:Maindyson}) soit non ($G^{4}_{(3)}$ dans \ref{fig:Maindyson}). De plus, si cette ligne s\'epare le graphe en deux composantes disconnect\'ees, le premier vertex peut soit appartenir \`a la composante \`a deux pattes ($G^{4}_{(1)}$ dans \ref{fig:Maindyson}) soit \`a la composante \`a quatre pattes ($G^{4}_{(2)}$ dans \ref{fig:Maindyson}). 

Cela est une simple classification des graphes: les composantes repr\'esent\'ees dans la figure \ref{fig:Maindyson} prennent en compte leurs facteurs combinatoires.
Nous pouvons donc \'ecrire l'\'equation de Dyson:
\bea
\label{eq:MainDyson}
 G^4=G^4_{(1)}(0,m,0,m)+G^4_{(2)}(0,m,0,m)+G^4_{(3)}(0,m,0,m)\, .
\eea

Le terme $G^{4}_{(2)}$ est z\'ero par renormalisation de masse. $G^{4}_{(1)}+G^{4}_{(3)}$ donne 
$\Gamma^4=\lambda (1-\partial \Sigma)^2$ 
apr\`es amputation des propagateurs externes.

Nous allons analyser ici en d\'etail la contribution $G^{4}_{(1)}$. 
Elle est de la forme:
\bea
G^4_{(1)}(0,m,0,m)=\lambda C_{0 m} G^{2}(0, m) G^{2}_{ins}(0,0;m)\,.
\eea
Mais, par l'identit\'e de Ward:
\bea
G^{2}_{ins}(0,0;m)&=&\lim_{\zeta\rightarrow 0}G^{2}_{ins}(\zeta ,0;m)=
\lim_{\zeta\rightarrow 0}\frac{G^{2}(0,m)-G^{2}(\zeta,m)}{\zeta}\nonumber\\
&=&-\partial_{L}G^{2}(0,m) \, .
\eea
En utilisant la forme explicite du propagateur on a
$\partial_L C^{-1}_{ab}=\partial_R C^{-1}_{ab}=\partial C^{-1}_{ab}=1$. En utilisant  l'\'equation (\ref{eq:MainG2Sigmarelation}), on conclut:
\bea\label{eq:Maing41}
G^4_{(1)}(0,m,0,m)&=&\lambda
C_{0m}\frac{C_{0m}C^2_{0m}[1-\partial_{L}\Sigma(0,m)]}{[1-C_{0m}\Sigma(0,m)]
(1-C_{0m}\Sigma(0,m))^2}\nonumber\\
&=&\lambda [G^{2}(0,m)]^{4}\frac{C_{0m}}{G^{2}(0,m)}[1-\partial_{L}\Sigma(0,m)]\, .
\eea
Le traitement de $G^4_{(3)}(0,m,0,m)$ est plus difficile et nous allons seulement l'esquisser ici. Nous "ouvrons" la face ``premi\`ere \`a droite''. 
Son index somm\'e est appel\'e $p$ (figure \ref{fig:Maindyson}). Cette op\'eration se traduit pour les fonctions de Green nues par:
\bea
\label{eq:Mainopening}
G^{4,bare}_{(3)}(0,m,0,m)=C_{0m}\sum_{ p} G^{4,bare}_{ins}(p,0;m,0,m)\, .
\eea
et pour les fonctions renormalis\'ees par:
\bea
\label{eq:MainOpen2}
G^4_{(3)}(0,m,0,m)&=& C_{0m}\sum_{p} G^{4}_{ins}(0,p;m,0,m)\nonumber\\
     &-&C_{0m}(CT_{lost})G^{4}(0,m,0,m)\,.
\eea

Si on utilise l'identit\'e de Ward pour la fonction $G^{4,bare}_{ins}(p,0;m,0,m)$ on peut n\'egliger l'un des deux termes. En combinant le deuxi\`eme avec le contre-terme manquant on obtient un terme qui, combin\'e avec (\ref{eq:Maing41}) nous donne
\ref{th:MainBeta}.

Nous arrivons ainsi \`a prouver que le flot de la constante de couplage $\lambda$ est born\'e au long de la trajectoire du groupe de renormalisation. Comme d\'ej\`a mentionn\'e, il faut \'etendre ce r\'esultat au niveau constructif. Les premiers pas dans cette direction ont \'et\'e faits par Rivasseau et Magnen. Leurs travaux nous permettent d'esp\'erer que l'extension de ce r\'esultat au niveau constructif ne va pas tarder.

L'invariance de jauge sous-jacente reste encore \`a comprendre. La transformation des $X$ mentionn\'ee ci-dessus est en relation directe avec les dif\'eomorphismes qui pr\'eservent l'aire. Nous esp\'erons qu'une telle \'etude nous aidera \`a formuler une th\'eorie invariante de fond et sera un pas vers une proposition de quantification de la gravitation.

\chapter{Repr\'esentation param\'etrique}

Un outil essentiel dans la th\'eorie commutative des champs est la repr\'esentation param\'etrique. Elle donne une repr\'esentation compacte des amplitudes. De plus, elle fournit des formules topologiques et canoniques. Elle est le point de d\'epart pour la r\'egularisation dimensionnelle, qui est la seule fa\c con de renormaliser une th\'eorie de jauge sans briser l'invariance de jauge.

Dans les th\'eories habituelles, la construction de cette repr\'esentation n'est pas trop compliqu\'ee. Au lieu de r\'esoudre les conservations d'impulsion au vertex on exprime les fonctions $\delta$ de conservation d'impulsion en transform\'ee de Fourier. L'amplitude d'un graphe $G$ avec impulsion externe $p$ en $D$ dimensions est, apr\`es int\'egration des variables internes:
\bea \label{eq:Mainsymanzik} 
{\cal A}_G (p) = \delta(\sum p)\int_0^{\infty} 
\frac{e^{- V_G(p,\alpha)/U_G (\alpha) }}{U_G (\alpha)^{D/2}} 
\prod_l  ( e^{-m^2 \alpha_l} d\alpha_l )\ .
\eea
Les deux polyn\^omes de Symanzik $U_G$ et $V_G$ sont:
\be\label{eq:Mainsymanzik1}
U_G = \sum_T \prod_{l \not \in T} \alpha_l \ ,
\ee
\be\label{eq:Mainsymanzik2}
V_G = \sum_{T_2} \prod_{l \not \in T_2} \alpha_l  (\sum_{i \in E(T_2)} p_i)^2 \, .
\ee 

Les deux polyn\^omes sont explicitement des sommes de termes positifs. Cette positivit\'e terme \`a terme est essentielle pour la renormalisation et la r\'egularisation dimensionnelle car il nous suffit d'\'evaluer une partie des termes et de borner le reste. Les sommes sont index\'ees par des objets topologiques: les arbres d'un graphe ou les paires d'arbres d'un graphe. On remarque aussi la d\'emocratie des arbres: ils contribuent chacun avec un terme de poids $1$.

La repr\'esentation param\'etrique du mod\`ele $\phi^{\star 4}_4$ fait intervenir des polyn\^omes dans les variables $t_{\ell}=\tanh\frac{\alpha_{\ell}}{2}$, appel\'es donc hyperboliques. Nous allons retrouver la positivit\'e explicite de la repr\'esentation et nous allons pouvoir isoler des termes dominants index\'es par des bi-arbres, qui sont des objets topologiques qui g\'en\'eralisent les arbres. A cause des complications techniques, nous allons pr\'esenter ici seulement le premier polyn\^ome.

A ce point nous rencontrons une subtilit\'e: dans la th\'eorie commutative nous ne 
pouvons pas int\'egrer sur toutes les positions internes. A cause de l'invariance par translation, l'int\'egrale sur toutes les positions internes nous fournit un facteur infini de volume. Nous int\'egrons sur toutes les positions sauf une! Les polyn\^omes sont ind\'ependants par rapport \`a ce point non int\'egr\'e, et ils restent donc canoniques. Dans notre th\'eorie l'invariance par translation est perdue. Nous pouvons int\'egrer sur toutes les positions internes. Nous pr\'ef\'erons par contre isoler un vertex (not\'e $\bar v$) et ne pas int\'egrer sur sa fonction $\delta$. Cela nous permettra de retrouver les r\'esultats commutatifs dans une certaine limite. Notons que les polyn\^omes d\'ependent de cette racine explicite, mais leurs termes dominants n'en d\'ependent pas.

Nous notons $\ell=(i,j)$ la ligne du graphe qui va du coin $i$ d'un vertex au coin $j$ d'un autre. Nous allons passer aux variables courtes et longues. Deux matrices sont cruciales par la suite:
\begin{itemize}
\item La matrice $\e_{\ell i}$ est la matrice d'incidence du graphe. Elle vaut $1$ si $\ell$ entre en $i$, $-1$ si $\ell$ sorte de $i$, et zero si $\ell$ ne touche pas $i$,
\item $\eta_{\ell i}=\vert\e_{\ell i}\vert $. 
\end{itemize}
Pour un graphe orientable on a $\e_{\ell i}=(-1)^{i+1}\eta_{\ell i}$.

En s'appuyant sur une propri\'et\'e de factorisation sur les dimensions de l'espace-temps, nous \'ecrivons l'amplitude d'un graphe comme:
\bea
{\cal A}_{G,{\bar v}}  (\{x_e\},\;  p_{\bar v}) = K'  \int_{0}^{1} \prod_{\ell=1}^L  [ dt_{\ell} (1-t_l^2)^{D/2-1} ]
\frac
{e^{-  \frac {HV_{G, \bar{v}} ( t , x_e , p_{\bar v})}{HU_{G, \bar{v}} ( t )}}}
{HU_{G, \bar{v}} ( t )^{D/2}} \; ,
\eea
ou $t_{\ell}=\tanh\frac{\alpha_{\ell}}{2}$. Nous d\'efinissons aussi $c_{\ell}=1/t_{\ell}$.

Le premier polyn\^ome est:
\bea
HU_{G,\bar v}= \left(\prod_{\ell=1}^Lt_{\ell}\right)\, \det(A+B) \;,
\eea
avec $A+B$ une matrice carr\'ee $2L+N-1$ dimmensionnelle. La matrice $A$ est la contribution (diagonale dans les variables courtes et longues) des propagateurs:
\bea
A=\begin{pmatrix} c_{\ell} & 0 & 0\\ 0  & t_{\ell} & 0 \\ 0&0&0\\
   \end{pmatrix} \; ,
\eea
et la matrice $B$ (antisym\'etrique) provient des oscillations des vertex et des fonctions $\delta$:
\bea
B= \begin{pmatrix}\frac{1}{\Omega} E & C \\
-C^t & 0 \\
\end{pmatrix}\ .
\eea
La contribution d\'etaill\'ee des fonctions $\delta$ est:
\bea
C_{v l}=\begin{pmatrix}
\sum_{i\in v}(-1)^{i+1}\epsilon_{l i} \\
\sum_{i\in v}(-1)^{i+1}\eta_{l i} \\
\end{pmatrix}\ ,
\eea
et celle des vertex:
\bea
E^{vv}_{l,l'}&=&\sum_v
\sum_{i\ne j ; \; i, j \in v}  (-1)^{i+j+1} \omega(i,j)\eta_{l i}\eta_{l' j},
\nonumber\\
E^{uu}_{l,l'}&=&\sum_v
\sum_{i\ne j ; \; i, j \in v}  (-1)^{i+j+1} \omega(i,j)\epsilon_{l i}\epsilon_{l' j},
\nonumber\\
E^{uv}_{l,l'}&=&\sum_v
\sum_{i\ne j ; \; i, j \in v}  (-1)^{i+j+1} \omega(i,j)\epsilon_{l i}\eta_{l' j},
\eea 
avec $\omega(i,j) = 1$ si $i < j$,  $\omega(i,j) = -1$ si $i > j$;
de plus $E^{vu}_{l,l'} = - E^{uv}_{l',l}$.

\section{La positivit\'e}

Le premier polyn\^ome posse\`ede une propri\'et\'e de positivit\'e. En effet, nous avons prouv\'e le lemme suivant:
\begin{lemma}
Soit $A,B\in {\cal M}(N\times N)$ des matrices avec $A=(a_i\delta_{i j})$ une matrice diagonale et $B=(b_{i j})$ une matrice arbitraire, avec $b_{ii}=0$ ($B$ n'est pas n\'ecessairement antisym\'etrique). Nous avons:
\be
\det(A+B)=\sum_{K\subset \{1,\dotsc ,N\}}\det(B_{\hat{K}}) \prod_{i\in K}a_i 
\ee
o\`u $B_{\hat{K}}$ est la matrice obtenue de $B$ en effa\c cant les lignes et 
colonnes ayant des indices dans le sous-ensemble $K$.
\end{lemma}
Si en plus $B$ est antisym\'etrique, $B_{\hat{K}}$ l'est aussi et son d\'eterminant est positif, car il est le carr\'e d'un Pfaffien.

Cette propri\'et\'e de positivit\'e est vraie aussi pour le mod\`ele r\'eel, car sa preuve ne fait pas appel \`a l'orientabilit\'e.

Choisissons $K=I\cup J$ avec $I$ un sous-ensemble d'indices choisis parmi les premiers  $L$, associ\'es aux variables courtes et $J$ un sous-ensemble choisi parmi les $L$ indices suivants associ\'es aux variables longues. Le premier polyn\^ome prend la forme:
\bea\label{eq:MainHU}
HU_{G,{\bar v}} (t) &=&  \sum_{I,J}  \left(\frac{1}{\Omega}\right)^{2g(G)-k_{I,J}} 
\ n_{I,J}^2 \prod_{\ell \not\in I} t_{\ell} \prod_{\ell' \in J} t_{\ell'}\ ,
\eea
avec $k_{I,J}=\vert I\vert+\vert J\vert - L(G) - F(G) +1$, et 
$n_{I,J}=\text{Pf} (B'_{\hat{I}\hat{J}})$ avec $B'=\Omega B$.
La matrice $B'$ a des entr\'ees enti\`eres et son pffafien est un nombre entier. Cette forme est g\'en\'erale, et la preuve, de nouveau, ne fait pas appel \`a l'orientabilit\'e. Parmi les termes de la somme du (\ref{eq:MainHU}) certains sont nuls. Pour trouver des termes dominants nous devons trouver des ensembles $I$ et $J$ tels que $n_{I,J}\neq 0$.

\section{Les termes dominants}

Dans la r\'egion UV les termes dominants sont ceux ayant un degr\'e minimal en $t$. Nous n'allons pas donner ici la liste exhaustive des termes dominants\footnote{Notamment, une cat\'egorie de termes dominants n\'ecessaires pour la r\'egularisation dimensionnelle sera explicit\'ee dans le chapitre suivant. Ces termes seront notament n\'ecessaires pour trouver le comportement des sous-graphes primitivement divergents.}. N\'eanmoins, les termes qu'on explicite suffisent pour trouver le comptage de puissance d'un graphe. 

Choisions $I=\{1\cdots L\}$. Cela nous garantit que le premier produit de (\ref{eq:MainHU}) est vide. Nous cherchons les sous-ensembles $J$ de cardinal minimal tel que $n_{I,J}\neq 0$. Nous verrons que ces termes ont $\vert J \vert =F-1$, ce qui justifie la normalisation de (\ref{eq:MainHU}).

Nous g\'en\'eralisons les mouvements de Filk \cite{Filk:1996dm}, pour d\'efinir trois op\'erations topologiques pour un graphe \`a rubans. Nous repr\'esentons les pfaffiens d'int\'er\^et comme des int\'egrales grassmanniennes.

\begin{figure}
\centerline{\epsfig{figure=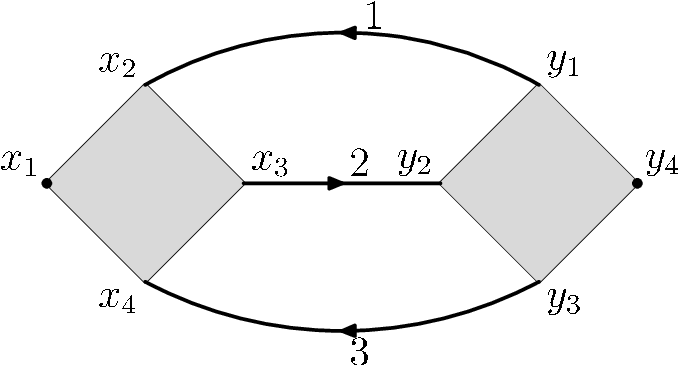,width=6cm} \hfil \epsfig{figure=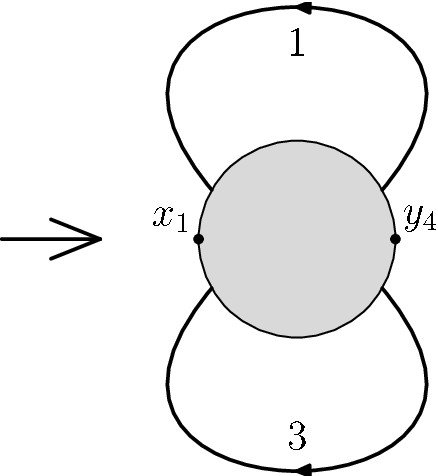,width=2cm}}
\caption{Le premier mouvement de Filk pour le Sunshine}
\label{fig:Mainfirstfilk}
\end{figure}

\begin{figure}
\centerline{\epsfig{figure=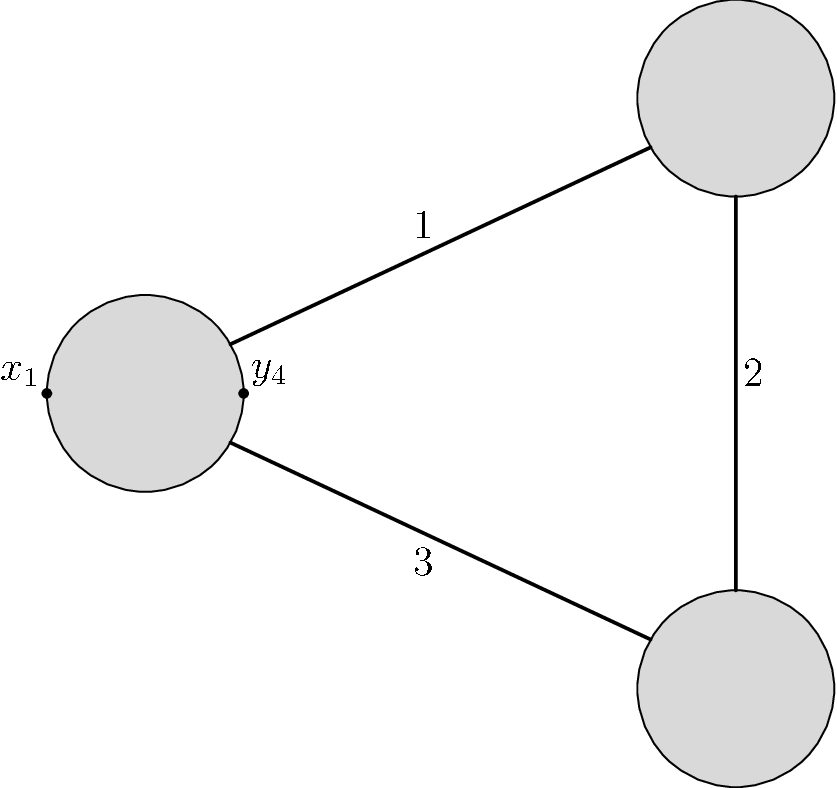,width=4cm} \hfil \epsfig{figure=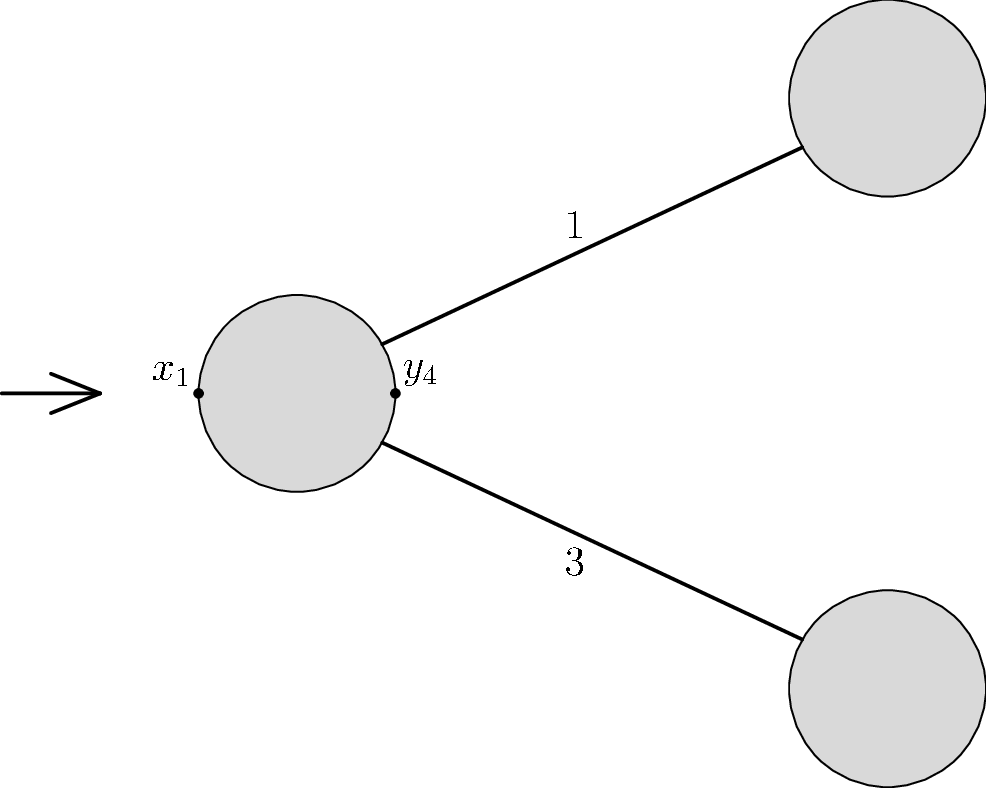,width=4cm}}
\caption{Le premier mouvement de Filk pour le dual du Sunshine}
\label{fig:Maindualsunshine}
\end{figure}

La premi\`ere est l'habituelle "premier mouvement de Filk" d\'ej\`a introduit, la r\'eduction d'une ligne d'arbre dans le graphe direct. Cela revient \`a coller les deux vertex touchant la ligne, de coordination $p$ et $q$ pour obtenir un grand vertex de coordination $p+q-2$. Le nouveau graphe ainsi obtenu a une ligne et un vertex en moins $N(G')=N(G)-1$, $L(G')=L(G)-1$, $F(G')=F(G)$. Le genre du graphe est conserv\'e sous cette op\'eration, \`a cause de la relation $2g(G)-2=N(G)-L(G)+F(G)$.
Dans le graphe dual cette op\'eration efface la ligne d'arbre du graphe direct.
La r\'eduction par cette op\'eration de la ligne centrale du graphe sunshine est pr\'esent\'ee dans la figure \ref{fig:Mainfirstfilk}, et son effet sur le graphe dual du sunshine est pr\'esent\'e dans la figure \ref{fig:Maindualsunshine}. En terme des variables grassmanniennes cette op\'eration apparie la variable de la ligne r\'eduite avec celle d'un des vertex coll\'es.

\begin{figure}
\centerline{\epsfig{figure=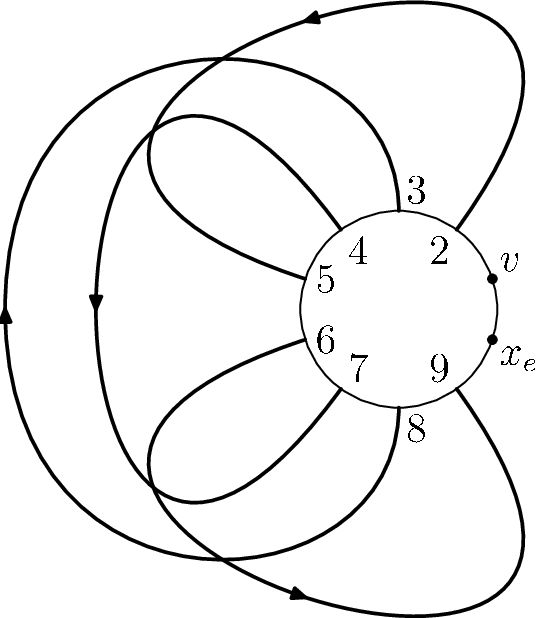,width=5cm}}
\caption{Rosette avec racine explicit\'ee}
\label{fig:Mainrosette}
\end{figure}

It\'erons cette op\'eration un maximum de fois. On peut toujours r\'eduire un arbre qui recouvre le graphe direct, ayant $N(G)-1$ lignes. Le r\'esultat de cette op\'eration est un graphe avec un seul vertex n'ayant que des lignes de tadpole et ayant le genre du graphe initial: la rosette associ\'ee au graphe. La rosette a une racine correspondant au vertex $\bar v$. De plus, la rosette a un ordre cyclique, trigonom\'etrique direct. Nous avons besoin de cet ordre cyclique pour pr\'eciser l'ordre des points sur la rosette.
Un exemple est donn\'e dans la figure \ref{fig:Mainrosette}. Les lignes conservent leurs orientations et pour les graphes orientables elles sortent des points pairs et entrent dans les points impairs.

Dans l'annexe \ref{sec:param} nous prouvons que cette op\'eration reproduit la forme du facteur d'oscillation d'un graphe. Par cons\'equent, pour \'evaluer un terme de la somme \ref{eq:MainHU} il suffit de regarder la rosette associ\'ee \`a un graphe et de lui associer le facteur:
\bea
-\sum_{i,j}\omega^{\bar v}(i,j)\e_{\ell i}\e_{\ell j}
\eea

\begin{figure}
\centerline{\epsfig{figure=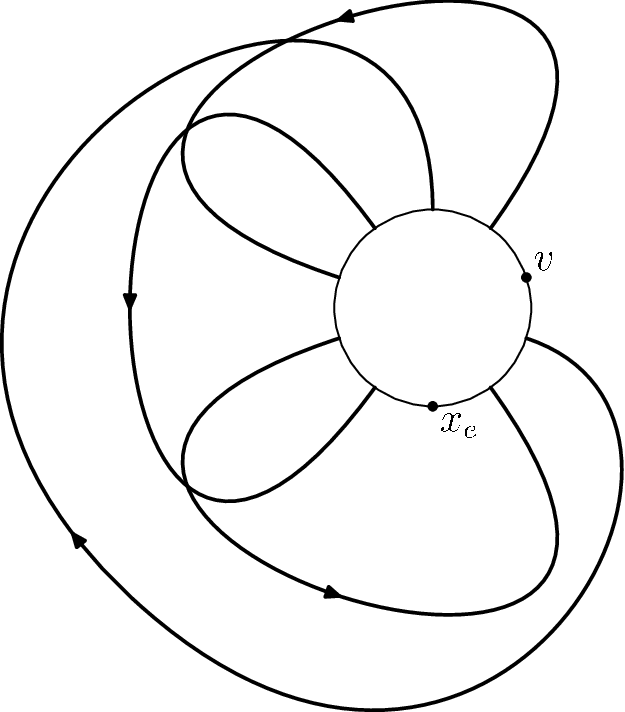,width=5cm}}
\caption{Une superrosette}
\label{fig:Mainsuperrosette}
\end{figure}

La seconde op\'eration topologique est la r\'eduction d'une ligne d'arbre dans le graphe {\it dual}. Cette op\'eration efface la ligne dans le graphe direct. Le genre du graphe (direct et dual) ne change pas sous cette op\'eration.
En it\'erant cette operation maximalement, nous r\'eduisons $F(G)-1$ lignes d'un arbre dans le graphe dual. Nous aboutissons \`a un graphe ayant $N(G)-1$ lignes d'un arbre $T(G)$ dans le graphe direct r\'eduites et $F(G)-1$ lignes d'un arbre ${\cal T}(G)$ du graphe dual r\'eduites. Ce graphe est une rosette avec une seule face, appel\'e {\it superrosette}.
Son dual est aussi une superrosette. Un exemple est pr\'esent\'e dans la figure 
\ref{fig:Mainsuperrosette}.

Cette op\'eration va simplement effacer les lignes et colonnes correspondantes aux lignes du graphe dual r\'eduites de la matrice $B'$.

La troisi\`eme op\'eration topologique est la r\'eduction du genre d'une rosette.
Un {\it croisement naturel} est un croisement des deux lignes sur une rosette tel que le point final de la premi\`ere ligne succ\`ede au point initial de la deuxi\`eme sur la rosette.
De telles lignes ont une face interne commune. Si une rosette a des croisements, elle aura certainement des croisements naturels.
Les lignes $2-5$ et $4-8$ dans la figure \ref{fig:Mainrosette} sont un exemple de croisement naturel.

\begin{figure}
\centerline{\epsfig{figure=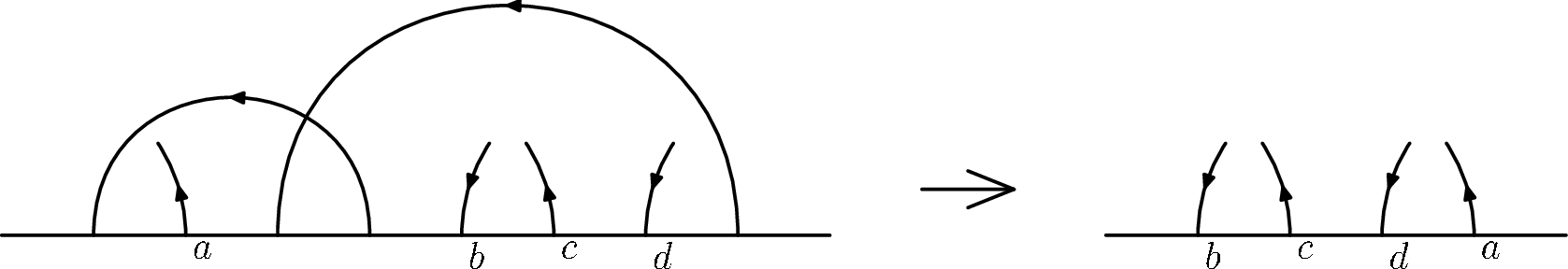,width=12cm,height=2cm}}
\caption{Troisi\`eme mouvement de Filk}
\label{fig:Mainthirdfilk}
\end{figure}

La r\'eduction du genre efface les lignes d'un croisement naturel et permute les points 
survol\'es par les deux lignes, comme dans la figure \ref{fig:Mainthirdfilk}. 
On appelle cette op\'eration le troisi\`eme mouvement de Filk. Elle diminue le nombre de lignes par $2$, recolle les faces d'une mani\`ere coh\'erente, et d\'ecro\^it le genre par $1$. En terme des variables grassmanniennes, cela correspond \`a apparier les deux variables des lignes d'un croisement naturel.

A l'aide de cette troisi\`eme op\'eration nous prouvons que
\begin{lemma}
Le d\'eterminant associ\'e \`a une rosette vaut
\begin{itemize}
\item  $0$ si $F(G)$ est plus grand que $1$
\item $2^{2g}$ si $F(G)=1$
\end{itemize}
\end{lemma}

Pour avoir un terme non nul il suffit donc de choisir $J$ admissible, ce qui veut dire que:
\begin{itemize}
\item $J$ contient un arbre ${\cal T}(G)$ dans le graphe dual,
\item le compl\'ement de $J$ contient un arbre $T(G)$ dans le graphe direct 
\item La rosette obtenue en effa\c cant les lignes de $J$ et en contractant les lignes de $T(G)$ est un superrosette, donc \`a une seule face.
\end{itemize}
Nous avons la borne:
\bea
HU_{G,\bar v}\ge \sum_{J admissible} \left( \frac{2}{\Omega} \right)^{2g-k_J} \prod_{\ell\in J} t_{\ell} \; ,
\eea
avec $k_J=\vert J \vert -F(G)+1$.

Les termes dominants sont ceux avec $\vert J \vert$ minimal, donc ceux pour lesquels $J$ est un arbre ${\cal T}(G)$ dans le graphe dual et l'on a:
\bea
HU_{G,\bar v} \ge \left(\frac{2}{\Omega}\right)^{2g}\sum_{{\cal T}(G)}\prod_{\ell \in {\cal T}(G)}t_{\ell} \; .
\eea

Nous allons utiliser cette borne pour \'evaluer le degr\'e de divergence d'un graphe. Nous nous pla\c cons \`a attribution d'\'echelle fix\'ee $0\le t_1\le t_2\dots \le t_{L(G)}$\footnote{Ce traitement est d\'etaill\'e dans la section suivante}. Nous changeons de variable dans $t_1=\lambda^2$, $t_i=\lambda^2 x_i$. La divergence ou convergence du graphe est gouvern\'ee par la nature de l'int\'egrale sur $\lambda$:
\bea
{\cal A} \le \int_0 d\lambda \lambda^{2L(G)-1+D[F(G)-1]}\sim \int_0 d\lambda \lambda^{-1} \lambda^{8g(G)+N_e(G)-4} \, ,
\eea
o\`u nous avons utilis\'e $N(G)-L(G)+F(G)=2-2g(G)$ et $2L(G)=4N(G)-N_e(G)$. Nous voyons tout de suite que l'int\'egrale au-dessus est certainement convergente si le graphe a plus de quatre pattes externes ou s'il est non-planaire. Par des traitements similaires nous pouvons conclure qu'un graphe est convergent aussi s'il a plus d'une face bris\'ee.

Dans ce chapitre nous avons introduit la repr\'esentation param\'etrique dans la TCNC.
La continuation naturelle de notre traitement est de proc\'eder \`a la r\'egularisation et renormalisation dimensionnelle de notre mod\`ele. Ce programme sera achev\'e dans le chapitre suivant. 

Une autre direction de recherche est d'approfondir les nouveaux polyn\^omes hyperboliques introduits dans notre travail. Les bi-arbres, ensembles des lignes form\'es d'un arbre dans le graphe direct et un arbre dans le graphe dual, sont la g\'en\'eralisation des arbres de Symanzick pour les TCNC sur le plan de Moyal. Leur \'etude semble \^etre n\'ecessaire pour la construction finale du mod\`ele. Nous pensons que des th\'eor\`emes g\'en\'eralisant celles du type "tree-matrix" (voir \cite{A}) pourront \^etre \'etablis.

La dualit\'e de Langmann-Szabo se traduit au niveau des polyn\^omes par une sym\'etrie $\Omega\to \Omega^{-1}$. Cette propri\'et\'e sera exploit\'ee pour \'etablir la liste compl\`ete des termes des polyn\^omes.

\newpage
\thispagestyle{empty}

\chapter{R\'egularisation dimensionnelle}

Dans le chapitre pr\'ec\'edent nous avons obtenu la repr\'esentation param\'etrique du mod\`ele $\Phi^{\star 4}_4$. Dans ce chapitre nous allons appliquer cette repr\'esentation afin d'achever sa r\'egularisation et la renormalisation dimensionnelle (RRD).

Une raison pour s'int\'eresser \`a la r\'egularisation dimensionnelle est qu'elle est le seul sch\'ema d'extraction des singularit\'es dans la th\'eorie des champs qui respecte les invariances de jauge. Il est vrai que l'on ne poss\`ede pas pour l'instant une th\'eorie de jauge non commutative vulcanis\'ee bien \'etablie; n\'eanmoins, une fois une telle th\'eorie obtenue, il sera naturel de la renormaliser \`a l'aide des techniques introduites dans ce chapitre.

Il y a une deuxi\`eme raison d'examiner la RRD. La localit\'e des contre-termes dans les th\'eories commutatives r\'epose sur des propri\'et\'es de factorisation des polyn\^omes de Symanzik. Ces propri\'et\'es sont faciles \`a prouver \`a partir de la forme explicite des polyn\^omes. La moyalit\'e, dans notre cas, va reposer sur le m\^eme type de propri\'et\'es. Cette fois-ci la factorisation est bien plus difficile \`a prouver. Nous devons nous appuyer sur des manipulations nontriviales au niveau des variables de Grassmann.

Une fois la factorisation des amplitudes \'etablie il est possible de r\'eformuler la RRD dans le langage des alg\`ebres de Connes-Kreimer. L'approche alg\'ebrique \`a la renormalisation peut ensuite \^etre exploit\'ee pour formuler et r\'esoudre des nouvelles classes de probl\`emes de Hilbert.

Pour effectuer la RRD nous devons combiner les techniques classiques du cas commutatif avec les r\'esultats nouveaux charact\'eristiques du monde non commutatif. La principale diff\'erence par rapport au cas commutatif vient du fait que nous avons une nouvelle cat\'egorie de graphes primitivement divergents. La preuve de la m\'eromorphie et l'extraction des p\^oles, par contre, est formellement identique dans les cas commutatif et non commutatif.

Comme d\'ej\`a mentionn\'e auparavant, nous avons besoin d'une am\'elioration par rapport aux r\'esultats obtenus dans le chapitre pr\'ec\'edent. En effet, l'am\'elioration du comptage des puissances d'un graphe dans le nombre de faces bris\'ees n'est apparente qu'en utilisant le deuxi\`eme polyn\^ome. Pour un sous-graphe d'un grand graphe, par contre, nous avons une am\'elioration d\'ej\`a au niveau du premier polyn\^ome. Pour la mettre en \'evidence l'on doit utiliser des termes dominants suppl\'ementaires de $G$, avec un scaling global en $t_G^{F(G)+2g(G)-1}$, mais avec un scaling dans les variables de $S$ en $t_S^{F(S)-2}$ si $B(S)\ge 2$. L'existence de tels termes sera prouv\'ee par la suite.

L'extraction des singularit\'es dans la RRD se fait de la mani\`ere suivante: Nous introduisons des secteurs de Hepp. Les int\'egrales sur les param\`etres de Hepp des sous-graphes $S$ du graphe $G$ qui ne sont pas primitivement divergents convergent. Pour les $S$ primitivement divergents nous obtenons des singularit\'es explicites que nous extrayons.

Pour prouver techniquement que les int\'egrales correspondantes aux sousgraphes planaires avec plus d'une face bris\'ee sont convergentes nous avons besoin de nous appuyer sur certains termes dominants du premi\`ere polyn\^ome. Ces termes existent, mais ils ne sont pas parmi ceux explicit\'es dans le chapitre pr\'ec\'edent.

\section{Les termes dominants suppl\'ementaires}
\label{sec:dominantssup}

Les nouveaux termes sont associ\'es \`a des ensembles $I$ et $J$ particuliers. Pour les sp\'ecifier nous devons introduire quelques d\'efinitions.

Appelons "ligne de genre" une ligne qui fait partie d'un croisement naturel.
Soit $S$ un sous-graphe planaire de $G$ avec plus d'une face bris\'ee. Nous appelons $\tilde \ell$ la ligne de $G$ qui brise la face de $S$.

Si l'on r\'eduit $S$ \`a une rosette il existe une ligne de boucle $\ell_2\in S$ qui soit croise $\tilde \ell$, soit la survole. Cette ligne $\ell_2$ s\'epare deux faces bris\'ees de $S$.

\begin{definition}
Soit $J_0$ un sous-ensemble de lignes internes du graphe $G$. $J_0$ est appel\'e
{\it pseudo-admissible} si:
\begin{itemize}
\item Son compl\'ement est l'union d'un arbre $T(G)$ de $G$ 
et $\ell_2$,
\item Ni $\tilde \ell$ ni $\ell_2$ n'appartienent pas \`a $T(G)$,
\end{itemize}
\end{definition}

Soit $I_0=\{\ell_1 \hdots \ell_L\} - \tilde \ell \equiv I- \tilde
\ell$. Nous avons $|I_0|=L(G)-1$ et $|J_0|=L(G)-N(G)=F(G)-2+2g(G)$.
\begin{theorem}
Le pfaffien associ\'e \`a $I_0$ et $J_0$ est
\beqa
n_{I_0, J_0}^2&=&4, \mbox{ si $\tilde \ell$ est une ligne de genre de $G$}\nonumber\\
&=& 16, \mbox{ si $\tilde \ell$ n'est pas une ligne de genre de $G$.}\nonumber
\eeqa
\end{theorem}

Nous utilisons le lemme pr\'ec\'edent de la fa\c con suivante: Choisissons pour $T(G)$ un arbre, qui restreint \`a $S$ est sous-arbre \footnote{L'existence de tels arbres est triviale. Nous commen\c cons par choisir un arbre en $S$ qui ne contient pas $l_2$ (ce qui est toujours possible car $S$ est une particule irr\'eductible) et ensuite nous le compl\'etons en un arbre de $G$ ne contenant pas $\tilde \ell$ (ce qui est de nouveau possible car $G$ est \'egalement une particule irr\'eductible).}. Un terme correspondant \`a un pfaffien du haut a un pr\'efacteur de la forme:
\bea
  t_{\tilde \ell}\prod_{\ell \notin T(G)\text{ et }\ell \neq \ell_2} t_{\ell} \, .
\eea

Le degr\'e global dans les variables $t_G$ est $F(G)+2g(G)-1$, donc il s'agit bien d'un terme dominant pour le graphe $G$.

Le degr\'e dans les variables $t_S$ est $L(S)-(N(S)-1)-1= L(S)-N(S)=F(S)-2$, car $S$ est planaire. Par cons\'equent, sous rescaling de tous les $t_S$ par $x_i^2$ le degr\'e minimal de $x_i$ est au plus $x_i^{2[L(S)-N(S)]}$.

\section{Factorisation}

Dans la repr\'esentation param\'etrique la localit\'e des contre-termes dans la th\'eorie commutative se traduit par les propri\'et\'es de factorisation des polyn\^omes. Au niveau du premier polyn\^ome, pr\'esent\'e dans l'\'equation (\ref{eq:Mainsymanzik1}):
\be\label{eq:Mainsymanzik1rapel}
U_G = \sum_T \prod_{l \not \in T} \alpha_l \ ,
\ee
cela se traduit par la propri\'et\'e suivante:
Soit $S$ un sous-graphe primitivement divergent de $G$ (un sous-graphe \`a deux ou \`a quatre pattes externes) et $G/S$ le graphe $G$ o\`u le sous-graphe $S$ a \'et\'e r\'eduit \`a un point. Le polyn\^ome des $G$ se s\'epare dans la somme des deux contributions suivantes:
\bea
U_G=\sum_{T, T\vert_{S} \text{ arbre en }S} \prod_{l \not \in T} \alpha_l+
    \sum_{T, T\vert_{S} \text{ n'est pas un arbre en }S} \prod_{l \not \in T} 
    \alpha_l \, ,
\eea
Sous un rescaling de tous les param\`etres $\alpha_S\mapsto \rho^2 \alpha_S$ l'ordre dominant en $\rho$ du polyn\^ome $U_G$ est donn\'e par la premi\`ere contribution du haut, $U_G^1$:
\bea
U_G^{1}&=&\rho^{2[L(S)-N(S)+1]}\sum_{T_1\text{ arbre en }S}\prod_{l \not \in T_1} \alpha_l \sum_{T_2, T_2 \text{ arbre en }G/S} \prod_{l \not \in T_2} \alpha_l \nonumber\\
&=&\rho^{2[L(S)-N(S)+1]} U_S U_{G/S} \, .
\eea

Nous allons prouver une propri\'et\'e similaire pour les polyn\^omes des TCNC. Comme la forme de ces polyn\^omes est beaucoup plus complexe que la forme des polyn\^omes de Simanzik, la preuve de cette factorisation est tr\`es technique. En effet, pour prouver la factorisation nous devons retourner \`a l'expression des polyn\^omes comme d\'eterminants et ensuite prouver que ces d\'eterminants se factorisent comme des produits des deux autres d\'eterminants.

\begin{figure}[ht]
\centerline{\epsfig{figure=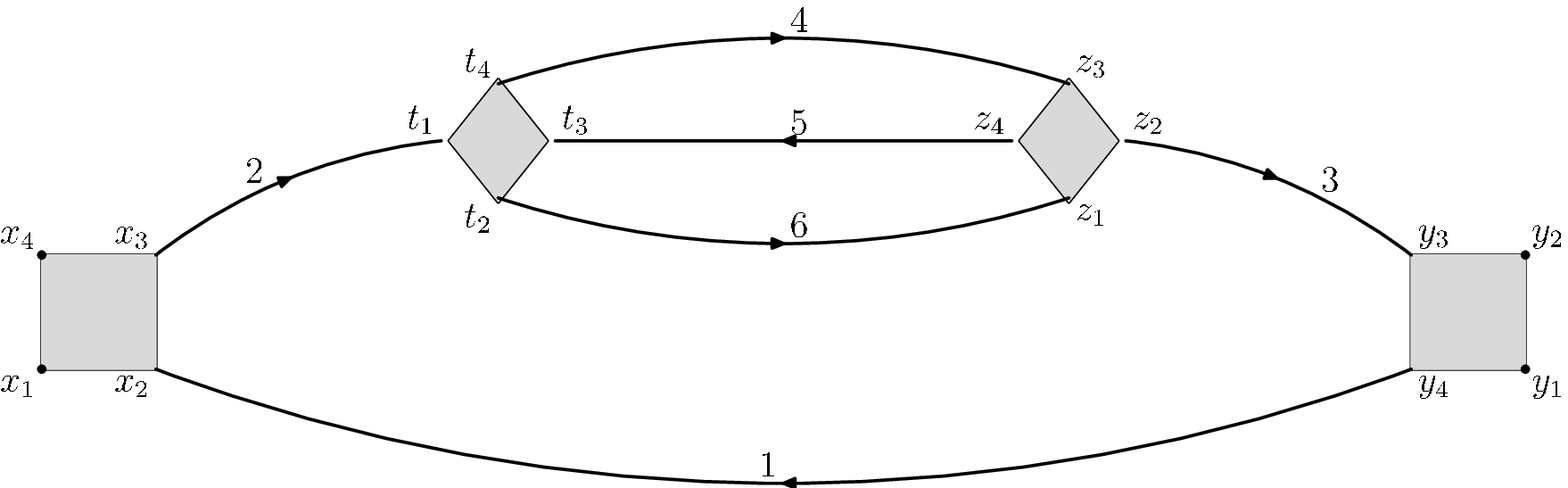,width=12cm}}
\caption{Un graphe avec un sous-graphe primitivement divergent ($\ell_4,\ell_5$ $\ell_6$)}
\label{fig:Mainhyper}
\end{figure}

\begin{figure}[ht]
\centerline{\epsfig{figure=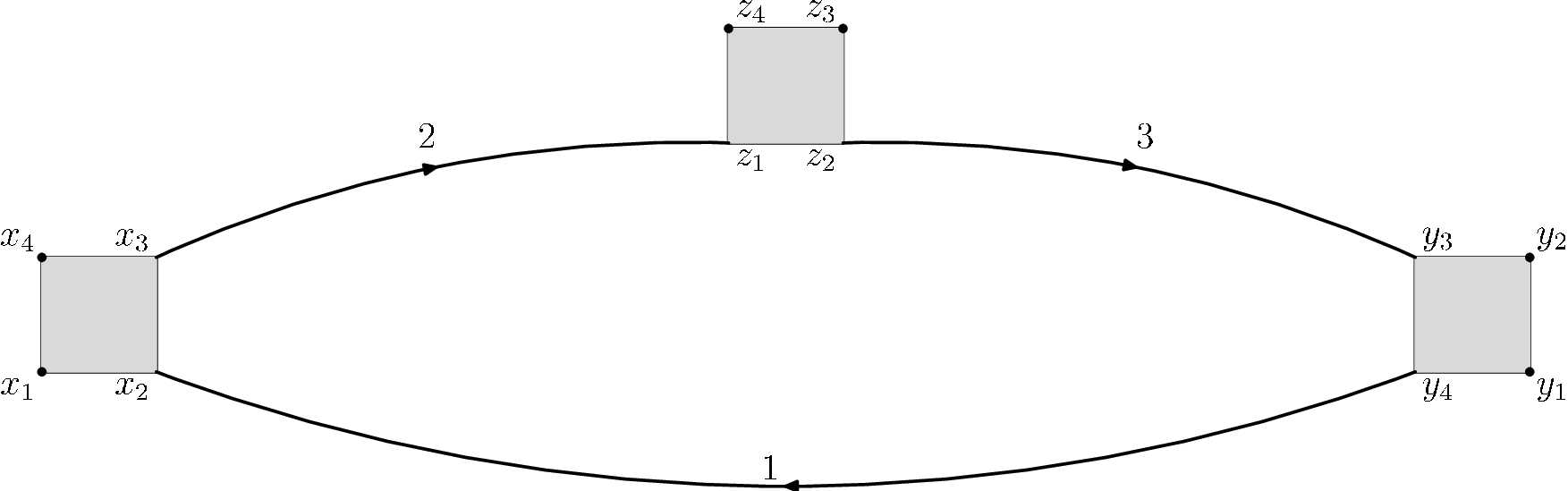,width=12cm}}
\caption{Le graphe $G/S$}
\label{fig:Mainbula-deformata}
\end{figure}

Soit $S$ un sous-graphe primitivement divergent de $G$. $S$ est donc planaire avec une seule face bris\'ee et \`a deux ou quatre pattes externes. $G/S$ est le graphe $G$ o\`u $S$ a \'et\'e r\'eduit \`a un vertex de Moyal (voir figures \ref{fig:Mainhyper}, et \ref{fig:Mainbula-deformata}).

Nous notons par $l(\rho)$ les termes dominants en $\rho$ dans la zone ultraviolette. Dans l'appendice \ref{chap:dimreg} nous prouvons les deux affirmations suivantes:
\begin{theorem}
Sous le rescaling
\beqa
t_{\alpha}\mapsto \rho^2 t_{\alpha}
\eeqa
des param\`etres correspondant \`a un sous-graphe divergent $S$ d'un graphe $G$, le premier polyn\^ome factorise comme:
\beqa
HU^{l(\rho)}_{G,\bar V}=HU^{l(\rho)}_{S,\bar V_S} HU_{G/S,\bar V} \; .
\eeqa
\end{theorem}
Pour la partie exponentielle on a le r\'esultat:
\begin{proposition}
   Sous le rescaling $t_{\alpha}\mapsto \rho^2 t_{\alpha}$ des param\`etres du 
   sous-graphe $S$ nous avons:
   \beqa
    \frac{HV_G}{HU_G}\Big{\vert}_{\rho =0}=\frac{HV_{G/S}}{HU_{G/S}}\, .
   \eeqa
\end{proposition}

Mis ensemble ces deux r\'esultats donnent la factorisation des int\'egrants des amplitudes:
\beqa\label{eq:Mainfactfinal}
\frac{e^{-\frac{HV_G(\rho)}{HU_G(\rho)}}}{HU_G(\rho)^{D/2}}=
\frac{1}{[HU^{l(\rho)}_{S}]^{D/2}}(1+\rho^2{\cal O}_S) \frac{e^{-\frac{HV_{G/S}}{HU_{G/S}}}}{HU_{G/S}^{D/2}} \, .
\eeqa
Pour la fonction \`a deux points nous prouvons en plus que l'op\'erateur ${\cal O}_S$ correspond \`a une renormalisation de masse et de fonction d'onde.

\section{M\'eromorphie}

Nous pr\'esentons ci-dessous en d\'etail la preuve de la m\'eromorphie des amplitudes de Feynman en TCNC. La proc\'edure de soustraction est similaire \`a celle des th\'eories commutatives et nous n'insistons pas sur ces d\'etails ici.

L'amplitude d'un graphe de Feynman s'\'ecrit: 
\beqa
\label{eq:MainHUGV-t}
{\cal A}_{G,{\bar V}}  (x_e,\;  p_{\bar V}, D) = \left(\frac{\tom}{2^{\frac
    D2 -1}}\right)^L  \int_{0}^{1} \prod_{\ell=1}^L  
 dt_\ell (1-t_\ell^2)^{\frac D2 -1} 
\frac{
e^{-  \frac {HV_{G, \bar{V}} ( t_\ell , x_e , p_{\bar v})}{HU_{G, \bar{v}} ( t )}}
}
{HU_{G, \bar{V}} ( t )^{D/2}} \, .
\eeqa
Nous analysons seulement les graphes connexes avec au moins deux pattes externes.
Nous prolongeons cette expression dans le plan complexe. On d\'efinit un secteur de Hepp $\sigma$ comme:
\beqa
\label{eq:Mainhepp-nc}
0\le t_1 \le \ldots \le t_L \, ,
\eeqa
et nous effectuons le changement de variables adapt\'e:
\beqa
\label{eq:Mainchange-nc}
t_\ell=\prod_{j=\ell}^L x_j^2,\ \ell =1,\ldots, L.
\eeqa
Soit $G_i$ le sous-graphe compos\'e par les lignes $t_1$ \`a $t_i$.
On note $L(G_i)=i$ le nombre des lignes de $G_i$, $g(G_i)$ son genre, $F(G_i)$ son nombre de faces, et ainsi de suite. L'amplitude est:
\beqa
{\cal A}_{G,\bar V}=&&\Big{(}\frac{\tilde \Omega}{2^{(D-4)/2}}\Big{)}^L
\int_{0}^{1}\prod_{i=1}^L 
\left(1-(\prod_{j=i}^L x_j^2)^2\right)^{\frac D2 -1}
dx_{i} 
\nonumber\\
&&\prod_{i=1}^{L}x_{i}^{2L(G_i)-1}
\frac{e^{-\frac{HV_{G,\bar V}(x^2)}{HU_{G,\bar V}(x^2)}}}{HU_{G,\bar V}(x^2)}\, .
\eeqa 
Dans l'\'equation ci-dessus nous factorisons dans $HU_{G,\bar V}$ le mon\^ome avec le plus petit degr\'e dans chaque variable $x_i$:
\beqa
\label{eq:Mainampli-x}
{\cal A}_{G,{\bar V}}  (x_e,\;  p_{\bar v}) = &&\left(\frac{\tom}{2^\frac
    D2}\right)^L  \int_{0}^{1} \prod_{\ell=1}^L  
 dx_\ell \left(1-(\prod_{j=\ell}^L x_j^2)^2\right)^{\frac D2 -1} 
\nonumber\\ 
&& x_i^{2L(G_i)-1-D b'(G_i)}
 \frac{e^{-\frac {HV_{G, \bar V}}{HU_{G,\bar V}}}}{(a s^b+ F (x^2))^\frac D2}.
\eeqa
Le dernier terme ci-dessus est toujours born\'e par une constante. Des divergences ne peuvent appara\^itre que dans la r\'egion $x_i$ proche de $0$. Nous retrouvons, comme auparavant, que cette th\'eorie n'a pas de divergences infrarouges m\^eme \`a masse z\'ero.
Le nombre $b'(G_i)$ est fix\'e par la topologie du graphe $G_i$. Il vaut:
\beqa
b'(G_i)=\begin{cases}
    {\displaystyle \le L(G_i)-[n(G_i)-1]-2g(G_i)} &\text{si } 
      g(G_i)>0
     \vspace{.3cm}\\
     {\displaystyle \le L(G_i)-n(G_i)} &\text{si }
      g(G_i)=0 \text{ et } B(G_i)>1
     \vspace{.3cm}\\
     {\displaystyle =L(G_i)-[n(G_i)-1]} &\text{si }
       g(G_i)=0 \text{ et } B(G_i)=1 \\
  \end{cases} .\nonumber
\eeqa
Pour prouver la premi\`ere et la derni\`ere ligne nous devons analyser des termes dominants du $HU_G$ avec $I=\{1 \hdots L \}$ et $J$ admissible en $G$, tel que l'arbre $T(G)$, inclut dans l'ensemble compl\'ementaire de $J$, est sous-arbre en $G_i$.

Pour la deuxi\`eme ligne nous devons choisir $I=\{1 \hdots L \}-\tilde\ell$ 
et $J$ {\it pseudo-admissible}, comme dans la section \ref{sec:dominantssup}.

Nous pouvons conclure que $b'(G_i)$ est au plus $L(G_i)-n(G_i)+1$ et qu'il prend cette valeur si et seulement si  $g(G_i)=0$ et $B(G_i)=1$. Par cons\'equent,
la convergence dans le r\'egime UV ($x_i\to 0$) est assur\'ee si:
\beqa
 \Re [ 2L(G_i)-Db'(G_i) ]>0 ,\ i=1\ldots L \, .
\eeqa
Comme
\beqa
\Re[2L(G_i)-Db'(G_i)]>\Re\Big{(}2L(G_i)-D[L(G_i)-n(G_i)+1]\Big{)} \, ,
\eeqa
on a toujours convergence pourvu que:
\beqa
\Re D<2\le \frac{4n(G_i)-N(G_i)}{n(G_i)-N(G_i)/2+1} \le \frac{2L(G_i)}{L(G_i)-n(G_i)+1} \, .
\eeqa
avec $N_e(G_i)$ le nombre des points externes de $G_i$
\footnote{On utilise ici la relation topologique $4N(G_i)-N_e(G_i)=2L(G_i)$}.
Donc ${\cal A}_{G, \bar V}(D)$ est analytique dans la bande
\beqa
\label{eq:Maindomain-nc}
{\cal D}^\sigma=\{ D\, | \, 0 < \Re\, D <  2 \}.
\eeqa
On prolonge maintenant ${\cal A}$ comme fonction de $D$ pour $2 \le \Re D \le 4$. Pourvu que:
\begin{itemize}
\item $g(G_i)>0$
\item $g(G_i)=0$ and $B(G_i)>1$
\item $N(G_i)>4$ \, ,
\end{itemize}
la bande d'analyticit\'e est imm\'ediatement prolong\'ee \`a:
\beqa
{\cal D}^\sigma=\{ D\, | \, 0 < \Re\, D <  4+\e_G \}.
\eeqa
pour $\e_G$ un petit nombre positif d\'ependent du graphe.

En effet, dans les premiers deux cas on a $b'(G_i)\le L(G_i)-n(G_i)$ donc l'int\'egrale sur $x_i$ converge pour:
\beqa
\Re D\le 4 < \frac{4n(G_i)-N(G_i)}{n(G_i)-N(G_i)/2}=\frac{2L(G_i)}{L(G_i)-n(G_i)}
\, ,
\eeqa
et dans le troisi\`eme cas, comme $N(G_i)>4$, l'int\'egrale sur $x_i$ converge pour:
\beqa
\Re D\le 4 < \frac{4n(G_i)-N(G_i)}{n(G_i)-N(G_i)/2+1}=
\frac{2L(G_i)}{L(G_i)-n(G_i)+1} \, .
\eeqa

Seuls les sous-graphes planaires avec une seule face bris\'ee avec deux ou quatre points pr\'esentent des divergences UV dans la bande $2, 4+\e$. Ils sont appel\'es primitivement divergents.
Soit $S$ un sous-graphe primitivement divergent. On note son param\`etre de Hepp $\rho$. 
En utilisant l'\'equation (\ref{eq:Mainfactfinal}), sa contribution \`a l'amplitude est: 
\beqa
\label{eq:Mainpolamp}
{\cal A}^{\rho}_{G,\bar V_G}\sim
\int_{0}d\rho \rho^{2L(S)-1-D[L(S)-n(S)+1]}
\frac{1}{HU^l_{S,\bar V_{S}}|_{\rho=1}}(1+\rho^2 {\cal O}_S)
\frac{e^{-\frac{HV_{G/S}}{HU_{G/S}}}}{HU_{G/S}}\, . \nonumber
\eeqa

L'int\'egrale sur $\rho$ est un op\'erateur m\'eromorphique en $D$ avec la partie divergente donn\'ee par:
\beqa
\frac{r_1}{2L(S)-D[L(S)-n(S)+1]}+
     \frac{r_2}{2L(S)-D[L(S)-n(S)+1]+2}{\cal O}_S \, . \nonumber
\eeqa
Nous avons un p\^ole isol\'e \`a $D=4$ si $S$ a quatre points. Si $S$ a deux points, nous avons deux p\^oles, l'un \`a $D=4-2/N(S)$ et l'autre \`a  $D=4$.
Comme toutes les singularit\'es sont de ce type, nous avons prouve que ${\cal A}_G$
est m\'eromorphe dans la bande:
\beqa
{\cal D}^\sigma=\{ D\, | \, 0 < \Re\, D < 4+\e_G \}.
\eeqa

\qed

Dans notre \'etude nous avons prouv\'e la factorisation des amplitudes de Feynman des TCNC. Nous avons r\'eussi \`a prouver que les amplitudes sont des fonctions holomorphes dans le plan complexe (la g\'en\'eralisation de notre preuve pour la bande $0,4$ pour tout le plan complexe \'etant triviale). Une fois la structure des singularit\'es \'etablie nous avons pu les extraire \`a l'aide des op\'erateurs de Taylor (voir appendice \ref{chap:dimreg}). 

Pour le futur, une voie de recherche sera de reprendre les m\^emes proc\'edures pour une th\'eorie de jauge vulcanis\'ee. Comme nous ne disposons pas pour l'instant d'une telle th\'eorie, nous ne pouvons pas d\'emarrer un tel programme pour l'instant.

Une autre voie de recherche qui devrait aboutir \`a une r\'eponse est de trouver lequel des traitements que nous avons effectu\'es peut se g\'en\'eraliser pour d'autres vari\'et\'es non commutatives.

Une autre question tr\`es int\'eressante est d'essayer d'\'etablir un nombre de conditions n\'ecessaires et suffisantes pour que les amplitudes de Feynman se factorisent. Si nous disposons d'une telle caract\'erisation, nous pourrons nous prononcer sur la question de la renormalisabilit\'e pour des th\'eories des champs plus g\'en\'eriques.

\newpage
\thispagestyle{empty}

\chapter{Le repr\'esentation de Mellin compl\`ete}

En s'appuyant sur la repr\'esentation param\'etrique des TCNC introduite dans \cite{gurauhypersyman,RivTan}, nous allons maintenant introduire la repr\'esentation de Mellin Compl\`ete (CM) du mod\`ele $\Phi^{\star 4}_4$. Cette repr\'esentation des amplitudes de Feynman est la plus compacte connue. De plus, elle est un nouveau point de d\'epart pour la renormalisation et pour l'\'etude des comportements asymptotiques des amplitudes sous rescaling arbitraire des invariants externes.

Dans les th\'eories commutatives le probl\`eme du d\'eveloppement asymptotique est formul\'e de la mani\`ere suivante. Etant donn\'ee une amplitude de Feynman nous rescalons l'un de ses invariants externes par un param\`etre $\alpha$. Nous cherchons \`a exprimer l'amplitude comme s\'erie asymptotique de puissances et puissances du logarithme de $\alpha$. La repr\'esentation de Mellin nous fournit ce d\'eveloppement dans certains cas.

Une propri\'et\'e des polyn\^omes de la repr\'esentation param\'etrique requise pour pouvoir prouver l'existence d'un d\'eveloppement asymptotique est la propri\'et\'e FINE. Un polyn\^ome est appel\'e FINE s'il se factorise dans tout secteur de Hepp. Pour les polyn\^omes FINE la transformation de Mellin inverse, qui nous donne l'amplitude en fonction de $\alpha$, a une structure m\'eromorphe et ses r\'esidus nous donne l'expansion asymptotique.

Si un polyn\^ome n'est pas FINE, nous pouvons le d\'ecomposer dans plusieurs morceaux, chacun avec la propri\'et\'e FINE. Nous pouvons ensuite associer un param\`etre de Mellin \`a tout sous-polyn\^ome FINE. Si nous poussons ce raisonnement \`a ses limites, nous pouvons associer un param\`etre de Mellin \`a tout mon\^ome. Cette repr\'esentation est appel\'ee "Mellin Compl\`ete" (CM).

De plus, dans le cadre de cette repr\'esentation, nous pouvons int\'egrer les param\`etres de Schwinger et nous pouvons reformuler la renormalisation comme l'\'etude des d\'eplacements des pieds des contours d'int\'egration des variables de Mellin dans des cellules du plan complexe. Cela nous permet de traiter au m\^eme niveau les amplitudes divergentes et convergentes dans l'ultraviolet.

A l'aide de la repr\'esentation CM nous prouvons le r\'esultat suivant.
Soit une amplitude de Feynman $G(s_{k})$ \'ecrite en fonction des ses invariants $s_{k}$ (on inclut aussi les masses des particules). Soit un r\'egime asymptotique d\'efini par:
\begin{equation}
s_{k}  \rightarrow \alpha ^{a_k} {s}_{k},  
\end{equation}
avec $a_k$ positif, n\'egatif, ou nul.

Nous laissons $\alpha $ tendre vers l'infini. $G(s_{k})$ a un d\'eveloppement asymptotique comme fonction de $\alpha$ du type:
\begin{equation}
G(\alpha;s_{k} )=\sum_{p=p_{\text{max}}}^{-\infty }\sum_{q=0}^{q_{\text{max}%
}(p)}G_{pq}(s_k)\alpha ^p\ln ^q\alpha ,
\end{equation}
o\`u $p$ est un nombre rationnel ($p_{max}$ son maximum) et $q$, \`a $p$ donn\'e, est un nombre naturel.

Notre but final est de retrouver un tel r\'esultat pour TCNC. Nous allons nous concentrer par la suite sur le d\'ebut, soit l'introduction de la CM pour TCNC. Nous trouvons une structure distributionnelle plus compliqu\'ee que dans le cas commutatif, mais nous obtenons un premier r\'esultat, celui de la m\'eromorphie des amplitudes de Feynman.

\section{La repr\'esentation CM dans la TCNC} 

Pour toute fonction $f(u)$, d\'erivable par morceaux pour $u>0$, si l'int\'egrale 
\be 
g(x)=\int_0^\infty du\,u^{-x-1}f(u) 
\ee
est absolument convergente pour $\alpha < \Re x<\beta $, alors pour $\alpha <\sigma <\beta $ 
\be
f(u)=\frac 1{2\pi i}\int_{\sigma -i\infty }^{\sigma +i\infty }dx\,g(x)\,u^x .
\ee
La fonction $g$ s'appelle la transform\'ee de Mellin de $f$.

Pour $f(u) = e^{-u}$, et $u=HV_k/HU$, la transform\'ee de Mellin inverse est:
\be
e^{-HV_k/HU}=\int_{\tau _k}\Gamma (-y_k)\left( \frac{HV_k}{HU}\right) ^{y_k}, 
\label{main:gamayk} 
\ee
o\`u $\int_{\tau _k}$ est une notation abr\'eg\'ee pour $\int_{-\infty }^{+\infty }
\frac{d(\Im y_k)}{2\pi }$, et  $\Re y_k$ est fix\'e \`a $\tau _k<0$.

Pour tout $u$ positif, en choissant $x$ avec $\Re x <0$, tel que $0<\Re x+u$, nous avons l'\'egalit\'e suivante:
\be
\Gamma (u)\left( A+B\right) ^{-u}=\int_{-\infty }^\infty \frac{d(\text{Im }x)%
}{2\pi }\Gamma (-x)A^x\Gamma (x+u)B^{-x-u} . \label{main:gamauamaisb} 
\ee

Nous prenons $A\equiv HU_1$ et $B\equiv HU_2+HU_3+\cdots $ et nous utilisons it\'erativement l'\'egalit\'e ci-dessus pour $u=\sum_k y_k+D/2$, pour obtenir:
\be
\Gamma \left( \sum_k y_k+\frac D2\right) HU^{-\sum_k y_k-\frac D2}=\int_\sigma 
\prod_j\Gamma (-x_j)HU_j^{x_j},  \label{main:gamayk2} 
\ee
avec $\Re x_j=\sigma _j<0$, $\Re \left( \sum_k y_k +\frac D2\right) =\sum_k\tau 
_k+\frac D2>0$, et $\int_\sigma $ veut dire $\int_{-\infty }^{+\infty }\prod_j
\frac{d(\Im x_j)}{2\pi }$ avec $\sum_jx_j+\sum_k y_k =-\frac D2$.

Pour utiliser les deux transformations pr\'esent\'ees ci-dessus nous devons s\'eparer les polyn\^omes hyperboliques dans des sommes des mon\^omes. Pour cela, nous commen\c cons par l'expression du premier polyn\^ome:
\beqa
HU_{\cG,{\bar V}} (t) &=&   \sum_{{K_U}= I\cup J, \  n + |{K_U}|\; {\rm impaire}}  s^{2g-k_{{K_U}}} \ n_{{K_U}}^2
\prod_{\ell \not\in I} t_\ell \prod_{\ell' \in J} t_{\ell'}\ 
\nonumber \\
&=&\sum_{{K_U}}a_{K_U} \prod_\ell t_\ell^{u_{\ell {K_U}}}\equiv 
\sum_{K_U} HU_{K_U} .
\eeqa
avec:
\begin{itemize}

\item $I$ sous-ensemble des premiers $L$ indices avec son cardinal $\vert I \vert $,
 $J$ sous-ensemble des  $L$ indices suivants de cardinal $\vert J\vert $,

\item $B$ est la matrice antisym\'etrique restante,

\item $n_{{K_U}}=\mathrm{Pf}(B_{\widehat{{K_U}}})$ le pfaffien de $B$ avec 
les index de ${K_U}= I \cup J$ omis,

\item $k_{{K_U}}$ est $ \vert {K_U}\vert - L(G) - F(G) +1$,

\item $ a_{K_U} = s^{2g-k_{{K_U}}}n_{{K_U}}^2$, $s=\Omega^{-1}$,
\begin{equation}
u_{\ell {K_U}}=\left\{  
\begin{array}{ll} 
0 & \mbox{ si } \ell \in I \mbox{ et } \ell\notin J \\  
1 & \mbox{ si } (\ell \notin I \mbox{ et } \ell\notin J) \mbox{ ou } (\ell \in I \mbox{ et } \ell\in J)\\
2 &  \mbox{ si } \ell \notin I \mbox{ et } \ell\in J
\end{array} 
\right. .
\end{equation}

\end{itemize}
Le second polyn\^ome $HV$ se d\'ecompose en partie r\'eelle $HV^\cR$ et imaginaire $HV^\cI$.
En plus des ensembles $I$ et $J$ du haut, nous introduisons une ligne particuli\`ere 
$\tau \notin I$, analogue  d'une coupure d'un arbre en deux.
D\'efinissons $\mathrm{Pf}(B_{\hat{{K_V}}\hat{\tau}})$ le pfaffien de la matrice 
obtenue de $B$ en effa\c cant les lignes et les colonnes dans $I$, $J$ et $\tau$. De plus, nous d\'efinissons $\epsilon_{I,\tau}$ la signature de la permutation:
\beqa
1,\dotsc ,d\rightarrow 1,\dotsc,\hat{i_1},\dotsc ,\hat{i_{\vert I\vert }},\dotsc
,\hat{i_{\tau}},\dotsc ,d, i_{\tau},i_{\vert I\vert }\dotsc, i_1\,.
\eeqa
avec $d=2L(G)+N(G)-1$. Alors:
\beqa
HV^\cR_{\cG,{\bar V}}&=&
 \sum_{{K_V}= I \cup J }\prod_{\ell \notin I} t_\ell
 \prod_{\ell' \in J} t_{\ell'}
\Big{[}\sum_{e_1}x_{e_1}\sum_{\tau\notin {K_V}}P_{e_1\tau}\epsilon_{{K_V}\tau}
\mathrm{Pf}(B_{\hat{{K_V}}\hat{\tau}})\Big{]}^2 \, .
\nonumber\\
&=&
\sum_{K_V} s^\cR_{K_V}\left( \prod_{\ell=1}^L t_\ell^{v_{\ell {K_V}}}\right) \equiv 
\sum_{K_V} HV_{K_V}^\cR
\eeqa
avec
\beqa s_{K_V}^\cR= \left(
\sum_e x_e \sum_{\tau\notin {K_V}} P_{e\tau} \e_{{K_V} \tau} 
\mathrm{Pf}(B_{\hat{{K_V}}\hat{\tau}})
\right)^2 \eeqa
et $v_{\ell {K_V}}$ sont donn\'es par la m\^eme formule que $u_{\ell K_U}$. 

La partie imaginaire fait intervenir de paires des lignes $\tau, \tau'$ et une signature correspondante (d\'etaill\'ee en \cite{gurauhypersyman}):
\beqa 
HV_{\cG,\bar V}^\cI&=&\sum_{{K_V}= I \cup J }\prod_{\ell \notin I} t_\ell
 \prod_{\ell' \in J} t_{\ell'}\nonumber \\
&&
\epsilon_{K_V}\mathrm{Pf}(B_{\hat{{K_V}}})
\Big{[}\sum_{e_1,e_2} \Big{(}
\sum_{\tau\tau'}P_{e_1\tau}\epsilon_{{K_V}\tau\tau'}\mathrm{Pf}(B_{\hat{{K_V}}\hat{\tau}\hat{\tau'}})
P_{e_2\tau'}\Big{)} x_{e_1}\wedge x_{e_2}\Big{]}\,.\nonumber
\\
&=&
\sum_{K_V} s^\cI_{K_V}\left( \prod_{\ell=1}^L t_\ell^{v_{\ell {K_V}}}\right) \equiv 
\sum_{K_V} HV_{K_V}^\cI
\eeqa
avec:
\beqa s_{K_V}^\cI=\epsilon_{K_V} \mathrm{Pf}(B_{\hat{{K_V}}})
\left(
\sum_{e, e'} (\sum_{\tau, \tau'}   P_{e \tau} \e_{{K_V} \tau \tau'} 
\mathrm{Pf}(B_{\hat{{K_V}}\hat{\tau}\hat{\tau'}}) P_{e'\tau'})x_e \wedge x_{e'}
\right) . 
\eeqa

Les principales diff\'erences entre la TCNC et la TCC sont:
\begin{itemize}
\item la pr\'esence des constantes $a_j$ dans $HU$,
\item la partie imaginaire $i\, HV^\cI$ de $HV$,
\item Le fait que $u_{\ell j}$ et $v_{\ell k}$ peuvent prendre la valeur $2$ (pas seulement $0$ et $1$).

\end{itemize}

Pour la partie r\'eelle $HV^\cR$ de $HV$ nous utilisons l'identit\'e:
\begin{equation}
e^{-HV^\cR_{K_V}/HU}=\int_{\tau_{K_V}^\cR}\Gamma (-y_{K_V}^\cR)\left( \frac{HV^\cR_{K_V}}{HU}\right) ^{y_{K_V}^\cR}, 
\end{equation} 
qui introduit l'ensemble des param\`etres de Mellin $y_{K_V}^\cR$.

Pour la partie imaginaire une identit\'e similaire n'est vraie qu'au sens des distributions. Nous avons, pour $HV^\cI_{K_V}/U >0$ et $-1 < \tau_{K_V}^\cI <0$: 
\begin{equation}
e^{-i\, HV^\cI_{K_V}/HU}=\int_{\tau_{K_V}^\cI}\Gamma (-y_{K_V}^\cI)\left( \frac{i\, HV^\cI_{K_V}}{HU}\right) ^{y_{K_V}^\cI}.
\end{equation} 
Nous introduisons ainsi un autre ensemble de param\`etres de Mellin $y_{K_V}^\cI$.
Le caract\`ere distributionnel de cette \'egalit\'e est l'une des principales diff\'erences avec le cas commutatif.

Pour $HU$ nous utilisons la repr\'esentation:
\begin{equation} 
\Gamma \left( \sum_{K_V}y_{K_V}+\frac D2\right) (HU)^{-\sum_{K_V} (y_{K_V}^\cR+ y_{K_V}^\cI) -\frac D2}=\int_\sigma 
\prod_{K_U}\Gamma (-x_{K_U})HU_{K_U}^{x_{K_U}},
\end{equation}
 avec $y_{K_V}=y_{K_V}^\cR + y_{K_V}^\cI$.

Une amplitude g\'en\'erale de Feynman s'\'ecrit comme:
\beqa
{\cal A}_G &=&{\rm K'}\int_\Delta \frac{\prod_{K_U} a_{K_U}^{x_{K_U}} \Gamma (-x_{K_U})}{\Gamma (-\sum_{K_U}x_{K_U})}
\left( \prod_{K_V} (s_{K_V}^\cR)^{y_{K_V}^\cR}\Gamma (-y_{K_V}^\cR) \right)
\nonumber\\
&&\left( \prod_{K_V} (s_{K_V}^\cI)^{y_{K_V}^\cI}\Gamma (-y_{K_V}^\cI) \right)\int_0^1 \prod_{\ell =1}^L dt_\ell (1-t_\ell^2)^{\frac D2 -1}
t_\ell^{\phi_\ell -1}
\eeqa
avec:
\begin{equation} 
\phi_\ell\equiv \sum_{K_U} u_{\ell {K_U}}x_{K_U}+\sum_{K_V}
(v^\cR_{\ell {K_V}}y_{K_V}^\cR+v^\cI_{\ell {K_V}}y_{K_V}^\cI) +1 .
\end{equation} 
Par $\int_\Delta $ nous entendons une int\'egrale sur $\frac{\text{Im }x_{K_U}}{2\pi i}$, $\frac{\text{Im }y_{K_V}^\cR}{2\pi i}$ et
$\frac{\text{Im }y_{K_V}^\cI}{2\pi i}$, o\`u
$\Delta $ est le domaine convexe
\begin{equation} 
\Delta =\left\{ \sigma ,\tau^\cR, \tau^\cI \left|  
\begin{array}{l} 
\sigma _{K_U}<0;\;\tau^\cR_{K_V}<0;\; -1 < \tau^\cI_{K_V}<0;\;
\\
\sum_{K_U}x_{K_U}+\sum_{K_V} (y_{K_V}^\cR+ y_{K_V}^\cI)=-\frac D2; \\  
\forall \ell,\;\text{Re }\phi_\ell\equiv \sum_{K_U}u_{\ell {K_U}}\sigma _{K_U}\\
+\sum_{K_V}
(v^\cR_{\ell {K_V}}\tau^\cR_{K_V}+v^\cI_{\ell {K_V}}\tau^\cI_{K_V})+1>0 
\end{array} 
\right. \right\}
\end{equation} 
et $\sigma $, $\tau^\cR $ et $\tau^\cI $ repr\'esentent $\text{Re }x_{K_U}$, $\text{Re }y_{K_V}^\cR$ et $\text{Re }y_{K_V}^\cI$.

Les int\'egrales $dt_\ell$ peuvent \^etre effectu\'ees en utilisant la repr\'esentation de la fonction b\^eta:
\beqa
\int_0^1 \prod_{\ell =1}^L dt_\ell (1-t_\ell^2)^{\frac D2 -1}
t_\ell^{\phi_\ell -1}=\frac 12 \beta (\frac{\phi_\ell}2, \frac D2).
\eeqa
De plus, nous avons:
\bea
  \beta (\frac{\phi_\ell}2, \frac D2) =\frac {\Gamma (\frac{\phi_\ell}2)
  \Gamma (\frac D2)}{\Gamma (\frac{\phi_\ell+D}2)}.
\eea
Cette repr\'esentation est convergente pour $0<\Re D<2$. Nous obtenons ainsi la repr\'esentation CM d'une amplitude de Feynman dans la TCNC:
\begin{theorem}
Toute amplitude de Feynman d'un graphe de $\phi^{\star 4}$ est analytique au moins dans la bande $0<\Re D<2$ o\`u elle a la repr\'esentation CM suivante:
\beqa 
{\cal A}_\cG &=&{\rm K'}\int_\Delta \frac{\prod_{K_U} a_{K_U}^{x_{K_U}} \Gamma (-x_{K_U})}{\Gamma (-\sum_{K_U}x_{K_U})}
\left( \prod_{K_V} (s_{K_V}^\cR)^{y_{K_V}^\cR}\Gamma (-y_{K_V}^\cR) \right)
\nonumber\\
&& \left( \prod_{K_V} (s_{K_V}^\cI)^{y_{K_V}^\cI}\Gamma (-y_{K_V}^\cI) \right)
\left( \prod_{\ell=1}^L \frac {\Gamma (\frac{\phi_\ell}2)
  \Gamma (\frac D2)}{2\Gamma (\frac{\phi_\ell+D}2)} \right).\nonumber\\
\eeqa
vraie en tant que distribution temp\'er\'ee des invariants externes.
\end{theorem}

A partir de cette formule nous avons obtenu la continuation de la repr\'esentation CM d'une amplitude \`a une distribution temp\'er\'ee d\'ependant m\'eromorphiquement de $D$ dans tout le plan complexe.

Une fois la repr\'esentation de Mellin Compl\`ete introduite et la m\'eromorphie prouv\'ee nous devrons essayer de trouver les g\'en\'eralisations des r\'esultats classiques obtenus \`a l'aide de cette repr\'esentation dans la th\'eorie commutative.
Le premier but est de reformuler la renormalisation dans le nouveau langage. Dans les th\'eories commutatives la renormalisation est \'equivalente au d\'eplacement des pieds des contours d'int\'egration entre diff\'erentes cellules de Mellin dans le plan complexe. La soustraction est \'equivalente avec le passage d'une repr\'esentation de la fonction $\Gamma(p)$ pour $0<p$ \`a une repr\'esentation pour $-1<p<0$. Il faut v\'erifier que ces soustractions sont donn\'ees par des contre-termes de la forme du Lagrangien. En tenant compte des travaux pr\'esent\'es dans le chapitre pr\'ec\'edent nous pensons que ce r\'esultat suivra.

Une fois la renormalisabilit\'e prouv\'ee, le deuxi\`eme but est de prouver l'existence des s\'eries asymptotiques sous un rescaling des invariants de l'amplitude. La combination de notre repr\'esentation de CM et des techniques classiques fournira tr\`es certainement ce r\'esultat.

\chapter{Conclusion}

Le long de cette th\`ese nous avons \'etudi\'e le mod\`ele $\Phi^{\star 4}_4$ de plusieurs points de vue. Nous avons \'etudi\'e le mod\`ele \`a l'aide de diverses techniques introduites le long de nos travaux. Ainsi, nous disposons aujourd'hui pour les TCNC d'une panoplie d'outils d'analyse presque tout aussi d\'evelopp\'ee que pour leurs partenaires commutatives.

Parmi les divers resultats pr\'esent\'es dans nos travaux, probablement le plus excitant est l'annulation de la fonction beta du mod\`ele. Les mod\`eles avec des flots born\'es sont tr\`es coh\'erents du point de vue math\'ematique, notamment nous pouvons esp\'erer les construire non perturbativement. Si cette caract\'eristique s'av\`ere g\'en\'erique pour les TCNC vulcanis\'ees elle fournira \`a elle seule un tr\`es fort argument en leur faveur.

Les identit\'es de Ward que nous avons utilis\'ees pour prouver ce r\'esultat ont un statut tr\`es particulier. Dans les th\'eories de champs commutatives les identit\'es de Ward sont la cons\'equence de l'invariance de jauge. En consid\'erant un mod\`ele invariant de jauge nous devons fixer la jauge. Ensuite, si nous agissons avec une transformation de jauge infinit\'esimale de la m\^eme fa\c con que dans nos travaux nous obtenons les identit\'es de Ward. 

Il semble que notre mod\`ele initial corresponde \`a une action d\'ej\`a fix\'ee de jauge! Nous nous demandons donc si il existe une forme invariante de jauge associ\'ee \`a notre mod\`ele. Une fois une telle invariance comprise nous devons construire une action de Yang Mills associ\'ee. Cette action sera diff\'erente des actions de Yang Mills d\'ej\`a propos\'ees par \cite{GrosseYM,WalletYM}.

Nous concluons que le sens des transformations unitaires qui intervient dans la preuve de l'annulation de la fonction b\^eta devrait \^etre approfondi. Si nous consid\'erons que les op\'erateurs associ\'es aux coordonn\'ees se transforment r\'eellement sous les unitaires nous voyons que ces transformations correspondent aux diff\'eomorphismes pr\'eservant l'aire. En effet, si dans une transformation:
\bea
x'^{\mu}=U^{\dagger}x^{\mu}U
\eea
nous prenons $U=e^{iH}$ avec $H$ une matrice herm\'etienne infinit\'esimale, nous voyons qu'au premier ordre en $H$ le jacobien de la transformation $x\to x'$ est $J(x,x')=1$. Quel est le sens physique de cette transformation? Elle appara\^it naturellement dans l'\'etude des fluides non commutatifs (voir \cite{Polych}), car les fluides sont incompressibles. Nous ne disposons pas pour l'instant d'une interpr\'etation convaincante de ces transformations pour nos mod\`eles.

Une direction de recherche future est la construction des mod\`eles fermioniques non commutatifs renormalisables. Des pas dans cette direction ont \'et\'e faits par Vignes-Tourneret. L'auteur a propos\'e une vulcanisation du mod\`ele de Gross-Neveu engendr\'e par un champ de fond. Le mod\`ele est renormalisable pour tout $\Omega\neq 1$, mais la fonction b\^eta du mod\`ele n'est pas nulle.
Ces travaux semble sugg\'erer que la finitude des flots n'est pas g\'en\'erique dans les th\'eories vulcanis\'ees.

Neanmoins, il existe d'autres fa\c cons de vulcaniser une th\'eorie fermionique. Une proposition pour le mod\`ele de Gross Neveu (et m\^eme le mod\`ele de Yukawa non commutatif) qui ne peut pas \^etre interpr\'et\'ee en terme d'un champ de fond est la suivante:
\bea
S=\int \bar{\psi} (\gamma^1\partial_1+\gamma^2\partial_2+\gamma^3\tilde x_1 + \gamma^4 \tilde x_2) \psi + \lambda\bar{\psi}\star\psi\star\bar{\psi}\star\psi \; ,
\eea
avec $\gamma$ des matrices de Dirac $4\times 4$. A premi\`ere vue il semble \'etrange qu'on ait besoin des spineurs quadri-dimensionnels pour d\'ecrire un syst\`eme de fermions \`a deux dimensions. Une telle th\'eorie peut-elle \^etre raisonnable? 

La r\'eponse est la suivante: la d\'eduction historique de l'\'equation de Dirac part de l'\'equation du mouvement pour les bosons. Dirac a trouv\'e la bonne fa\c con d'extraire la racine carr\'ee du d'Alembertien:
\bea
\gamma^{\mu} \partial_{\mu} \gamma^{\nu} \partial_{\nu} = g^{\mu\nu} \partial_{\mu} \partial_{\nu} \; ,
\eea
Dans les TCNC le bon hamiltonien libre pour les bosons \`a deux dimensions est:
\bea
H_0=\partial_1^2+\partial_2^2+\tilde x_1^2+\tilde x_2^2 \; .
\eea 
Le hamiltonien libre pour les fermions devrait \^etre donc la racine carr\'ee de cet op\'erateur, qui fait intervenir des spineurs quadri-dimensionnels. Une proposition similaire a \'et\'e faite ind\'ependament en  \cite{GrosseFin}. Les auteurs introduisent un op\'erateur similaire en quatre dimensions  (qui fait intervenir par cons\'equent des spineurs en seize dimensions) et l'ont utilis\'e pour calculer une action spectrale du type Connes-Lott d'un champ de jauge coupl\'e \`a un boson de Higgs de la forme du mod\`ele $\phi_4^{\star 4}$. 

Nous nous proposons d'\'etudier par la suite cette th\'eorie et notamment sa fonction b\^eta.

L'\'etude des th\'eories non commutatives, et en particulier $\Phi^{\star 4}_4$ au niveau constructif est d'un grand int\'er\^et. Si le mod\`ele est construit nous disposons d'un deuxi\`eme argument de coh\'erence math\'ematique en sa faveur. 

L'absence du fant\^ome de Landau est un bon signe qui va dans cette direction. De plus, les travaux r\'ecents de Rivasseau et Magnen \cite{constructiv1,constructiv2} ont abouti \`a la construction du mod\`ele dans une tranche. Il nous reste n\'eanmoins encore plusieurs points d\'elicats \`a traiter. Nous avons besoin de prouver que la fonction beta est nulle constructivement, ce qui repr\'esente un travail technique non n\'egligeable. Ensuite ces r\'esultats devrait se combiner en une analyse constructive multi tranche de notre mod\`ele. Il est int\'eressant de mentionner que la sym\'etrie entre les faces droites et gauches intervient d'une fa\c con cruciale dans les preuves constructives monotranche. C'est une nouvelle indication que la sym\'etrie sous-jacente du mod\`ele est cruciale pour sa coh\'erence.

Nous avons d\'ej\`a mentionn\'e les fluides non-commutatifs. Les diff\'eomorphismes pr\'eservant l'aire intervient d'une fa\c con tr\`es naturelle comme groupe d'invariance d'une telle th\'eorie. Cependant nous ne savons pas encore appliquer notre groupe de renormalisation \`a ces th\'eories. Cela est principalement d\^u aux difficult\'es techniques li\'ees aux d\'eg\'en\'erescences du propagateur. Ces singularit\'es sont la cons\'equence de la structure des niveaux de Landau. A cause de l'espacement entre les niveaux nous ne pouvons pas d\'ecouper le spectre du hamiltonien libre en tranches arbitrairement \'etroites. Nous avons test\'e diff\'erentes r\'egularisations possibles, mais nous ne disposons pas pour l'instant d'une recette d\'efinitive du traitement des ces singularit\'es.

Y-a-t il un lien entre nos travaux et la quantification de la gravitation? Nous proposons le sc\'enario suivant:

Nous voulons traiter les op\'erateurs de position comme des champs quantiques et int\'egrer sur leurs fluctuations. En fin de compte \`a l'\'echelle de Planck l'espace temps lui m\^eme devient dynamique. Nous pouvons choisir toujours de nous placer dans un syst\`eme de r\'ef\'erence inertiel. Cela veut dire que pour toute g\'eom\'etrie, localement l'espace-temps tangent est plat. Nous pensons que les effets des fluctuations des op\'erateurs de position devraient \^etre ressentis aussi dans ce contexte plat. Dans un tel syst\`eme, en premi\`ere approximation la quantification de l'espace-temps prendra la forme d'un espace de Moyal. Int\'egrer sur les fluctuations des op\'erateurs de position nous donnerait ainsi le cadre g\'en\'erique pour la quantification de la gravitation. Il est en plus possible que les configurations des champs avec des op\'erateurs de position unitairement \'equivalents soient sur la m\^eme section de jauge, et que nous ayons \`a effectuer une fixation de jauge.

Les auteurs de \cite{Langmann:2003if} ont trait\'e le mod\`ele $\Phi_4^{\star 4}$ avec des techniques des mod\`eles int\'egrables. Dans ces travaux les auteurs ont notamment cherch\'e une limite thermodynamique non triviale du mod\`ele pour certaines valeurs des param\`etres. 
L'annulation de la fonction beta au point de dualit\'e ($\Omega=1$) indique l'existence de cette limite. Une direction de recherche future est donc la construction explicite de la fonction de partition du mod\`ele.
De plus, les cons\'equences de nos travaux pour des mod\`eles de matrices g\'en\'eraux, m\'erite aussi une \'etude d\'etaill\'ee. En effet dans la base matricielle au point de dualit\'e nous avons \`a faire \`a un mod\`ele de matrices non identiquement distribues, sa covariance \'etant $\frac{1}{m+n}$. Si nous arrivons \`a construire explicitement la fonction de partition, nous pourrons essayer de le faire aussi pour d'autres covariances.

Une autre direction de recherche future est d'adapter nos m\'ethodes pour des g\'eom\'etries non commutatives plus g\'en\'erales. Nous nous proposons de mener une \'etude de ces mod\`eles sur des espaces-temps courbes munis de produits de Moyal adapt\'es. A la lumi\`ere de r\'ecents travaux sur ce sujet nous mentionnons que la Moyalit\'e des contretermes est une caract\'eristique tr\'es g\'en\'erale, associ\'e aux propri\'et\'es d'associativit\'e et de tracialit\'e de la d\'eformation.

Ensuite nous pourrons tester si la fonction b\^eta des ces mod\`eles est nulle. La repr\'esentation param\'etrique peut probablement \^etre introduite d'une fa\c con similaire au cas plat. Dans ces conditions la r\'egularisation et renormalisation dimensionnelle peuvent \^etre envisag\'ees. 
Nous pouvons aussi esp\'erer appliquer nos techniques aux vari\'et\'es non commutatives plus g\'en\'erales, comme les sph\`eres floues par exemple, mais nous ne savons pas encore comment vulcaniser les th\'eories de champs sur ces vari\'et\'ees.

Le nouveau groupe de renormalisation que nous avons d\'evelopp\'e dans notre th\`ese devrait aussi trouver une application dans la reformulation du mod\`ele standard par Connes et Chamseddine. Il est possible que la non-commutativit\'e tr\`es simple mise en \'evidence par les deux auteurs (mais limit\'ee \`a un espace interne) devienne plus compliqu\'ee au fur et \`a mesure que nous montons en \'energie et envahisse l'espace-temps $\R^4$ ordinaire. Dans ces conditions le groupe de renormalisation commutatif devrait \^etre remplac\'e par notre groupe de renormalisation non commutatif \`a partir d'une certaine \'echelle, entrainant une modification des flots du mod\`ele standard.

Comme la dualit\'e de Langmann-Szabo est la raison profonde qui tue le fant\^ome de Landau, on peut se poser la question de savoir si elle peut jouer un r\^ole similaire \`a la supersym\'etrie pour dompter les divergences UV. En effet l'id\'ee de la supersym\'etrie souffre du fait que nous n'avons encore trouv\'e aucun partenaire supersym\'etrique. Si dans des exp\'eriences futures nous trouvons de tels  partenaires, ils seront de toute fa\c con de plusieurs ordres de grandeur plus lourds, ce qui devrait \^etre expliqu\'e. Les partenaires des \'electrons sont d\'ej\`a bien plus lourds que ceux-ci, mais la situation est bien pire pour les neutrinos. Cet \'ecart \'enorme a du mal \`a s'interpr\'eter avec les th\'eories actuelles. Il est donc raisonnable de chercher des alternatives \`a la supersym\'etrie.

Nous trouvons dans nos travaux que la dualit\'e de Langmann Szabo joue un r\^ole similaire \`a la supersym\'etrie et am\'eliore les comportements UV des th\'eories de champs. La non-commutativit\'e peut donc \^etre une alternative \`a la super sym\'etrie. Pour \^etre juste il est vrai que tandis que le flot de la constante de couplage $\lambda$ est fini dans nos th\'eories, les masses divergent encore quadratiquement. Pour pouvoir consid\'erer les th\'eories vulcanis\'ees comme des vraies alternatives aux th\'eories supersym\'etriques nous devons premi\`erement comprendre mieux le statut de ces divergences.
Une telle \'etude aboutira en fin de compte \`a la question cl\'e:
Est-ce qu'on peut donner un sens physique \`a la dualit\'e de Langmann Szabo entre position et impulsion et si oui, est-ce qu'elle peut jouer un r\^ole dans le mod\`ele standard et dans le m\'ecanisme de brisure spontan\'ee de sym\'etrie?

M\^eme si la TCNC ne s'applique pas \`a l'au del\`a du mod\`ele standard elle fournit un formalisme qui g\'en\'eralise la th\'eorie quantique des champs habituelle. Par cons\'equent elle doit nous permetre de mieux comprendre des probl\`emes ouverts de la th\'eorie commutative, comme des probl\`emes en champ fort, tel l'effet Hall quantique ou le confinement. C'est une motivation suppl\'ementaire profonde pour poursuivre l'\'etude de ces mod\`eles.

\chapter{Appendice Technique}

La premi\`ere section de ce chapitre est d\'edi\'ee au plan de Moyal, et la seconde a la renormalisation.

\section{Le plan de Moyal}
\label{sec:planMoyal}
\vskip 0.5cm

Le plan de Moyal peut \^etre introduit de plusieurs fa\c con. On va opter ici pour une d\'efinition "utilitaire" en sacrifient les subtilit\'es math\'ematiques en faveur de la simplicit\'e.

Soit $\cal S$ l'espace des fonctions Schwarz (\`a d\'ecroissance rapide) d\'efinies sur $\RR^D$ avec $D$ pair. Soit $\theta$ une matrice antisym\'etrique $D\times D$ aux entr\'ees r\'eelles. Entre deux fonctions $f,g\in \cal S$ on d\'efinit un produit d\'eform\'e (appel\'ee produit de Moyal et not\'e $\star$) par l'int\'egrale:
\bea
\label{eq:prodMoyal}
  (f\star g)(x)=\int\frac{d^Dk}{(2\pi)^D}~d^Dy~
  f(x+\frac{1}{2}\theta k)~g(x+y)~e^{iky} \, .
\eea
La condition que $f$ et $g$ sont Schwarz assure la convergence de l'int\'egrale 
(\ref{eq:prodMoyal}). De plus on voit que la valeur du produit de Moyal en un point prend en compte les valeurs des fonctions dans tout l'espace: on \`a a faire avec un produit non local.

On peut \'etendre le produit de Moyal pour des fonctions plus g\'en\'erales, notamment si $f$ est une fonction coordonn\'e, $f(x)=x^{\mu}$, l'int\'egrale est encore convergente.
\bea
  \label{eq:coordMoyal}
  (x^{\mu}\star g)(x)&=&\int\frac{d^Dk}{(2\pi)^D}~d^Dy~
  (x^{\mu}+\frac{1}{2}\theta^{\mu\nu}k_{\nu})g(x+y)e^{iky}\nonumber\\
   &=&x^{\mu}g(x)+\frac{1}{2}\theta^{\mu\nu}\int\frac{d^Dk}{(2\pi)^D}~d^Dy~
   g(x+y)\frac{\partial}{\imath\partial y^{\nu}}e^{iky}\nonumber\\
   &=&x^{\mu}g(x)+\frac{\imath}{2}\theta^{\mu\nu}\partial_{\nu}g(x) \, ,
\eea
analogue:
\bea
  (g\star x^{\mu})(x)=x^{\mu}g(x)-\frac{\imath}{2} \theta^{\mu\nu} \partial_{\nu} 
   g(x) \, .
\eea
Les deux relations d'avant nous fournisent une repr\'esentation du produit habituel et de la d\'eriv\'ee a l'aide du produit de Moyal:
\bea
   \label{eq:derivMoyal}
  \theta_{\lambda\mu}~[x^{\mu},g]_{\star}=\imath~\partial_{\lambda}g
  \, , \, \{x^{\mu},g\}_{\star}=2~x^{\mu}g\, .
\eea

On note que sous le signe d'int\'egrale on peut remplacer un produit $\star$ par un produit habituel. En effet:
\bea
\label{eq:integralMoyal}
\int d^Dx f\star g&=&\int d^Dx \frac{d^Dk}{(2\pi)^D}~d^Dy~
  f(x+\frac{1}{2}\theta k)~g(x+y)~e^{iky}\nonumber\\
    &=&\int d^Dx' \frac{d^Dk}{(2\pi)^D}~d^Dy~
  f(x'-y+\frac{1}{2}\theta k)~g(x')~e^{iky}\nonumber\\
   &=&\int d^Dx' \frac{d^Dk}{(2\pi)^D}~d^Dy'~
  f(x'-y')~g(x')~e^{iky'}\nonumber\\
   &=&\int d^Dx'~d^Dy'~
  f(x'-y')~g(x')~\delta(y')\nonumber\\
    &=&\int d^Dx'~f(x')g(x') \,.
\eea

Cette propri\'et\'e est essentielle car elle nous permettra par la suite de transformer tous les produits des champs dans la partie quadratique des actions en produits de Moyal.

Pour r\'ef\'erence ult\'erieure on calcule ici aussi la forme de l'int\'egrale du produit de Moyal dans l'espace direct entre quatre fonctions:
\bea
\label{eq:4fonctonMoyal}
\int d^Dx (f_1\star f_2\star f_3\star f_4)(x)
\eea

Un simple changement de variable dans l'eq. (\ref{eq:prodMoyal}) nous fournit la repr\'esentation suivante:
\bea
(f\star g)(x)=\frac{1}{(2\pi)^D\det(2\theta^{-1})}\int f(x_1)g(x_2)
e^{2\imath x_2\theta^{-1}x_1-2\imath x\theta^{-1}(x_1-x_2)}
\eea
et on obtient:
\bea
\label{eq:prod4Moyal}
&&\int f_1\star f_2\star f_3\star f_4=\frac{\det(2\theta^{-1})^2}{(2\pi)^{2D}}\int d^Dx\prod_{i=1,4}d^Dx_i
f_i(x_i)\nonumber\\
&&e^{2\imath x_2\theta^{-1}x_1+2\imath x_4\theta^{-1}x_3-2\imath x\theta^{-1}(x_1-x_2+x_3-x_4)}
=\frac{\det(2\theta^{-1})}{(2\pi)^{D}} \nonumber\\
&&\int \prod_{i=1,4}d^Dx_i
f_i(x_i) \delta(x_1-x_2+x_3-x_4)e^{-2\imath(x_1\theta^{-1}x_2+x_3\theta^{-1}x_4)}
\eea

\subsection{La base matricielle}

Un outil tr\`es important dans l'\'etude du plan de Moyal est la base matricielle. Pour simplicit\'e on se limite \`a deux dimensions de l'espace, la g\'en\'eralisation en toute dimension arbitraire paire \'etant triviale. En deux dimensions on peut effectuer un changement de coordonn\'ees tel que la matrice $\theta$ prend la forme de Jordan:
\bea
\label{eq:Jordan}
\theta=\begin{pmatrix}
              0 & \theta \\
	      -\theta & 0\\
           \end{pmatrix}\equiv \theta \wedge \;.
\eea

Appellons les coordonn\'ees de l'espace $x^1$ et $x^2$. On introduit les variables holomorphe et anti holomorphe:
\bea
\label{eq:abara}
 a&=&\frac{1}{\sqrt{2}}(x^1+\imath x^2) \; , \bar{a}=\frac{1}{\sqrt{2}}(x^1-\imath x^2)\nonumber\\
\partial_a&=&\frac{1}{\sqrt{2}}(\partial_{1}-\imath \partial_{2})\; ,
\partial_{\bar{a}}=\frac{1}{\sqrt{2}}(\partial_{1}+\imath \partial_{2})\, .
\eea
Ils respectent les propriet\'es suivantes:
\bea
\label{eq:abaraprop}
(a\star f)(x)&=&a(x)f(x)+\frac{\theta}{2}\partial_{\bar{a}}f(x)\; ,
(f\star a)(x)=a(x)f(x)-\frac{\theta}{2}\partial_{\bar{a}}f(x)\nonumber\\
(\bar{a}\star f)(x)&=&\bar{a}(x)f(x)-\frac{\theta}{2}\partial_{a}f(x)\; ,
(f\star \bar{a})(x)=\bar{a}(x)f(x)+\frac{\theta}{2}\partial_{a}f(x)
\eea
et la mesure de Lebesgue sur le plan devient $dx^1\wedge dx^2=\imath da\wedge d\bar{a}$

Soit la fonction
\bea
\label{eq:gaussienne}
 f_{0}(a,\bar{a})=2e^{-\frac{2}{\theta}a\bar{a}}\, .
\eea
Elle est invariante par le produit de Moyal:
\bea
\label{eq:f0starf0}
 f_{0}\star f_{0}&=&4\int d^2y\frac{d^2k}{(2\pi)^2}
 e^{-\frac{1}{\theta}(2x^2+y^2+2xy+\theta x\wedge k+\frac{\theta^2}{4}k^2+\imath ky)}\nonumber\\
 &=&\frac{4e^{-\frac{2x^2}{\theta}}}{(2\pi)^2}\int d^2yd^2k 
   e^{-\frac{1}{2}\begin{pmatrix}
                             y &k \\
                            \end{pmatrix}
                          \begin{pmatrix}
	        	   \frac{2}{\theta} & -\imath \\
			   -\imath & \frac{\theta}{2}\\
                           \end{pmatrix}
			   \begin{pmatrix}
			   y \\
			  k\\
		  \end{pmatrix}+
		  \begin{pmatrix}
		  y & k\\
		  \end{pmatrix}
		  \begin{pmatrix}
		  -\frac{2}{\theta}x \\
		  \frac{1}{2}\wedge x\\
		  \end{pmatrix}
     }\nonumber\\
     &=& 2 e^{-\frac{2}{\theta}x^2}
     e^{-\frac{1}{2}
                         \begin{pmatrix}
			   -\frac{2}{\theta}x &-\frac{1}{2} x \wedge\\
                         \end{pmatrix}
			 \begin{pmatrix}
			 \frac{\theta}{4} & \frac{\imath}{2}\\
			 \frac{\imath}{2} & \frac{1}{\theta}\\
			 \end{pmatrix}
                  \begin{pmatrix}
		  -\frac{2}{\theta}x \\
		  \frac{1}{2}\wedge x\\
		  \end{pmatrix}
     }=f_{0}\, .
\eea

Soit l'ensemble des focntons $\{f_{mn}\}$ (polyn\^omes d'Hermite en deux dimensions) d\'efini par:
\bea
 f_{mn}=\frac{1}{\sqrt{m!n!\theta^{m+n}}}\bar{a}^{\star m}\star f_{0}\star a^{\star n}\, .
\eea

Notons que $\bar{a}^{\star m}\star f_0=2^m\bar{a}^m f_0$ et $f_0\star a^{\star n}=2^na^nf_0$, $\bar{f_{mn}}=f_{nm}$.De plus:
\bea
a\star f_{mn}=\sqrt{m\theta}f_{m-1n}\; , f_{mn}\star \bar{a}=\sqrt{n\theta}f_{mn-1}\, ,
\eea
ce qui nous donne:
\bea
f_{mn}f_{kl}=\delta_{nk}f_{ml}\; ,\int f_{mn}=2\pi\theta\delta_{mn}\, .
\eea

Toute fonction peur \^etre d\'eveloppe sur cette base:
\bea
\phi(x)=\sum_{mn}\phi_{mn}f_{mn}\; ,\phi_{mn}=\int dx f_{nm}\star \phi(x)\, .
\eea

Le produit de Moyal entre deux fonctions est repr\'esente dans cette base par le produit matricielle. De plus la matrice associe a la complexe conjugu\'ee d'une fonction est la herm\'etique conjugu\'ee de la matrice associe a la fonction.

\appendix

\chapter{Propagators for NCQFT}
\label{sec:Propagators}

\begin{center}

  Razvan \textsc{Gurau}, 
  Vincent \textsc{Rivasseau} and
  Fabien \textsc{Vignes-Tourneret}

  \vskip 3ex  

  \textit{Laboratore de Physique Th\'eorique, B\^at.\ 210, 
    Universit\'e Paris XI \\ F-91405 Orsay Cedex, France}
  \\
  e-mail: \texttt{razvan.gurau@th.u-psud.fr}, 
  \texttt{vincent.rivasseau@th.u-psud.fr}, 
  \texttt{fabien.vignes@th.u-psud.fr}
\end{center}

\vskip 5ex

  In this paper we provide exact expressions for propagators of noncommutative
  Bosonic or Fermionic field theories after adding terms of the Grosse-Wulkenhaar
  type in order to ensure Langmann-Szabo covariance. We emphasize 
  the new Fermionic case and we give in particular all necessary bounds for the multiscale
  analysis and renormalization of the noncommutative Gross-Neveu model. 


\section{Introduction}

This paper is the first of a series in which we plan to extend the proof of perturbative 
renormalizability of noncommutative $\phi^4_4$ field theory
\cite{GrWu03-1,GrWu04-3,Rivasseau2005bh} to other noncommutative models 
(see \cite{DN} for a general review on noncommutative field theories).

We have in mind in particular Fermionic field theories either of the relativistic type,
such as the Gross-Neveu model in Euclidean two dimensional space \cite{Mitter:1974cy,Gross:1974jv}, 
or of the type used in condensed matter for many body theory, in
which there is no symmetry between time and space. In the commutative case, these non-relativistic theories 
are just renormalizable in any dimension \cite{FT,Benfatto:1996ngIN,Salmhofer}. 
Their noncommutative version should be relevant 
for the study of Fermions in 2 dimensions in magnetic fields, hence for the quantum Hall effect.
Of course a future goal is also to find the right extension of the Grosse-Wulkenhaar method to
gauge theories. 

In this paper we generalize the computation of the Bosonic $\phi^4_4$ propagator
of \cite{GrWu04-3} and provide the exact expression of the propagators of Fermionic 
noncommutative field theories on the Moyal plane. We will restrict our analysis to one pair 
of noncommuting coordinates, as the generalization to several pairs are 
trivial. These propagators are not the ordinary commutative propagators: 
they have to be modified to obey Langmann-Szabo duality, according to the 
pioneering papers \cite{GrWu04-3,LaSz}. We propose to call {\it vulcanization}
this modification of the theory, and to call {\it vulcanized} the resulting theory and propagators.

Like in \cite{Rivasseau2005bh} we can slice the corresponding noncommutative heat kernels 
according to the Schwinger parameters in order to derive a multiscale analysis. In this framework
the  theory with a finite number of slices has a cutoff, and the removal of the cutoff 
(ultraviolet limit) corresponds to summing over infinitely many slices. 
However for the Fermionic propagators treated in this paper 
this multiscale analysis is harder than in the Bosonic case.
In $x$-space the propagator end-terms oscillate rather than decay as in 
the  Bosonic $\phi^4_4$ case. In matrix basis the behavior of the propagator
is governed by a non-trivial critical point in parameter space and in contrast
with the Bosonic case there is no general scaled decay for all indices.
The main result of this paper is the detailed analysis of this critical point,
leading to Theorem \ref{maintheorem}, namely to the bounds required for the multiscale analysis of the 
2 dimensional Euclidean Non-Commutative Gross-Neveu model. 
These bounds are slightly worse than in the $\phi^4_4$ case. 
However they should suffice for a complete proof of renormalizability of the model 
to all orders (see the discussion after Theorem \ref{maintheorem}). This complete proof
is postponed to a future paper \cite{RenNCGN05}.

\section{Conventions}
\setcounter{equation}{0}

The two dimensional Moyal space ${\mathbb R}^2_\theta$
is defined by the following associative non-commutative star product
\begin{equation}
  (a\star b)(x) = \int \frac{d^2k}{(2\pi)^2} \int d^2 y \; a(x{+}\tfrac{1}{2}
  \theta {\cdot} k)\, b(x{+}y)\, \mathrm{e}^{\mathrm{i} k \cdot y}\;.
\label{starprod}
\end{equation}
which corresponds to a constant commutator:
\begin{equation}
  [ x^i , x^j ]  = i  \Theta ^{ij} \ \ ,  {\rm where } 
 \ \Theta=\begin{pmatrix}\phantom{-}0&\theta\\-\theta&0\end{pmatrix}. 
\end{equation}

In order to perform the second quantization one must first identify 
the Hilbert space of states of the first quantization. If we deal with a 
real field theory (for instance the $\phi^4$ theory which is treated in 
detail in \cite{GWR2x2}) one considers a real Hilbert space, 
whereas for a complex field theory (e.g. any Fermionic field theory) one has
to use a complex Hilbert space.

A basis in this Hilbert space is given by the functions 
$f_{mn}$ defined in \cite{GWR2x2,Gracia-Bondia1987kw}.  

Any complex-valued function defined on the plane can be decomposed in 
this basis as:
\begin{equation}
  \chi(x)=\sum_{m,n} \chi_{mn}f_{mn}(x).
\end{equation}

A crucial observation is that $\bar{f}_{mn}(x)=f_{nm}(x)$ (which can 
be verified on the explicit expression for $f_{mn}$). A real function 
in this basis obeys:
\begin{equation}
  \bar{\chi}(x)=\chi(x)\Rightarrow\overline{\sum\chi_{mn}f_{mn}(x)}=
  \sum\chi_{pq}f_{pq(x)}\Rightarrow\bar{\chi}_{mn}=\chi_{nm} .
\end{equation}

The scalar product (which must be sesquilinear for a 
complex Hilbert space) is then defined as:
\begin{eqnarray}
  \langle\phi,\chi\rangle&=&\int\frac{d^2x}{2\pi\theta}
  ~\bar{\phi}(x)\chi(x)=\sum\bar{\phi}_{pq}
  \chi_{rs}\int\frac{d^2x}{2\pi\theta}\bar{f}_{pq}f_{rs}\nonumber\\
  &=&
  \sum\bar{\phi}_{pq}\chi_{rs}\delta_{pr}\delta_{qs}=\sum\bar{\phi}_{pq}\chi_{pq} .
\end{eqnarray}

Notice that if $\phi$ is a real field we have:
\begin{equation}
  \langle\phi,\chi\rangle=\sum\phi_{qp}\chi_{pq},
\end{equation}
so that our conventions restrict to those in \cite{GWR2x2} for 
the real $\phi^4$ theory.
With this convention $\langle f_{kl},\chi\rangle=\chi_{mn}\int \bar{f}_{kl}f_{mn}=\chi_{kl}$.
A linear operator on this space acts like:
\begin{equation}
  [A\phi]_{kl}=\langle f_{kl},A\phi\rangle=\sum_{m,n}\langle f_{kl},Af_{mn}\rangle\phi_{mn}.
\end{equation}

At this point the convention consistent with that for the real 
$\phi^4$ theory found in \cite{GWR2x2} is to note:
\begin{equation}
  \langle f_{kl},Af_{mn}\rangle=\int\frac{d^2x}{2\pi\theta}\bar{f}_{kl}(x)
  \int d^2y A(x,y)f_{mn}(y):=A_{l,k;m,n} .
\end{equation}
With this convention the product of operators is
\begin{equation}
  \lsb AB\rsb_{p,q;r,s}=\langle f_{qp},ABf_{rs}\rangle=\sum_{t,u}\langle f_{qp},Af_{tu}\rangle
  \langle f_{tu},Bf_{rs}\rangle=\sum_{t,u}A_{p,q;t,u}B_{u,t;r,s}
\end{equation}
and the identity operator $I$ has the matrix elements:
\begin{equation}
  \phi_{mn}=\sum_{p,q}\langle f_{mn},If_{pq}\rangle\phi_{pq}
  \Rightarrow I_{n,m;p,q}=\delta_{pm}\delta_{nq}.
\end{equation}

We pass now to the second quantization.
The quadratic part of the action is generically (in $x$ space):
\begin{equation}
  S=\int d^2x~\bar{\chi}(x)~H(x,y)~\chi(y) .
\end{equation}

In the matrix basis one has the action:
\begin{equation}
  S=2\pi\theta\sum_{m,n,k,l}\frac{1}{2} \bar{\chi}_{pq}\Big{(}2\int\frac{d^2x}{2\pi\theta}\bar{f}_{pq}(x)
  H(x,y)f_{kl}(y)\Big{)} \chi_{kl}.
\end{equation}
We define the Hamiltonian in the matrix basis as:
\begin{equation}
  H_{q,p;k,l}=2\int\frac{d^2x}{2\pi\theta}\bar{f}_{pq}(x)
  H(x,y)f_{kl}(y)
\end{equation}
where the $2$ has been included in order to maintain the conventions 
in \cite{GWR2x2} for the case of a real field.

\section{Non-Commutative Schwinger Kernels}
\setcounter{equation}{0}

In this section we provide the explicit formulas for the Schwinger representation
of the non-commutative kernels or propagators of free scalar Bosonic and 
spin 1/2 Fermionic theories on the Moyal plane. These formulas are 
essential for a multiscale analysis based on slicing the Schwinger parameter.

The different propagators of interest are expressed  
via the Schwinger parameter trick as:
\begin{equation}
  H^{-1}=\int_0^{\infty}dt~e^{-tH} .
\end{equation} 

\subsection{Bosonic $\boldsymbol{x}$ Space Kernel}

We define $x\wedge x'=x_0x'_1-x_1x'_0$ and $x\cdot x'=x_0x'_0+x_1x'_1$.
The following lemma generalizes the Mehler kernel \cite{simon79:funct}:
\begin{lemma}
  \label{HinXspace}Let H be:
  \begin{equation}
    H=\frac{1}{2}\big{[}-\partial_0^2-\partial_1^2+
    \Omega^2x^2-2\imath B(x_0\partial_1-x_1\partial_0)\Big{]}.
  \end{equation}     
  The integral kernel of the operator $e^{-tH}$ is:
  \begin{equation}
    e^{-tH}(x,x')=\frac{\Omega}{2\pi\sh\Omega t}e^{-A},
  \end{equation}
  \begin{equation}
    A=\frac{\Omega\ch\Omega t}{2\sh\Omega t}(x^2+x'^2)-
    \frac{\Omega\ch Bt}{\sh\Omega t}x\cdot x'-\imath
    \frac{\Omega\sh Bt}{\sh\Omega t}x\wedge x'.
  \end{equation}
\end{lemma}
\prf
We note that the kernel is correctly normalized: as    
$\Omega =B\rightarrow 0$ we  
have
\begin{equation}
  e^{-tH}(x,x')\rightarrow\frac{1}{2\pi t}e^{-\frac{|x-x'|^2}{2t}},
\end{equation}
which is the normalized heat kernel.\\
We must then check the equation
\begin{equation}\label{diffbasic}
  \frac{d}{dt}e^{-tH}+He^{-tH}=0.
\end{equation}
In fact
\begin{eqnarray}
  \frac{d}{dt}e^{-tH}&=&\frac{\Omega e^{-A}}{2\pi\sh\Omega t}
  \Big{\{}-\Omega\coth\Omega t +
  \frac{\Omega^2}{2\sh^2\Omega t}(x^2+x'^2)
  \nonumber\\
  &+&\Omega\frac{B\sh\Omega t\sh Bt-\Omega\ch\Omega t\ch Bt}
  {\sh^2\Omega t}x\cdot x' 
  \nonumber\\
  &+&\imath\Omega\frac{B\ch Bt\sh\Omega t-\Omega\sh Bt\ch\Omega t}
  {\sh^2\Omega t}x\wedge x'
  \Big{\}}.
\end{eqnarray}
Moreover
\begin{eqnarray}  
  \frac{(-\partial^2_1-\partial^2_2)}{2}e^{-tH}
  &=& \frac{\Omega e^{-A}}{2\pi\sh\Omega t}\Big{\{}
  \Omega\coth\Omega t -\frac{1}{2}
  \Big{[}\frac{\Omega\ch\Omega t}{\sh\Omega t}x-\frac{\Omega\ch 
    Bt}{\sh\Omega t}x'\Big{]}^2
  \nonumber\\
  &+&\frac{\Omega^2\sh^2 Bt}{2\sh^2\Omega t}x'^2
  +\imath\frac{\Omega^2\ch\Omega t\sh Bt}{\sh^2\Omega t}x\wedge x'
  \Big{\}}
\end{eqnarray}
and
\begin{equation}
  \imath B(x_1\partial_2-x_2\partial_1)= \frac{\Omega}{2\pi\sh\Omega 
    t}e^{-A}\Big{\{}
  (-\imath\frac{B\Omega\ch Bt}{\sh\Omega t}x\wedge x'+
  \frac{B\Omega\sh Bt}{\sh\Omega t}x\cdot x')
  \Big{\}} .
\end{equation}
It is now straightforward to verify the 
differential equation (\ref{diffbasic}). \hfill\qed

\begin{corollary}
  Let H be:
  \begin{equation}
    H=\frac{1}{2}\big{[}-\partial_0^2-\partial_1^2+
    \Omega^2x^2-2\imath\Omega(x_0\partial_1-x_1\partial_0)\Big{]}.
  \end{equation}
  The integral kernel of the operator $e^{-tH}$ is:
  \begin{equation}
    e^{-tH}(x,x')=\frac{\Omega}{2\pi\sh\Omega t}e^{-A},
  \end{equation}
  \begin{equation}
    A=\frac{\Omega\ch\Omega t}{2\sh\Omega t}(x^2+x'^2)-
    \frac{\Omega\ch \Omega t}{\sh\Omega t}x\cdot x'-\imath
    \Omega x\wedge x'.
  \end{equation}
\end{corollary}

\subsection{Fermionic $\boldsymbol{x}$ Space Kernel}

The two-dimensional free commutative Fermionic field theory is defined by the 
Lagrangian
\begin{equation}
\label{eq:LGN}
{\cal L}=\psib(x)\lbt\ps+\mu\rbt\psi(x). 
\end{equation}
The propagator of the theory $\lbt\ps + \mu\rbt^{-1}(x,y)$ can be calculated thanks to
the usual heat kernel method as
\begin{align}
  \lbt\ps
  +\mu\rbt^{-1}(x,y)&=\lbt-\ps+\mu\rbt\lbt\lbt\ps+\mu\rbt\lbt-\ps+\mu\rbt\rbt^{-1}(x,y)\nonumber\\
  &=\lbt-\ps+\mu\rbt\lbt p^{2}+\mu^{2}\rbt^{-1}(x,y)\\
  &=\lbt-\ps+\mu\rbt\int_{0}^{\infty}\frac{dt}{4\pi t}\,
  e^{-\frac{(x-y)^{2}}{4t}-\mu^{2}t}\\
  &=\int_{0}^{\infty}\frac{dt}{4\pi
    t}\,\lbt\frac{-\imath}{2t}(\xs-\ys)+\mu\rbt e^{-\frac{(x-y)^{2}}{4t}-\mu^{2}t}.
\end{align}
In the noncommutative case we have to modify the free action, adding a Grosse-Wulkenhaar term to 
implement Langmann-Szabo duality. This shall prevent ultra-violet infrared mixing in theories with 
generic interaction of the Gross-Neveu type and allow consistent renormalization to all orders of perturbation. 

The free action becomes after vulcanization
\begin{equation}\label{sfree}
S_{free} = \int d^2 x \psib^a(x)\lbt\ps +\mu+\Omega\xts\rbt\psi^a  (x)
\end{equation}
where $\xt=2\Theta^{-1}x$ and $\Theta=\begin{pmatrix}\phantom{-}0&\theta\\-\theta&0\end{pmatrix}$ and
$a$ is a color index which takes values $1$ to $N$. The corresponding propagator $G$ is 
diagonal in this color index, so we omit it in this section\footnote{There
is no star product in these formulas, since we reduce quadratic expressions in the fields
in non-commutative theory to usual integrals with ordinary products.}.
To compute this propagator we write as in the
commutative case:
\begin{eqnarray}
G&=& \lbt\ps+\mu+\Omega\xts\rbt^{-1}=\lbt-\ps+\mu-\Omega\xts\rbt  . Q^{-1},
\nonumber\\
Q&=& \lbt\ps+\mu+\Omega\xts\rbt\lbt-\ps+\mu-\Omega\xts\rbt
  \nonumber\\
  &=&\mathds{1}_{2}\otimes\lbt
  p^{2}+\mu^{2}+\frac{4\Omega^{2}}{\theta^{2}}x^{2}\rbt
  +\frac{4\imath\Omega}{\theta}\gamma^{0}\gamma^{1}\otimes{\text{Id}}
  \nonumber\\
&+&\frac{4\Omega}{\theta}\mathds{1}_{2}\otimes  L_{2},
\end{eqnarray}
where $L_{2}=x^{0}p_{1}-x^{1}p_{0}$. 

To invert $Q$ we use again the Schwinger trick and obtain:
\begin{lemma}
\label{FermioXspace} We have:
\begin{eqnarray}
G(x,y) &=& -\frac{\Omega}{\theta\pi}\int_{0}^{\infty}\frac{dt}{\sinh(2\Ot
t)}\, e^{-\frac{\Ot}{2}\coth(2\Ot t)(x-y)^{2}+\imath\Ot
x\wedge y}
\\ 
&&    \lb\imath\Ot\coth(2\Ot t)(\xs-\ys)+\Omega(\xts-\yts)- \mu \rb
e^{-2\imath\Ot t\gamma^{0}\gamma^{1}}e^{-t\mu^{2}}  
\nonumber
\end{eqnarray}
It is also convenient to write $G$ in terms of commutators:
\begin{eqnarray}    
G(x,y)  &=&-\frac{\Omega}{\theta\pi}\int_{0}^{\infty}dt\,\lb \imath\Ot\coth(2\Ot
t)\lsb\xs, \Gamma^t  \rsb(x,y) \right.
\nonumber\\
&&
\left. +\Omega\lsb\slashed{\tilde{x}}, \Gamma^t \rsb(x,y)  -\mu \Gamma^t (x,y)  \rb
e^{-2\imath\Ot t\gamma^{0}\gamma^{1}}e^{-t\mu^{2}}, 
\label{xfullprop}
\end{eqnarray} 
where
\begin{eqnarray}
\Gamma^t (x,y)  &=&
\frac{1}{\sinh(2\Ot t)}\,
e^{-\frac{\Ot}{2}\coth(2\Ot t)(x-y)^{2}+\imath\Ot x\wedge y}
\end{eqnarray}
with $\Ot=\frac{2\Omega}{\theta}$ and $x\wedge y=x^{0}y^{1}-x^{1}y^{0}$.
\end{lemma}
\noindent {\proof} The proof follows along the same lines than for Lemma \ref{HinXspace}
and is given in detail in Appendix \ref{app2}. Note that the constant term 
$e^{-2\imath\Ot t\gamma^{0}\gamma^{1}}$ is developped in (\ref{expgamma01}).
\qed
\medskip

\subsection{Bosonic Kernel in the Matrix Basis}

Let $H$ be as in Lemma \ref{HinXspace}, with $\Omega\rightarrow 
\frac{2\Omega}{\theta}$ and $B\rightarrow\frac{2B}{\theta}$. A 
straightforward computation shows that in the matrix basis we have:
\begin{align}
  H_{m,m+h;l+h,l}=&\frac{2}{\theta}
  (1+\Omega^2)(2m+h+1)\delta_{m,l}-\frac{4Bh}{\theta}~\delta_{m,l}\label{eq:HBini}
  \\
  &\hspace{-1cm}-\frac{2}{\theta}(1-\Omega^2)[\sqrt{(m+h+1)(m+1)}~\delta_{m+1,l}+
  \sqrt{(m+h)m}~\delta_{m-1,l}].\nonumber
  \label{HinMat}
\end{align}
Notice that in the limiting case $\Omega=B=1$ the operator becomes diagonal.

The corresponding propagator in the matrix basis for $B=0$ 
can be found in \cite{GrWu04-3} and \cite{GWR2x2}. The result is
that the only non zero matrix elements of the 
exponential are:  
\begin{eqnarray} 
    &&[e^{-\frac{t\theta}{8\Omega}H}]_{m,m+h;l+h,l}=\\
    \nonumber 
    &&\sum^{\min(m,l)}_{u=\max(0,-h)}
    \Big{(}\frac{4\Omega}{(1+\Omega)^2}\Big{)}^{h+2u+1}
    \Big{(}\frac{1-\Omega}{1+\Omega}\Big{)}^{m+l-2u}
    {\cal E}(m,l,h,u){\cal A}(m,l,h,u)
\end{eqnarray}     
with
\begin{equation}\label{eq:A}
{\cal A}(m,l,h,u)=\frac{\sqrt{m!(m+h)!l!(l+h)!}}{(m-u)!(l-u)!(h+u)!u!},
\end{equation}
\begin{equation}
{\cal E}(m,l,h,u)=\frac{e^{-t(\frac{h+1}{2}+u)}(1-e^{-t})^{m+l-2u}}
{(1-(\frac{1-\Omega}{1+\Omega})^2)e^{-t})^{m+l+h+1}}.
\end{equation}

Having in mind the slicing of the propagators needed to 
carry out the renormalization (see \cite{Rivasseau2005bh} for the 
$\phi^4$ case), we will use a slightly different 
representation of the propagator:
\begin{equation}
  H^{-1}=\frac{\theta}{8\Omega}\int_{0}^{1}\frac{d\alpha}
  {1-\alpha}(1-\alpha)^{\frac{\theta}{8\Omega}H}.
\end{equation} 
One has then the lemma:
\begin{lemma}\label{lemma:B0bis}
  Let $H$ be given by equation (\ref{eq:HBini}) with $B=0$. We have:
  \bea 
    [(1-\alpha)^{\frac{\theta}{8\Omega}H}]_{m,m+h;l+h,l}
    &=&\sum_{u=max(0,-h)}^{min(m,l)}{\cal A}(m,l,h,u)
    \Big{(}C\frac{1+\Omega}{1-\Omega}\Big{)}^{m+l-2u}\nonumber\\
    &&{\cal E}(m,l,h,u)
  \eea 
  with ${\cal A}(m,l,h,u)$ as before, $C=\frac{(1-\Omega)^2}{4\Omega}$, and
  \begin{equation}
    {\cal E}(m,l,h,u)=\frac{(1-\alpha)^{\frac{h+2u+1}{2}}\alpha^{m+l-2u}}
    {(1+C\alpha)^{m+l+h+1}}.
  \end{equation}
\end{lemma}
The proof is given in Appendix \ref{app3} below. Extending to the  $B\ne 0$ case, we get
easily the following corollary, useful for studying the Gross-Neveu model:
\begin{corollary}
  Let $B\neq 0$. Denote $H_0=H|_{B=0}$ We have:
  \begin{equation}
    [(1-\alpha)^{\frac{\theta}{8\Omega}H}]_{m,m+h;l+h,l}=
    [(1-\alpha)^{\frac{\theta}{8\Omega}H_0}]_{m,m+h;l+h,l}
    (1-\alpha)^{-\frac{4B}{8\Omega}h}.
  \end{equation}
\end{corollary}

\subsection{Fermionic Kernel in the Matrix Basis}

Let $L_{2}=-\imath(x^{0}\partial_{1}-x^{1}\partial_{0})$. The inverse of the quadratic form
\begin{equation}
\Delta= Q- \frac{4 \imath \Omega} {\theta} \gamma^0\gamma^1= p^{2}+\mu^{2}+\frac{4\Omega^{2}}{\theta^2} x ^{2} +\frac{4B}{\theta}L_{2}
\end{equation}
is given by the previous section:
\begin{align}
  \label{eq:propinit}
  \Gamma_{m, m+h; l + h, l} 
  &= \frac{\theta}{8\Omega} \int_0^1 d\alpha\,  
  \dfrac{(1-\alpha)^{\frac{\mu^2 \theta}{8 \Omega}-\frac{1}{2}}}{  
    (1 + C\alpha )} 
  \Gamma^{\alpha}_{m, m+h; l + h, l}\;,
  \nonumber\\
  \Gamma^{(\alpha)}_{m, m+h; l + h, l}
  &= \left(\frac{\sqrt{1-\alpha}}{1+C \alpha} 
  \right)^{m+l+h}\left( 1-\alpha\right)^{-\frac{Bh}{2\Omega}} \\
  &
  \sum_{u=0}^{\min(m,l)} {\cal A}(m,l,h,u)\ 
  \left( \frac{C \alpha (1+\Omega)}{\sqrt{1-\alpha}\,(1-\Omega)} 
  \right)^{m+l-2u}\;,
  \label{eq:propinit-b}
\end{align}
where ${\cal A}(m,l,h,u)$ is given by (\ref{eq:A}) and $C$ is defined in Lemma
\ref{lemma:B0bis}.

The Fermionic propagator $G$ (\ref{xfullprop}) in matrix space
can be deduced from this kernel. One should simply take $B= \Omega$,
add the missing $\gamma^0 \gamma^1$ term, and compute the action of
$-\ps-\Omega\xts+\mu$ on $\Gamma$. Hence we have to compute $\lsb x^{\nu},\Gamma\rsb$ in the
matrix basis. It is easy to express the multiplicative operator $x^{\nu}$ in
this matrix basis. Its commutator with $\Gamma$ follows from
\begin{align}
  \lsb x^{0},\Gamma\rsb_{m,n;k,l}=&2\pi\theta\sqrt\frac{\theta}{8}\lb\sqrt{m+1}
  \Gamma_{m+1,n;k,l}-\sqrt{l}\Gamma_{m,n;k,l-1}+\sqrt{m}\Gamma_{m-1,n;k,l}
\right.\nonumber\\
&-\sqrt{l+1}\Gamma_{m,n;k,l+1}+\sqrt{n+1}\Gamma_{m,n+1;k,l}-\sqrt{k}
\Gamma_{m,n;k-1,l}\nonumber\\
&\left.+\sqrt{n}\Gamma_{m,n-1;k,l}-\sqrt{k+1}
  \Gamma_{m,n;k+1,l}\rb,\label{x0Gamma}\\
  \lsb
  x^{1},\Gamma\rsb_{m,n;k,l}=&2\imath\pi\theta\sqrt\frac{\theta}{8}\lb\sqrt{m+1}
  \Gamma_{m+1,n;k,l}-\sqrt{l}\Gamma_{m,n;k,l-1}-\sqrt{m}
  \Gamma_{m-1,n;k,l} \right.
\nonumber\\
&+\sqrt{l+1}\Gamma_{m,n;k,l+1}
-\sqrt{n+1}\Gamma_{m,n+1;k,l}+\sqrt{k}\Gamma_{m,n;k-1,l}
\nonumber\\
&\left.+\sqrt{n}\Gamma_{m,n-1;k,l}-\sqrt{k+1}\Gamma_{m,n;k+1,l}\rb.
  \label{x1Gamma}
\end{align}
This leads to the formula for $G$ in matrix space:
\begin{lemma}Let $G_{m,n;k,l}$ be the matrix basis kernel of the operator\\
$\lbt\ps+\Omega\xts+\mu\rbt^{-1}$. We have:
\begin{eqnarray}
G_{m,n;k,l}&=& 
-\frac{2\Omega}{\theta^{2}\pi^{2}} \int_{0}^{1} 
d\alpha\, G^{\alpha}_{m,n;k,l}  
\nonumber\\
G^{\alpha}_{m,n;k,l}&=&\lbt\imath\Ot\frac{2-\alpha}{\alpha}\lsb\xs,
\Gamma^{\alpha}\rsb_{m,n;k,l}
+\Omega\lsb\slashed{\tilde{x}},\Gamma^{\alpha}\rsb_{m,n;k,l} - \mu\,\Gamma^{\alpha}_{m,n;k,l}\rbt
\nonumber\\
&&\times\lbt\frac{2-\alpha}{2\sqrt{1-\alpha}}
\mathds{1}_{2}-\imath\frac{\alpha}{2\sqrt{1-\alpha}}\gamma^{0}\gamma^{1}
\rbt.\label{eq:matrixfullprop}
\end{eqnarray}
where $\Gamma^{\alpha}$ is given by (\ref{eq:propinit-b}) and the commutators 
by formulae (\ref{x0Gamma}) and (\ref{x1Gamma}).
\end{lemma}
The first two terms in (\ref{eq:matrixfullprop}) contain commutators and are grouped together
under the name $G^{\alpha, {\rm comm}}_{m,n;k,l}$. The last term is called 
$G^{\alpha, {\rm mass}}_{m,n;k,l}$.  Hence
\begin{eqnarray}\label{commterm}
G^{\alpha, {\rm comm}}_{m,n;k,l}&=& \lbt\imath\Ot\frac{2-\alpha}{\alpha}\lsb\xs,
\Gamma^{\alpha}\rsb_{m,n;k,l} +\Omega\lsb\slashed{\tilde{x}},\Gamma^{\alpha}\rsb_{m,n;k,l} \rbt  \nonumber\\
&&\times\lbt\frac{2-\alpha}{2\sqrt{1-\alpha}}
\mathds{1}_{2}-\imath\frac{\alpha}{2\sqrt{1-\alpha}}\gamma^{0}\gamma^{1} \rbt.
\end{eqnarray}
\begin{eqnarray}\label{massterm}
G^{\alpha, {\rm mass}}_{m,n;k,l}&=& - \mu\, \Gamma^{\alpha}_{m,n;k,l}
\times\lbt\frac{2-\alpha}{2\sqrt{1-\alpha}}
\mathds{1}_{2}-\imath\frac{\alpha}{2\sqrt{1-\alpha}}\gamma^{0}\gamma^{1} \rbt.
\end{eqnarray}

\section{NC Gross-Neveu Model in the MB}
\setcounter{equation}{0}

The Euclidean two-dimensional Gross-Neveu model is written in terms of $N$ pairs of conjugate
Grasmmann fields $\psib^a , \psi^a$, $a=1, ... N$. In the commutative case 
the interaction has the following "$N$-vector" form:
\begin{equation}
\lambda\lbt \sum_a \psib^a  \psi^a \rbt^{2}(x).
\end{equation}

There are several non-commutative generalizations of this interaction, 
depending on how to put the star product with respect to color 
and conjugation. The most general action involving a vertex with two colors,
each present on one $\psib$ and one $\psi$ takes the form (using cyclicity of the integral trace
of star products):
\begin{align}\label{completeaction}
S&= S_{free} + \int \Tr   V\lbt\psib,\psi\rbt \ ,
\nonumber\\
 V\lbt\psib,\psi\rbt&=\sum_{a,b}\lambda_{1}\psib^a\star\psi^{a}\star\psib^b\star\psi^{b}
      +\lambda_{2}\psib^a\star\psi^{b}\star\psib^b\star\psi^{a}\nonumber\\
      &+\lambda_{3}\psib^a\star\psib^b\star\psi^{a}\star\psi^{b}
      +\lambda_{4}\psib^a\star\psib^b\star\psi^{b}\star\psi^{a}
\end{align}
where $S_{free}$ is defined in (\ref{sfree}).
Recall that in the matrix basis the Grasmmann fields 
$\psib^a , \psi^a$, $a=1, ... N$ are Grassmann matrices 
$\psib^a_{mn}, \psi^a_{mn}$, $a=1, ... N$, $m,n \in {\mathbb N}$.

\begin{figure}
\begin{center}
\includegraphics[width=6cm]{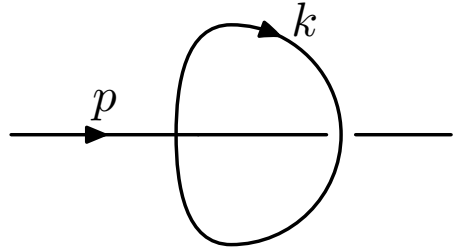}
\end{center}
\caption{The non-planar tadpole}
\label{figtadpole}
\end{figure}

In the cases $\lambda_3 =\lambda_4 = 0$, there are no non planar tadpole of the type of
figure \ref{figtadpole}
which lead to  infrared-ultraviolet mixing in $\phi^4_4$. 
Hence one may superficially conclude that there is no need
to vulcanize the free action, hence to use (\ref{sfree}). However even for
$\lambda_3 =\lambda_4 = 0$, four point functions lead
to "logarithmic" IR/UV  mixing,
hence to "renormalons" effects (large graphs with amplitudes of size $n!$ at order $n$).
Since we want anyway to renormalize generic Gross-Neveu actions in which 
$\lambda_3 \ne 0$ or $\lambda_4 \ne 0$, we shall always use the vulcanized free action
and propagator.

A proof of the BPHZ renormalization theorem according to the multiscale analysis \cite{Rivasseau2005bh}
decomposes into two main steps. First one has to prove bounds on the sliced propagator;
then using these bounds one has to prove that irrelevant operators in the multiscale analysis
give rise to convergent sums, and that this is also the case for marginal and relevant operators
{\it after} subtracting a singular part of the same form than the original action.

In this paper we provide the first part of this proof, namely the appropriate bounds.
The rest of the proof of renormalizability to all orders of the generic model (\ref{completeaction})
is postponed to a future paper \cite{RenNCGN05}.
However we illustrate on a particular example below why the bounds of 
this paper (which are optimal in a certain sense) should be sufficient for this task, 
which is however significantly more difficult
than the correpsonding task for $\phi^4_4$.

The multiscale  slice decomposition is performed as in \cite{Rivasseau2005bh}
\begin{equation}
  \label{eq:slices}
  \int_{0}^1 d\alpha = \sum_{i=1}^\infty \int_{M^{-i}}^{M^{-i+1}} 
  d\alpha
\end{equation}
and leads to the following propagator for the $i^{\text{th}}$ slice: 
\begin{align}
  \Gamma^i_{m,m+h,l+h,l} 
  &=\frac{\theta}{8\Omega}  \int_{M^{-i}}^{M^{-i+1}} d\alpha\; 
  \dfrac{(1-\alpha)^{\frac{\mu_0^2 \theta}{8 \Omega}-\frac{1}{2}}}{  
    (1 + C\alpha )} 
  \Gamma^{(\alpha)}_{m, m+h; l + h, l}\;.
  \label{prop-slice-i}
\end{align}
\begin{eqnarray}
G_{m,n;k,l}&=& \sum_{i=1}^\infty G^i_{m,n;k,l} \ ; \ G^i_{m,n;k,l} = 
-\frac{2\Omega}{\theta^{2}\pi^{2}} \int_{M^{-i}}^{M^{-i+1}} 
d\alpha\, G^{\alpha}_{m,n;k,l}  
\label{eq:matrixfullpropsliced}
\end{eqnarray}

We split $G$ according to (\ref{commterm}) and  (\ref{massterm}),
and we now bound $\vert G^i_{m,n;k,l}\vert$. We define $h= n-m$ and $p=l-m$.
By obvious symmetry of the integer indices we can assume $h \ge 0 $, and 
$p \ge 0$, so that   the smallest  of the four integers $m,n,k,l$ is $m$ and
the largest is $k=m+h+p$. The main result of this paper is the following bound:

\begin{theorem}\label{maintheorem}
Under the assumed conditions $h =n-m\ges 0 $ and $p=l-m \ges 0$
the Gross-Neveu propagator in a slice $i$ obeys the bound
\begin{eqnarray}\label{mainbound1}  
\vert G^{i,{\rm comm}}_{m,n;k,l}\vert&\les&   
K M^{-i/2} \bigg(  \frac{\exp \{- \frac{c p ^2  }{1+ kM^{-i}}
- \frac{ c M^{-i}}{1+k} (h - \frac{k}{1+C})^2  )  \}}{(1+\sqrt{ kM^{-i}}) }  
\nonumber\\
&&+ e^{- c k M^{-i} - c  p }\bigg).
\end{eqnarray}
for some (large) constant $K$ and (small) constant $c$ which depend only on $\Omega$.
Furthermore the part with the mass term has a slightly different bound:
\begin{equation} \label{mainbound2}  
\vert  G^{i,{\rm mass}}_{m,n;k,l}\vert\les   
K M^{-i} \bigg(  \frac{\exp \{- \frac{c p ^2  }{1+ kM^{-i}}
- \frac{ c M^{-i}}{1+k} (h - \frac{k}{1+C})^2  )  \}}{1+\sqrt{ kM^{-i}}}  
+ e^{- c k M^{-i} - c  p }\bigg).
\end{equation}
\end{theorem}

The rest of the paper is devoted to the proof of this theorem.
We give this proof only for $i\gg 1$, the ``first slices'' 
being unimportant for renormalization.

In the rest of this section we indicate
how this bound leads to efficient power counting estimates
for renormalization.

Recall first that for any non-commutative Feynman graph $G$ we can define
the genus of the graph, called $g$ and the number of faces ``broken by external legs", called $B$
as in \cite{GrWu04-3,Rivasseau2005bh}.
We have $g \ge 0$ and $B\ge 1$. 
The power counting established for $\phi^4_4$ in 
\cite{GrWu04-3,Rivasseau2005bh} 
involves the superficial degree of divergence of a graph
\begin{equation}
\omega (G) = (2 -N/2)  - 4 g  -2(B-1) \ ,
\end{equation}
and it is positive only for $N=2$ and $N=4$ subgraphs
with $g=0$ and $B=1$. These are the only non-vacuum graphs
that have to be renormalized. We expect the same conclusion for the two-dimensional
Gross-Neveu model, since this holds for the commutative counterpart of these models.

Let us sketch now why the multislice analysis based on bound (\ref{maintheorem}) proves that a 
graph  with internal propagators in slice $i >> 1$ and external legs in slice 1 
with $N\ge 6$ or $N=4$ and $g\ge 1$ does not require renormalization.
First remark that the second term in bound (\ref{maintheorem})  
gives exactly the same decay proved in \cite{Rivasseau2005bh}.
The $O(1)$ decay in $p$ means that the model is quasi-local in the sense
of \cite{GWR2x2}. Hence all indices except those of independent faces cost $O(1)$ to sum.
Each main "face index" is summed with the scale decay $k M^{-i}$, hence each face sum costs $M^i$
in two dimensions. Combined with the $M^{-i/2}$ scaling factors of the propagators 
in (\ref{maintheorem}) one recovers the usual power counting in $\omega$.

\begin{figure}
  \begin{center}
    \includegraphics[scale=.7]{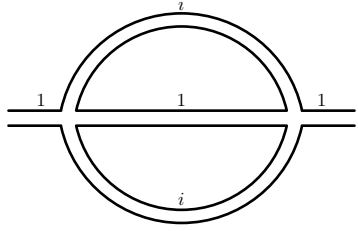}
  \end{center}
  \caption{The Sunset Graph}
  \label{figsunset}
\end{figure}

Hence let us concentrate on the more difficult case
of the first part of bound (\ref{maintheorem}), and  for instance 
consider the ``sunset" graph $G$ of Figure \ref{figsunset}

When the scales of the three internal lines are roughly identical, this graph should be
renormalized as a two-point subgraph and nothing particularly new
happens. But something new occurs when the two {\it exterior} lines have scales $i>>1$ and the interior line
has scale $1$ (like the external legs), as shown on the figure.
In the usual multiscale analysis we do not have a divergent two point 
subgraph in the traditional sense.The subgraph made of the two external lines
 is ``dangerous", i.e. has all internal scales above all external ones. But this graph 
has  $N=4$ and $g=0$, $B=2$, and it should {\it not} and in fact {\it cannot } be renormalized.
However applying bound (\ref{mainbound1})  the sum over $i$ diverges logarithmically! 
Indeed the sums over the $p$
indices cost only $O(1)$ as usual for a quasi-matrix model, 
but the two internal faces sums, together with their lines prefactor give
\begin{equation}
\sum_{k, \delta h = 0}^{\infty} 
  M^{-i/2} M^{-i/2}   e^{-M^{-i} k} \frac{e^{ -  (\delta h )^2 /k }}{1 + \sqrt k} = O(1) .
\end{equation}

What is the solution to this riddle?
In this case it is the full {\it two-point } subgraph $G$
which has to be renormalized. This works because the renormalization improvement 
brings modified ``composite propagators"  {\it solely}
on the exterior face of the graph \cite{GrWu04-3}. 
These improved propagators have scale $i$, hence they bring a factor $M^{-i}$, 
and the sum over $i$ converges.

One has to generalize this argument, and show that all the counterterms are of the right form
to complete the BPHZ theorem for this kind of models \cite{RenNCGN05}.
In short all dangerous subgraphs with $N=2,4$, $g=0$, $B=1$,  {\it and} $N=4$, $g=0$, $B=2$
should be renormalized, the last ones being renormalized by the corresponding 
{\it two-point} function counterterms.
This subtlety makes the multislice formulation of renormalization in the non-commutative 
Gross-Neveu model more complcated  than in the non-commutative $\phi^4_4$.

\section{Proof of The Main Theorem}
\setcounter{equation}{0}

We cast the propagator for $B=\Omega$ in the following form:
\begin{equation}
  \Gamma=\int_{0}^{1}~d\alpha\frac{(1-\alpha)^{-1/2}}{1+C\alpha}\Gamma^{\alpha}
\end{equation}
with:
\begin{equation}\label{expressgamma}
  \Gamma^{\alpha}=\Big{(}\frac{\sqrt{1-\alpha}}{1+C\alpha}\Big{)}^{2m+p}
  \frac{1}{(1+C\alpha)^h}\sum_{u=o}^{m}
  \Big{(}\frac{\alpha\sqrt{C(1+C)}}
  {\sqrt{(1-\alpha)}}\Big{)}^{2m+p-2u}{\cal A}(m,m+p,h,u).
\end{equation}
We have:
\begin{equation}
\Gamma^{\alpha}=e^{(2m+p)\ln\frac{\sqrt{1-\alpha}}{1+C\alpha}-
h\ln(1+C\alpha)}\sum_{0\les v=m-u\les m}e^{(2v+p)\ln
\frac{\alpha\sqrt{C(1+C)}}{\sqrt{1-\alpha}}}{\cal A}(m,m+p,h,u).
\end{equation}
We consider the regime $\alpha \ll C \ll 1$ hence we limit ourselves 
as usually to a parameter $\Omega$ close to 1. We use Stirling's formula to write:
\begin{equation}
  \Gamma^{\alpha}\les K
  \sum_{0\les v\les k - h - p }e^{f(m,l,v)}\frac{\Big{(}
    (2\pi)^4  k    (k-h) (k-p) (k - h -p ) 
    \Big{)}^{1/4}}
  {\sqrt{(2\pi)^4v(v+p)(k-v-h -p )(k-v - p)}}
\end{equation}
with:
\begin{eqnarray}
  f&=&(2k -2h - p)\ln\frac{\sqrt{1-\alpha}}{1+C\alpha}-h\ln(1+C\alpha)+
  (2v+p)\ln\frac{\alpha\sqrt{C(1+C)}}{\sqrt{1-\alpha}}
  \nonumber\\
  &&+\frac{k - h -p }{2}\ln (k - h -p )
  +\frac{k-h}{2}\ln(k-h)+\frac{k- p }{2}
  \ln(k - p )
  \nonumber\\
  &&+\frac{k}{2}\ln k
  -v\ln v-(v+p)\ln(v+p) - (k-v- p )\ln(k-v - p )
  \nonumber\\
  &&-(k-v-h - p )\ln(k-v-h - p ).
\end{eqnarray}
We define the reduced variables $x = v/k$,  $y =h/k$, 
$z=p/k$. These parameters live in the compact
simplex $0 \les x, y, z \les 1$,  $0 \les x+ y+ z \les 1$. In
our propagator bound we can replace the Riemann sum over $v$ with the 
integral. Since we have a bounded function on a compact interval, 
this is a rigorous upper bound (up to some inessential overall constant)
for $k$ large, which is the case of interest:
\begin{equation}\label{integralinx}
  \Gamma^{\alpha}\les \int_0^{1-y-z} dx  \frac{[(1-y)(1-z)(1-z-y)]^{1/4}}
  {[x(x+z)(1-x-z) (1-x-y-z)]^{1/2}} e^{k g(x,y,z)}
\end{equation}
with the function $g$ defined by:
\begin{eqnarray}
  g&=&(2-2y-z)\ln\frac{\sqrt{1-\alpha}}{1+C\alpha}
  +(2x+z)\ln\frac{\alpha\sqrt{C(1+C)}}{\sqrt{1-\alpha}}- y \ln(1+C\alpha)
  \nonumber\\
  &&+ \frac{1-y}{2}\ln(1-y) +\frac{1-z}{2}\ln(1-z)  +\frac{1-y-z}{2}\ln(1-y-z)
  \nonumber\\
  &&-x\ln x-(x+z)\ln(x+z)  -(1-x-z)\ln(1-x-z) 
  \nonumber\\
  &&-(1-x-y-z)\ln(1-x-y-z).
\end{eqnarray}
The differential is 
\begin{eqnarray}\label{diferential}
  dg &=& dx \left\{  \ln\frac{\alpha^2 C(1+C) (1-x-z) (1-x-y-z) }
    {(1-\alpha)x(x+z)}  \right\}\\
  &&+ dy \left\{  \ln\frac{(1+C\alpha)(1-x-y-z)}{(1-\alpha)\sqrt{(1-y)(1-y-z)}} 
  \right\}
\nonumber\\
&&+ dz \left\{  \ln\frac{\alpha\sqrt{C(1+C)}(1+C\alpha)(1-x-z)(1-x-y-z)}{(1-\alpha)(x+z)\sqrt{(1-z)(1-y-z)}} 
\right\}\nonumber.
\end{eqnarray}
The second derivatives are
\begin{eqnarray}\label{seconderivatives}
  \frac{\partial ^2g}{\partial x^2}&=&-\frac{1}{x}-\frac{1}{x+z}-\frac{1}{1-x-z}
  -\frac{1}{1-x-y-z},\nonumber\\
  \frac{\partial^2g}{\partial y^2}&=&\frac{1}{2(1-y)}+ \frac{1}{2(1-y-z)}- \frac{1}{1-x-y-z},\nonumber\\
  \frac{\partial^2g}{\partial z^2}&=&\frac{1}{2(1-z)}+ \frac{1}{2(1-y-z)}- \frac{1}{x+z} 
  - \frac{1}{1-x-z} - \frac{1}{1-x-y-z},\nonumber\\
  \frac{\partial^2g}{\partial x \partial y }&=&-\frac{1}{1-x-y-z},\nonumber\\
  \frac{\partial^2g}{\partial x \partial z }&=&-\frac{1}{x+z}-\frac{1}{1-x-z}-\frac{1}{1-x-y-z},\nonumber\\
  \frac{\partial^2g}{\partial y \partial z}&=&\frac{1}{2(1-y-z)}-\frac{1}{1-x-y-z}. 
\end{eqnarray}

\begin{lemma}
The function $g$ is concave in the simplex.
\end{lemma}
\prf
We have to prove that the quadratic form $-Q$
defined by the 3 by 3 symmetric matrix of the second derivatives is negative in the whole simplex.
In others words we should prove that the oppposite quadratic form
\begin{eqnarray}
  \label{concave}
  Q&=&[\frac{1}{x} +\frac{1}{x+z}+  \frac{1}{1-x-z}+\frac{1}{1-x-y-z}]u^2
  \nonumber\\
  &+& [\frac{1}{1-x-y-z} - \frac{1}{2(1-y)} - \frac{1}{2(1-y-z)}] v^2
  \nonumber\\
  &+& [ \frac{1}{x+z} 
  +\frac{1}{1-x-z} +\frac{1}{1-x-y-z} - \frac{1}{2(1-z)}- \frac{1}{2(1-y-z)} ] w^2
  \nonumber\\
  &+& \frac{2}{1-x-y-z} uv + [ \frac{2}{x+z} + \frac{2}{1-x-z} + \frac{2}{1-x-y-z} ] uw
  \nonumber\\
  &+& [\frac{2}{1-x-y-z}-\frac{1}{1-y-z}] vw 
\end{eqnarray}
\newpage
is positive. But we have
\begin{eqnarray}
  \label{concave1}
  Q&=& \frac{(u+v+w)^2}{1-x-y-z} + \frac{(u+w)^2}{(x+z)(1-x-z)}  +
  \frac{u^2}{x}
  - \frac{v^2}{2(1-y)}  - \frac{(v+w)^2}{2(1-y-z)}
  \nonumber\\
  &\ge&
  \frac{(u+v+w)^2}{1-x-y-z} + \frac{(u+w)^2}{2(x+z)} +
  \frac{u^2}{2 x} - \frac{v^2}{2(1-y)} - \frac{(v+w)^2}{2(1-y-z)}
  \nonumber\\
  &=& \frac{1}{2}\left( \left[ \frac{(u+v+w)^2}{1-x-y-z} - \frac{v^2}{1-y} + \frac{(u+w)^2}{x+z}\right] \right.
  \nonumber\\&& \quad \quad \quad + \left. \left[ \frac{(u+v+w)^2}{1-x-y-z} - \frac{(v+w)^2}{1-y-z} + \frac{u^2}{x}\right]\right)
  \nonumber\\
  &=& \frac{1}{2}\left(\frac{1}{1-y}\left[\sqrt{\frac{x+z}{1-x-y-z}}(u+v+w) + \sqrt{\frac{1-x-y-z}{x+z}}(u+w)\right]^2\right.
  \nonumber\\
  &+& \frac{1}{1-y-z}\left. \left[ \sqrt{\frac{x}{1-x-y-z}}(u+v+w) + \sqrt{\frac{1-x-y-z}{x}}u\right]^2\  \right)
  \nonumber\\
  &\ge& 0.
\end{eqnarray}
\qed

\begin{lemma}\label{uniquemaximum}
  The only critical point of the function in the closed simplex is at $x_0 = \frac{C\alpha}{1 + C\alpha}$, $y_0= \frac{1}{1 + C}$, $z=0$,
  where the function $g=0$.
\end{lemma} 
\prf 
One can easily check that:
\begin{equation}
  g(x_0,y_0,0)= \partial_x g(x_0,y_0,0)= \partial_{y}g(x_0,y_0,0)=
  \partial_{z}g(x_0,y_0,0)=0.
\end{equation}
The unicity follows from the concavity of the function $g$.
\hfill\qed
\bigskip\\
Our bound on $g$ will be inspired by the steepest descent method around the critical point.
We divide now the simplex into:
\begin{itemize}
\item the neighborhood of the maximum. We call this region the ``mountain top''.
  It corresponds to $\delta x=\vert x-x_0 \vert\ll\alpha $, $\delta y=\vert
  y-y_0\vert\ll O(1)$, $z\ll\alpha$.
  For aesthetic reasons we prefer to use a reference quadratic form $Q_0$
  to define a smooth border of this region. Hence putting $X=(\delta x, \delta y , z)$ we
  define the mountain top by the condition
  \begin{equation}\label{defmountaintop}
    XQ_0^{\phantom{0}t}\!X  = \frac{(\delta x)^2  + z^2}{\alpha^2} + (\delta y)^2 \les \eta
  \end{equation}
  where $\eta$ is a small constant,

\item the rest of the simplex. This region is defined by $XQ_0^{\phantom{O}t}\!X\ge \eta $. 
\end{itemize}

\subsection{The ``Mountain Top''}
In this region we use the Hessian approximation and check that the cubic
correction terms are small with respect to this leading order.

\begin{lemma} In the mountain top region, for some small constant $c$ 
  (which may depend on $C$, hence on $\Omega$):
  \begin{equation}\label{mainhessianbound}
    g(X) \le  - c \alpha XQ_0^{\phantom{O}t}\!X  =  - c \alpha [\frac{(\delta x)^2  + z^2}{\alpha^2} + (\delta y)^2 ]
  \end{equation}
\end{lemma}

\prf
>From (\ref{seconder}) we evaluate the second order derivatives of $g$ at leading order 
in $\alpha$ at the maximum: 
\begin{align}
  \frac{\partial ^2g}{\partial x^2} &\approx -\frac{2}{\alpha C
  };&\frac{\partial^2g}{\partial x \partial z }&\approx -\frac{1}{\alpha
    C},\nonumber\\
  \frac{\partial^2g}{\partial z^2} &\approx -\frac{1}{\alpha C};& \frac{\partial^2g}{\partial x \partial y } &\approx
  -\frac{1+C}{C},\nonumber\\  
  \frac{\partial^2g}{\partial y^2} &\approx -\alpha\frac{1+C}{C};&  
  \frac{\partial^2g}{\partial y \partial z}&\approx -\frac{1+C}{2C}.\label{seconder}
\end{align}

It is easy to diagonalize
the corresponding 3 by 3 quadratic form, and to check that
in the neighborhood of the maximum it is smaller than $- c \alpha  Q_0$ 
for some small constant $c$:
\begin{equation} 
  \label{hessianbound}
  g_{Hessian}(\delta x,\delta y,z)= X  Q_{Hessian}^{\phantom{Hessian}t}\!X  
  \les -\alpha c XQ_0^{\phantom{O}t}\!X  =-\alpha c
  \Big{(} \frac{(\delta x)^2 + z^2}{\alpha^2}
  +(\delta y)^2 \Big{)},
\end{equation}
where $g_{Hessian}$ is the Hessian approximation to the function $g$.
It is easy to check from the expression (\ref{diferential}) of the differential $dg$
that the third order derivatives scale in the appropriate way so that choosing 
the constant $\eta$ small enough in (\ref{defmountaintop}), 
the function $g$ obeys the same bound than (\ref{hessianbound}) with 
a slightly different constant $c$. 
\hfill\qed

\subsection{The Rest of the Simplex}

To bound the function $g$ in the whole simplex we use the previous notation $X=(\delta x, \delta y, z)$. 
Drawing the segment from point $X$ to the origin (i.e. the mountain top), we
cross the border of the mountain top at $X_0= \lambda X$ with $X_{0}Q_0^{\phantom{O}t}\!X_{0} =\eta$. We
define $X_1= X_0/2= (\lambda/2) X $.  $X$ out of the mountain top means that
$\lambda = \sqrt{\eta/XQ_0^{\phantom{O}t}\!X}\les 1$. 

\begin{lemma}\label{firstoutbound}
Out of the mountain top region the function $g(X) = g(\delta x, \delta y, z)$ 
obeys the bound, for some small enough constant $c$:
\begin{equation}\label{goodlargefieldbound}
g(\delta x, \delta y, z) \les - c (\alpha   + \vert \delta x \vert + z).
\end{equation}
\end{lemma}
\prf
We use concavity of the function on the segment considered, which means that the function $g$ 
is below its first order Taylor approximation at $X_1$:
\begin{equation} 
  g(X) \les g(X_1) +  \langle dg(X_1),X-X_1\rangle. 
\end{equation}
At $X_1$ the Hessian approximation of $g$ is valid, say up to a factor 2. Hence
\begin{equation} 
  g(X) \les (1/2)[ X_1 Q_{Hessian}^{\phantom{Hessian}t}\!X_1  +  2(X-X_1) Q_{Hessian}^{\phantom{Hessian}t}\!X_1 ].
\end{equation} 
Using (\ref{hessianbound}) we can relate $Q_{Hessian}$ to our reference quadratic form 
$Q_0$ up to a constant and get for some small $c$:
\begin{eqnarray} 
  g(X) &\les& - c \alpha [ X_1 Q_{0}^{\phantom{0}t}\!X_1  +  2(X-X_1) Q_{0}^{\phantom{0}t}\!X_1 ] 
  \nonumber\\ &=& - c \alpha  [(\eta/4)  + \lambda(1-\lambda/2)XQ_0^{\phantom{0}t}\!X ] 
  \nonumber\\ &=& - c' \alpha  [1  + \sqrt{XQ_0^{\phantom{0}t}\!X}] 
\end{eqnarray} 
for some constant $c'$ smaller than $c$. In the last line we used $1-\lambda/2\ges 1/2$ and
$\lambda = \sqrt{\eta/XQ_0^{\phantom{0}t}\!X}$.
Finally 
\begin{equation}
  \alpha \sqrt{X Q_0 ^tX} \ge  \sqrt{(\delta x)^2  + z^2}\ges (\vert \delta x \vert  + z)/\sqrt{2}
\end{equation}
completes the proof of (\ref{goodlargefieldbound}).
\hfill\qed

\subsection{Integration on $\boldsymbol{x}$}
It remains now to prove some explicit decay of the function
$G$ in the variable $z$ after integration in $x$ in (\ref{integralinx}). 
The decay in $z$ is necessary to prove that the model is a quasi-local matrix model 
in the sense of \cite{GrWu03-1}. For the mountain top region,
we do not have any decay in $k$ so we want also to exhibit the decay in y.
\begin{lemma}\label{goodbound}
  For some large constant  $K$ and small constant $c$, under the condition $\alpha k \ge 1$
  we have
  \begin{equation}\label{auxbound2} 
    \Gamma^{\alpha}\les  K \bigg(  \frac{\exp \{- \frac{c}{\alpha k} p^2  
      - \frac{ c \alpha}{k} (h - \frac{k}{1+C})^2  ) \}}{\sqrt{\alpha k}}  
    + e^{- c \alpha k - c  p }   \bigg).
  \end{equation}
\end{lemma}
\prf
In the integration on $x$ in (\ref{integralinx}) we can insert 
$1 = \chi + (1-\chi)$ where $\chi$ is the characteristic function of the mountain top.
In the first term we apply the bound (\ref{mainhessianbound}) and in the
second the bound (\ref{goodlargefieldbound}). In this second case 
we use a better estimation of the prefactors in front of $e^{kg}$ in
(\ref{integralinx}). Actually their expression in (\ref{integralinx}) leads to
a spurious logarithmic divergence due to the bad behaviour of the Stirling
approximation close to 0. We will use
\begin{eqnarray}
  K\sqrt{n+1}\,n^{n}e^{-n}\les&n!&\les K'\sqrt n\,n^{n}e^{-n},
\end{eqnarray}
wich leads to an integral of the type 
\begin{align}
  \int_{0}^{1}\frac{dx}{x+1/k}e^{-ckx}&\les
  K\int_{1/k}^{\infty}\frac{dx}{x}e^{-ckx},
\end{align}
which is bounded by a constant. Scaling back to the original variables
completes the proof of Lemma \ref{goodbound}.
\hfill\qed

\subsection{The Region $\boldsymbol{\alpha k\les 1}$}
In this region we do not need Stirling's formula at all.
It is easier to derive a direct simple bound on $\Gamma^{\alpha}_{m,m+h;k,m+p}$ (recall that $k=m+h+p$):
\begin{lemma}\label{lemma:boundakpetit}
  For $M$ large enough and $\alpha k< 1$, there exists constants $K$ and $c$  such that
  \begin{equation}
    \label{eq:propppetit}
    \Gamma^{\alpha}_{m,m+h;k,m+p}\les Ke^{-c(\alpha k+p)}.
  \end{equation}
\end{lemma}
{\proof} We assume $h\ges 0$ and $p\ges 0$. For $\alpha k\les 1$, 
the following crude bound on $\Gamma^{\alpha}$ follows easily from (\ref{expressgamma}):
\begin{equation}\label{eq:kpetit1}
  \Gamma^{\alpha}_{m,m+h;k, m+p}\les 
  e^{-\alpha(C+1/2)(2m+p)-(C/2)\alpha h}
  \sum_{u=0}^{m}\frac{X^{m-u}}{(m-u)!}\frac{Y^{m+p-u}}{(m+p-u)!}
\end{equation}
where $X=\frac{\alpha\sqrt{C(C+1)}}{\sqrt{1-\alpha}}\sqrt{m(m+h)}$ and $Y=\frac{\alpha\sqrt{C(C+1)}}{\sqrt{1-\alpha}}\sqrt{(m+p)(m+p+h)}$. 
\begin{align}
  \label{eq:kpetit2}
  \Gamma^{\alpha}_{m,m+h,l+h,l}&\les e^{-\alpha(C+1/2)(2m+p)-(C/2)\alpha h}
  \nonumber\\  
  &\frac{\big( \frac{\alpha k \sqrt{C(C+1)}}{\sqrt{1-\alpha}} \big)^{p}}{p!}
  \sum_{u=0}^{m}\frac{\big( \frac{\alpha k \sqrt{C(C+1)}}{\sqrt{1-\alpha}} \big)^{2(m-u)}}{(m-u)!^{2}}
\end{align}
The sum over $u$ is bounded by a constant. For $C$ small (i.e. $\Omega $ close to 1) we have certainly
$\frac{\sqrt{C(C+1)}}{\sqrt{1-\alpha}}\les 1/2$ hence we get the
desired result.\qed
\bigskip\\
Combining with (\ref{auxbound2}) we conclude that Lemma \ref{goodbound} always holds:
\begin{lemma}
  \label{goodbound1}
  For some large constant  $K$ and small constant $c$
  we have
  \begin{equation}\label{auxbound3} 
    \Gamma^{\alpha}\les  K \bigg(  \frac{\exp \{- \frac{c}{1+\alpha k} p^2  
      - \frac{ c \alpha}{1+ k} (h - \frac{k}{1+C})^2  ) \}}{1 + \sqrt{\alpha k}}  
    + e^{- c \alpha k - c  p }\bigg).
  \end{equation}
\end{lemma}

\subsection{Numerator Terms}

In this section we check that the numerators in the Gross-Neveu propagator
bring the missing power counting factors, hence we complete the proof of
Theorem \ref{maintheorem}.

The bound (\ref{mainbound2}) is nothing but the direct consequence of
multiplying the bound of Lemma \ref{goodbound1} by the width $M^{-i}$
of the integration interval over $\alpha$. Hence we now prove (\ref{mainbound1}).

In the commutator terms (\ref{commterm})
the $\Omega$ and the $\imath\frac{\alpha}
{2\sqrt{1-\alpha}}\gamma^{0}\gamma^{1}$
are smaller by at least an $\alpha$ factor. Therefore the largest piece
is the $O(1/\alpha)[\tilde x , \Gamma]$ term. The $1/alpha$ factor 
compensates the  width $M^{-i}$
of the integration interval over $\alpha$. Hence we need to prove that a $[\tilde x , \Gamma]$
numerator adds  a factor $\sqrt{\alpha}$ to the bound of Lemma \ref{goodbound1}:

The $\Omega$ and the $\imath\frac{\alpha}{2\sqrt{1-\alpha}}\gamma^{0}\gamma^{1}$
terms in (\ref{commterm}) have an additional factor $\alpha$, hence are much easier 
to bound and left to the reader. The bound for the mass term $\mu \Gamma$ just
involves multpliying the bound of Lemma \ref{goodbound1} by the appropriate weight
$M^{-i}$ coming from the $\alpha$ integration, which leads easily to (\ref{mainbound2}).

The commutator  $[\slashed x , \Gamma]$ involves terms like  
\begin{align}
&\sqrt{m+1}\Gamma_{m+1,n;k,l}-\sqrt{l}\Gamma_{m,n;k,l-1}\nonumber\\
=&\lbt\sqrt{m+1}-   \sqrt{l}\rbt\Gamma_{m,n;k,l-1} 
+ \sqrt{m+1}\lbt\Gamma_{m+1,n;k,l}-\Gamma_{m,n;k,l-1}\rbt. \label{eq:typnum}
\end{align}

\subsubsection{The first term}
The first  term is the easiest to bound. It is zero unless $p=l-m-1 \ges 1$.
In this case,  we have 
\begin{equation}
\sqrt{m+1}-   \sqrt{l} \les  \frac{2p}{1 + \sqrt l}.
\end{equation}

Using Lemma \ref{lemma:dl}: 
\begin{itemize}
\item On the mountain top we have $l =k-h \simeq \frac{C}{1+C} k$,
hence $\frac{2p}{1 + \sqrt l} \les O(1) \frac{p}{1 + \sqrt k}  $. An additional
factor $\alpha$ comes from (\ref{inboundd}), hence we have a bound in $O(1) \frac{\alpha  p}{1 + \sqrt k}$.
Using a fraction of the decay $ e^{-c \alpha p^2/k}$ from Lemma \ref{goodbound1} bouinds
$O(1) \frac{\alpha  p}{1 + \sqrt k}$ by $\sqrt{\alpha}$.

\item Out of the mountain top we have a factor $\alpha \sqrt{kl}$ which comes from 
(\ref{outboundd}), and we can use $p e^{-cp} \les  e^{-c'p}$ and 
$\sqrt{\alpha k} e^{-c\alpha k} \les  e^{-c'\alpha k}$. Hence the desired factor $\sqrt\alpha$
is obtained.

\end{itemize}

It remains to prove
\begin{lemma}\label{lemma:dl}
Let $(m,l,h)\in\mathds{N}^{3}$ with $p=l-m-1 \ges 1$. We have 
\begin{itemize}
\item On the mountain top
\begin{equation}\label{inboundd}
      \Gamma_{m,m+h;k,m+p}\les K\alpha\,\Gamma_{m,m+h;m+p-1+h,m+p-1},
    \end{equation}
  \item if $\Gamma$ is out the mountain top region
    \begin{equation}\label{outboundd}
      \Gamma_{m,m+h;k,m+p}\les K\alpha \sqrt{kl}\,\Gamma_{m,m+h;m+p-1+h,m+p-1}.  
    \end{equation}
  \end{itemize}
\end{lemma}
{\proof} Let $p\ges 1$. The kernel $\Gamma_{m,m+h;m+p+h,m+p}$ is given by
\begin{align}
  \Gamma_{m,m+h;m+p+h,m+p}=&\lbt\frac{\sqrt{1-\alpha}}{1+C\alpha}
  \rbt^{2m+p}(1+C\alpha)^{-h}\nonumber\\
  &\sum_{u=0}^{m}(\alpha D)^{2(m-u)+p}{\cal A}(m,m+p,h, u)  
\end{align}
with ${\cal A}$ defined by (\ref{eq:A}) and $D(\alpha)=\sqrt{\frac{C(C+1)}{1-\alpha}}$.

The scaling factor $(\alpha D)^{2(m-u)+p}$ reaches its maximum at
$u=m$. There we have a factor $(\alpha D)^{p}$. For $p\ges 1$,
we factorize $\alpha D$ to get
\begin{align}
  \Gamma_{m,m+h;m+p+h,m+p}=&\alpha
  D\lbt\frac{\sqrt{1-\alpha}}{1+C\alpha}\rbt^{2m+p}(1+C\alpha)^{-h}\nonumber\\
  &\sum_{u=0}^{m}(\alpha D)^{2(m-u)+p-1}{\cal A}(m,m+p,h, u).
\end{align}
Using $\binom{m+p}{m+p-u}=\frac{m+p}{m+p-u}\binom{m+p-1}{m+p-1-u}$ and $\binom{m+p+h}{m+p-u}=\frac{m+p+h}{m+p-u}\binom{m+p-1+h}{m+p-1-u}$, we have
\begin{align}
  \Gamma_{m,m+h;m+p+h,m+p}&\les K\alpha\lbt\frac{\sqrt{1-\alpha}}{1+C\alpha}\rbt^{2m+p-1}
  (1+C\alpha)^{-h}\sum_{u=0}^{m}(\alpha
  D)^{2(m-u)+p-1}\nonumber\\
  &\hspace{1cm}\times{\cal A}(m,m+p-1,h, u)\frac{\sqrt{(m+p)(m+p+h)}}{m+p-u}.
\end{align}
On the ``mountain top'' we express $m+p-u$ in term of $k$ and get
\begin{align}
  \frac{1}{m+p-u}&=\frac{1}{k-u}\ \frac{1}{1-\frac{h}{k-u}}.
\end{align}
Moreover $u\simeq\alpha k$ and $(k-u)^{-1}\simeq
k^{-1}$, which leads to  (\ref{inboundd}).

Out of the mountain top we use simply $m+p-u \ges 1$ and\\$\sqrt{(m+p)(m+p+h)}= \sqrt{kl}$,
and (\ref{outboundd}) follows.
\qed

\subsubsection{The second term}

We now focus on the terms involving differences of $\Gamma$'s. For this we need
some identities on the combinatorial factor ${\cal A}$: 
\begin{align}
  {\cal A}(m,l,h,u)&=\frac{\sqrt{ml}}{u}{\cal A}(m-1,l-1,h+1,u-1),\text{ for
    $u\ges 1$},\label{eq:A-1}\\
  {\cal A}(m,l,h,u)&=\frac{\sqrt{m(m+h)}}{m-u}{\cal A}(m-1,l,h,u),\label{eq:Am-1}\\
  {\cal A}(m,l,h,u)&=\frac{\sqrt{m(m+h)l(l+h)}}{(m-u)(l-u)}
  {\cal A}(m-1,l-1,h,u).\label{eq:Aml-1}.
\end{align}
Let us recall $h=n-(m+1)$ and $p=l-(m+1)$. Then
$\Gamma_{m+1,n;k,l}=\Gamma_{m+1,m+1+h;l+h,l}$ and
$\Gamma_{m,n;k,l-1}=\Gamma_{m,m+h+1;l+h,l-1}$. 

Using Lemma (\ref{thm:numerator-terms}) we get:
\begin{equation}
  \sqrt{m+1}\lbt\Gamma_{m+1,n;k,l}-\Gamma_{m,n;k,l-1}\rbt
        \les K\sqrt{\alpha}\Gamma^{\alpha},
\end{equation}
  on the mountain top by (\ref{numtermmountop}) and the sequel, as well as 
out of the critical region by (\ref{numtermouttop}). The factors thus 
obtained together with Lemma (\ref{goodbound1})
complete the proof of (\ref{maintheorem}).

\begin{lemma}\label{thm:numerator-terms}
For $M$ large enough and $\Omega$ close enough to 1, there exist
$K$ such that 
\begin{equation}
\label{eq:fullprop-bound}
    G^{\alpha}_{m,n;k,l}\les K\sqrt{\alpha}\Gamma^{\alpha}_{m-1,n; k-1, l-2}.
  \end{equation}
\end{lemma}
{\proof} It remains to bound the second term in (\ref{eq:typnum}),\\ 
namely $\sqrt{m+1}\lbt\Gamma_{m+1,n;k,l}-\Gamma_{m,n;k,l-1}\rbt$. 
\begin{align}
  &\Gamma_{m+1,m+1+h;l+h,l}-\Gamma_{m,m+h+1;l+h,l-1}\\
  =&\lbt\frac{\sqrt{1-\alpha}}{1+C\alpha}\rbt^{m+1+l}(1+C\alpha)^{-h}
  \sum_{u_{1}=0}^{m+1}(\alpha D)^{m+1+l-2u_{1}}{\cal A}(u_{1};m+1,l,h)\nonumber\\
  &-\lbt\frac{\sqrt{1-\alpha}}{1+C\alpha}\rbt^{m+l-1}(1+C\alpha)^{-h-1}
  \sum_{u_{2}=0}^{m}(\alpha
  D)^{m+l-1-2u_{2}}{\cal A}(u_{2};m,l,h+1)\nonumber.
\end{align}

Thus we conclude, again up to boundary terms (treated in Subsection \ref{subsubboundary} below)
\begin{align}\label{boundaryterm1}
  &\Gamma_{m+1,m+1+h;l+h,l}-\Gamma_{m,m+h+1;l+h,l-1}\\
  &\les \lbt\frac{\sqrt{1-\alpha}}{1+C\alpha}\rbt^{m+l+1}(1+C\alpha)^{-h}
  \sum_{u_{1}=1}^{m+1}(\alpha D)^{m+1+l-2u_{1}}{\cal A}(u_{1};m+1,l,h)\nonumber\\
  &-\lbt\frac{\sqrt{1-\alpha}}{1+C\alpha}\rbt^{m+l-1}(1+C\alpha)^{-h-1}
  \sum_{u_{2}=0}^{m}(\alpha
  D)^{m+l-1-2u_{2}}{\cal A}(u_{2};m,l-1,h+1)\nonumber
\end{align}
Let $u_{1}=u+1$. Thanks to (\ref{eq:A-1}) we write the sum over $u_{1}$
as 
\begin{align}
  &\sum_{u_{1}=1}^{m+1}(\alpha D)^{m+1+l-2u_{1}}{\cal
    A}(u_{1};m+1,l,h)\nonumber\\
  =&\sum_{u=0}^{m}(\alpha D)^{m+l-1-2u}{\cal A}(u;m,l-1,h+1)\frac{\sqrt{l(m+1)}}{u+1}.
\end{align}
The difference in (\ref{boundaryterm1}) is now written as 
\begin{align}
  {\cal D}\defi&\lbt\frac{\sqrt{1-\alpha}}{1+C\alpha}\rbt^{m+l-1}(1+C\alpha)^{-h-1}
  \sum_{u=0}^{m}(\alpha
  D)^{m+l-1-2u}{\cal A}(u;m,l-1,h+1)\nonumber\\
  &\times\lb\lbt\frac{\sqrt{1-\alpha}}{1+C\alpha}\rbt^{2}(1+C\alpha)\frac{\sqrt{l(m+1)}}
  {u+1}-1\rb. \label{bondaryterm2}
\end{align}
The factor between braces is expressed as
$\lb\frac{\sqrt{l(m+1)}}{u+1}-1-\alpha\frac{1+C}{1+C\alpha}\frac{\sqrt{l(m+1)}}{u+1}\rb$.
The scaling factor $(\alpha D)^{m+l-1-2u}$ is maximum for $u=m$ where it
reaches $(\alpha D)^{p}$. We can consider up to bondary terms that
the $u$-sum goes only up to $m-1$. We can factorize $\alpha^{2}$ and prove that
the remaining terms are smaller than $1/\alpha$ on the ``mountain top'' or
$k$ outside this critical region.

For the $u$-sum, with $u$ running only to $m-1$ we have:
\begin{align}
  {\cal D} = i&\lbt\frac{\sqrt{1-\alpha}}{1+C\alpha}\rbt^{m+l-1}
  (1+C\alpha)^{-h-1}\sum_{u=0}^{m-1}(\alpha
  D)^{m+l-1-2u}{\cal A}(u;m,l-1,h+1)\nonumber\\
  &\times\lb\frac{\sqrt{l(m+1)}}{u+1}-1-\alpha\frac{1+C}{1+C\alpha}
  \frac{\sqrt{l(m+1)}}{u+1}\rb,\\
  {\cal D}\les&K\alpha^{2}\lbt\frac{\sqrt{1-\alpha}}{1+C\alpha}\rbt^{m+l-3}
  (1+C\alpha)^{-h-1}\nonumber\\
  &\times\sum_{u=0}^{m-1}(\alpha D)^{m+l-3-2u}{\cal A}(u;m-1,l-2,h+1)\nonumber\\
  &\times\frac{\sqrt{m(m+h+1)(l-1)(l+h)}}{(m-u)(l-1-u)}
  \lb\frac{\sqrt{l(m+1)}}{u+1}-1-c_{1}\alpha\frac{\sqrt{l(m+1)}}{u+1}\rb\nonumber\\
  = &K  \sum_{u=0}^{m-1} \Gamma (u,m-1,l-2, h+1)  E(u, m,l,h)
\end{align}   
where    $c_1 =\frac{1+C}{1+C\alpha} $, and
\begin{eqnarray}   
E (u, m,l,h) = \alpha^{2} \frac{\sqrt{m(m+h+1)(l-1)(l+h)}}{(m-u)(l-1-u)(u+1)(1+ C \alpha)}
\times\nonumber\\   
\lb(1- \alpha) \sqrt{l (m+1)} -(1+ C\alpha )  (u+1) \rb\nonumber .
\end{eqnarray}
Once more the subsequent procedure depends on the value of the indices
in the configuration space $\N^{3}$. If they stand on the mountain top, we pass to the variables
$x = v/k = C\alpha /(1+C\alpha )  + \delta x $, $  y = h+1 /k  = 1/(1+C)  + \delta y$, 
$z = p/k = \delta z$. We can assume $\alpha k\ges 1$ and we can evaluate 
$E$ as:
\begin{eqnarray}
E &\les &   O(1) [  (1- \alpha)  \sqrt{(1-y) (1-y-z)}  -   (1+ C \alpha) (1-x-y-z) ]    +   O(1/k) 
\nonumber\\ 
& \les & O( \vert \alpha \delta y \vert  + \vert \delta x \vert  + \vert \delta z \vert )  +   O(1/k) 
\label{numtermmountop}
\end{eqnarray}
The rest of the sum recontstructs the usual mountain bound of lemma \ref{goodbound1}.
Using the Hessian estimates $\delta y  \simeq 1/ \sqrt{\alpha k}$, $\delta x \simeq \delta z \simeq \sqrt {\alpha k} $, 
the $E$ correction simply provides an additional
factor $\sqrt {\alpha /k}$ to the estimate of Lemma \ref{goodbound1}. 
Adding the $\sqrt{m} \les \sqrt{k}$ factor we recover the desired $\sqrt{\alpha}$ factor.

If the indices are outside the critical region,
\begin{eqnarray}
{\cal D}& \les&  K\alpha^{2}\Gamma_{m-1,m+h;l+h-1,l-2} \max_{u\les
    m-1}\lbt\frac{\sqrt{m(m+h+1)(l-1)(l+h)}}{(m-u)(l-1-u)}\right.\\
&\times&\left.\lb\frac{l-1-u}{u+1}-c_{1}\alpha\frac{\sqrt{l(m+1)}}{u+1}\rb\rbt \les K\alpha^{2}\Gamma_{m-1,m+h;l+h-1,l-2}\,k\sqrt{1+\frac pm}\nonumber\\
 && \sqrt{m+1}\ {\cal D}\ \les \  K\sqrt\alpha\, e^{-c(p+\alpha k)}
\label{numtermouttop}
\end{eqnarray}
This is a better bound.\qed

\subsubsection{Boundary Terms}
\label{subsubboundary}
For purists we add the treatement of the boounary terms, and show that they do not perturb the previous computations. We start with the $u_{1}=0$ term in (\ref{boundaryterm1}). We have:
\begin{align}
  {\cal O}\defi&\lbt\frac{\sqrt{1-\alpha}}{1+C\alpha}\rbt^{m+1+l}(1+C\alpha)^{-h}(\alpha
  D)^{m+1+l}{\cal A}(0;m+1,l,h)\nonumber\\
  =&\lbt\frac{\sqrt{1-\alpha}}{1+C\alpha}\rbt^{m+1+l}(1+C\alpha)^{-h}(\alpha
  D)^{m+1+l}\sqrt{\binom{m+1+h}{m+1}\binom{l+h}{l}}\nonumber\\ 
  \les&(\alpha D)^{m+1+l}e^{-\alpha(C+1/2)(m+1+l)-\frac{C'}{2}
  \alpha h}\max_{h}e^{-\frac{C'}{4}\alpha
    h}\frac{(m+1+h)^{(m+1)/2}}{\sqrt{(m+1)!}}\nonumber\\
  &\hspace{1cm}\times\max_{h}e^{-\frac{C'}{4}\alpha h}\frac{(l+h)^{l/2}}{\sqrt{l!}}\\
  \les&\alpha^{(m+1+l)/2}\left(\frac{D\sqrt
      2}{\sqrt{C'}}\right)^{m+1+l}e^{-\alpha(C+1/2-\frac{C'}{4})(m+1+l)-\frac{C'}{2}\alpha
    h}\nonumber\\
  \les&(9\alpha)^{(m+1+l)/2}e^{-\frac 12\alpha(m+1+l)-\frac{C'}{2}
  \alpha h}\les(9\alpha)^{p/2}e^{-c\alpha k}
\end{align}
Then $\sqrt{m+1}{\cal O}\les K\sqrt\alpha\,e^{-c(\alpha k+p)}$ which is the
desired result.

\vspace*{\stretch{2}}

We now study the $u=m$ term in (\ref{bondaryterm2})and prove that it obeys (\ref{eq:fullprop-bound}):
\begin{align}
  {\cal M}\defi&\lbt\frac{\sqrt{1-\alpha}}{1+C\alpha}\rbt^{m+l-1}(1+C\alpha)^{-h-1}(\alpha
  D)^{p}{\cal A}(m;m,l-1,h+1)\nonumber\\
  &\times\lb\lbt\frac{\sqrt{1-\alpha}}{1+C\alpha}\rbt^{2}(1+C\alpha)
  \frac{\sqrt{l(m+1)}}{m+1}-1\rb\nonumber\\
  \les&\,\Gamma_{m,m+h+1;l+h,l-1}\times\lb\frac{\sqrt{l(m+1)}}{m+1}-1-
  \alpha\frac{1+C}{1+C\alpha}\frac{\sqrt{l(m+1)}}{m+1}\rb\nonumber\\
  \les&\,\Gamma_{m,m+h+1;l+h,l-1}\times\lb\frac{l}{m+1}-1-
  \alpha\frac{1+C}{1+C\alpha}\sqrt{1+\frac{p}{m+1}}\rb\nonumber\\
  \les&\,K\Gamma_{m,m+h+1;l+h,l-1}\times\lb\frac{p}{m+1}+\alpha\rb.\label{eq:um3}
\end{align}
Remark that the $u=m$ term is not on the mountain top so that for the function
$\Gamma_{m,m+h+1;l+h,l-1}$ appearing in the equations (\ref{eq:um3}) we use
the bounds outside the mountain top region. Then we have to treat two
different cases according to the value of $\alpha k$. If $\alpha k\les 1$,
thanks to (\ref{eq:kpetit1}) 
\begin{align}
  {\cal M}\les& e^{-c\alpha k}\lbt\frac{\alpha
    \sqrt{(m+p)(m+1+p+h)}}{2}\rbt^{p}\times\lb\frac{p}{m+1}+\alpha\rb,\nonumber\\
  \sqrt{m+1}{\cal M}\les&\alpha p\sqrt{\frac{(m+p)(m+1+p+h)}{m+1}}\lbt\frac{\alpha
    k}{2}\rbt^{p-1}e^{-c\alpha k}\nonumber\\
  &+\alpha\sqrt{m+1}e^{-c\alpha k-c'p}\nonumber\\
  \les&\lbt\alpha p\sqrt{k}\sqrt{1+\frac{p}{m+1}}+\sqrt\alpha\rbt
  e^{-c(p+\alpha k)}\les K\sqrt\alpha e^{-c(p+\alpha k)}.
\end{align}
If $\alpha k\ges 1$,
\begin{align}
  \sqrt{m+1}{\cal M}\les&\lbt\frac{p}{\sqrt{m+1}}+\alpha\sqrt{m+1}\rbt
  e^{-c(p+\alpha k)}\les K\sqrt\alpha e^{-\frac c2(p+\alpha k)}.
\end{align}
\bigskip


\section{The ordinary B=0 bound}
\setcounter{equation}{0}
\label{app1}

In this section we use the above analysis to revisit
the bounds on the $\phi^4_4$ propagator given in \cite{Rivasseau2005bh}.
By convention the indices of the $\phi^{4}$ propagator are two-dimensionnal indices.
Using obvious notations
\begin{align}
  G^{\phi^{4}}_{m,m+h;l+h,l}&=\int_{0}^{1}d\alpha\,
  \frac{(1-\alpha)^{\frac{\mu^{2}\theta}{8\Omega}}}{(1+C\alpha)^{2}}\,
  G^{\alpha,\phi^{4}}_{m,m+h;l+h,l},\label{eq:propphi4}\\
  G^{\alpha,\phi^{4}}_{m,m+h;l+h,l}&=\Gamma^{\alpha}_{m,m+h;l+h,l}\,
  (1-\alpha)^{h/2}\les\Gamma^{\alpha}_{m,m+h;l+h,l}\,e^{-\frac{\alpha
      h}{2}}.
\end{align}
We have then the following result (recalling $p=l-m$)
\begin{theorem}\label{thm:phi4}
The $\phi^4_4$ propagator in a slice obeys the bound
\begin{equation}\label{eq:boundphi4}  
    G^i_{m,n;k,l}\les K M^{-i}\min\lbt 1,(\alpha k)^{p}\rbt e^{-c(\alpha k+p)}
  \end{equation}
  for some (large) constant $K$ and (small) constant $c$ which depend only on $\Omega$.
\end{theorem}
{\proof} In this context the analysis of Lemmas \ref{lemma:boundakpetit} and
\ref{goodbound1} leads to the modified bound
\begin{eqnarray}
  G^\alpha_{m,n;k,l}&\les& K e^{-\frac{\alpha h}{2}}
  \Big{(}\frac{\exp\{-\frac{c}{\alpha k}p^2  
    - \frac{c\alpha}{k} (h - \frac{k}{1+C})^2\}}{\sqrt{\alpha k}}
   \nonumber\\
   &+& \min\lbt 1,(\alpha k)^{p}\rbt e^{-c(\alpha k+p)}
   \Big{)}.
   \label{eq:phi4proof1}
\end{eqnarray}
The second term is already of the desired form in (\ref{eq:boundphi4}). Moreover:
\begin{align}
  \exp\lb-\frac{c\alpha}{k}\lbt h - \frac{k}{1+C}\rbt^2-\frac{\alpha
    h}{2}\rb&\les\exp\lb -\frac{c}{1+C}\alpha k-\alpha h\lbt \frac 12-\frac{2c}{1+C}\rbt\rb.
\end{align}
Then choosing $c$ small enough, the first term in (\ref{eq:phi4proof1}) is bounded by
\begin{align}
  K \exp\lb -c\alpha k-\frac{c}{\alpha k}p^{2}\rb&\les K\exp\lb -\frac
  c2\alpha k -c\sqrt2\,p\rb.
\end{align}
This completes the proof.\qed
\bigskip\\
Remark that in contrast with the Gross-Neveu case, the $\phi^{4}$ propagator has
no critical point in index space. The bound is the same for all $(m,l,h)\in\N^{3}$ and looks
like the bound (\ref{goodlargefieldbound}). Note also that the bound (\ref{eq:boundphi4}) allows to recover the
propositions 1, 2, 3 and 4 of \cite{Rivasseau2005bh} in a very direct manner.

\section{Fermionic Kernel}
\setcounter{equation}{0}
\label{app2}

 Let us compute first
\begin{eqnarray}
  Q^{-1}&=&\int_{0}^{\infty}dt\, e^{-tQ},\nonumber\\
  Q^{-1}(x,y)&=&\frac{\Omega}{\theta\pi}\int_{0}^{\infty}dt\frac{e^{-t\mu^{2}}}{\sinh(2\Ot t)}\,
  e^{-\frac{\Ot}{2}\coth(2\Ot t)(x-y)^{2}+\imath\Ot x\wedge y}e^{-2\imath\Ot
    t\gamma^{0}\gamma^{1}}\nonumber\\
  &\defi&\frac{\Omega}{\theta\pi}\int_{0}^{\infty}dt\,e^{-t\mu^{2}}e^{-2\imath\Ot
    t\gamma^{0}\gamma^{1}}\Gamma^t(x,y), \nonumber\\
\Gamma^t(x,y)  &=&
\frac{1}{\sinh(2\Ot t)}\,
e^{-\frac{\Ot}{2}\coth(2\Ot t)(x-y)^{2}+\imath\Ot x\wedge y}
\end{eqnarray}
with $\Ot=\frac{2\Omega}{\theta}$ and $x\wedge y=x^{0}y^{1}-x^{1}y^{0}$.

We only have to check that $e^{-tQ}$ is a solution of 
\begin{equation}
  \label{eq:expQ}
  \frac{dP}{dt}+QP=0.
\end{equation}
\vspace*{\stretch{2}}
The constant is fixed by the requirement that in the limit $\Omega\to 0$, $Q^{-1}$
goes to the usual heat kernel.
\begin{eqnarray}
  \frac{d\Gamma^t}{dt}&=&\frac{e^{-\frac{\Ot}{2}\coth(2\Ot t)(x-y)^{2}
  +\imath\Omega x\wedge y}}{\sinh(2\Ot t)}
    \nonumber\\
    &&\lbt-\frac{2\Ot\cosh(2\Ot t)}{\sinh(2\Ot
    t)}+\frac{\Ot^{2}}{\sinh^{2}(2\Ot t)}(x-y)^{2}\rbt,\nonumber\\
  \partial^{\nu}\Gamma^t&=&\frac{e^{-\frac{\Ot}{2}\coth(2\Ot
      t)(x-y)^{2}-\imath\Omega x\cdot\yt}}{\sinh(2\Ot t)}\lbt-\Ot\coth(2\Ot
  t)(x-y)^{\nu}-\imath\Omega\yt^{\nu}\rbt,\nonumber\\
  \Delta\Gamma^t&=&\frac{e^{-\frac{\Ot}{2}\coth(2\Ot t)(x-y)^{2}-\imath\Omega
      x\cdot\yt}}{\sinh(2\Ot t)}\times\nonumber\\
  &&\lbt -2\Ot\coth(2\Ot t)+\lbt -\Ot\coth(2\Ot
  t)(x-y)^{\mu}-\imath\Omega\yt^{\mu}\rbt^{2}\rbt\nonumber\\
  &=&\frac{e^{-\frac{\Ot}{2}\coth(2\Ot t)(x-y)^{2}-\imath\Omega
      x\cdot\yt}}{\sinh(2\Ot t)}\lbt -2\Ot\coth(2\Ot t)+\Ot^{2}\coth^{2}(2\Ot
  t)(x-y)^{2}\right.\nonumber\\
  &&\left.-\Omega^{2}\yt^{2}+ 2\imath\Omega\Ot\coth(2\Ot t)x\cdot\yt\rbt. 
\end{eqnarray}
The operator $L_{2}=x^{0}p_{1}-x^{1}p_{0}=-\imath x^{0}\partial_{1}+\imath
x^{1}\partial_{0}$ acting on $\Gamma^t$ gives:
\begin{eqnarray}
  L_{2}\Gamma^t&=&-\imath\frac{x^{0}}{\sinh(2\Ot t)}\lbt -\Ot\coth(2\Ot
  t)(x-y)^{1}-\imath\Omega\yt^{1}\rbt
  \nonumber\\
  &+&\imath\frac{x^{1}}{\sinh(2\Ot t)}\lbt -\Ot\coth(2\Ot
  t)(x-y)^{0}-\imath\Omega\yt^{0}\rbt
  e^{-\frac{\Ot}{2}\coth(2\Ot t)(x-y)^{2}-\imath\Omega
    x\cdot\yt}
  \nonumber\\
  &=&\frac{e^{-\frac{\Ot}{2}\coth(2\Ot t)(x-y)^{2}-\imath\Omega
    x\cdot\yt}}{\sinh(2\Ot t)}\lbt\imath\coth(2\Ot t)\Omega x\cdot\yt-\Ot x\cdot
  y\rbt. 
\end{eqnarray}
\begin{eqnarray}
  \lbt-\Delta+2\Ot L_{2}+\Ot^{2}x^{2}\rbt\Gamma^t&=&\frac{e^{-\frac{\Ot}{2}\coth(2\Ot t)(x-y)^{2}-\imath\Omega
    x\cdot\yt}}{\sinh(2\Ot t)}\lbt2\Ot\coth(2\Ot t)\right.\nonumber\\ 
&&\hspace{-2cm}-\Ot^{2}\coth^{2}(2\Ot  t)(x-y)^{2}+\Ot^{2}y^{2}
-2\imath\Omega\Ot\coth(2\Ot t)x\cdot\yt\nonumber\\
&&\hspace{-2cm}\left. +2\imath\Omega\Ot\coth(2\Ot t)x\cdot\yt-2\Ot^{2}x\cdot
  y+\Ot^{2}x^{2}\rbt\nonumber\\
  =\frac{e^{-\frac{\Ot}{2}\coth(2\Ot t)(x-y)^{2}
  -\imath\Omega  x\cdot\yt}}{\sinh(2\Ot t)}&\times&\big( 2\Ot\coth(2\Ot t)-\frac{\Ot^{2}}{\sinh^{2
    }(2\Ot t)}(x-y)^{2}\big)
\end{eqnarray}
With the $\mu^{2}$ and $\gamma^{0}\gamma^{1}$ terms, eq. (\ref{eq:expQ}) is satisfied.

For the full propagator it remains to compute
$e^{-2\imath\Ot t\gamma^{0}\gamma^{1}}$ and the action of
$\lbt-\ps+\mu-\Omega\xts\rbt$ on $e^{-tQ}$.
\begin{align}\label{expgamma01}
  e^{-2\imath\Ot t\gamma^{0}\gamma^{1}}&=\sum_{n\ges 0}\frac{(-2\imath\Ot
    t)^{2n}}{(2n)!}(-\mathds{1}_{2})^{n} + \sum_{n\ges 1}\frac{(-2\imath\Ot
    t)^{2n+1}}{(2n+1)!}(-\gamma^{0}\gamma^{1})^{n}\nonumber\\
  &=\cosh(2\Ot t)\mathds{1}_{2}-\imath\sinh(2\Ot t)\gamma^{0}\gamma^{1}.
\end{align}
With the convention $\slashed{p}=-\imath\ds$ we have
$-\slashed{p}=\imath\gamma^{\nu}\partial_{\nu}$ and
\begin{equation}
  -\slashed{p}\Gamma^t =\frac{\imath\gamma^{\nu}e^{-\frac{\Ot}{2}\coth(2\Ot t)(x-y)^{2}
  -\imath\Omega x\cdot\yt}}{\sinh(2\Ot t)}\big(-\Ot\coth(2\Ot
  -t)(x-y)_{\nu}-\imath\Omega\yt_{\nu}\big).
\end{equation}
which completes the proof of the Lemma.

\section{Matrix basis Kernel}
\label{app3}
\setcounter{equation}{0}

We will restrict our attention to the case $h\ge 0$. We see that:
\begin{equation}
  \lim_{\alpha\to 0} {\cal E}=\delta_{mu}\delta_{lu}\rightarrow
  (1-\alpha)^{\frac{\theta}{8\Omega}H}|_{\alpha=O}=I
\end{equation}
so that we have the correct normalization.
One must now only check the differential equation:
\begin{equation}
  (1-\alpha)\frac{d}{d\alpha}(1-\alpha)^{\frac{\theta}{8\Omega}H}+
  \frac{\theta}{8\Omega}H(1-\alpha)^{\frac{\theta}{8\Omega}H}=0.
\end{equation}

A straightforward computation yields the result:
$$
-(1-\alpha)\frac{d}{d\alpha}[(1-\alpha)^{\frac{\theta}
  {8\Omega}H}]_{m,m+h;l+h,l}=\sum_{u=0}^{min(m,l)}{\cal A}(m,l,h,u)
\Big{(}C\frac{1+\Omega}{1-\Omega}\Big{)}^{m+l-2u}$$
$$\times\Big{[}
(2m+h+1)\Big{(}\frac{1}{2}+C\frac{1-\alpha}{1+C\alpha}\Big{)}-
(m-u)\Big{(}1+\frac{1-\alpha}{\alpha}+C\frac{1-\alpha}
{1+C\alpha}\Big{)}$$
\begin{equation}
  -(l-u)\Big{(}\frac{1-\alpha}{\alpha}-C\frac{1-\alpha}
  {1+C\alpha}\Big{)}
  \Big{]}{\cal E}(m,l,h,u).
\end{equation}
We will treat the last term in the above sum. Using the equality:
\begin{eqnarray}
  (l-u){\cal A}(m,l,h,u)&=&{\cal A}(m,l,h,u+1)\times \nonumber\\
  &&\hspace{-3cm}\Big{[}
  \frac{(m+1)(m+h+1)}{m-u}-(2m+h+1)+(m-u-1)
  \Big{]}
\end{eqnarray}
and changing the dummy variable from $u$ to $v=u+1$, the term 
rewrites as:
\begin{eqnarray}
  && \frac{\alpha C^2(\frac{1+\Omega}{1-\Omega})^2}{1+C\alpha}
  \sum_{v}
  \Big{[}
  \frac{(m+1)(m+h+1)}{m+1-v}-(2m+h+1)+(m-v)
  \Big{]}
  \nonumber\\   
  && \hspace{2cm} {\cal A}(m,l,h,v)
  \Big{(}C\frac{1+\Omega}{1-\Omega}\Big{)}^{m+l-2v}
  {\cal E}(m,l,h,v).
\end{eqnarray}
Coupling the identical terms in the two sums we get the coefficients 
of:
\begin{align} 
  2m+h+1&\rightarrow 
  \frac{1}{2}+C\frac{1-\alpha}{1+C\alpha}+
  \frac{\alpha C^2(\frac{1+\Omega}{1-\Omega})^2}{1+C\alpha}=
  C+\frac{1}{2}=\frac{1+\Omega^2}{4\Omega}\\
  m-u&\rightarrow (-1)
  \Big{[}\frac{1}{\alpha}+C\frac{1-\alpha}{1+C\alpha}+\frac{\alpha 
    C^2(\frac{1+\Omega}{1-\Omega})^2}{1+C\alpha}\Big{]}=
  -\frac{1+C\alpha}{\alpha}.
\end{align}
The complete sum is then:
\begin{eqnarray}
  &&\sum_{u}\Big{(}C\frac{1+\Omega}{1-\Omega}\Big{)}^{m+l-2u}
  {\cal A}(m,l,h,u){\cal E}(m,l,h,u)\times\\
  \nonumber
  &&\hspace{-.8cm} \Big{[}
  (2m+h+1)\frac{1+\Omega^2}{4\Omega}-(m-u)\frac{1+C\alpha}{\alpha}
  -\frac{(m+1)(m+h+1)}{m+1-u}\frac{\alpha 
    C^2(\frac{1+\Omega}{1-\Omega})^2}{1+C\alpha}\Big{]} .
\end{eqnarray}
Using 
\begin{eqnarray} (m-u){\cal A}(m,l,h,u)&=&\sqrt{m(m+h)}{\cal A}(m-1,l,h,u),\\
  \frac{(m+1)(m+h+1)}{(m+1-v)}{\cal A}(m,l,h,u)&=&
  \sqrt{(m+1)(m+h+1)}{\cal A}(m+1,l,h,u), \nonumber
\end{eqnarray}
one can cast the result into the form:
\begin{eqnarray}
  &&-(1-\alpha)\frac{d}{dt}[(1-\alpha)^{\frac{\theta}{8\Omega}H}
  ]_{m,m+h;l+h,l}=\\
  &&\frac{1+\Omega^2}{4\Omega}(2m+h+1)
  [(1-\alpha)^{\frac{\theta}{8\Omega}H}
  ]_{m,m+h;l+h,l}\nonumber\\
  &&-C\frac{1+\Omega}{1-\Omega}\sqrt{m(m+h)}
  [(1-\alpha)^{\frac{\theta}{8\Omega}H}]_{m-1,m-1+h;l+h,l}\nonumber\\
  &&-\sqrt{(m+1)(m+h+1)}C\frac{1-\Omega}{1+\Omega}
  [(1-\alpha)^{\frac{\theta}{8\Omega}H}]_{m+1,m+1+h;l+h,l} .
\end{eqnarray}
On the other hand:
\begin{eqnarray}
  &&\frac{\theta}{8\Omega}H_{m,m+h;p+h,p}=
  \frac{1+\Omega^2}{4\Omega}(2m+h+1)\delta_{mp}
  \nonumber\\
  &&\hspace{-.5cm} -\frac{1-\Omega^2}{4\Omega}
  [\sqrt{(m+h+1)(m+1)}~\delta_{m+1,p}+\sqrt{(m+h)m}~\delta_{m-1,p}]
\end{eqnarray}
and the differential equation is checked.

\paragraph{Acknowledgement}
We thank S.~Al~Jaber, V.~Gayral, J.~Magnen and R.~Wulkenhaar for discussions at
various stages of this work. We also thank our anonymous referee for
interesting comments which led to this improved version.



\chapter{Renormalization of $\Phi^{\star 4}_4$ in $x$ Space}

\begin{center} 

\vskip 4ex

Razvan \textsc{Gurau}, Jacques \textsc{Magnen$^*$}, \\
Vincent \textsc{Rivasseau} and
Fabien \textsc{Vignes-Tourneret}

\vskip 3ex  

\textit{Laboratoire de Physique Th\'eorique, B\^at.\ 210, CNRS UMR 8627\\
Universit\'e Paris XI,  F-91405 Orsay Cedex, France\\
$^*$Centre de Physique Th\'eorique, CNRS UMR 7644\\
Ecole Polytechnique F-91128 Palaiseau Cedex, France}
\\
e-mail: \texttt{razvan.gurau@th.u-psud.fr}, 
\texttt{magnen@cpht.polytechnique.fr}, 
\texttt{vincent.rivasseau@th.u-psud.fr}, 
\texttt{fabien.vignes@th.u-psud.fr}
\end{center}

\vskip 5ex

In this paper we provide a new proof that the Grosse-Wulkenhaar
non-commutative scalar $\Phi^4_4$ theory
is renormalizable to all orders in perturbation theory, and extend it to more
general models with covariant derivatives. 
Our proof relies solely on 
a multiscale analysis in $x$ space.
We think this proof is simpler and could be more adapted to
the future study of these theories (in particular at the non-perturbative or constructive level). 

\section{Introduction}
\setcounter{equation}{0}

In this paper we recover the proof of perturbative 
renormalizability of non-com\-mutative $\Phi^4_4$ field theory
\cite{GrWu03-1,GrWu04-3,Rivasseau2005bh} by a method solely based on $x$ space.
In this way we avoid completely the sometimes tedious use of the matrix basis and 
of the associated special functions of \cite{GrWu03-1,GrWu04-3,Rivasseau2005bh}.
We also extend the corresponding BPHZ theorem to the 
more general complex Langmann-Szabo-Zarembo $\bar\varphi \star \varphi \star \bar\varphi \star \varphi $ model 
with covariant derivatives, hereafter called the LSZ model.
This model has a slightly more complicated propagator, and is exactly solvable in a certain limit \cite{Langmann:2003if}.

Our method builds upon previous work of Filk and Chepelev-Roiban
\cite{Filk:1996dm,Chepelev:2000hm}. These works however remained inconclusive \cite{CheRoi}, since these 
authors used the right interaction but not the right propagator, hence
the problem of ultraviolet/infrared mixing prevented them from obtaining a finite 
renormalized perturbation series. The Grosse Wulkenhaar breakthrough was to realize
that the right propagator in non-commutative field theory 
is not the ordinary commutative propagator, but has to be modified to obey Langmann-Szabo duality \cite{GrWu04-3,LaSz}.

Non-commutative field theories (for a general review see \cite{DN}) deserve a thorough 
and systematic investigation. Indeed they may be relevant for physics beyond the standard model.
They are certainly effective models for certain limits of string theory
\cite{SeiWitt,a.connes98:noncom}. Also they form almost certainly the
correct framework for a microscopic {\it ab initio} understanding 
of the quantum Hall effect which is currently lacking.
We think that $x$ space-methods are probably more powerful for the future 
systematic study of  the noncommutative Langmann-Szabo covariant field theories.

Fermionic theories such as the two dimensional Gross-Neveu model can be shown to be
renormalizable to all orders in their Langmann-Szabo covariant versions, using either the matrix basis
or the direct space version developed here \cite{RenNCGN05}. However the $x$-space version seems the most 
promising for a complete non perturbative construction, using Pauli's principle to controll the
apparent (fake) divergences of perturbation theory. 
In the case of $\phi^4_4$, recall that although the commutative version is until now fatally flawed 
due to the famous Landau ghost, there is some hope that the non-commutative field theory treated 
at the perturbative level in this paper may also exist at the constructive level \cite{GrWu04-2,Rivasseau:2004az}. 
Again the $x$-space formalism is probably better than
the matrix basis for a rigorous investigation of this question.

In the first section of this paper we establish the $x$-space power counting 
of the theory using the Mehler kernel form of the propagator in direct space given in \cite{Propaga}.
In the second section we prove that the divergent subgraphs can be renormalized
by counterterms of the form of the initial Lagrangian.
The LSZ models are treated in the Appendix.

\paragraph{Acknowledgment}
We thank V.~Gayral and R.~Wulkenhaar for useful discussions on this work.

\vspace*{\stretch{2}}

\section{Power Counting in $x$-Space}
\setcounter{equation}{0}

\subsection{Model, Notations}

The simplest noncommutative $\varphi^4_4$ theory is defined on ${\mathbb R}^4$ equipped
with the associative and noncommutative Moyal product
\begin{align}
  (a\star b)(x) &= \int \frac{d^4k}{(2\pi)^4} \int d^4 y \; a(x{+}\tfrac{1}{2}
  \theta {\cdot} k)\, b(x{+}y)\, \mathrm{e}^{\mathrm{i} k \cdot y}\;.
\label{starprodRAZ}
\end{align}

The renormalizable action functional introduced in \cite{GrWu04-3} is
\begin{equation}\label{action}
S[\varphi] = \int d^4x \Big( \frac{1}{2} \partial_\mu \varphi
\star \partial^\mu \varphi + \frac{\Omega^2}{2} (\tilde{x}_\mu \varphi )
\star (\tilde{x}^\mu \varphi ) + \frac{1}{2} \mu_0^2
\,\varphi \star \varphi 
+ \frac{\lambda}{4!} \varphi \star \varphi \star \varphi \star
\varphi\Big)(x)\;,
\end{equation}
where $\tilde{x}_\mu=2(\theta^{-1})_{\mu\nu} x^\nu$ and the Euclidean
metric is used.

In four dimensional $x$-space the propagator is \cite{Propaga}
\begin{equation}
C(x,x')=\frac{\Omega ^2}{[2\pi\sh\Omega t ]^2}
e^{-\frac{\Omega\coth\Omega t}{2}(x^2+x'^2)-
\frac{\Omega}{\sh\Omega t}x \cdot x'    - \mu_0^2 t }
\end{equation}
and the (cyclically invariant) vertex is \cite{Filk:1996dm}
\begin{equation}\label{vertex}
V(x_1, x_2, x_3, x_4) = \delta(x_1 -x_2+x_3-x_4 )e^{i
\sum_{1 \les i<j \les 4}(-1)^{i+j+1}x_i \theta^{-1}  x_j}
\end{equation}
where we note\footnote{Of course two different $\theta$ parameters could be used for the two 
symplectic pairs of variables of  ${\mathbb R}^4$.}
$x \theta^{-1}  y  \equiv  \frac{2}{\theta} (x_1  y_2 -  x_2  y_1 +  x_3  y_4 - x_4  y_3 )$.

The main result of this paper is a new proof in configuration space of
\begin{theorem}
\label{BPHZ1}
The theory defined by the action (\ref{action}) 
is renormalizable to all orders of perturbation theory.
\end{theorem} 

Let $G$ be an arbitrary connected graph. The amplitude associated with this graph is 
(with selfexplaining notations):
\begin{eqnarray}
&&A_G=\int \prod_{v,i=1,...4} dx_{v,i} \prod_l dt_l   \nonumber  \\
&& \prod_v \left[ \delta(x_{v,1}-x_{v,2}+x_{v,3}-x_{v,4})e^{\imath
\sum_{i<j}(-1)^{i+j+1}x_{v,i}\theta^{-1} x_{v,j}} \right]
  \\
&&  \prod_l 
\frac{\Omega^2}{[2\pi\sinh(\Omega t_l)]^2}e^{-\frac{\Omega}{2}\coth(\Omega 
t_l)(x_{v,i(l)}^{2}+x_{v',i'(l)}^{2})
+\frac{\Omega}{\sinh(\Omega t_l)}x_{v,i(l)} . x_{v',i'(l)}   - \mu_0^2 t_l} .
\nonumber
\label{amplitude}
\end{eqnarray} 

For each line $l$ of the graph joining positions $x_{v,i(l)}$ and $x_{v',i'(l)}$, 
we choose an orientation and we define 
the ``short" variable $u_l=x_{v,i(l)}-x_{v',i'(l)}$ and the ``long" variable $v_l=x_{v,i(l)}+x_{v',i'(l)}$. 
With these notations, defining $\Omega t_l=\alpha_l$, the propagators in our graph can be 
written as:
\begin{equation}\label{tanhyp}
\int\prod_l \frac{\Omega d\alpha_l}{[2\pi\sinh(\alpha_l)]^2}
e^{-\frac{\Omega}{4}\coth(\frac{\alpha_l}{2})u_l^2-
\frac{\Omega}{4}\tanh(\frac{\alpha_l}{2})v_l^2  - \frac{\mu_0^2}{\Omega} \alpha_l}\; .
\end{equation} 

\subsection{Orientation and Position Routing}

A rule to solve the $\delta$ functions at every vertex is a ``position routing"
exactly analog to a momentum routing in the ordinary commutative case, except for the
additional difficulty of the cyclic signs which impose to orient the lines.
It is well known that there is no canonical such routing but 
there is a routing associated to any choice of a spanning tree in $G$. 
Such a tree choice is also useful to orient the lines of the graph, 
hence to fix the exact sign definition of the ``short" line variables $u_l$, and to 
optimize the multiscale power counting bounds below.

\begin{figure}[htbp]
\centering
\includegraphics[scale=1]{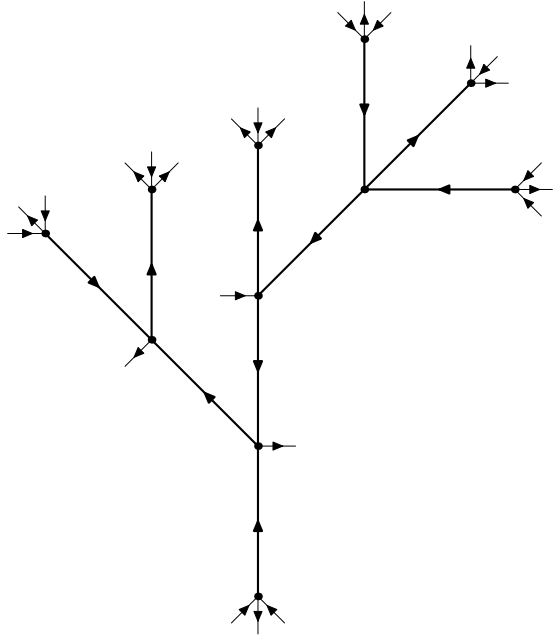}
\caption{Orientation of a tree}
\label{fig:treeorient}
\end{figure}

Let $n$ be the number of vertices of $G$, $N$ 
the number of its external fields, and $L$ the number of internal lines of $G$. We have $L= 2n-N/2$. 
Let $T$ be a rooted tree in the graph (when the graph is not a vacuum graph
it is convenient to choose for the root a vertex with external fields but this is not essential). 
We orient first all the lines of the tree and all the remaining 
half-loop lines or ``loop fields", following the cyclicity of the vertices.
This means that starting from an arbitrary orientation of a first field at the root and 
inductively climbing into the tree, at each vertex we follow the cyclic order to alternate entering 
and exiting lines as in Figure \ref{fig:treeorient}. 

Every line of the tree by definition of this orientation has one end exiting a vertex and an other
entering another one. This may not be true for the loop lines, which join two ``loop fields". 
Among these, some exit one vertex and enter another; they are called  
well-oriented. But others may enter or exit at both ends. These loop lines
are subsequently referred to as ``clashing lines". If there are no clashing lines, 
the graph is called orientable. If not, it is called non-orientable. 

We will see below that non-orientable graphs are irrelevant in the renormalization group sense. 
In fact they do not occur at all in some particular models such as
the LSZ model treated in the Appendix, or in the most natural noncommutative Gross-Neveu models
\cite{RenNCGN05}. 

For all the well-oriented lines (hence all tree propagators plus some of the 
loop propagators) we define in the natural way $u_l=x_{v,i(l)}-x_{v',i'(l)}$ if the 
line enters at $x_{v,i(l)}$ and exits from $x_{v',i'(l)}$. 
Finally we fix an additional (completely arbitrary) auxiliary orientation for all
the clashing loop lines, and fix in the same way $u_l=x_v-x_{v'}$ with respect to this auxiliary orientation.
 
It is also convenient to define the set of ``branches" associated to the rooted tree $T$.
There are $n-1$ such branches $b(l)$, one for each of the $n-1$ lines $l$ of the tree,
plus the full tree itself, called the root branch, and noted $b_0$.
Each such branch is made of the subgraph $G_b$ containing all the vertices ``above $l$"
in $T$, plus the tree lines and loop lines joining these vertices. It has also ``external fields" 
which are the true external fields hooked to $G_b$, plus the loop fields in $G_b$ for the
loops with one end (or ``field") inside and one end 
outside $G_b$, plus the upper end of the tree line $l$ itself to which $b$ is associated.
In the particular case of the root branch, $G_{b_0} = G$ and
the external fields for that branch are simply  all true external fields. 
We call $X_b$ the set of all external fields $f$ of $b$.

We can now describe the position routing associated to $T$.
There are $n$ $\delta$ functions in (\ref{amplitude}), hence $n$ linear equations for the $4n$ positions,
one for each vertex. The momentum routing associated to the tree $T$ solves this system by passing 
to another equivalent system of  $n$ linear equations, one for each branch of the tree.
This equivalent system is obtained by summing the arguments of the 
$\delta$ functions of the vertices in each branch. Obviously the Jacobian of this transformation is 1,
so we simply get another equivalent set of $n$ $\delta$ functions, one for each branch.

Let us describe more precisely the positions summed in these branch equations, using the orientation.
Fix a particular branch $G_b$, with its subtree $T_b$. In the branch sum we find a sum 
over all the $u_l$ short parameters of the lines $l$ in $T_b$ and no $v_l$ long parameters
since $l$ both enters and exits the branch. This is also true for the set 
$L_b$ of well-oriented loops lines with both fields in the branch.
For the set $L_{b,+}$ of clashing loops lines with both fields entering the branch,
the short variable disappears and the long variable remains; the same is true but with a minus sign
for the set $L_{b,-}$ of clashing loops lines with both fields exiting the branch.
Finally we find the sum of positions of all external fields for the branch (with the signs
according to entrance or exit). For instance in the particular case of Figure \ref{fig:exbranch}, the delta function
is
\begin{equation}
\delta\lbt u_{l_{1}}+u_{l_{2}}+u_{l_{3}}+u_{L_{1}}+u_{L_{3}}-v_{L_{2}}+X_{1}-X_{2}+X_{3}+X_{4} 
\rbt \; .
\end{equation} 
\begin{figure}[htbp]
  \centering
  \includegraphics[scale=0.8]{xphi4-fig.2}
  \caption{A branch}
  \label{fig:exbranch}
\end{figure}

The position routing is summarized  by:

\begin{lemma}[Position Routing]
We have, calling $I_G$ the remaining integrand in (\ref{amplitude}):
\begin{eqnarray}
A_G &=& \int \Big[ \prod_v  \big[ \delta(x_{v,1}-x_{v,2}+x_{v,3}-x_{v,4})\big] \, \Big]
I_G(\{x_{v,i}  \}  )   \\
&=& \int \prod_{b}
\delta \left(   \sum_{l\in T_b \cup L_b }u_{l} + \sum_{l\in L_{b,+}}v_{l}-\sum_{l\in L_{b,-}}v_{l}
+\sum_{f\in X_b}\epsilon(f) x_f \right) I_G(\{x_{v,i}  \}), \nonumber 
\end{eqnarray} 
where $\epsilon(f)$ is $\pm 1$ depending on whether the field $f$ enters or exits the branch.
\end{lemma}

Using the above equations one can at least solve all the long tree variables $v_l$ 
in terms of external variables, short variables and long loop variables, using the
$n-1$ non-root branches. There remains then the root branch $\delta$ function.
If $G_b$ is orientable, this $\delta$ function of branch $b_0$
contains only short and external variables, since $L_{b,+}$ and $ L_{b,-}$ are empty. 
If $G_b$ is non-orientable one can solve for an additional ``clashing" long loop variable.
We can summarize these observations in the following lemma:

\begin{lemma}\label{lemmarouting}
The position routing solves any long tree variable $v_l$ as a function 
of:
\begin{itemize}
\item the short tree variable $u_l$ of the line $l$ itself,
\item the short tree and loop variables with both ends in $G_{b(l)}$,
\item the long loop variables of the clashing loops with both ends in $G_{b(l)}$ (if any),
\item the short and long variables of the loop lines 
with one end inside $G_{b(l)}$ and the other outside.
\item the true external variables $x$ hooked to  $G_{b(l)}$.
\end{itemize}
The last equation corresponding to the root branch is particular. 
In the orientable case it does not contain any long variable, 
but gives a linear relation among the short variables and the external positions.
In the non-orientable case it gives a linear relation between 
the long variables $w$ of all the clashing loops in the graph 
some short variables $u$'s and all the external positions.
\end{lemma}

{}From now on, each time we use this lemma to solve the long tree variables $v_l$ 
in terms of the other variables, we shall call $w_l$ rather than $v_l$ the remaining $n+1 - N/2$ 
independent long loop variables. Hence looking at the long variables names the reader can check
whether Lemma \ref{lemmarouting} has been used or not.
\label{filkreduc1}

\subsection{Multiscale Analysis and Crude Power Counting}

In this section we follow the standard procedure of multiscale analysis \cite{Riv1}.
First the parametric integral for the propagator is sliced in the usual way :
\begin{equation}
C(u,v)=C^0(u,v)+\sum_{i=1}^{\infty}C^i(u,v),
\end{equation} 
with
\begin{equation}
C^i(u,v)=\int_{M^{-2i}}^{M^{-2(i-1)}}\frac{\Omega d\alpha} 
{[2\pi\sinh\alpha]^{2}}e^{-\frac{\Omega}{4}\coth\frac{\alpha}{2} 
u^2-\frac{\Omega}{4}\tanh\frac{\alpha}{2}v^2  -\frac{\mu_0^2}{\Omega} \alpha_l }
\end{equation}  

\begin{lemma} For some constants $K$ (large) and $c$ (small):
\begin{equation}\label{eq:propbound-phi4}
C^i (u,v) \les K M^{2i}e^{-c [ M^{i}\Vert u \Vert + M^{-i}\Vert v\Vert ] }
\end{equation} 
(which a posteriori justifies the terminology of ``long" and ```short" variables).
\end{lemma}
The proof is elementary, as it 
relies only on second order approximation of the hyperbolic functions near the origin.

Taking absolute values, hence neglecting all oscillations, leads to the following crude bound:
\begin{equation}\label{assignsum}
\vert A_G \vert \les \sum_{\mu}\int du_ldv_l\prod_l  C^{i_l}(u_l,v_l)
\prod_v \delta_v \ ,
\end{equation}
where $\mu$ is the standard assignment of an integer index $i_l$ to each propagator 
of each internal line $l$ of the graph $G$, which represents its ``scale". We will 
consider only amputated graphs. Therefore we have no 
external propagators, but only external vertices of the graph; 
in the renormalization group spirit, the convenient convention is to assign 
all external indices of these external fields to a  fictitious $-1$ ``background" scale. 

To any assignment $\mu$ and scale $i$
are associated the standard connected components $G_k^i$, $k=1,... ,k(i)$ 
of the subgraph $G^i$ made of all lines with scales $j\ges i$. These tree components 
are partially ordered according to their inclusion relations and the (abstract) tree
describing these inclusion relations is called the Gallavotti-Nicol\`o tree \cite{GaN}; 
its nodes are the $G_k^i$'s and its root is the complete graph $G$ (see Figure
\ref{multiscale-tools}).
\begin{figure}[htbp]
  \centering
  \subfloat[A $\varphi^{4}$ graph]{\label{fig:exgraph}\includegraphics[scale=0.8]{xphi4-fig.3}}
  \subfloat[Example of scale
  attribution]{\label{fig:exscale}\includegraphics[scale=0.5]{xphi4-fig.4}}\\
  \vspace*{1cm}
  \subfloat[The ``Gallavotti-Nicol\`o'' tree]{\label{fig:GNtree}\includegraphics[scale=0.7]{xphi4-fig.5}}
  \caption{}
  \label{multiscale-tools}
\end{figure}

\vspace*{\stretch{1}}

More precisely for an arbitrary subgraph $g$ one defines: 
\begin{equation}
 i_g(\mu)=\inf_{l\in g}i_l(\mu) \quad , \quad 
 e_g(\mu)=\sup_{l \mathrm{~external~line~of~} g}i_l(\mu)\ .
\end{equation}
The subgraph $g$ is a $G_k^i$ for a given $\mu$ if and 
only if $i_g(\mu)\ges i> e_g(\mu)$. 
As is well known in the commutative field theory case, the key to optimize the bound over 
spatial integrations is to choose the real tree $T$ compatible with the abstract Gallavotti-Nicol\`o tree,
which means that the restriction $T_k^{i}$ of  $T$ to any $G_k^{i}$ must still span $G_k^{i}$.
This is always possible (by a simple induction from leaves to root).
We pick such a compatible tree $T$ and use it both to orient the graph as in the previous section and
to solve the associated branch system of $\delta$ functions according to Lemma \ref{lemmarouting}
We obtain: 
\begin{eqnarray}\label{bound1}
\vert A_{G,\mu} \vert &\les& K^n\prod_l M^{2i_l}\int du_ldv_l \prod_l
e^{-c [ M^{i_l}\Vert u_l \Vert  + M^{-i_l} \Vert v_l \Vert ]}
\prod_b\delta_b   \; .
\nonumber\\
&\les&  K^n\prod_l M^{2i_l}\int du_l dw_l \prod_l
e^{-c [ M^{i_l}\Vert u_l \Vert  + M^{-i_l} \Vert v_l (u,w,x) \Vert ]}  \delta_{b_0} \; .
\end{eqnarray} 

The key observation is to remark that any
long variable integrated at scale $i$ costs $KM^{4i}$ whereas any short 
variable integrated at scale $i$ brings $KM^{-4i}$, and the variables ``solved" by the $\delta$ functions
bring or cost nothing. For an orientable graph the optimal solution is easy:
we should solve the $n-1$ long variables $v_l$'s of the tree 
propagators in terms of the other variables, because this is the maximal number of long
variables that we can solve, and they have highest possible indices because
$T$ has been chosen compatible with the Gallavotti-Nicol\`o tree structure.
Finally  we still have the last $\delta_{b_0}$ function (equivalent to the overall momentum conservation
in the commutative case). It is optimal to use it
to solve one external variable (if any ) in terms of all the 
short variables and the external ones. Since external variables are typically smeared
against unit scale test functions, this leaves power counting invariant\footnote{In the case
of a vacuum graph, there are no external variables and we must 
therefore use the last $\delta_{b_0}$ function to solve 
the lowest possible short variable in terms of all others. In this way,
we loose the $M^{-4i}$ factor for this short integration. This is why the power counting
of a vacuum graph at scale $i$ is not given by the usual formula $M^{(4-N)i}= M^{4i}$ below at $N=0$,
but is in $M^{8i}$, hence worse by $M^{4i}$. 
This is of course still much better than the commutative case, because in that case and
in the analog conditions, that is without a fixed internal point, vacuum graphs
would be worse than the others by an ... 
infinite factor, due to translation invariance! In any case 
vacuum graphs are absorbed in the normalization of the theory, 
hence play no role in the renormalization.}.

The non-orientable case is slightly more subtle.
We remarked that in this case the system of branch equations allows 
to solve $n$ long variables as a functions of all the others. Should we always choose these $n$ 
long variables as the $n-1$ long tree variables plus one long loop variable? 
This is {\it not always} the  optimal choice.
Indeed when several disjoint $G^i_k$ subgraphs are non-orientable it is better
to solve more long clashing loop variables, essentially one per disjoint non-orientable $G^i_k$,
because they spare higher costs than if tree lines were chosen instead.
We now describe the optimal procedure, using words rather than equations 
to facilitate the reader's understanding.

Let ${\cal C}$ be the set of all the clashing loop lines. Each clashing
loop line has a certain scale $i$, therefore belongs to one and only one 
$G^i_k$ and consequently to all $G^j_{k'}\supset G^i_k$. We now define the 
set $S$ of $n$ long variables to be solved via the $\delta$ functions. First we 
put in $S$ all the $n-1$ long tree variables $v_l$. Then we scan 
all the connected components $G^i_k$ starting from the leaves towards the root,
and we add a clashing line to $S$ each time some new non-orientable component $G^i_k$ appears.
We also remove $p-1$ tree lines from $S$ each time $p\ges 2$ 
non-orientable components merge into a single one. In the end we obtain a new set
$S$ of exactly $n$ long variables.

More precisely suppose some $G^i_k$ at scale $i$ is a ``non-orientable leaf",
which means that is contains some clashing lines at scale $i$ but none at scales $j>i$. We 
then choose one (arbitrary) such clashing line and put it in the set $S$. Once a clashing
line is added to $S$ in this way it is never removed and no
other clashing line is chosen in any of the $G^j_k$ at 
lower scales $j<i$ to which the chosen line belongs. (The reader should be 
aware that this process allows nevertheless several clashing lines of $S$ to belong to a single $G^i_k$,
provided they were added to different connected components 
at upper scales.) When $p\ges 2$ non-orientable components merge at scale $i$
into a single non-orientable $G^i_k$, we can find $p-1$ lines 
in the part of the tree $T^i_k$ joining them together,
(e.g. taking them among the first lines on the unique paths in $T$ from these $p$ components
towards the root) and remove them from $S$.

We see that if we have added in all $q$ clashing lines to the set $S$, we 
have eliminated $q-1$ tree lines. The final set $S$ thus obtained in the end has 
exactly $n$ elements. The non trivial statement is that thanks to inductive use of Lemma \ref{lemmarouting}
in each $G^i_k$, we can solve all the long variables in the set $S$ with the branch system 
of $\delta$ functions associated to $T$. 

We perform now all remaining integrations. This spares the 
corresponding $M^{4i}$ integration cost for each long variable in $S$. 
For any line not in $S$ we see that the net power counting is 1, since the cost of the long variable
integration exactly compensates the gain of the short variable integration. 
But for any line in $S$ we earn the $M^{-4i}$ power counting of the 
corresponding short variable $u$ without paying the $M^{4i}$ cost of the long variable.

Gathering all the corresponding factors together 
with the propagators prefactors $M^{2i}$ leads to the 
following bound:
\begin{equation}
\vert A_{G,\mu}\vert  \  \les \  K^n
\prod_l M^{2i_l}\prod_{l\in S}M^{-4i_l } \ .
\end{equation} 
Remark that if the graph is well-oriented this formula remains true but the set $S$ consists of only 
the $n-1$ tree lines.

In the usual way of \cite{Riv1} we write 
\begin{equation}
\prod_{l}M^{2i_l}=\prod_{l}\prod_{i=1}^{i_l}M^2=
\prod_{i,k}\prod_{l\in G^i_k}M^2=\prod_{i,k}M^{2l(G^i_k)}
\end{equation} 
and
\begin{equation}
\prod_{l\in S}\prod_{i=1}^{i_l}M^{-4i_l}=
\prod_{i,k}\prod_{l\in G^i_k\cap S}M^{-4} ,
\end{equation} 
and we must now only count the number of elements in $G^i_k\cap S$.

If $G^i_k$ is orientable, it contains no clashing lines, hence 
$G^i_k\cap S=T^i_k$, and the cardinal of $T^i_k$ is $n(G^i_k)-1$. 

If $G^i_k$ contains one or more clashing lines and
$p$ clashing lines $l_1$, ... , $l_p$ in $G^i_k$ have been chosen to belong to $S$, then 
$p-1$ tree variables in $T^i_k$ have also been removed from $S$ and 
$G^i_k\cap S=T^i_k\cup\{l_1, ... \; , l_p\}-\{\mathrm{p-1~tree~variables}\}$, hence
the cardinal of $G^i_k\cap S$ is $n(G^i_k)$.

Using the fact that $2l(G^i_k)-4n(G^i_k)=-N(G^i_k)$ we can summarize 
these results in the following lemma:

\begin{lemma}\label{crudelemma}
The following bound holds for a connected graph (with external arguments integrated
against fixed smooth test functions):
\begin{equation}
\vert A_{G,\mu} \vert \les  K^n \prod_{i,k}M^{-\omega(G^i_k)}
\end{equation} 
for some (large) constant $K$, with $\omega(G^i_k)=N(G^i_k)-4$ if $G^i_k$ 
is orientable and $\omega(G^i_k)=N(G^i_k)$ if $G^i_k$ is non-orientable.
\end{lemma}
This lemma is optimal {\it if vertices oscillations are not taken into account}, and proves that 
non-orientable subgraphs are irrelevant.
But it is not yet sufficient for a renormalization theorem to all orders of perturbation. 

\subsection{Improved Power Counting}

Recall that for any non-commutative Feynman graph $G$ we can define
the genus of the graph, called $g$ and the number of faces ``broken by external legs", called $B$
\cite{GrWu04-3,Rivasseau2005bh}.
We have $g \ges 0$ and $B\ges 1$. 
The power counting established with the matrix basis in \cite{GrWu04-3,Rivasseau2005bh}, rewritten in the language
of this paper \footnote{Beware that the factor $i$ in \cite{Rivasseau2005bh} is now $2i$, and that the $\omega$
used here is the convergence rather than divergence degree. Hence there is both a sign change and a factor 2
of difference between the $\omega$'s of this paper and the ones of \cite{Rivasseau2005bh}.} is:
\begin{equation}\label{truepowercounting}
\omega (G) = N -4  + 8 g  + 4(B-1) \ ,
\end{equation}
hence we must (and can) renormalize only 2 and 4 point subgraphs
with $g=0$ and $B=1$, which we call \emph{planar regular}. They are the only non-vacuum graphs with $\omega \les 0$.

In the previous section we established that
\begin{equation}
\omega (G) \ges  N -4 \ , \ {\rm if}\   G\  {\rm orientable}\ , \ \   \omega (G) \ges  N  \ , \ 
{\rm if}\    G  \ {\rm non\ orientable}\ .
\end{equation} 

It is easy to check that planar regular subgraphs are orientable, but the 
converse is not true. Hence to prove that {\it orientable non-planar} 
subgraphs or {\it orientable planar} subgraphs with $B\ges 2$ are irrelevant 
requires to use a bit of the vertices oscillations to improve Lemma \ref{crudelemma}
and get:

\begin{lemma}\label{improvedbound}
For orientable subgraphs with $g\ges 1$ we have
\begin{equation}\label{improvednonplanar}
\omega (G) \ges  N + 4 \; .  
\end{equation} 
For orientable subgraphs with $g = 1$ and $B\ges 2$ we have
\begin{equation}\label{improvedbrokenfaces}
\omega (G) \ges  N  \; .
\end{equation} 
\end{lemma}
This lemma although still not giving (\ref{truepowercounting}) is sufficient for the purpose of this paper. 
For instance it implies directly that graphs which contain only irrelevant subgraphs in the sense of (\ref{truepowercounting}) have finite amplitudes uniformly bounded by $K^n$,
using the standard method of \cite{Riv1} to bound the assignment sum over $\mu$ in (\ref{assignsum}).

The rest of this subsection is essentially devoted to the proof of this Lemma \ref{improvedbound}.

We return before solving $\delta$ functions, hence to the $v$ variables. 
We will need only to compute in a precise way the oscillations which are 
quadratic in the long variables $v$'s to prove (\ref{improvednonplanar})
and the linear oscillations in $v \theta^{-1} x$ to prove (\ref{improvedbrokenfaces}). 
Fortunately an analog problem 
was solved in momentum space by Filk and Chepelev-Roiban \cite{Filk:1996dm,Chepelev:2000hm},
and we need only a slight adaptation of their work to position space.
In fact in this subsection short variables are quite inessential but it is convenient 
to treat on the same footing the long $v$  and the external $x$ variables, so we introduce a new 
global notation $y$ for all these variables. The vertices rewrite as
\begin{equation}
\prod_v\delta(y_1-y_2+y_3-y_4+\epsilon^iu_i)
e^{\imath \big(\sum_{i<j}(-1)^{i+j+1}y_i\theta^{-1}y_j + yQu + uRu \big)} \ .
\end{equation} 
for some inessential signs $\epsilon^i$ and some symplectic
matrices $Q$ and $R$.

Since we are not interested in the precise oscillations in the short $u$ variables 
we will note in the sequel quite sloppily $E_u$ any linear combination of 
the $u$ variables. Let's consider the first Filk reduction \cite{Filk:1996dm}, which contracts tree lines
of the graph. It creates progressively generalized vertices
with even number of fields. At a given induction step and for a tree line joining two 
such generalized vertices with respectively $p$ and $q-p+1$
fields ($p$ is even and $q$ is odd), 
we assume by induction that the two vertices are
\begin{eqnarray}\label{firstfilk1}
&& \delta(y_1-y_2+y_3...-y_p+E_u)
\delta(y_p-y_{p+1}+...-y_q+E_u)
\\
&& \hskip-1cm e^{  \imath \big(\sum_{1\les i<j\les p} (-1)^{i+j+1}y_i\theta^{-1}y_j+
\sum_{p\les i<j\les q} (-1)^{i+j+1}y_i\theta^{-1}y_j+yQu+ uRu   \big) } \ .\nonumber
\end{eqnarray} 
Using the second $\delta$ function we see that:
\begin{equation}\label{solveyp}
y_p=y_{p+1}-y_{p+2}+....+y_q-E_u \ .
\end{equation} 
Substituting this expression in the first $\delta$ function we get:
\begin{eqnarray}\label{firstfilk2}
  &&\delta(y_1-y_2+...-y_{p+1}+..-y_q+E_u)
  \delta(y_p-y_{p+1}+...-y_q+E_u)
 \\
 && \hskip-1cm e^{\imath\big( \sum_{1\les i<j\les p} (-1)^{i+j+1}y_i\theta^{-1}y_j+
\sum_{p\les i<j \les q} (-1)^{i+j+1}y_i\theta^{-1}y_j+yQu+ uRu \big)} \ . \nonumber
\end{eqnarray} 
 
The quadratic terms which include $y_p$ in the exponential are (taking 
into account that $p$ is an even number):
\begin{equation}
\sum_{i=1}^{p-1}(-1)^{i+1}y_i\theta^{-1}y_p+\sum_{j=p+1}^q
  (-1)^{j+1}y_p\theta^{-1}y_j \ .
\end{equation} 
Using the expression (\ref{solveyp}) for $y_p$ we see that the second term gives only 
terms in $yLu$. The first term yields:
\begin{equation}
\sum_{i=1}^{p-1}\sum_{j=p+1}^q (-1)^{i+1+j+1}y_i\theta^{-1}y_j=
  \sum_{i=1}^{p-1}\sum_{k=p}^{q-1}(-1)^{i+k+1}y_i\theta^{-1}y_j \ ,
\end{equation} 
which reconstitutes the crossed terms, and we have recovered the inductive form
of the larger generalized vertex.
 
\begin{figure}[htbp]
\centering
\includegraphics[scale=1.3]{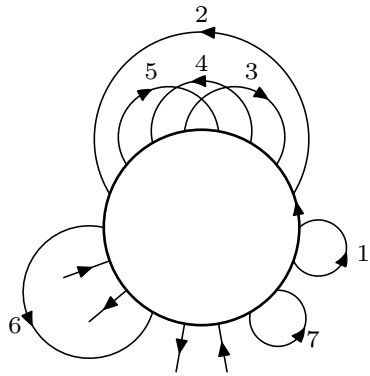}
\caption{A typical rosette}
\label{fig:exrosette}
\end{figure}

One should be aware that $y_p$ has disappeared from the final result, but 
that all the subsequent $y_{s>p}$ have changed sign.
This complication arises because of the cyclicity of the vertex.
As $p$ was chosen to be even (which implies $q$ odd)  
we see that $q-1$ is even as it should. Consequently by 
this procedure we will always treat only even vertices.
We finally rewrite the product of the two vertices as:
\begin{eqnarray}
&&\delta(y_1-y_2+...+y_{p-1}-y_{p+1}+..-y_q+E_u) \delta(y_p-y_{p-1}+...-y_q+E_u)
\nonumber\\
&&  e^{\imath\big( \sum_{1\les i<j\les q}(-1)^{i+j+1}y_i\theta^{-1}y_j+yQu+ uRu \big)}
\end{eqnarray} 
where the exponential is written in terms of the {\it reindexed} vertex 
variables. In this way we can contract all lines of a spanning tree $T$
and reduce $G$ to a single vertex with ``tadpole loops" called a ``rosette graph"
\cite{Chepelev:2000hm}.  
In this rosette to keep track of cyclicity is essential so rather than the ``point-like"
vertex of \cite{Chepelev:2000hm} we prefer to draw the rosette as a cycle 
(which is the border of the former tree) bearing loops lines on it 
(see Figure \ref{fig:exrosette}).
Remark that the rosette can also be considered as a big vertex, with $r=2n+2$ fields, on which $N$
are external fields with external variables $x$ and $2n+2-N$ are loop fields for the corresponding
$n+1-N/2$ loops. When the graph is orientable (which is the case to consider in Lemma
\ref{improvedbound}, 
the fields alternatively enter and exit, and correspond
to the fields on the border of the tree $T$, which we meet turning around counterclockwise
in Figure \ref{fig:treeorient}.
In the rosette the long variables $y_l$
for $l$ in $T$ have disappeared. Let us call $z$ the set of remaining long loop and external variables. Then the
rosette vertex factor is
\begin{equation}\label{rosettefactor}
\delta(z_1-z_2+...-z_r+E_u)
e^{\imath\big(\sum_{1\les i<j\les r}(-1)^{i+j+1}z_i\theta^{-1}z_j+zQu+ uRu\big)} \ .
\end{equation} 

The initial product of $\delta$ functions has not disappeared
so we can still write it as a product over branches like in the previous section and use it to solve
the $y_l$ variables in terms of the $z$ variables and the short $u$ variables. The net effect
of the Filk first reduction was simply to rewrite the
root branch $\delta$ function and the combination of all vertices oscillations 
(using the other $\delta$ functions) as the new big vertex or rosette factor
(\ref{rosettefactor}).

The second Filk reduction \cite{Filk:1996dm} further simplifies the rosette factor by erasing the
loops of the rosette which do not cross any other loops or arch over external fields. 
Here again the same operation is possible. 
Consider indeed such a rosette loop $l$
(for instance loop 2 in Figure \ref{fig:exrosette}). This means that
on the rosette cycle there is an even number of vertices in betwen the two ends of that loop and 
moreover that the sum of $z$'s in betwen these two ends must be zero, since they are loop variables
which both enter and exit between these ends. 
Putting together all the terms in the exponential which contain $z_l$ we 
conclude exactly as in \cite{Filk:1996dm} that these long $z$ variables 
completely disappears from the rosette oscillation factor, which simplifies as in \cite{Chepelev:2000hm}
to 
\begin{equation}\label{rosettefactorsimp}
\delta(z_1-z_2+...-z_r+E_u)
e^{\imath\lbt z {\cal I}z+zQu+ uRu\rbt} \;  ,
\end{equation} 
where ${\cal I}_{ij}$ is the antisymmetric ``intersection matrix" of \cite{Chepelev:2000hm} (up to a different sign convention). Here ${\cal I}_{ij}= +1$ if oriented loop line $i$ crosses oriented loop line $j$  coming from its right,
${\cal I}_{ij} = -1$ if $i$ crosses $j$  coming from its left,
and ${\cal I}_{ij} = 0$ if $i$ and $j$ do not cross. These formulas are also true for
$i$ external line and $j$ loop line or the converse, provided one extends the
external lines from the rosette circle radially to infinity to
see their crossing with the loops. Finally when $i$ and $j$ are external lines
one should define ${\cal I}_{ij} = (-1)^{p+q+1}$ if $p$ and  $q$ are the numbering
of the lines on the rosette cycle (starting from an arbitrary origin).

If a node $G^i_{k}$ of the Gallavotti-Nicol\`o tree is orientable but non-planar ($g \ges 1$), 
there must therefore exist two intersecting loop lines in the rosette corresponding 
to this $G^i_k$, with long variables $w_1$ and $w_2$. 
Moreover since $G^i_{k}$ is orientable, none of the long loop variables associated 
with these two lines belongs to the set $S$ of long variables eliminated by the $\delta$
constraints. Therefore, after 
integrating the variables in $S$ the basic mechanism to improve the
power counting of a single non planar subgraph is the following. The integral
\bea
\int dw_1dw_2 e^{-M^{-2i_1}w_1^2-M^{-2i_2}w_2^2
- iw_1\theta^{-1}w_2+w_1 . E_1(x,u)+w_2 E_2(x,u)} \, ,
\eea
becomes:
\bea
\label{gainoscill}
&&\int dw'_1dw'_2 e^{-M^{-2i_1}(w_1')^2
-M^{-2i_2}(w'_2)^2 +iw'_1\theta^{-1}w'_2 + (u,x)Q(u,x)}
\nonumber\\
&=&  K  M^{4i_1} \int dw'_2
e^{- (M^{2i_1}+ M^{-2i_2})(w'_2)^2 }=
K M^{4i_1}M^{-4i_2} \; .
\eea 
In these equations we used for simplicity $M^{-2i}$ 
instead of the correct but more complicated factor $(\Omega /4) \tanh (\alpha /2 )$
(see \ref{tanhyp}) (of course this does not change the argument) and we performed
a unitary linear change of variables $w'_1 = w_1 + \ell_1 (x, u)$, $w'_2 = w_2 + \ell_2 (x, u)$
to compute the oscillating $w'_1$ integral. The gain in (\ref{gainoscill}) is
$M^{-8i_2}$, which is the difference between $M^{-4i_2}$ and 
the normal factor $M^{4i_2}$ that the $w_2$ integral would have cost if
we had done it with the regular $e^{-M^{-2i_2}w_2^2}$ factor for long variables. 
To maximize this gain we can assume $i_1 \les i_2$.

This basic argument must then be generalized to each non-planar leaf in the Gallavotti-Nicol\`o
tree. This is done exactly in the same way as the inductive definition of the set $A$
of clashing lines in the non-orientable case.
In any orientable non-planar `primitive" $G^i_k$ node (i.e. not containing sub non-planar nodes)
we can choose an arbitrary pair of crossing loop lines 
which will be integrated as in (\ref{gainoscill}) using this oscillation.
The corresponding improvements are independent.

This leads to an improved amplitude bound:
\begin{equation}
\vert A_{G,\mu} \vert \les K^n \prod_{i,k}M^{-\omega(G^i_k)}\ ,
\end{equation} 
where now $\omega(G^i_k)=N(G^i_k) + 4$ if $G^i_k$ 
is orientable and non planar (i.e. $g \ges 1$).
This bound proves (\ref{improvednonplanar}).

Finally it remains to consider the case of nodes $G^i_k$ which are planar orientable
but with $B \ges 2$. In that case there are no crossing loops in the rosette but
there must be at least one loop line arching over a non trivial subset
of external legs in the $G^i_k$ rosette (see line 6 in Figure \ref{fig:exrosette}). We have then a non trivial integration
over at least one external variable, called $x$, of at least one long loop variable called $w$.
This ``external" $x$ variable without the oscillation improvement 
would be integrated with a test function of scale 1 (if it is a true external line of scale $1$)
or better (if it is a higher long loop variable)\footnote{Since the loop line arches 
over a non trivial (i.e. neither full nor empty) subset
of external legs of the rosette, the variable $x$ cannot be the full combination 
of external variables in the ``root" $\delta$ function.}. But we get now
\begin{eqnarray}\label{gainoscillb}
&&\int dx dw e^{-M^{-2i}w^2
- iw\theta^{-1}x  +w.E_1(x',u)}
\nonumber\\
&=&  K  M^{4i} \int dx 
e^{-M^{+2i} x^2 }=
K' \ ,
\end{eqnarray} 
so that a factor $M^{4i}$ in the former bound becomes $O(1)$ hence is improved by $M^{-4i}$.
This proves (\ref{improvedbrokenfaces}) hence completes the proof of Lemma \ref{improvedbound}.
\qed
\bigskip\\
This method could be generalized to get the true power counting (\ref{truepowercounting}).
One simply needs a better description of the rosette oscillating factors when $g$ or $B$ increase.
It is in fact possible to ``disentangle" the rosette by some kind of ``third Filk move". Indeed the
rank of the long variables quadratic oscillations is exactly the genus, and the rank of the linear 
term coupling these long variables to the external ones is exactly $B-1$. So one can through a unitary
change of variables on the long variables inductively disentangle adjacent crossing pairs 
of loops in the rosette. This means that it is possible to diagonalize the 
rosette symplectic form through explicit moves of the loops along the rosette. 
Once oscillations are factorized in this way, the single improvements shown in this section
generalize to one improvement of $M^{-8i}$ per genus and one improvement of $M^{-4i}$ per broken face.
In this way the exact power counting (\ref{truepowercounting}) should be
recovered by pure $x$-space techniques which never require the use of the matrix basis. 
This study is more technical and not really necessary for the BPHZ theorem proved in
this paper.

\section{Renormalization}
\setcounter{equation}{0}

In this section we need to consider only divergent subgraphs, namely the planar
two and four point subgraphs with a single external face ($g=0$, $B=1$, $N=2$ or 4).
We shall prove that they can be renormalized by appropriate counterterms of the form
of the initial Lagrangian. We compute first
the oscillating factors $Q$ and $R$ of the short variables in (\ref{rosettefactorsimp}) for these graphs. 
This is not truly necessary for what follows, but is a good exercise.

\subsection{The Oscillating Rosette Factor}
In this subsection we define another more precise representation
for the rosette factor obtained after applying the first Filk moves
to a graph of order $n$.  We rewrite in terms of $u_l$ and $v_l$
the coordinates of the ends of the tree lines $l$, $l=1,\dots,n-1$ (those contracted in the first Filk moves),
but keep as variables called $s_1,\dots,s_{2n+2}$ the positions of all external fields and all ends of loop 
lines (those not contracted in the first Filk moves). 

\begin{figure}[htbp]
  \centering
  \includegraphics[scale=1.3]{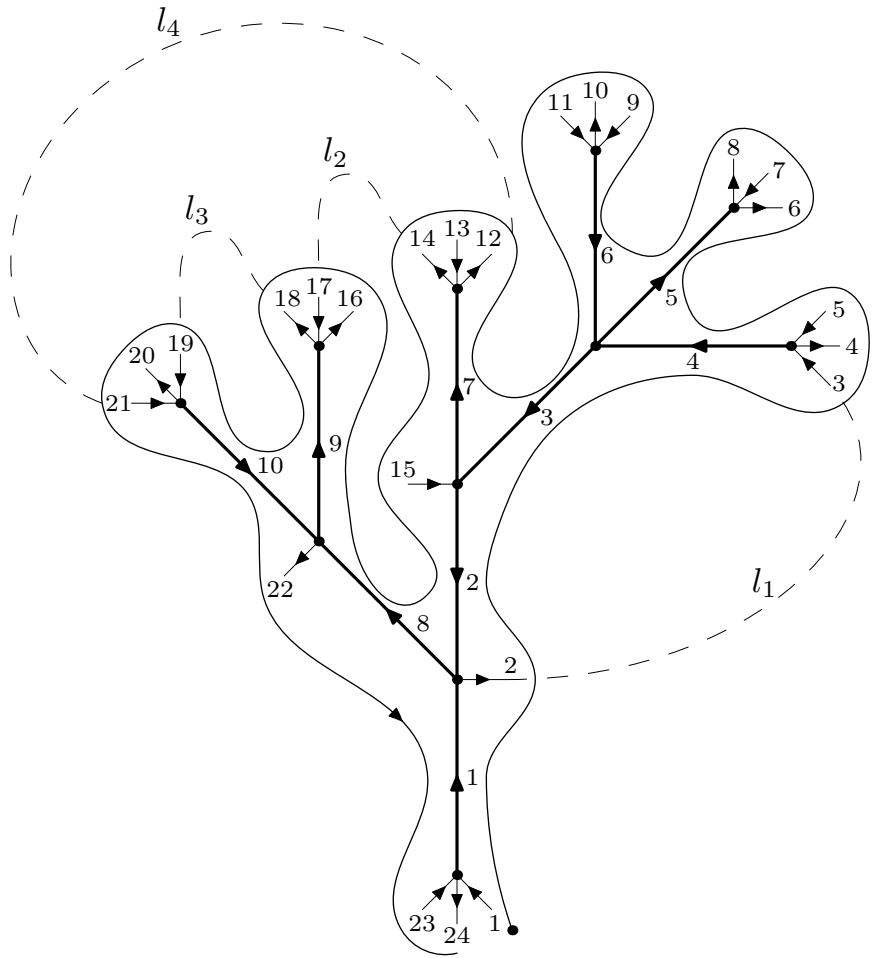}
  \caption{Total ordering of the tree lines and fields}
  \label{fig:turn-around-tree}
\end{figure}

We start from the root and turn around the tree in the trigonometrical sense. We number
separately all the fields as $1,\dots,2n+2$ and all the tree lines as $1,\dots,n-1$ in the order they are met,
but we also define a global ordering $\prec$ on the set of all the fields and
tree lines according to the order in which they are met (see Figure
\ref{fig:turn-around-tree}). In this way we know whether field number $p$ is
met before or after tree line number $q$. For example, in Figure
\ref{fig:turn-around-tree}, field number $8\prec$ tree line number $6$.

\begin{lemma}
The rosette contribution after a complete first Filk reduction is exactly:
\begin{eqnarray}
&&  \delta(s_1-s_2+\dots-s_{2n+2}+\sum_{l\in T}u_l)
e^{i\sum_{0\les i<j\les r}(-1)^{i+j+1}s_i\theta^{-1} s_j}
\nonumber \\
&& e^{-i\sum_{l \prec l'}u_l\theta^{-1} u_{l'}}e^{-i\sum_l \epsilon(l)\frac{u_l\theta^{-1} v_l}{2}} 
e^{i\sum_{l,i \prec l}(-1)^{i} s_i\theta^{-1}u_l+i\sum_{l,i \succ l}u_l \theta^{-1}(-1)^{i} s_i}\ ,
\end{eqnarray}
where $\epsilon(l)$ is $-1$ if the tree line $l$ is oriented towards the root and $+1$ if it is not. 
\end{lemma}

\noindent{\bf Proof:}
We proceed by induction. We contract the tree lines according to their ordering. 
In this way, at any step $k$ we contract a generalized vertex with $2k+2$
external fields corresponding to the contraction of the $k-1$ first lines with 
a usual four-vertex with $r=4$, and obtain a new generalized vertex with $2k+4$
fields.

We suppose inductively that the generalized vertex has the above form and prove that it keeps
this form after the contraction. We denote the external coordinates of this vertex as $s_1,\dots,s_{2k+2}$ 
and the coordinates of the four-vertex as $t_1,\dots,t_4$. We contract the propagator $(s_p,t_q)$ with 
associated variables $v=s_p+t_q$ and $u=(-1)^{p+1}s_p+(-1)^{q+1}t_q$. We also 
note that, since the tree is orientable, $p+q$ is odd.

Adding the arguments of the two $\delta$ functions gives the global 
$\delta$ function. We have the two equations:
\begin{equation}
s_1-s_2+\dots-s_{2k+2}+\sum u_s=0 \quad, \quad t_1-t_2+t_3-t_4=0 \; .
\end{equation}
Using the invariance of the $t$ vertex we can always eliminate 
the contribution of $t_q$ in the phase factor. We therefore have:
\begin{eqnarray}
\varphi&=& [s_1-s_2+\dots+(-1)^{p}s_{p-1}]\theta^{-1}
(-1)^{p}s_p\nonumber\\
&&+(-1)^{p}s_p\theta^{-1}[(-1)^{p+2}s_{p+1}+\dots-s_{2k+2}]
\nonumber\\
&=& [s_1-s_2+\dots+(-1)^ps_{p-1}]\theta^{-1} [-u+(-1)^{q+1}t_q] \nonumber\\
 &&+[-u+(-1)^{q+1}t_q]\theta^{-1}[(-1)^{p+2}s_{p+1}+\dots.-s_{2k+2}] .
\end{eqnarray}
As $(-1)^{q+1}t_q=\sum_{i=1,i\neq q}^4(-1)^it_i$ we see that the $s\theta^{-1} t_q$ terms
in the above expression reproduce exactly the crossed terms needed to complete the first exponential.
We rewrite the other terms as:
\begin{eqnarray}
&&[s_1-s_2+\dots+(-1)^{p}s_{p-1}]\theta^{-1}(-u)+(-u)\theta^{-1} 
[(-1)^{p+2}s_{p+1} \dots-s_{2k+2}] \nonumber \\
&=&[s_1-s_2+\dots+(-1)^{p}s_{p-1}]\theta^{-1}(-u)\nonumber\\
&&+(-u)\theta^{-1}[-s_1+s_2\dots+(-1)^p s_p-\sum_s u_s]\nonumber \\
&=&2[s_1-s_2+\dots+(-1)^{p}s_{p-1}]\theta^{-1}(-u)+(-u)\theta^{-1}(-1)^p s_p 
+u\theta^{-1}\sum_s u_s \nonumber\\
&=&2\sum_{i \prec l}(-1)^{i}s_i\theta^{-1}u+(-1)^{p+1}\frac{u\theta^{-1}v}{2}+
\sum_s u\theta^{-1}u_s\; .
\end{eqnarray}
where we have used $(-1)^p s_p=(-1)^p (v-u)/2$. 

Note that further contractions will not involve $s_1\dots s_{p-1}$. After collecting all the 
contractions and using the global delta function we write:
\begin{equation}
2 \sum_{l,i \prec l}(-1)^i s_i\theta^{-1}u_l=\sum_{l,i \prec  l}(-1)^is_i\theta^{-1}u_l+
\sum_{l,i \succ  l} u_l \theta^{-1} (-1)^i s_i+\sum_{l,l'}u_l\theta^{-1}u_{l'} ,
\end{equation}
and the last term is zero by the antisymmetry of $\theta^{-1}$.
\qed

We note ${\cal L}$ the set of loop lines, and analyze now further the rosette contribution for planar graphs. We 
call now $x_{i},\,i=1,\dots, N$ the $N$ external positions. We choose as first external field $1$ 
an arbitrary entering external line. We define an ordering among the set of all lines,
writing $l' \prec l$ if both ends 
of $l'$ are before the first end of $l$ when turning around the tree as in
Figure \ref{fig:turn-around-tree} where $l_{1}\prec l_{2}$. Analogously we define 
$l \prec j$ when $j$ is an external vertex ($l_{1}\prec x_{4}$ in Figure
\ref{fig:turn-around-tree}). We define $l'\subset  l$ if both ends of $l'$ lie
in between the ends of $l$ on the rosette ($l_{2}\subset l_{4}$ in Figure \ref{fig:turn-around-tree}). 
We count a loop line as positive if it turns in the trigonometric sense like the rosette
and negative if it turns clockwise. Each loop line $l \in {\cal L}$ has now a sign 
$\epsilon(l)$ associated with this convention, and we now explicit its end
variables in terms of $u_l$ and $w_l$.

With these conventions we prove the following lemma:
\begin{lemma}\label{exactoscill}
The vertex contribution for a planar regular graph is exactly:
\begin{eqnarray}
&&\delta(\sum_{i}(-1)^{i+1}x_{i}+\sum_{l\in T\cup {\cal L}} u_l)
e^{\imath\sum_{i,j}(-1)^{i+j+1}x_{i}\theta^{-1} x_{j}} 
\nonumber \\ 
&&e^{\imath\sum_{l\in T \cup {\cal L},\;   l \prec j}u_l\theta^{-1} (-1)^{j}x_j
+\imath\sum_{l\in T \cup {\cal L},\; l \succ j }(-1)^j x_{j}\theta^{-1} u_l}
\nonumber \\
&&e^{-\imath\sum_{l,l'\in T \cup {\cal L},\; l \prec l' }u_l\theta^{-1} u_{l'}
-\imath\sum_{l\in  T}\frac{u_l\theta^{-1} v_l}{2}\epsilon(l)
-\imath\sum_{l\in {\cal L}}\frac{u_l\theta^{-1} w_l}{2}\epsilon(l)}
\nonumber \\
&&e^{-\imath\sum_{l\in{\cal L},\, l' \in {\cal L} \cup T;\; l'\subset  l}u_{l'}\theta^{-1} w_l \epsilon(l)} \ .
\end{eqnarray}  
\end{lemma}

\prf We see that the global root $\delta$ function has the argument:
\begin{equation}
\sum_i (-1)^{i+1}x_i+\sum_{l\in {\cal L}\cup T}u_l .
\end{equation}
Since the graph has one broken face we always have an even number of vertices on the external face between 
two external fields. We express all the internal loop variables as 
functions of $u$'s and $w$'s. Therefore the quadratic term in the external 
vertices can be written as:
\begin{equation}
\sum_{i<j}(-1)^{i+j+1}x_i\theta^{-1} x_j\ .
\end{equation}

The linear term in the external vertices is:
\begin{align}
&\sum_{i<j}(-1)^{i+1}s_i\theta^{-1} (-1)^jx_j+\sum_{i>j}(-1)^jx_j\theta^{-1} 
(-1)^{i+1}s_i\nonumber\\
&+\sum_{l\in T, l \succ j}(-1)^{j}x_{j}\theta^{-1} u_l +\sum_{l\in T, l  \prec j} u_l \theta^{-1}(-1)^{j}x_{j}\nonumber
\end{align}
and equals
\begin{align}
&\sum_{l'\in {\cal L}, l' \succ j}u_{l'}\theta^{-1} (-1)^jx_j+\sum_{l'\in {\cal L}, l' \succ j}(-1)^j x_j \theta^{-1} u_{l'}
\nonumber\\
&+ \sum_{l\in T, l \succ j}(-1)^{j}x_{j}\theta^{-1} u_l+\sum_{l\in T, l \prec j} u_l \theta^{-1}(-1)^{j}x_{j}\ .
\end{align}

Consider a loop line from $s_p$ to $s_q$ with $p<q$. Its contribution to the 
vertex amplitude decomposes in a "loop-loop" term and a "loop-tree" term. The first one is:
\begin{align}
&\sum_{i<p}(-1)^{i+1}s_i\theta^{-1} (-1)^p s_p+\sum_{\substack{p<i\\i\neq q}}(-1)^ps_p\theta^{-1}
(-1)^{i+1}s_i+s_{p}\theta^{-1}s_{q}\nonumber\\
&+\sum_{\substack{i<q\\i\neq p}}(-1)^{i+1}s_i\theta^{-1} (-1)^q s_q+\sum_{q<i}(-1)^ps_q\theta^{-1}
(-1)^{i+1}s_i\nonumber\\
=&\sum_{i<p}(-1)^{i+1}s_i\theta^{-1} [(-1)^{p}s_p+(-1)^qs_q] \nonumber\\
&+\sum_{q<i}[(-1)^ps_p+(-1)^{q}s_q]\theta^{-1} (-1)^{i+1}s_i\nonumber\\
&+\sum_{p<i<q}(-1)^{i+1}s^i\theta^{-1} [(-1)^{p+1}s_p+(-1)^qs_q]+
s_p\theta^{-1} s_q \ .
\end{align}

Taking into account that $(-1)^{i+1}s_i+(-1)^{j+1}s_j=u_{l'}$ if $s_i$ and 
$s_j$ are the two ends of the loop line $l'$, we can rewrite the above 
expression as:
\begin{eqnarray}
\sum_{l' \prec l}u_{l'}\theta^{-1} (-u_l)+\sum_{l' \succ l}(-u_l)\theta^{-1} u_{l'}+
\sum_{l'\subset  l}u_{l'}\theta^{-1} (-1)^{p+1}w_l 
\nonumber\\
+ (-1)^{p+1}\frac{u_l\theta^{-1} w_l}{2}+\sum_{l',l\subset  l'}u_l\theta^{-1} 
(-1)^{i+1}w_{l'}\ ,
\end{eqnarray}
where $l$ is fixed in all the above expressions.
Summing the contributions of all the lines (being careful not to count the same term twice) we get the 
final result:
\begin{equation}
-\sum_{l'  \prec l}u_{l'}\theta^{-1} u_l-\sum_{l,l'\subset  l}u_{l'}\theta^{-1} 
w_l~\epsilon(l)-\sum_l\frac{u_l\theta^{-1} w_l~\epsilon(l)}{2}\ .
\end{equation}

We still have to add the "loop-tree" contribution. It reads:
\begin{align}
&\sum_{l'\in T,l'\prec p}u_{l'}\theta^{-1}(-1)^{p}s_{p}+\sum_{l'\in T,l' \succ p}(-1)^{p} s_p \theta^{-1}
u_{l'}\nonumber\\
&+\sum_{l'\in T,l'  \prec q}u_{l'}\theta^{-1}(-1)^{q}s_{q}+\sum_{l'\in T,l' \succ q}(-1)^{q} s_q \theta^{-1}u_{l'}\nonumber
\end{align}
and equals
\begin{align}
&\sum_{l'\in T;l'  \prec p,q}u_{l'}\theta^{-1}[(-1)^{p}s_p+(-1)^q s_q]+\sum_{l'\in T;l' \succ p,q} [(-1)^{p}s_p+(-1)^q s_q]\theta^{-1}u_{l'}\nonumber\\
&+\sum_{l'\in T;p\prec  l' \prec q}u_{l'}\theta^{-1}[(-1)^{p+1}s_p+(-1)^{q}s_q]\nonumber\\
=&\sum_{l'\in T;l'\prec l}u_{l'}\theta^{-1}(-u_l)+\sum_{l' \in T;l' \succ  l}(-u_l)\theta^{-1}u_{l'}+
\sum_{l'\in T;l'\subset  l}u_{l'}\theta^{-1}(-1)^{p+1}w_l\ .
\end{align}
Collecting all the factors proves the lemma.
\qed

\subsection{Renormalization of the Four-point Function}\label{ren4pt}

Consider a 4 point subgraph which needs to be renormalized,
hence is a node of the Gallavotti-Nicol\`o tree.
This means that there is $(i,k)$ such that $N(G^{i}_{k})=4$. The four external positions
of the amputated graph are labeled $x_{1},x_{2},x_{3}$ and $x_{4}$. We also define $Q$, $R$ and
$S$ as three skew-symmetric matrices of respective sizes $4\times l(G^{i}_{k})$,
$l(G^{i}_{k})\times l(G^{i}_{k})$
and $[n(G^{i}_{k})-1]\times l(G^{i}_{k})$, where we recall 
that $n(G)-1$ is the number of loops of a 4 point graph with $n$ vertices. The amplitude associated to the
connected component $G^{i}_{k}$ is then
\begin{eqnarray}
A(G^{i}_{k})(x_{1},x_{2},x_{3},x_{4})&=&\int
\prod_{\ell\in T^{i}_{k}}  du_{\ell}   C_{\ell}(x, u, w) 
\prod_{l \in G^{i}_{k},\,l\not\in T}  du_{l}  d w_{l} C_{l}(u_l, w_l) 
\nonumber\\
&&\hskip-4cm\delta\Big(x_{1}-x_{2}+x_{3}-x_{4}+\sum_{l\in G^{i}_{k}} u_l\Big)
e^{\imath \lbt \sum_{p<q}(-1)^{p+q+1}x_{p}\theta^{-1}
x_{q}+XQU+URU+USW \rbt } . \label{eq:4pt-ini}
\end{eqnarray}
The exact form of the factor $\sum_{p<q} (-1)^{p+q+1}x_{p}\theta^{-1}
x_{q}$  follows from Lemma \ref{exactoscill}. From this Lemma and (\ref{exactvvalue}) below 
would also follow exact expressions for
$Q$, $R$ and $S$, but we wont need them. The important fact is that 
there are no quadratic oscillations in $X$ times $W$ (because $B=1$) nor in $W$ times $W$ (because $g=0$).
$C_{l}$ is the propagator of the
line $l$. For loop lines $C_{l}$ is expressed in terms of $u_l$ and $w_l$
by formula (\ref{tanhyp}), (with $v$ replaced by our notation $w$ for long variables of loop lines). 
But for tree lines $\ell \in T^i_k$
recall that the solution of the system of branch $\delta$ functions for $T$ has reexpressed the
corresponding long variables $v_\ell$ in terms of the short variables $u$,
and the external and long loop variables of the branch graph $G_\ell$ 
which lies ``over" $\ell$ in the rooted tree $T$.
This is the essential content of the subsection \ref{filkreduc1}. 
More precisely consider a line $\ell \in T^i_k$ with scale $i(\ell)\ges i$;
we can write
\begin{equation}\label{exactvvalue}
v_\ell =  X_\ell + W_\ell + U_\ell 
\end{equation} 
where 
\begin{equation}\label{xvalue}
X_\ell = \sum_{e\in E(\ell)}  \epsilon_{\ell,e} x_{e}
\end{equation}
is a linear combination on the set of external variables of the branch graph $G_\ell$ with 
the correct alternating signs $\epsilon_{\ell,e}$, 
\begin{equation}\label{wvalue}
W_\ell = \sum_{l \in {\cal L} (\ell)} 
\epsilon_{\ell,l} w_{l}
\end{equation}
is a linear combination over the set ${\cal L} (\ell)$ of long loop variables 
for the external lines of $G_\ell$ (and $\epsilon_{\ell,l}$ are other signs), and
\begin{equation}\label{uvalue}
U_\ell = \sum_{l' \in S (\ell)} \epsilon_{\ell, l'} u_{l'}
\end{equation}
is a linear combination over a set $S_\ell$ 
of short variables that we do not need to know explicitly.
The tree propagator for line $\ell$ then is
\begin{equation}\label{propatre}
C_{\ell}(u_\ell, X_\ell, U_\ell, W_\ell) = \int_{M^{-2i(\ell)}}^{M^{-2(i(\ell)-1)}} 
\frac{\Omega d\alpha_\ell e^{-\frac{\Omega}{4} 
\{ \coth(\frac{\alpha_\ell}{2})u_l^2  + \tanh(\frac{\alpha_\ell}{2}) [X_\ell + W_\ell + U_\ell ]^2 \} }}{[2\pi\sinh(\alpha_\ell)]^2}
\ .
\end{equation}

To renormalize, let us call $e= \max e_p$, $p=1,...,4 $ the highest external index 
of the subgraph $G^{i}_{k}$.  We have $e<i$ since $G^{i}_{k}$
is a node of the Gallavotti-Nicol\`o tree. We evaluate $A(G^{i}_{k})$ on external 
fields\footnote{For the external index to be exactly $e$ the external smearing factor should be in fact 
$\prod_{p} \varphi^{\les e}(x_p) - \prod_{p} \varphi^{\les e-1}(x_p)$ but this subtlety is inessential.}
$\varphi^{\les e}(x_p)$ as:
\begin{eqnarray}
A(G^{i}_{k})&=&\int\prod_{p=1}^{4}dx_{p}\varphi^{\les e}(x_{p})\,
A(G^{i}_{k})(x_{1},x_{2},x_{3},x_{4})\nonumber\\
&=&\int\prod_{p=1}^{4}dx_{p } \varphi^{\les e}(x_{p})\ e^{\imath\text{Ext}}
\prod_{\ell\in T^{i}_{k} }  du_{\ell}   C_{\ell}(u_\ell, tX_\ell, U_\ell, W_\ell) 
\\
\prod_{l \in G^{i}_{k} \, \ l \not \in T} & du_{l}&  d w_{l} C_{l}(u_l, w_l) \ 
\delta\Big(\Delta+t\sum_{l\in G^{i}_{k}}u_{l}\Big)
e^{\imath tXQU+\imath URU+\imath USW}\  \Bigg|_{t=1}\ . \nonumber
\end{eqnarray}
with $\Delta = x_{1}-x_{2}+x_{3}-x_{4}$ and $\text{Ext}=\sum_{p<q=1}^{4}(-1)^{p+q+1}x_{p}\theta^{-1} x_{q}$. 

This formula is designed so that at $t=0$ all dependence on the external variables $x$
factorizes out of the $u,w$ integral in the desired vertex form for renormalization
of the $\varphi \star \varphi \star \varphi \star \varphi$ interaction in the action (\ref{action}).
We now perform a Taylor expansion to first order with respect to the $t$ variable and prove 
that the remainder term is irrelevant. Let $\mathfrak{U}=\sum_{l\in G^{i}_{k}}u_{l}$, and
\begin{eqnarray}
{\mathfrak R}(t) &=&   -\sum_{\ell\in T^{i}_{k} }\frac{\Omega}{4}\tanh(\frac{\alpha_\ell}{2}) \Bigg\{
t^2 X_\ell ^2 + 2t X_\ell \big[ W_\ell + U_\ell \big] \Bigg\} 
\nonumber\\
&\equiv& - t^2  {\cal A} X . X  -  2t {\cal A}X . (W + U )
\ .
\end{eqnarray} 
where ${\cal A}_\ell = \frac{\Omega}{4}\tanh(\frac{\alpha_\ell}{2})$, and $X\cdot Y$ means
$\sum_{\ell\in T^{i}_{k} } X_\ell  . Y_\ell $.
We have
\begin{eqnarray}
A(G^{i}_{k})&=&\int\prod_{p=1}^{4}dx_{p}\varphi^{\les e}(x_{p})\, e^{\imath\text{Ext}}
\prod_{\ell\in T^{i}_{k} }  du_{\ell}   \ C_{\ell}(u_\ell, U_\ell, W_\ell) 
\nonumber\\
&&
\bigg[ \prod_{l \in G^{i}_{k} \, \ l \not \in T}  du_{l}  d w_{l} C_{l}(u_l, w_l) \bigg]
\ e^{\imath URU+\imath USW}
\\
&&\hspace{-2cm}
\Bigg\{ \delta(\Delta)
+ \int_{0}^{1}dt\bigg[ \mathfrak{U}\cdot \nabla \delta(\Delta+t\mathfrak{U})
+\delta(\Delta+t\mathfrak{U}) [\imath XQU  + {\mathfrak R}' (t)]  \bigg] 
e^{\imath tXQU + {\mathfrak R}(t)}  \Bigg\} \ .\nonumber
\end{eqnarray}
where
$C_{\ell}(u_\ell, U_\ell, W_\ell) $ is given by (\ref{propatre})
but taken at $X_\ell=0$.

The first term, denoted by $\tau A$, is of the desired form (\ref{vertex}) times a number
independent of the external variables $x$. It is asymptotically constant in
the slice index $i$, hence the sum over $i$
at fixed $e$ is logarithmically divergent: this is the divergence
expected for the four-point function. It remains only to check that $(1-\tau)A$ converges
as $i-e \to \infty$.  But we have three types of terms in  $(1-\tau)A$, 
each providing a specific improvement over the regular, log-divergent power counting of $A$:

\begin{itemize}

\item The term $\mathfrak{U}\cdot \nabla \delta(\Delta+t\mathfrak{U})$. For this term, integrating by parts over
external variables, the $\nabla $ acts on external fields $\varphi^{\les e}$, 
hence brings at most $M^e$ to the bound,
whether the $\mathfrak{U}$ term brings at least $M^{-i}$.

\item  The term $XQU$. Here $X$ brings at most $M^e$ and $U$ brings at least $M^{-i}$.

\item The term ${\mathfrak R}' (t)$. It decomposes into terms in ${\cal A} X \cdot X$, ${\cal A} X\cdot U$
and ${\cal A} X\cdot W$. Here
the ${\cal A}_\ell$ brings at least $M^{-2 i(\ell)}$, $X$ brings at worst $M^{e}$,
$U$ brings at least $M^{-i}$ and $X_\ell  W_\ell$ brings at worst $M^{e+i(\ell)}$.
This last point is the only subtle one:
if $\ell \in T^i_k$, remark that because $T^i_k$ is a sub-tree within each
Gallavotti-Nicol\`o subnode of $G^i_k$, in particular all parameters
$w_{l'}$ for $l' \in{\cal L} (\ell)$ which appear in $W_\ell$
must have indices lower or equal to $i(\ell)$ (otherwise they would have been chosen instead of $\ell$
in $T^i_k$).

\end{itemize}

In conclusion, since $i(\ell) \ges i$, the Taylor remainder term $(1-\tau)A$
improves the power-counting of the connected component
$G_{k}^{i}$ by a factor at least $M^{-(i-e)}$.
This additional $M^{-(i-e)}$ factor makes $(1-\tau)A(G^{i}_{k})$ convergent 
and irrelevant as desired. 

\vspace*{\stretch{0.5}}

\subsection{Renormalization of the Two-point Function}\label{Ren2pt}

We consider now the nodes such that $N(G^{i}_{k})=2$.  We use the same notations than in the previous subsection.  The two external points are labeled $x$ and $y$. Using the global 
$\delta$ function,  which is now 
$\delta\Big(x-y + {\mathfrak U}\Big)$,
we remark that the external oscillation $e^{\imath x \theta^{-1} y}$
can be absorbed in a redefinition of the term $e^{\imath tXQU}$, which we do from now on.
Also we want to use expressions symmetrized over $x$ and $y$.
The full amplitude is
\begin{eqnarray}\label{2point1}
A(G^{i}_{k}) &=&\int dx dy 
\varphi^{\les e}(x)\varphi^{\les e}(y) \delta\Big(x-y + {\mathfrak U}\Big)
\\
&&  \prod_{l \in G^{i}_{k} ,\; l \not \in T}  du_{l}  d w_{l} C_{l}(u_l, w_l)
\nonumber\\
&&
\prod_{\ell\in T^{i}_{k} }  du_{\ell}   C_{\ell}(u_\ell, X_\ell, U_\ell, W_\ell) 
\ e^{\imath XQU+\imath URU+\imath USW}\; . \nonumber
\end{eqnarray}
First we write the identity
\begin{eqnarray}
\varphi^{\les e}(x)\varphi^{\les e}(y) &=& 
\frac 12 \bigg[ [\varphi^{\les e}(x)]^2 + [\varphi^{\les e}(y) ]^2 \; -\; 
[ \varphi^{\les e}(y) - \varphi^{\les e}(x)]^2 \bigg]\; ,
\label{eq:2pt-sym}
\end{eqnarray}  
we develop it as
\begin{eqnarray}\label{symdev}
\varphi^{\les e}(x)\varphi^{\les e}(y) &=& 
\frac 12 \Bigg\{ [\varphi^{\les e}(x)]^2 + [\varphi^{\les e}(y) ]^2 
- \bigg[   (y-x)^\mu \cdot \nabla_\mu \varphi^{\les e}(x)
\\&&
\hskip-1cm  + 
\int_{0}^{1}ds (1-s)  (y-x)^\mu (y-x)^\nu \nabla_\mu \nabla_\nu
\varphi^{\les e}(x + s(y-x))  \bigg]^2 \Bigg\}\; ,
\nonumber
\end{eqnarray}  
and substitute into (\ref{2point1}).
The first term $A_0$ is a symmetric combination 
with external fields at the same argument. Consider the case with the two external legs
at $x$, namely the term in $[\varphi^{\les e}(x)]^2 $. 
For this term we integrate over $y$. This uses the $\delta$ function. We 
perform then a Taylor expansion in $t$ at order $3$ of the  remaining function
\begin{equation}
  \label{eq:f}
  f(t)=  e^{\imath tXQ U   + {\mathfrak R}(t)}\; ,
\end{equation} 
where we recall that
${\mathfrak R}(t)= - [ t^2  {\cal A} X . X + 2t {\cal A}X . (W + U )]$. We get 
\begin{eqnarray}\label{2point2}
A_0 &=& \frac 12 \int dx  [\varphi^{\les e}(x)]^2    \,
e^{\imath  (URU+ USW)} 
\nonumber\\
&& \prod_{l \in G^{j}_{k} , \; l \not \in T}  du_{l}  d w_{l} C_{l}(u_l, w_l)
 \prod_{\ell\in T^{i}_{k}}  du_{\ell}   C_{\ell}(u_\ell, U_\ell, W_\ell) 
\nonumber\\
&& 
\lbt f(0)+f'(0)+\frac 12f''(0)+\frac 12\int_{0}^{1}dt\,(1-t)^{2}f^{(3)}(t)\rbt\
\ . 
\end{eqnarray}

In order to evaluate that expression, let $A_{0,0},A_{0,1},A_{0,2}$ be the
zeroth, first and second order terms in this Taylor
expansion, and $A_{0,R}$ be the remainder term. First,
\begin{eqnarray}
A_{0,0}&=&\int dx\, [\varphi^{\les e}(x)]^2  \; e^{\imath (URU+ USW)}
\prod_{l \in G^{i}_{k} \; , l \not \in T}  du_{l}  d w_{l} C_{l}(u_l, w_l)
\nonumber\\
&&
\prod_{\ell\in T^{i}_{k} }  du_{\ell}   C_{\ell}(u_\ell, U_\ell, W_\ell)  
\end{eqnarray}
is quadratically divergent and exactly of the expected form for the mass counterterm. Then
\begin{eqnarray}
A_{0,1}&=&
\frac 12 \int dx [\varphi^{\les e}(x)]^2 \ e^{\imath (URU+ USW)}
\prod_{l \in G^{i}_{k}  ,\; l \not \in T}  du_{l}  d w_{l} C_{l}(u_l, w_l)
\nonumber\\
&&
\prod_{\ell\in T^{i}_{k} }  du_{\ell}   C_{\ell}(u_\ell, U_\ell, W_\ell)     
\bigg(  \imath XQU  +  {\mathfrak R}' (0)   \bigg) 
\end{eqnarray} 
vanishes identically. Indeed all the terms are odd 
integrals over the $u,w$ variables. $A_{0,2}$ is more complicated:
\begin{eqnarray}\label{eqtwo2}
A_{0,2}&=& \frac 12 \int dx [\varphi^{\les e}(x)]^2 \ e^{\imath (URU+ USW)}
\prod_{l \in G^{i}_{k}  ,\; l \not \in T}  du_{l}  d w_{l} C_{l}(u_l, w_l)\nonumber\\
&&
\prod_{\ell\in T^{i}_{k} }  du_{\ell}   C_{\ell}(u_\ell, U_\ell, W_\ell)  \ \Bigg( -( XQU)^2   
\nonumber\\
&& \hskip -1cm- 4\imath XQU {\cal A}X \cdot (W + U )  -2 {\cal A}X \cdot X +4  [{\cal A}X \cdot (W + U )] ^2 
 \Bigg) .
\end{eqnarray}

The four terms in $(XQU)^2$, $XQU {\cal A}X \cdot W$ ${\cal A}X \cdot X$ and  $[{\cal A}X \cdot W ] ^2 $
are logarithmically divergent and contribute to the renormalization of the  harmonic frequency term
$\Omega$ in (\ref{action}). (The terms in $x^\mu x^\nu$ with  $\mu \ne \nu$ do not survive by parity
and the terms in $(x^\mu )^2$  have obviously the same coefficient. 
The other terms in $XQU {\cal A}X \cdot U$,
 $({\cal A}X \cdot U)({\cal A}X \cdot W)$ and  $[{\cal A}X \cdot U ] ^2 $ are irrelevant. Similarly the terms 
 in $A_{0,R}()x$ are all irrelevant.
 
For the term in $A_{0}(y)$ in which we have $\int dx [\varphi^{\les e}(y)]^2 $ we have to perform a similar
computation, but beware that it is now $x$ which is integrated with the $\delta$ function so that 
$Q$, $S$, $R$ and ${\mathfrak R}$ change, but not the conclusion.
 
Next we have to consider the term in $\bigg[ (y-x)^\mu \cdot \nabla_\mu \varphi^{\les e}(x) \bigg]^2 $
in (\ref{symdev}), for which we need to develop the
$f$ function only to first order. 
Integrating over $y$ replaces each $y-x$ by a ${\mathfrak U}$ factor so that we get a term
\begin{eqnarray}
A_{1}&=&  \frac 12\int dx\, \bigg[ {\mathfrak U}^\mu \cdot \nabla_\mu \varphi^{\les e}(x) \bigg]^2 
\; e^{\imath (URU+ USW)}\prod_{l \in G^{i}_{k} \; , l \not \in T}  du_{l}  d w_{l} C_{l}(u_l, w_l)
\nonumber\\
&&
\prod_{\ell\in T^{i}_{k} }  du_{\ell}   C_{\ell}(u_\ell, U_\ell, W_\ell) \bigg( f(0)+\int_0^1 dt  f'(t) dt \bigg)
\end{eqnarray}

The first term is
\begin{eqnarray}
A_{1,0} &=& \frac 12\int dx\, \bigg[ {\mathfrak U}^\mu \cdot \nabla_\mu \varphi^{\les e}(x) \bigg]^2 
\; e^{\imath (URU+ USW)}\prod_{l \in G^{i}_{k} \; , l \not \in T}  du_{l}  d w_{l} C_{l}(u_l, w_l)
\nonumber\\
&&
\prod_{\ell\in T^{i}_{k} }  du_{\ell}   C_{\ell}(u_\ell, U_\ell, W_\ell) 
\end{eqnarray}
The terms with $\mu \ne \nu$ do not survive by parity. The other ones reconstruct a counterterm proportional
to the Laplacian. The power-counting of this factor $A_{1,0}$ is improved, with respect to $A$, by
a factor $M^{-2(i-e)}$ which makes it only logarithmically divergent, as should 
be for a wave-function counterterm. 

The remainder term in $A_{1,R}^{x}$ has an additional factor at worst $M^{-(i-e)}$
coming from the $\int_0^1 dt  f'(t) dt $ term,
hence is irrelevant and convergent.

Finally the remainder terms $A_{R}$ with three or four gradients in (\ref{symdev})
are also irrelevant and convergent. Indeed  we have terms of various types:

\begin{itemize}

\item There are terms in $U^3$ with $\nabla^3 $. The  $\nabla $ 
act on the variables $x$, hence on external fields, hence bring at most $M^{3e}$ to the bound,
whether the $\mathfrak{U}^3$ brings at least $M^{-3i}$.

\item Finally there are terms with 4 gradients which are still smaller.
\end{itemize}

Therefore for the renormalized amplitude $A_{R}$
the power-counting is improved, with respect to $A_{0}$, by a factor $M^{-3(i-e)}$, 
and becomes convergent.

Putting together the results of the two previous section, we have
proved that the usual effective series  which expresses any connected function of the theory
in terms of an infinite set of effective couplings, related one to each other by a discretized flow \cite{Riv1},
have finite coefficients to all orders. Reexpressing these effective series in terms of the 
renormalized couplings would reintroduce in the usual way the Zimmermann's forests of ''useless" counterterms
and build the standard ``old-fashioned" renormalized series.
The most explicit way to check finiteness of these renormalized series in order to complete the ``BPHZ theorem"
is to use the standard ``classification of forests"  which distributes  Zimmermann's forests into packets
such that the sum over assignments in each packet is finite \cite{Riv1}\footnote{One could also use the popular
inductive scheme of Polchinski, which however does not extend yet to non-perturbative 
``constructive" renormalization}. This part is completely standard and identical
to the commutative case. Hence the proof of Theorem \ref{BPHZ1} is completed.

\section{The LSZ Model}
\setcounter{equation}{0}

In this section we prove the perturbative renormalizability of a generalized\\
Langmann-Szabo-Zarembo model \cite{Langmann:2002ai}. It consists in a bosonic complex scalar field theory in
a fixed magnetic background plus an harmonic oscillator. The quartic interaction is of the Moyal type. The
action functional is given by 
\begin{align}
  S=&\int\frac 12\bar{\varphi}\lbt -D^{\mu}D_{\mu}+\Omega^{2}x^{2}+\mu_{0}^{2}\rbt\varphi
 +\lambda\,\bar{\varphi}\star\varphi\star\bar{\varphi}\star\varphi\label{eq:lszaction}
\end{align}
where $D_{\mu}=\partial_{\mu}-\imath B_{\mu\nu}x^{\nu}$ is the
covariant derivative. The $1/2$ factor is somewhat unusual in a complex
theory but it allows us to recover exactly the results given in
\cite{Propaga} with $\Omega^{2}\rightarrow\omega^{2}=\Omega^{2}+B^{2}$.
By expanding the quadratic part of the action, we
get a $\Phi^{4}$-like kinetic part plus an angular momentum term:
\begin{align}
  \bar{\varphi}D^{\mu}D_{\mu}\varphi
  +\Omega^{2}x^{2}\bar{\varphi}\varphi=&\,\bar{\varphi}\lbt\Delta -\omega^{2}x^{2}-2BL_{5}\rbt\varphi
\end{align}
with $L_{5}=x^{1}p_{2}-x^{2}p_{1}+x^{3}p_{4}-x^{4}p_{3}= x \wedge \nabla$. Here the
skew-symmetric matrix $B$ has been put in its canonical form
\begin{equation}
  \label{eq:Bform}
  B=\begin{pmatrix}\begin{matrix}0&-1\\1&\phantom{-}0\end{matrix}&(0)\\
    (0)&\begin{matrix}0&-1\\1&\phantom{-}0\end{matrix}
    \end{pmatrix}.
\end{equation}
In $x$ space, the interaction term is exactly the same as (\ref{vertex}). The complex conjugation of the fields only selects the orientable graphs.\\
At $\Omega=0$, the model is similar to the Gross-Neveu theory. This will be
treated in a future paper \cite{RenNCGN05}. If we additionally set
$B=\theta^{-1}$ we recover the integrable LSZ model \cite{Langmann:2002ai}.

\subsection{Power Counting}
\label{sec:lszpowcount}

The propagator corresponding to the action (\ref{eq:lszaction}) has been
calculated in \cite{Propaga} in the two-dimensional case. The
generalization to higher dimensions e.g. four, is straightforward:
\begin{align}
  C(x,y)=&\int_{0}^{\infty}dt\,\frac{\omega^{2}}{(2\pi\sinh\omega
    t)^{2}}\ \exp-\frac\omega 2\lbt\frac{\cosh Bt}{\sinh\omega
    t}(x-y)^{2}\right.\label{eq:lszprop}\\
&\left.+\frac{\cosh\omega t-\cosh Bt}{\sinh\omega
    t}(x^{2}+y^{2})+\imath\frac{\sinh Bt}{\sinh\omega t}x\theta^{-1} y\rbt.\nonumber
\end{align}
Note that the sliced version of (\ref{eq:lszprop}) obeys the same bound
(\ref{eq:propbound-phi4}) as the $\varphi^{4}$ propagator. Moreover the additional
oscillating phases $\exp\imath x\theta^{-1} y$ are of the form $\exp\imath\, u_{l}\theta^{-1} v_{l}$.
Such terms played no role in the power counting of the $\Phi^{4}$ theory. They
were bounded by one. This allows to conclude that Lemmas \ref{crudelemma} and \ref{improvedbound} hold
for the generalized LSZ model. Note also that in this case, the theory
contains only orientable graphs due to the use of complex fields.

\subsection{Renormalization}
\label{lszrenorm}

As for the noncommutative $\Phi^{4}$ theory, we only need to renormalize the
planar ($g=0$) two and four-point functions with only one external face.\\
Recall that the oscillating factors of the propagators are
\begin{equation}
  \label{ eq:osc-prop}
  \exp\imath\frac{\sinh Bt}{2\sinh\omega t}u_{l}\theta^{-1} v_{l}.
\end{equation}
After resolving the $v_{\ell},\,\ell\in T$ variables in terms of $X_{\ell}$,
$W_{\ell}$ and  $U_{\ell}$, they can be included in the vertices oscillations
by a redefinition of the $Q$, $S$ and $R$ matrices (see
(\ref{eq:4pt-ini})). For the four-point function, we can then perform the same
Taylor subtraction as in the $\Phi^{4}$ case.\\
The two-point function case is more subtle. Let us consider the generic
amplitude
\begin{eqnarray}
  A(G^{i}_{k})&=&\int dx dy 
  \bar{\varphi}^{\les e}(x)\varphi^{\les e}(y) \delta\big(x-y + {\mathfrak U}\big)
  \\
  &&\prod_{l \in G^{i}_{k},\, l\not\in T}du_{l}dw_{l} C_{l}(u_l, w_l)
  \nonumber\\
  &&\prod_{\ell\in T^{i}_{k}}du_{\ell}C_{\ell}(u_\ell,X_\ell,U_\ell,W_\ell) 
  \ e^{\imath XQU+\imath URU+\imath USW}\, .\nonumber
\end{eqnarray}
The symmetrization procedure
(\ref{eq:2pt-sym}) over the external fields is not possible anymore, the
theory being complex. Nevertheless we can decompose
$\bar{\varphi}(x)\varphi(y)$ in a symmetric and an anti-symmetric part:
\begin{align}
  \bar{\varphi}(x)\varphi(y)=&\,\frac
  12\lbt\bar{\varphi}(x)\varphi(y)+\bar{\varphi}(y)\varphi(x)+\bar{\varphi}(x)\varphi(y)-\bar{\varphi}(y)\varphi(x)\rbt\nonumber\\
  \defi&\,\lbt{\cal S}+{\cal A}\rbt\bar{\varphi}(x)\varphi(y).
\end{align}
The symmetric part of $A$, called $A_{s}$, will lead to the same
renormalization procedure as the $\Phi^{4}$ case. Indeed,
\begin{align}
  {\cal S}\bar{\varphi}(x)\varphi(y)=&\,\frac
  12\lbt\bar{\varphi}(x)\varphi(y)+\bar{\varphi}(y)\varphi(x)\rbt\nonumber\\
  =&\,\frac 12\lb\bar{\varphi}(x)\varphi(x)+\bar{\varphi}(y)\varphi(y)-\lbt\bar{\varphi}(x)-\bar{\varphi}(y)\rbt\lbt\varphi(x)-\varphi(y)\rbt\rb
\end{align}
which is the complex equivalent of (\ref{eq:2pt-sym}).\\
In the anti-symmetric part of $A$, called $A_{a}$, the linear terms
$\bar{\varphi}\nabla\varphi$ do not compensate:
\begin{align}
  {\cal A}\bar{\varphi}(x)\varphi(y)=&\,\frac
  12\lbt\bar{\varphi}(x)\varphi(y)-\bar{\varphi}(y)\varphi(x)\rbt\nonumber\\
  =&\,\frac
  12\Big(\bar{\varphi}(x)(y-x)\cdot\nabla\varphi(x)-(y-x)\cdot\nabla\bar{\varphi}(x)\varphi(x)\nonumber\\
    &+\frac 12\bar{\varphi}(x)((y-x)\cdot\nabla)^{2}\varphi(x)-\frac
  12((y-x)\cdot\nabla)^{2}\bar{\varphi}(x)\varphi(x)\nonumber\\
  &+\frac
  12\int_{0}^{1}ds(1-s)^{2}\bar{\varphi}(x)((y-x)\cdot\nabla)^{3}\varphi(x+s(y-x))\nonumber\\
  &-((y-x)\cdot\nabla)^{3}\bar{\varphi}(x+s(y-x))\varphi(x)\Big).\label{eq:taylor-as}
\end{align}

\newpage

We decompose $A_{a}$ into five parts following the Taylor expansion (\ref{eq:taylor-as}):
\begin{align}
  A_{a}^{1+}&=\int dxdy\,\bar{\varphi}(x)(y-x)\cdot\nabla\varphi(x)\delta\big(x-y + {\mathfrak U}\big)\\
  &\prod_{l \in G^{i}_{k},\, l\not\in T}du_{l}dw_{l} C_{l}(u_l, w_l)
  \nonumber\\
  &\prod_{\ell\in T^{i}_{k}}du_{\ell}C_{\ell}(u_\ell,X_\ell,U_\ell,W_\ell) 
  \ e^{\imath XQU+\imath URU+\imath USW}\nonumber
\end{align}
which is
\begin{align}
  &\int dx\,\bar{\varphi}(x)\,\mathfrak{U}\cdot\nabla\varphi(x)
  \prod_{l \in G^{i}_{k},\, l\not\in T}du_{l}dw_{l} C_{l}(u_l, w_l)
  \nonumber\\
  &\prod_{\ell\in T^{i}_{k}}du_{\ell}C_{\ell}(u_\ell,X'_\ell,U'_\ell,W_\ell) 
  \ e^{\imath XQ'U+\imath URU+\imath USW}\nonumber
\end{align}
where we performed the integration over $y$ thanks to the delta function. The
changes have been absorbed in a redefinition of $X_{\ell}$, $U_{\ell}$ and
$Q$. From now on $X_{\ell}$ (and $X$) contain only $x$ (if $x$ is hooked to the branch
$b(l)$) and we forget the primes for $Q$ and $U_{\ell}$. We expand the
function $f$ defined in (\ref{eq:f}) up to order 2: 
\begin{align}
  A_{a}^{1+}&=\int\bar{\varphi}(x)\,\mathfrak{U}
  \cdot\nabla\varphi(x)\prod_{l \in G^{i}_{k},\, l\not\in T}du_{l}dw_{l} C_{l}(u_l, w_l)
  \nonumber\\
  &\prod_{\ell\in T^{i}_{k}}du_{\ell}C_{\ell}(u_\ell,U_\ell,W_\ell) 
  \ e^{\imath URU+\imath USW}\nonumber\\
  &\lbt f(0)+f'(0)+\int_{0}^{1}dt\,(1-t)f^{''}(t)\rbt.
\end{align}
The zeroth order term vanishes thanks to the parity of the integrals with
respect to the $u$ and $w$ variables. The first order term contains 
\begin{align}
  \bar{\varphi}(x)\,\mathfrak{U^{\mu}}\nabla_{\mu}\varphi(x)\lbt\imath XQU+\mathfrak{R}'(0)\rbt.
\end{align}
The first term leads to
$(\mathfrak{U}^{1}\nabla_{1}+\mathfrak{U}^{2}\nabla_{2})\varphi(x^{1}U^{2}-x^{2}U^{1})$
with the same kind of expressions for the two other dimensions. Due to the
odd integrals, only the terms of the form
$(U^{1})^{2}x^{2}\nabla_{1}-(U^{2})^{2}x^{1}\nabla_{2}$ survive. We are left with integrals like
\begin{align}
\int&(u_{\ell}^{1})^{2}\prod_{l \in G^{i}_{k},\, l\not\in T}du_{l}dw_{l} C_{l}(u_l, w_l)
\prod_{\ell\in T^{i}_{k}}du_{\ell}C_{\ell}(u_\ell,U_\ell,W_\ell) 
\ e^{\imath URU+\imath USW}\; .
\end{align}
To prove that these terms give the same coefficient (in order to reconstruct a
$x\wedge \nabla$ term), note that, apart from the $(u_{\ell}^{1})^{2}$, the involved integrals are actually invariant
under an overall rotation of the $u$ and $w$ variables. Then by performing
rotations of $\pi/2$, we prove that the counterterm is of the form of the
Lagrangian. The $\mathfrak{R}'(0)$ and the remainder term in $A^{1+}_{a}$ are
irrelevant.

Let us now study the other terms in $A_{a}$.
\begin{align}
  A_{a}^{1-}&=-\int dx\,\mathfrak{U}\cdot\nabla\bar{\varphi}(x)\,\varphi(x)
  \prod_{l \in G^{i}_{k},\, l\not\in T}du_{l}dw_{l} C_{l}(u_l, w_l)
  \nonumber\\
  &\prod_{\ell\in T^{i}_{k}}du_{\ell}C_{\ell}(u_\ell,X_\ell,U_\ell,W_\ell) 
  \ e^{\imath XQU+\imath URU+\imath USW} \; .
\end{align}
Once more we decouple the external variables form the internal ones by Taylor
expanding the function $f$. Up to irrelevant terms, this only doubles the
$x\wedge\nabla$ term in $A_{a}^{1+}$.
\begin{align}\label{thelastone}
A_{a}^{2+}&=\frac 12\int\bar{\varphi}(x)\,(\mathfrak{U}\cdot\nabla)^{2}
\varphi(x)\prod_{l \in G^{i}_{k},\, l\not\in T}du_{l}dw_{l} C_{l}(u_l, w_l)\\
&\prod_{\ell\in T^{i}_{k}}du_{\ell}C_{\ell}(u_\ell,U_\ell,W_\ell) 
\ e^{\imath URU+\imath USW}\lbt f(0)+\int_{0}^{1}dt\,f^{'}(t)\rbt.
\nonumber
\end{align}
The $f(0)$ term renormalizes the wave-function. The remainder term in (\ref{thelastone})
is irrelevant. $A_{a}^{2-}$ doubles the $A_{a}^{2+}$ contribution. Finally the
last remainder terms (the last two lines in (\ref{eq:taylor-as})) are irrelevant
too. This completes the proof of the perturbative renormalizability of the LSZ
models.

Remark that if we had considered a real theory with a covariant derivative
which corresponds to a neutral scalar field in a magnetic background, the angular momentum
term wouldn't renormalize. Only the harmonic potential term would. 
It seems that the renormalization ``distinguishes''  the true theory 
in which a \emph{charged} field should couple to a magnetic field. It would be 
interesting to study the renormalization group flow of these kind of models 
along the lines of \cite{GrWu04-2}.

\chapter{Beta Function of $\Phi^{\star 4}_4$ to all orders}

\begin{center}Margherita Disertori, Razvan Gurau\\ 
Jacques Magnen, Vincent Rivasseau
\end{center}

\medskip

The simplest non commutative renormalizable field theory, the $\phi^4$ 
model on four dimensional Moyal space with harmonic potential 
is asymptotically safe up to three loops, as shown by 
H. Grosse and R. Wulkenhaar, M. Disertori and V. Rivasseau. 
We extend this result to all orders.

\noindent
{PACS numbers: 11.10.Nx, 11.10.Gh}

\noindent
{Work supported by ANR grant NT05-3-43374 ``GenoPhy".}

\section{Introduction} 

Non commutative (NC) quantum field theory (QFT) may be important for physics 
beyond the standard model and for understanding the quantum
Hall effect \cite{DN}.
It also occurs naturally as an effective regime of string theory \cite{SeiWitt,a.connes98:noncom}.

The simplest NC field theory is the $\phi_4^4$ 
model on the Moyal space. Its perturbative renormalizability 
at all orders has been proved by 
Grosse, Wulkenhaar and followers \cite{GrWu03-1,GrWu04-3,Rivasseau2005bh,GMRV}. 
Grosse and Wulkenhaar solved the difficult problem of ultraviolet/infrared
mixing by introducing a new harmonic potential term 
inspired by the Langmann-Szabo (LS)
duality \cite{LaSz} between positions and momenta. 

Other renormalizable models of the same kind, including the orientable
Fermionic Gross-Neveu model \cite{RenNCGN05}, have been recently also shown renormalizable at all orders 
and techniques such as the parametric representation have
been extended to NCQFT \cite{gurauhypersyman}.
It is now tempting to conjecture that
commutative renormalizable theories in general have NC renormalizable
extensions to Moyal spaces which imply new parameters. However
the most interesting case, namely the one of gauge theories, still remains elusive.

Once perturbative renormalization is understood, the next problem is to
compute the renormalization group (RG) flow.
It is well known that the ordinary commutative $\phi_4^4$
model is not asymptotically free in the ultraviolet regime. 
This problem, called the Landau ghost or triviality problem affects also quantum electrodynamics.
It almost killed quantum field theory, which was resurrected by the discovery of ultraviolet 
asymptotic freedom in non-Abelian gauge theory \cite{thooft}.

An amazing discovery was made in \cite{GrWu04-2}:
the non commutative $\phi_4^4$ model does not exhibit any Landau ghost 
at one loop. It is not asymptotically free either. 
For any renormalized Grosse-Wulkenhaar harmonic potential parameter $\Omega_{ren} >0$, 
the running $\Omega$ tends to the special LS dual point $\Omega_{bare} =1$ in the ultraviolet. As a result
the RG flow of the coupling constant is simply bounded \footnote{The Landau ghost can be recovered
in the limit $\Omega_{ren}\to 0$.}. This result was extended up to three loops in \cite{DR}.

In this paper we compute the flow at the special LS dual point $\Omega =1$, and check that 
the beta function vanishes at all orders using a kind of Ward identity inspired by 
those of the Thirring or Luttinger models \cite{MdC,BM1,BM2}.
Note however that in contrast with these models, the model we treat has 
quadratic (mass) divergences. 

The non perturbative construction of the model should combine
this result and a non-perturbative multiscale analysis \cite{Riv1,GJ}.
Also we think the Ward identities discovered here might be important for the
future study of more singular models such as Chern-Simons or Yang Mills theories,
and in particular for those which have been advocated in connection with 
the Quantum Hall effect \cite{Susk,Polych,Raam}.

In this letter we give the complete argument of the vanishing 
of the beta function at all orders in the renormalized coupling, but we assume
knowledge of renormalization and effective expansions as described e.g. in \cite{Riv1},
and of the basic papers for renormalization of NC $\phi^4_4$ 
in the matrix base \cite{GrWu03-1,GrWu04-3,Rivasseau2005bh}.

\section{Notations and Main Result}

We adopt simpler notations than those of \cite{GrWu04-2,DR}, and normalize so that $\theta =1$,
hence have no factor of $\pi$ or $\theta$.

The  bare propagator in the matrix base at $\Omega=1$ is
\be \label{propafixed}
C_{m n;k l} = C_{m n} \delta_{m l}\delta_{n k} \ ; \ 
C_{m n}= \frac{1}{A+m+n}\  ,
\ee
where $A= 2+ \mu^2 /4$, $m,n\in \mathbb{N}^2$ ($\mu$ being the mass)
and we used the notations
\be
\delta_{ml} = \delta_{m_1l_1} \delta_{m_2l_2}\ , \qquad m+n = m_1 + m_2 + n_1 + n_2 \ .
\ee

There are two version of this theory, the real and complex one. We focus on the complex case, the result
for the real case follows easily \cite{DR}.

The generating functional is:
\bea
&&Z(\eta,\bar{\eta})=\int d\phi d\bar{\phi}~e^{-S(\bar{\phi},\phi)+F(\bar{\eta},\eta,;\bar{\phi},\phi)}\nonumber\\
&&F(\bar{\eta},\eta;\bar{\phi},\phi)=  \bar{\phi}\eta+\bar{\eta}\phi \nonumber\\
&&S(\bar{\phi},\phi)=\bar{\phi}X\phi+\phi X\bar\phi+A\bar{\phi}\phi+
\frac{\lambda}{2}\phi\bar{\phi}\phi\bar{\phi}
\eea
where traces are implicit and the matrix $X_{m n}$ stands for $m\delta_{m n}$. $S$ is the action and $F$ the external sources. 

We denote $\Gamma^4(a,b,c,d)$ the amputated one particle irreducible four point function with external indices set to $a,b,c,d$. 
Furthermore we denote $\Sigma(a,b)$ the amputated one particle irreducible two point function 
with external indices set to $a,b$ (also called the self-energy). The wave function renormalization is $1-\partial \Sigma(0,0)$ where
$\partial\Sigma(0,0)\equiv\partial_L \Sigma = \partial_R \Sigma = \Sigma (1,0) - \Sigma (0,0)$ is the derivative of the self-energy with respect to one of the two indices $a$ or $b$ \cite{DR}.
Our main result is:

\medskip
\noindent{\bf Theorem}
\medskip
The equation:
\bea\label{beautiful}
\Gamma^{4}(0,0,0,0)=\lambda (1-\partial\Sigma(0,0))^2
\eea
holds up to irrelevant terms \footnote{Irrelevant terms include in particular all non-planar or planar with more than one broken face contributions.}
 to {\bf all} orders of perturbation, either as a bare equation with fixed ultraviolet cutoff,
or as an equation for the renormalized theory. In the latter case $\lambda $ should still be understood 
as the bare constant, but reexpressed as a series in powers of $\lambda_{ren}$.

\section{Ward Identities}

We orient the propagators from a $\bar{\phi}$ to a $\phi$.
For a field $\bar{\phi}_{a b}$ we call the index $a$ a 
{\it left index} and the index, $b$ a {\it right index}. The first (second) index of a $\bar{\phi}$ {\it allways} contracts with 
the second (first) index of a $\phi$.  Consequently for $\phi_{c d}$, $c$ is a {\it right index} and $d$ is a {\it left index}.

Let $U=e^{\imath B}$ with $B$ a small hermitian matrix. We consider the ``left" (as it acts only on the left indices) change of variables\footnote{There is a similar ``right" change of variables, acting only on the right indices.}:
\bea
\phi^U=\phi U;\bar{\phi}^U=U^{\dagger}\bar{\phi} \ .
\eea
The variation of the action is, at first order:
\bea
\delta S&=&\phi U X U^{\dagger}\bar{\phi}-\phi X \bar{\phi}\approx
\imath\big{(}\phi B X\bar{\phi}-\phi X B \bar{\phi}\big{)}\nonumber\\
&=&\imath B\big{(}X\bar{\phi}\phi-\bar{\phi}\phi X \big{)}
\eea
and the variation of the external sources is:
\bea
\delta F&=&U^{\dagger}\bar{\phi}\eta-\bar{\phi}\eta+\bar{\eta}\phi U-\bar{\eta}\phi 
        \approx-\imath B \bar{\phi}\eta+\imath\bar{\eta}\phi B\nonumber\\
	&=&\imath B\big{(}-\bar{\phi}\eta+\bar{\eta}\phi{)} .
\eea
We obviously have:
\bea
&&\frac{\delta \ln Z}{\delta B_{b a}}=0=\frac{1}{Z(\bar{\eta},\eta)}\int d\bar{\phi} d\phi
   \big{(}-\frac{\delta S}{\delta B_{b a}}+\frac{\delta F}{\delta B_{b a}}\big{)}e^{-S+F}\\
   &&=\frac{1}{Z(\bar{\eta},\eta)}\int d\bar{\phi} d\phi  ~e^{-S+F}
\big{(}-[X \bar{\phi}\phi-\bar{\phi}\phi X]_{a b}+
       [-\bar{\phi}\eta+\bar{\eta}\phi]_{a b}\big{)} \  .\nonumber
\eea

We now take $\partial_{\eta}\partial_{\bar{\eta}}|_{\eta=\bar{\eta}=0}$ 
on the above expression. As we have at most two insertions we get only the connected components of the correlation functions.
\bea
0=<\partial_{\eta}\partial_{\bar{\eta}}\big{(}
-[X \bar{\phi}\phi-\bar{\phi}\phi X]_{a b}+
       [-\bar{\phi}\eta+\bar{\eta}\phi]_{a b}\big{)}e^{F(\bar{\eta},\eta)} |_0>_c \ ,
\eea
which gives:
\bea
&&<\frac{\partial(\bar{\eta}\phi)_{a b}}{\partial \bar{\eta}}\frac{\partial(\bar{\phi}\eta)}{\partial \eta}
-\frac{\partial(\bar{\phi}\eta)_{a b}}{\partial \eta}\frac{\partial (\bar{\eta}\phi)}{\partial \bar{\eta}}
- [X \bar{\phi}\phi-\bar{\phi}\phi X]_{a b}
\frac{\partial(\bar{\eta}\phi)}{\partial \bar{\eta}}\frac{\partial (\bar{\phi}\eta)}{\partial\eta}>_c\nonumber\\
&&=0 .
\eea
Using the explicit form of $X$ we get:
\bea
&&(a-b)<[ \bar{\phi}\phi]_{a b}
\frac{\partial(\bar{\eta}\phi)}{\partial \bar{\eta}}\frac{\partial (\bar{\phi}\eta)}{\partial\eta}>_c\nonumber\\
&&=<\frac{\partial(\bar{\eta}\phi)_{a b}}{\partial \bar{\eta}}\frac{\partial(\bar{\phi}\eta)}{\partial \eta}>_c
-<\frac{\partial(\bar{\phi}\eta)_{a b}}{\partial \eta}\frac{\partial (\bar{\eta}\phi)}{\partial \bar{\eta}}> \ ,
\nonumber
\eea
and for $\bar{\eta}_{ \beta \alpha} \eta_{ \nu \mu}$ we get:
\bea
(a-b)<[ \bar{\phi}\phi]_{a b} \phi_{\alpha \beta} 
\bar{\phi}_{\mu \nu }>_c=
<\delta_{a \beta}\phi_{\alpha b} \bar{\phi}_{\mu \nu}>_c
-<\delta _{b \mu }\bar{\phi}_{a \nu} \phi_{\alpha \beta}>_c
\eea

We now restrict to terms in the above expressions which are planar with a single external face,
as all others are irrelevant. Such terms have $\alpha=\nu$, $a=\beta$ and $b=\mu$. 
The Ward identity for $2$ point function reads:
\bea\label{ward2point}
(a-b)<[ \bar{\phi}\phi]_{a b} \phi_{\nu a} 
\bar{\phi}_{b \nu }>_c=
<\phi_{\nu b} \bar{\phi}_{b \nu}>_c
-<\bar{\phi}_{a \nu} \phi_{\nu a}>_c
\eea
(repeated indices are not summed). 

\begin{figure}[hbt]
\centerline{
\includegraphics[width=100mm]{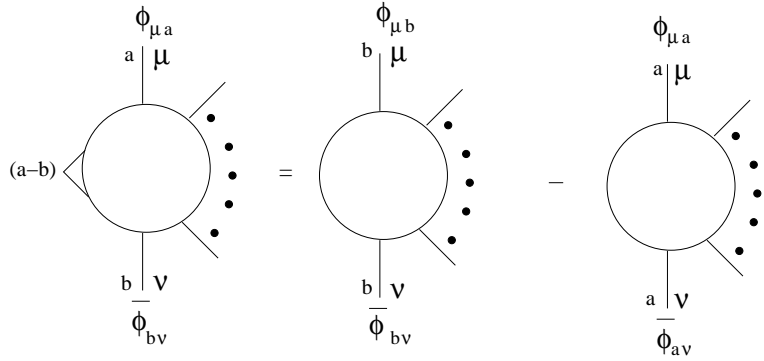}
}
\caption{The Ward identity for a 2p point function with insertion on the left face}\label{fig:Ward}
\end{figure}

Derivating further we get:
\bea
&&(a-b)<[\bar{\phi}\phi]_{a b}\partial_{\bar{\eta}_1}(\bar{\eta}\phi)
\partial_{\eta_1}(\bar{\phi}\eta) \partial_{\bar{\eta}_2}(\bar{\eta}\phi)
\partial_{\eta_2}(\bar{\phi}\eta) >_c=\\
&&<\partial_{\bar{\eta}_1}(\bar{\eta}\phi)
\partial_{\eta_1}(\bar{\phi}\eta)\big{[}
 \partial_{\bar{\eta_2}}
 (\bar{\eta}\phi)_{ab}\partial_{\eta_2}(\bar{\phi}\eta)-\partial_{\eta_2}(\bar{\phi}\eta)_{a b}
 \partial_{\bar{\eta}_2}(\bar{\eta}\phi) \big{]}>_c+1 \leftrightarrow 2 \ .\nonumber
\eea
Take $\bar{\eta}_{1~\beta \alpha}$, $\eta_{1~ \nu\mu}$, $\bar{\eta}_{2~\delta \gamma}$ and $\eta_{2~\sigma \rho}$.
We get:
\bea
&&(a-b)<[\bar{\phi}\phi]_{ab}\phi_{\alpha \beta}\bar{\phi}_{\mu \nu}\phi_{\gamma \delta}
\bar{\phi}_{\rho \sigma}>_c\\
&&=<\phi_{\alpha \beta}\bar{\phi}_{\mu \nu}  \delta_{a \delta}\phi_{\gamma b}\bar{\phi}_{\rho \sigma}>_c
-<\phi_{\alpha \beta}\bar{\phi}_{\mu \nu}\phi_{\gamma \delta}\bar{\phi}_{a \sigma}\delta_{b \rho}>_c+
\nonumber\\
&&<\phi_{\gamma \delta}\bar{\phi}_{\rho \sigma}  \delta_{a \beta}\phi_{\alpha b}\bar{\phi}_{\mu \nu}>_c
-<\phi_{\gamma \delta}\bar{\phi}_{\rho \sigma}\phi_{\alpha \beta}\bar{\phi}_{a \nu}\delta_{b \mu}>_c \ .
\nonumber
\eea
Again neglecting all terms which are not planar with a single external face leads to
\bea\label{ward4point}
&&(a-b) <\phi_{\alpha a}[\bar{\phi}\phi]_{ab}\bar{\phi}_{b\nu}\phi_{\nu \delta}\bar{\phi}_{\delta \alpha}>_c\nonumber\\
&&=
<\phi_{\alpha b}\bar{\phi}_{b \nu}\phi_{\nu \delta}\bar{\phi}_{\delta\alpha}>_c-
<\phi_{\alpha a}\bar{\phi}_{a \nu}\phi_{\nu \delta}\bar{\phi}_{\delta\alpha}>_c \ .
\nonumber
\eea
Clearly there are similar identities for $2p$ point functions for any $p$.

The indices $a$ and $b$ are left indices, so that we have the Ward identity with an insertion on a left face\footnote{There is a similar Ward identity obtained with the ``right" transformation, consequently with the insertion on a right face.}  as represented in  Fig. \ref{fig:Ward}.

We conclude this section by several remarks on the real theory. If $\phi(x)$ is a real function then $\phi_{ab}$ is a hermitian matrix. The action and the sources are:
\bea
 S=\phi X\phi+\frac{\lambda}{4}\phi^4\, , F=\phi\eta \,.
\eea
 We perform the change of variables (preserving the hermitian character of $\phi$):
\bea
\phi^U=U\phi U^{\dagger}
\eea
with constant $U$ a unitary matrix. A straightforward computation shows that the Jacobian of this change of variables is $1$ and the reader can check that the method above gives Ward identities identical with those of the complex model. 

\section{Proof of the Theorem}

We start this section by some definitions:
we will denote $G^{4}(m,n,k,l)$ the connected four point function restricted to the planar one broken face case, where $m,n,k,l$  are the indices of the external face in the correct cyclic order. The first index $m$ allways represents a left index.

\begin{figure}[hbt]
\centerline{
\includegraphics[width=60mm]{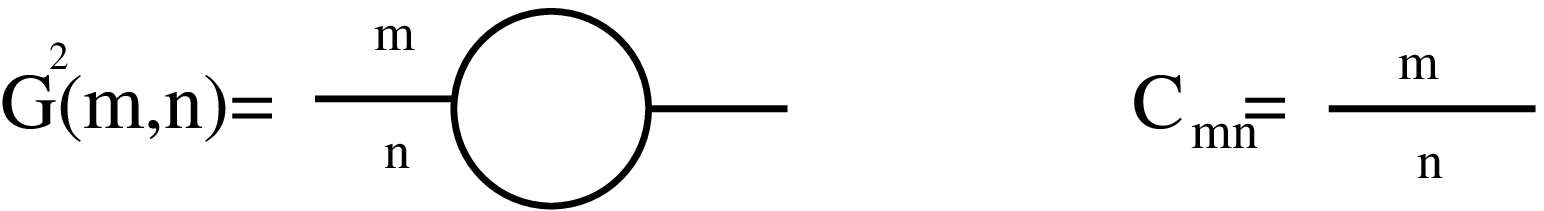}
}
\caption{The {\bf dressed} and the bare propagators}\label{fig:propagators}
\end{figure}

Similarely, $G^{2}(m,n)$ is the connected planar one broken face two point function with $m,n$ the indices on the external face (also called the {\bf dressed} propagator, see Fig. \ref{fig:propagators}). $G^{2}(m,n)$ and $\Sigma(m,n)$ are related by:
\bea
\label{G2Sigmarelation}
  G^{2}(m,n)=\frac{C_{m n}}{1-C_{m n}\Sigma(m,n)}=\frac{1}{C_{m n}^{-1}-\Sigma(m,n)} \, .
\eea 

$G_{ins}(a,b;...)$ will denote the planar one broken face 
connected function with one insertion on the left border where the matrix index jumps from $a$ to $b$. With this notations the Ward identity (\ref{ward2point}) writes:
\bea
(a-b) ~ G^{2}_{ins}(a,b;\nu)=G^{2}(b,\nu)-G^{2}(a,\nu)\, .
\eea

All the identities we use, either Ward identities or the Dyson equation of motion
can be written either for the bare  theory or for the theory with complete mass renormalization, which is the one considered in \cite{DR}. In the first case the parameter $A$ in (\ref{propafixed}) is the bare one, $A_{bare}$
and there is no mass subtraction. In the second case the parameter $A$ in (\ref{propafixed}) 
is $A_{ren}= A_{bare} - \Sigma(0,0)$, and every two point 1PI subgraph is subtracted at 0 external indices\footnote{These mass subtractions need not be rearranged into forests 
since 1PI 2point subgraphs never overlap non trivially.}. Troughout this paper $\partial_{L}$ will denote the derivative with respect to a left index and $\partial_{R}$ the one with respect to a right index. When the two derivatives are equal we will employ the generic notation $\partial$. 

Let us prove first the Theorem in the mass-renormalized case, then in the next subsection
in the bare case. Indeed the mass renormalized theory used is free from any quadratic divergences, and remaining logarithmic subdivergences in the ultra violet  cutoff can be removed easily by going, for instance, to the ``useful" renormalized effective series, 
as explained in \cite{DR}. 

\begin{figure}[hbt]
\centerline{
\includegraphics[width=110mm]{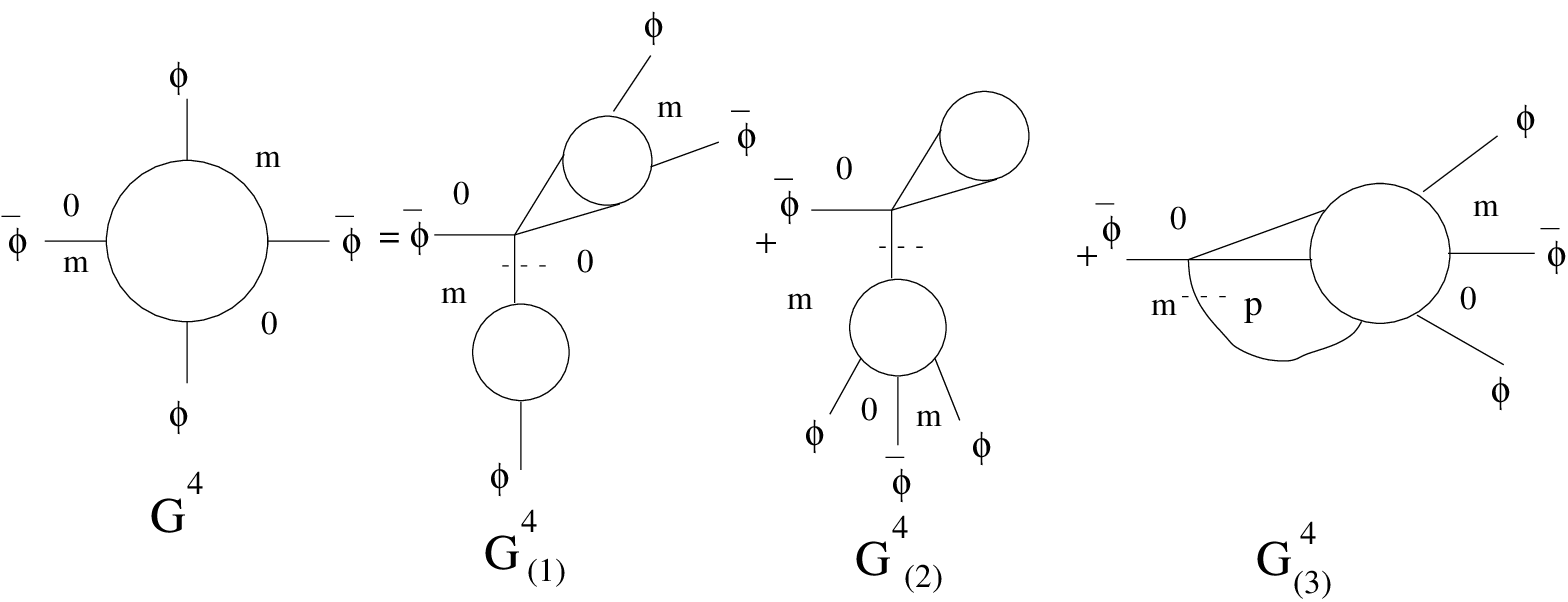}
}
\caption{The Dyson equation}\label{fig:dyson}
\end{figure}

We analyze a four point connected function $G^4(0,m,0,m)$ with index $m \ne 0$ on the right borders. 
This explicit break of left-right symmetry is adapted to our problem.

Consider a $\bar{\phi}$ external line and the first vertex hooked to it. 
Turning right on the $m$ border at this vertex we meet a new line (the slashed line in Fig. \ref{fig:dyson}). The slashed line either separates the graph into two disconnected components ($G^{4}_{(1)}$ and $G^{4}_{(2)}$ in Fig. \ref{fig:dyson}) or not 
($G^{4}_{(3)}$ in Fig. \ref{fig:dyson}). Furthermore, if the slashed line separates the graph into two disconnected components the first vertex may either belong to a four point component ($G^{4}_{(1)}$ in Fig. \ref{fig:dyson}) or to a two point component
($G^{4}_{(2)}$ in Fig. \ref{fig:dyson}). 

We stress that this is a {\it classification} of graphs: the different components depicted in Fig. \ref{fig:dyson} take into account all the combinatoric factors. Furthermore, the setting of the external indices to $0$ on the left borders and $m$ on the right borders distinguishes the $G^{4}_{(1)}$ and $G^{4}_{(2)}$ from their counterparts ``pointing upwards": indeed, the latter are classified in $G^{4}_{(3)}$!

We have thus the Dyson equation:
\bea
\label{Dyson}
 &&G^4(0,m,0,m)\\
&&=G^4_{(1)}(0,m,0,m)+G^4_{(2)}(0,m,0,m)+G^4_{(3)}(0,m,0,m)\, . \nonumber
\eea    

The second term,  $G^{4}_{(2)}$, is zero. Indeed the mass renormalized two point insertion is zero, as it has the external left index set to zero. Note that this is an insertion exclusively on the left border. The simplest case of such an insertion is a
 (left) tadpole. We will (naturally) call a general insertion touching only the left border a ``generalized left tadpole" and denote it by 
 $T^L$. 

We will prove that $G^{4}_{(1)}+G^{4}_{(3)}$ yields 
$\Gamma^4=\lambda (1-\partial \Sigma)^2$ after amputation of the four external propoagators.

We start with $G^{4}_{(1)}$. It is of the form:
\bea
G^4_{(1)}(0,m,0,m)=\lambda C_{0 m} G^{2}(0, m) G^{2}_{ins}(0,0;m)\,.
 \eea

By the Ward identity we have:
\bea
G^{2}_{ins}(0,0;m)&=&\lim_{\zeta\rightarrow 0}G^{2}_{ins}(\zeta ,0;m)=
\lim_{\zeta\rightarrow 0}\frac{G^{2}(0,m)-G^{2}(\zeta,m)}{\zeta}\nonumber\\
&=&-\partial_{L}G^{2}(0,m) \, .
\eea
Using the explicit form of the bare propagator we have $\partial_L C^{-1}_{ab}=\partial_R C^{-1}_{ab}=\partial C^{-1}_{ab}=1$. Reexpressing $G^{2}(0,m)$ by eq.  (\ref{G2Sigmarelation}) we conclude that:
\bea\label{g41}
G^4_{(1)}(0,m,0,m)&=&\lambda
C_{0m}\frac{C_{0m}C^2_{0m}[1-\partial_{L}\Sigma(0,m)]}{[1-C_{0m}\Sigma(0,m)]
(1-C_{0m}\Sigma(0,m))^2}\nonumber\\
&=&\lambda [G^{2}(0,m)]^{4}\frac{C_{0m}}{G^{2}(0,m)}[1-\partial_{L}\Sigma(0,m)]\, .
\eea
The self energy is (again up to irrelevant terms (\cite{GrWu04-3}):
\bea
\label{PropDressed}
\Sigma(m,n)=\Sigma(0,0)+(m+n)\partial\Sigma(0,0) 
\eea 
Therefore up to irrelevant terms ($C^{-1}_{0m}=m+A_{ren}$) we have:
\bea
\label{G2(0,m)}
G^{2}(0,m)=\frac{1}{m+A_{bare}-\Sigma(0,m)}=\frac{1}{m[1-\partial\Sigma(0,0)]+A_{ren}}
\, ,
\eea
and
\bea \label{cdressed}
\frac{C_{0m}}{G^{2}(0,m)}=1-\partial\Sigma(0,0)+\frac{A_{ren}}{m+A_{ren}}\partial\Sigma(0,0) \, .
\eea
Inserting eq. (\ref{cdressed}) in eq. (\ref{g41}) holds:
\bea
\label{g41final}
G^4_{(1)}(0,m,0,m)&=&\lambda [G^{2}(0,m)]^{4}\bigl( 
1-\partial\Sigma(0,0)+\frac{A_{ren}}{m+A_{ren}}\partial\Sigma(0,0)
\bigr) \nonumber\\
&&[1-\partial_{L}\Sigma(0,m)]\, .
\eea

\begin{figure}[hbt]
\centerline{
\includegraphics[width=140mm]{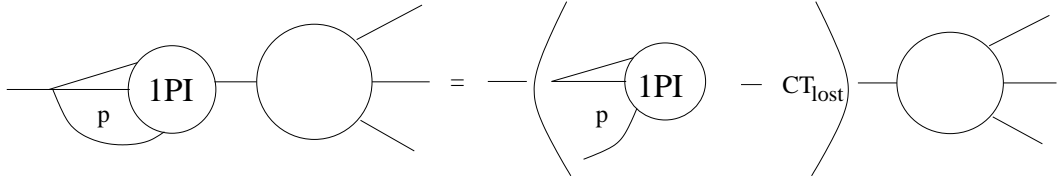}
}
\caption{Two point insertion and opening of the loop with index $p$}\label{fig:insertion}
\end{figure}

For the $G^4_{(3)}(0,m,0,m)$ one starts by ``opening" the face which is ``first on the right''. The summed index of this face is called  $p$ (see Fig. \ref{fig:dyson}).  For bare Green functions this reads:
\bea
\label{opening}
G^{4,bare}_{(3)}(0,m,0,m)=C_{0m}\sum_{ p} G^{4,bare}_{ins}(p,0;m,0,m)\, .
\eea
When passing to mass renormalized Green functions one must be cautious. It is possible that the face $p$ belonged to a  1PI two point insertion in $G^{4}_{(3)}$ (see the left hand side in Fig. \ref{fig:insertion}).
Upon opening the face $p$ this 2 point insertion disappears (see right hand side of Fig. \ref{fig:insertion})! 
When renormalizing, the counterterm  corresponding to this kind of two point insertion will be substracted on the left hand side of  eq.(\ref{opening}), but not on the right hand side. In the equation for $G^{4}_{(3)}(0,m,0,m)$ one must 
therefore \textit{add its missing counterterm}, so that:
\bea
\label{Open2}
G^4_{(3)}(0,m,0,m)&=& C_{0m}\sum_{p} G^{4}_{ins}(0,p;m,0,m)\nonumber\\
     &-&C_{0m}(CT_{lost})G^{4}(0,m,0,m)\,.
\eea

\begin{figure}[hbt]
\centerline{
\includegraphics[width=90mm]{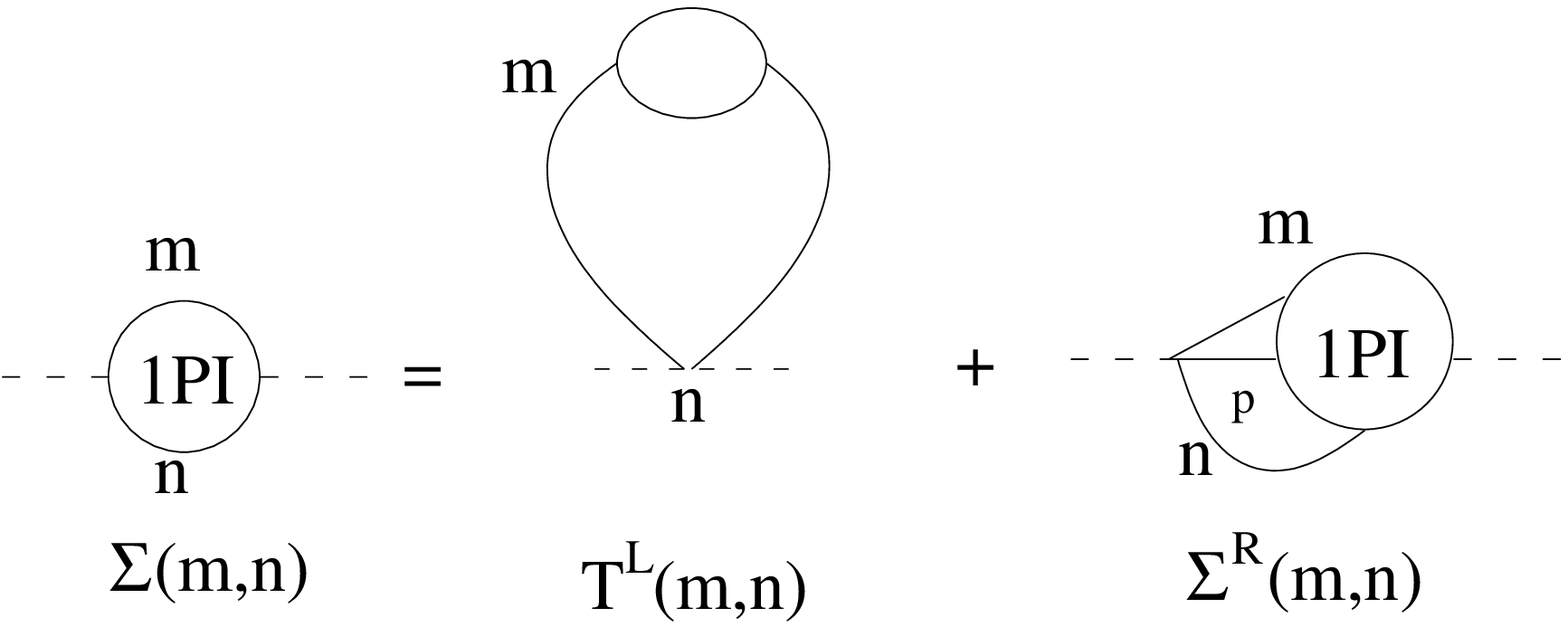}
}
\caption{The self energy}\label{fig:selfenergy}
\end{figure}

It is clear that not all 1PI 2 point insertions on the left hand side of Fig. \ref{fig:insertion} will be ``lost" on the right hand side. If the insertion is a ``generalized left tadpole" it is not ``lost" by opening the face $p$ (imagine a tadpole pointing upwards in Fig.\ref{fig:insertion}: clearely it will not be opened by opening the line). We will call the 2 point 1PI insertions ``lost" on the right hand side $\Sigma^R(m,n)$. Denoting the generalized left tadpole $T^{L}$ we can write (see Fig .\ref{fig:selfenergy}):
\bea
\label{eq:leftright}
  \Sigma(m,n)=T^{L}(m,n)+\Sigma^R(m,n)\, .
\eea
Note that as $T^{L}(m,n)$ is an insertion exclusively on the left border, it does not depend upon the right index $n$. We therefore have $\partial\Sigma(m,n)=\partial_R\Sigma(m,n)=\partial_R\Sigma^R(m,n)$.

The missing mass counterterm writes:
\bea\label{lostct}
CT_{lost}=\Sigma^R(0,0)=\Sigma(0,0)-T^{L}\, .
\eea
In order to evaluate $\Sigma^{R}(0,0)$ we proceed
by opening its face $p$ and using the Ward identity (\ref{ward2point}), to obtain:
\bea
\label{S2}
\Sigma^R(0,0)&=&\frac{1}{G^{2}(0,0)}\sum_{p}G^2_{ins}(0,p;0)\nonumber\\
    &=&\frac{1}{G^{2}(0,0)}\sum_{p}\frac{1}{p}[G^2(0,0)-G^2(p,0)]\nonumber\\
    &=&\sum_{p}\frac{1}{p} \biggl(1 -\frac{G^{2}(p,0)}{G^{2}(0,0)}\biggr) \, .
\eea

Using eq. (\ref{Open2}) and eq. (\ref{S2}) we have:
\bea\label{S3}
G^4_{(3)}(0,m,0,m)&=& C_{0m}\sum_{p} G^{4}_{ins}(0,p;m,0,m)\nonumber\\
&-&C_{0m} G^{4}(0,m,0,m) \sum_{p}\frac{1}{p}  
\biggl( 1- \frac{G^{2}(p,0)}{G^2(0,0)}  \biggr)  \, .
\eea

But by the Ward identity (\ref{ward4point}): 
\bea
\label{Ward4}
&&C_{0m} \sum_{p} G^{4}_{ins}(0,p;m,0,m) \nonumber\\
&&=C_{0m} \sum_{p} \frac{1}{p}\biggl( G^{4}(0,m,0,m)-G^{4}(p,m,0,m) \biggr)\, ,
\eea
The second term in eq. (\ref{Ward4}), having at least three denominators 
linear in $p$, is irrelevant \footnote{Any perturbation order of $G^4(p,m,0,m)$ is a polynomial in $\ln(p)$ divided by $p^2$. Therefore the sums over $p$ above are allways convergent.}
. Substituing eq. (\ref{Ward4}) in eq . (\ref{S3}) we have:
\bea
\label{G3}
G^4_{(3)}(0,m,0,m) =C_{0m}\frac{G^{4}(0,m,0,m)}{G^2(0,0)}\sum_{p}
\frac{G^{2}(p,0)}{p} \, .
\eea
To conclude we must evaluate the sum in eq. (\ref{G3}). Using eq. (\ref {G2(0,m)}) we have:
\bea
\label{derivee}
\sum_{p}\frac{G^{2}(p,0)}{p}=\sum_{p}\frac{G^{2}(p,0)}{p}\bigl( \frac{1}{G^{2}(0,1)}-\frac{1}{G^{2}(0,0)}\bigr)
\frac{1}{1-\partial\Sigma(0,0)}
\eea

In order to interpret the two terms in the above equation we start by performing the same manipulations as in eq (\ref{S2}) for $\Sigma^R(0,1)$. We get:
\bea
\label{S2new}
\Sigma^R(0,1)&=&\sum_{p}\frac{1}{p} 
\biggl(1 -\frac{G^{2}(p,1)}{G^2(0,1)}\biggr) =\sum_{p}\frac{1}{p} 
\biggl(1 -\frac{G^{2}(p,0)}{G^2(0,1)}\biggr)\, .
\eea
where in the second equality the we have neglected an irrelevant term. 

Substituting  eq. (\ref{S2}) and eq. (\ref{S2new}) in eq. (\ref{derivee}) we get:
\bea
\sum_{p}\frac{G^{2}(p,0)}{p}=\frac{\Sigma^R(0,0)-\Sigma^R(0,1)}{1-\partial\Sigma(0,0)}
=-\frac{\partial_{R}\Sigma^R(0,0)}{1-\partial\Sigma(0,0)}
=-\frac{\partial\Sigma(0,0)}{1-\partial\Sigma(0,0)}\, .
\eea
as $\partial_R\Sigma^R=\partial\Sigma$. Hence:
\bea\label{g43}
G^4_{(3)}(0,m,0,m)&=&-C_{0m}G^{4}(0,m,0,m)\frac{1}{G^{2}(0,0)}\frac{\partial\Sigma(0,0)}{1-\partial\Sigma(0,0)}
\nonumber\\
&=&-G^{4}(0,m,0,m)
\frac{A_{ren} \; \partial\Sigma(0,0)}{(m+A_{ren}) [1-\partial\Sigma(0,0)]} \ .
\eea
Using (\ref{g41final}) and (\ref{g43}), equation (\ref{Dyson}) rewrites as:
\bea
\label{Betafinal}
&&G^4(0,m,0,m)\Big{(}1+
\frac{A_{ren}\; \partial\Sigma(0,0)}{(m+A_{ren}) \; [ 1-\partial\Sigma(0,0)] }\Big{)}
\\
&&=\lambda (G^{2}(0,m))^{4}\Big{(}1-\partial\Sigma(0,0)+\frac{A_{ren}}{m+A_{ren}}\partial\Sigma(0,0)\Big{)}
[1-\partial_{L}\Sigma(0,m)]\, .\nonumber
\eea
We multiply (\ref{Betafinal}) by $[1-\partial\Sigma(0,0)]$ and amputate four times. As the differences $\Gamma^4(0,m,0,m,)-\Gamma^4(0,0,0,0)$ and $\partial_L\Sigma(0,m)-\partial_L\Sigma(0,0)$ are irrelevant we get:
\bea
\Gamma^{4}(0,0,0,0)=\lambda (1-\partial\Sigma(0,0))^2\, .
\eea
\qed 

\subsection{Bare identity}

Let us explain now why the main theorem is also true as an identity between bare functions, without
any renormalization, but with ultraviolet cutoff.

Using the same Ward identities, all the equations go through with only few differences:

- we should no longer add the lost mass counterterm in (\ref{Open2})

- the term $G^{4}_{(2)}$ is no longer zero.

- equation (\ref{cdressed}) and all propagators now involve the bare $A$ parameter.

But these effects compensate. Indeed the bare  $G^{4}_{(2)}$ term is the left generalized
tadpole $\Sigma - \Sigma^R$, hence
\begin{equation} \label{newleft}
G^{4}_{(2)}  (0,m,0,m) = C_{0,m} \bigl(  \Sigma(0,m) - \Sigma^R (0,m) \bigr) G^4(0,m,0,m)\; .
\end{equation}
Equation (\ref{cdressed}) becomes up to irrelevant terms
\bea \label{cdressedbare}
\frac{C^{bare}_{0m}}{G^{2,bare}(0,m)}=1-\partial_{L}\Sigma(0,0)+
\frac{A_{bare}}{m+A_{bare}}\partial_{L}\Sigma(0,0) 
- \frac{1}{m+A_{bare}}\Sigma(0,0) 
\eea
The first  term proportional to $ \Sigma(0,m) $ in (\ref{newleft})  combines with 
the new term in (\ref{cdressedbare}), and the second term proportional to $ \Sigma^R(0,m) $ in (\ref{newleft})
is exactly the former ``lost counterterm" contribution in (\ref{Open2}). This proves (\ref{beautiful}) in the bare case.

\section{Conclusion}

Since the main result of this paper is proved up to irrelevant 
terms which converge at least like a power of the ultraviolet cutoff, as this ultraviolet cutoff is lifted towards infinity,
we not only get that the beta function vanishes in the ultraviolet regime, but that it 
vanishes fast enough so that the total flow of the coupling constant is bounded. 
The reader might worry whether this conclusion is still true for the full model which has 
$\Omega_{ren} \ne 1$, hence no exact conservation of matrix indices along faces.
The answer is yes, because the flow of $\Omega $ towards its ultra-violet limit
$\Omega_{bare}=1$ is very fast (see e.g. \cite{DR}, Sect II.2). 

The vanishing of the beta function is a step towards a full non perturbative construction 
of this model without any cutoff, just like e.g. the one of the Luttinger model \cite{BM1,BGPS}.
But NC $\phi^4_4$ would be the first such \textit{four dimensional} model, and the only one
with non logarithmic divergences. Tantalizingly, quantum field theory might actually behave 
better and more interestingly on non-commutative than on commutative spaces.

\subsubsection*{Acknowledgment}
We thank Vieri Mastropietro for very useful discussions.

\newpage
\thispagestyle{empty}

\chapter{NCQFT Parametric Representation}
\label{sec:param}
\begin{center}
Razvan Gurau and Vincent Rivasseau\footnote{e-mail: Razvan.Gurau@th.u-psud.fr; Vincent.Rivasseau@th.u-psud.fr}\\
Laboratoire de Physique Th\'eorique, CNRS UMR 8627\\
Universit\'e Paris XI\footnote{Work supported by ANR grant NT05-3-43374 ``GenoPhy".}
\end{center}

In this paper we investigate the Schwinger parametric
representation for the Feynman amplitudes of the recently discovered
renormalizable $\phi^4_4$ quantum field theory on the Moyal non commutative
${\mathbb R^4}$ space. This representation involves new {\it hyperbolic} polynomials
which are the non-commutative analogs of the usual ``Kirchoff"
or ``Symanzik" polynomials of commutative field theory, 
but contain richer topological information.

\section{Introduction}

Non-commutative field theories (for a general review see \cite{DN}) deserve a thorough 
and systematic investigation. Indeed they may be relevant for physics beyond the standard model.
They are certainly effective models for certain limits of string theory
\cite{SeiWitt}, \cite{a.connes98:noncom}. What is often less emphasized is that they
can also describe effective physics in our ordinary standard world but
with non-local interactions, such as the physics of the quantum Hall effect
\cite{Susk}.

In joint work with J. Magnen and F. Vignes-Tourneret \cite{GMRV},
we provided recently a new proof that the Grosse-Wulkenhaar
scalar $\Phi^4$ theory on the Moyal space ${\mathbb R}^4$, hereafter 
called NC$\Phi^4_4$, is renormalizable to all orders in perturbation theory using direct space multiscale analysis.

The Grosse-Wulkenhaar breakthrough \cite{GrWu03-1,GrWu04-3}
found that the right propagator in non-commutative field theory 
is not the ordinary commutative propagator, but has to be modified to obey Langmann-Szabo duality \cite{GrWu04-3,LaSz}. Grosse and Wulkenhaar added an harmonic potential which
can be interpreted as a piece of the covariant Laplacian in a constant magnetic field. They computed the
corresponding ``vulcanized" propagator in the ``matrix base" which transforms the Moyal product 
into a matrix product. They use this representation to prove 
perturbative renormalizability of the theory up to some estimates 
which were finally proven in \cite{Rivasseau2005bh}.

Our direct space method builds upon the previous work of Filk \cite{Filk:1996dm} 
who introduced clever simplifications,
also called ``Filk moves", to treat the combination of 
oscillations and $\delta$ functions which characterize non commutative interactions. 
Minwalla, van Raamsdonk and Seiberg \cite{MvRS} computed a Schwinger parametric 
representation for the ``not-vulcanized" $\Phi^4_4$ non-commutative theory.
Subsequently Chepelev and Roiban computed also such a Schwinger parametric 
representation for this theory in \cite{Chepelev:2000hm}
and used it in \cite{CheRoi} to analyze power counting. 
These works however remained inconclusive, since they 
worked with the vertex but not the right propagator of NC$\Phi^4_4$, where
ultraviolet/infrared mixing prevents from obtaining a finite 
renormalized perturbation series. We have also been unable to find up to now 
the proofs for the formulas in \cite{MvRS,Chepelev:2000hm}, which in fact disagree.

The parametric representation introduced in this work is completely 
different from the ones of \cite{Chepelev:2000hm} or \cite{MvRS},
since it corresponds to the renormalizable vulcanized theory.
It no longer involves direct polynomials in the Schwinger parameters
but new polynomials of hyperbolic functions of these
Schwinger parameters. This is because the propagator of NC$\Phi^4_4$ is based on
the Mehler kernel rather than on the ordinary heat kernel.
These hyperbolic polynomials contain richer topological information
than in ordinary commutative field theory. Based on ribbon graphs, they 
contain information about their invariants, such
as the genus of the surface on which these graphs live.

This new parametric representation is a compact tool for the study of non commutative field theory which has the advantages
(positivity, exact power counting) but not the drawbacks 
(awkwardness of the propagator) of the matrix base representation.
It can be used as a starting point to work out 
the renormalization of the model directly in parametric space, as can be done
in the commutative case \cite{BL}. It is also a good starting point
to define the regularization and minimal dimensional
renormalization scheme of NC$\Phi^4_4$. This  dimensional scheme in the ordinary field theory case better preserves continuous symmetries such as gauge symmetries, hence played a historic
role in the proof of `tHooft and Veltman that non Abelian gauge theories on commutative ${\mathbb R^4}$ are renormalizable. It is also 
used extensively in the works of Kreimer and Connes \cite{K,CK} which recast 
the recursive BPHZ forest formula of perturbative renormalization into 
a Hopf algebra structure and relate it to a new class of Riemann-Hilbert problems; 
here the motivations to use dimensional renormalization
rather than e.g. subtraction at zero momentum come 
at least in part from number theory rather than from physics.

Following these works, renormalizability has also attracted considerable 
interest in the recent years as a pure mathematical structure. 
The renormalization group ``ambiguity" reminds mathematicians of
the Galois group ambiguity for roots of algebraic equations \cite{CoM}.
Finding new renormalizable theories may therefore be 
important for the future of pure mathematics as well as for physics.

This paper is organized as follows. In Section II we introduce notations
and define our new polynomials $HU$ and $HV$ which generalize the Symanzik
polynomials $U$ and $V$ of commutative field theory. In section III 
we prove the basic positivity property of the first polynomial $HU$
and compute leading ultraviolet terms which allow to recover 
the right power counting in the parametric 
representation, introducing a technical trick which we call
the ``third Filk move"
\footnote{For technical reasons
exact power counting was not fully established in \cite{GMRV}.}.
Section IV establishes the positivity properties 
and computes such leading terms for the second polynomial
$HV$, the one which gives the dependence in the external arguments.
Finally examples of these polynomials for various graphs 
are given in section V.


\section{Hyperbolic Polynomials}
\subsection{Notations}

The NC$\Phi^4_4$ theory is defined on ${\mathbb R}^4$ equipped
with the associative and noncommutative Moyal product
\begin{align}
(a\star b)(x) &= \int \frac{d^4k}{(2\pi)^4} \int d^4 y \; a(x{+}\tfrac{1}{2}
\theta {\cdot} k)\, b(x{+}y)\, \mathrm{e}^{\mathrm{i} k \cdot y}\;.
\label{paramstarprod}
\end{align}

The renormalizable action functional introduced in \cite{GrWu04-3} is
\begin{equation}\label{paramaction}
S[\phi] = \int d^4x \Big( \frac{1}{2} \partial_\mu \phi
\star \partial^\mu \phi + \frac{\Omega^2}{2} (\tilde{x}_\mu \phi )
\star (\tilde{x}^\mu \phi ) + \frac{1}{2} \mu_0^2
\,\phi \star \phi 
+ \frac{\lambda}{4!} \phi \star \phi \star \phi \star
\phi\Big)(x)\;,
\end{equation}
where the Euclidean
metric is used. In what follows the mass $\mu_0$ does not play any role so we put it to zero\footnote{This does not lead in this model to any infrared divergences.
Beware that our definition of $\Omega$ is different from the one of \cite{GrWu04-3}
by a factor $4\theta^{-1}$}.

In four dimensional $x$-space the propagator is \cite{Propaga}
\begin{equation}\label{paramtanhyp}
\int\prod_l \frac{\Omega d\alpha_l}{[2\pi\sinh(\alpha_l)]^{D/2}}
e^{-\frac{\Omega}{4}\coth(\frac{\alpha_l}{2})u_l^2-
\frac{\Omega}{4}\tanh(\frac{\alpha_l}{2})v_l^2}\; .
\end{equation} 
and the (cyclically invariant) vertex is:
\begin{equation}\label{paramvertex}
V(x_1, x_2, x_3, x_4) = \delta(x_1 -x_2+x_3-x_4 )e^{i
\sum_{1 \le i<j \le 4}(-1)^{i+j+1}x_i \theta^{-1}  x_j}\ ,
\end{equation}
where we note $x \theta^{-1}  y  \equiv  \frac{2}{\theta} (x_1  y_2 -  x_2  y_1 +  
x_3  y_4 - x_4  y_3 )$.

Permutational symmetry of the fields at all vertices, which characterizes commutative field theory,
is replaced by the more restricted cyclic symmetry. Hence
the ordinary Feynman graphs of $\Phi^4_4$ really become ribbon 
graphs in NC$\Phi^4_4$. For such a ribbon 
graph $G$, we call $n$, $L$, $N$, $F$, and $B$
respectively the number of vertices, of internal lines, of external half-lines, of faces and of
faces broken by some external half-lines. The Euler characteristic
is $2-2g = n-L+F$, where $g$ is the genus of the graph.
To each graph $G$ is associated a dual graph ${\cal G}$ of same genus by
exchanging faces and vertices.

In ordinary commutative field theory, in order to obtain Symanzik's polynomials it is not convenient
to solve the momentum conservation at the vertices through a momentum routing, because
this is not canonical. It is better to express 
these $\delta$ functions through their Fourier transform. After integration
over internal variables, the amplitude of an 
amputated graph $G$ with external momenta $p$ is, up to a normalization,
in space time dimension $D$ (of course the main case of interest 
in this paper is $D=4$):
\bea \label{symanzik} 
A_G (p) = \delta(\sum p)\int_0^{\infty} 
\frac{e^{- V_G(p,\alpha)/U_G (\alpha) }}{U_G (\alpha)^{D/2}} 
\prod_l  ( e^{-m^2 \alpha_l} d\alpha_l )\ .
\eea
The first and second Symanzik polynomials $U_G$ and $V_G$ are
\begin{equation}\label{symanzik1}
U_G = \sum_T \prod_{l \not \in T} \alpha_l \ ,
\end{equation}
\begin{equation}\label{symanzik2}
V_G = \sum_{T_2} \prod_{l \not \in T_2} \alpha_l  (\sum_{i \in E(T_2)} p_i)^2 \ , 
\end{equation}
where the first sum is over spanning trees $T$ of $G$
and the second sum  is over two trees $T_2$, i.e. forests separating the graph
in exactly two connected components $E(T_2)$ and $F(T_2)$; the corresponding
Euclidean invariant $ (\sum_{i \in E(T_2)} p_i)^2$ is, by momentum conservation, also
equal to $ (\sum_{i \in F(T_2)} p_i)^2$.

The topological formulas (\ref{symanzik1}) and (\ref{symanzik2}) are
a field-theoretic instance of the tree-matrix theorem of Kirchoff et al; 
for a recent review of this kind of theorems see \cite{A}.

In the non commutative case,
momentum routing is replaced by position routing \cite{GMRV}.
However this position routing 
is again non-canonical, depending on the choice of a particular tree. 
Therefore to compute the parametric representation we prefer to perform
a new level of Fourier transform: we represent the ``position conservation" 
rules as integrals
over new ``hypermomenta" $p_v$ associated to each of the vertices:
\begin{equation}
\delta(x_1 -x_2+x_3-x_4 ) = \int  \frac{d p_v}{(2 \pi)^D}
e^{ip_v(x_1^v-x_2^v+x_3^v-x_4^v)} = \int  \frac{d p_v}{(2 \pi)^D}
e^{p_v \sigma (x_1^v-x_2^v+x_3^v-x_4^v)}
\end{equation}
where $\sigma $ is the $D$ by $D$ matrix defined by $D/2$ matrices $\sigma_y$
on the diagonal (we assume $D$ even):
\bea
\sigma=\begin{pmatrix}
\sigma_y & \cdots & 0 \\
\cdots & \cdots & \cdots \\
0 & \cdots & \sigma_y
\end{pmatrix}  \ \ {\rm where} \ \ \sigma_y=\begin{pmatrix}
0 & -i \\
i & 0 \end{pmatrix} \ .
\eea

There is here a subtle difference with the commutative case.
The first commutative polynomial (\ref{symanzik1}) is not the determinant of the quadratic
form in the internal position variables, since this determinant vanishes by translation invariance.
It is rather the determinant of the quadratic form integrated over all internal positions of 
the graph {\it save one} (remark the overall 
momentum conservation in (\ref{symanzik})). This is a canonical object which does
not depend of the choice of the particular vertex whose position is not integrated 
(this can be seen explicitly on
the form (\ref{symanzik1}), which depends only on $G$).

In the non commutative case translation invariance is lost. This allows to define the amplitude of a graph
as a function of the external positions by integrating over all internal positions 
and hypermomenta, since the corresponding 
determinant no longer vanishes. In this way one can 
define {\it canonical} polynomials $HU_G$ and $HV_G$
which only depend on the ribbon graph $G$. 

But in practice it is often more convenient (for instance for renormalization or for understanding the
limit towards the commutative case)  to define the amplitude of a graph 
by integrating all the internal positions and hypermomenta save one, $p_{\bar{v}}$;
this helps to factorize an overall approximate ``position conservation" for the whole graph. 
However precisely because there is no translation invariance, the corresponding
polynomials $HU_{G,\bar{v}}$ and $HV_{G,\bar{v}}$ explicitly depend
on the ``rooted graph" $G,\bar{v}$, i.e. on the choice of $\bar{v}$ 
(although their leading ultraviolet terms do not depend on this choice, see below).

Consider a graph $G$ with $n$ vertices, $N$ external positions 
and a set $L$ of $2n-N/2$ internal lines or propagators.
Each vertex in NC$\phi^4$ is made of four ``corners",
bearing either a halfline or an external field, 
numbered as $1,2,3,4$ in the cyclic order given by the Moyal product. 
To each such corner is associated a position, noted $x_i$. The set $I$ of internal corners 
has $4n - N$ elements, labeled usually as $i,j,...$; the set $E$ of 
external corners has $N$ elements labeled as $e,e',...$.
A line $l$ of the graph joins two corners in $I$, with positions $(x^l_i,x^l_j)$ 
(which in general do not belong to the same vertex).

The amplitude of such a NC$\phi^4$ graph $G$ is then given, up to some inessential normalization $K$, by:
\bea
{\cal A}_G(\{x_e\})&=& K
\int \prod_{l} \frac{d\alpha_l}{\sh\alpha_l^{D/2}} \int \prod_{i \in I}d x_i
\prod_{v} d p_v\\
&&\prod_l e^{-\frac{\Omega}{4}\coth(\frac{\alpha_l}{2})(x_i^l-x_j^l)^2-
\frac{\Omega}{4}\tanh(\frac{\alpha_l}{2})(x_i^l+x_j^l)^2}\nonumber\\
&&\prod_v  e^{\frac{i}{2}\sum_{1\le i<j\le 4}(-1)^{i+j+1}x_i^v\theta^{-1} x_j^v+
p_v \sigma (x_1^v-x_2^v+x_3^v-x_4^v)}\nonumber
\eea
or, for some fixed root vertex $\bar {v}$ by
\bea
&&\int \prod_l \frac{d\alpha_l}{\sh\alpha_l^{D/2}} \int \prod_{i \in I}d x_i
\prod_{v\neq\bar{v}} d p_v 
\prod_l e^{-\frac{\Omega}{4}\coth(\frac{\alpha_l}{2})(x_i^l-x_j^l)^2-
\frac{\Omega}{4}\tanh(\frac{\alpha_l}{2})(x_i^l+x_j^l)^2}\nonumber\\
&&\prod_v  e^{\frac{i}{2}\sum_{1\le i<j\le 4}(-1)^{i+j+1}x_i^v\theta^{-1} x_j^v+p_v \sigma (x_1^v-x_2^v+x_3^v-x_4^v)}\ .
\eea

\subsection{Definition of $HU$ and $HV$}
The fundamental observation is that the integrals to perform being Gaussian, the result is
a Gaussian in the external variables divided by a determinant. This gives the definition of 
our hyperbolic parametric representation.
We introduce the notations $c_l = \coth(\frac{\alpha_l}{2}) = 1/t_l $
and $t_l = \tanh(\frac{\alpha_l}{2}) $. 
Using $\sinh \alpha_l = 2 t_l  / (1-t_l^2)  $ we obtain
\bea\label{hyperpcan}
{\cal A}_G (\{x_e\}) = K  \int_{0}^{\infty} \prod_l  [ d\alpha_l (1-t_l^2)^{D/2} ]
HU_G ( t )^{-D/2}   e^{-  \frac {HV_G ( t , x_e )}{HU_G ( t )}},
\eea
\bea\label{hyperpnoncan}
{\cal A}_{G,{\bar v}}  (\{x_e\},\;  p_{\bar v}) = K'  \int_{0}^{\infty} \prod_l  [ d\alpha_l (1-t_l^2)^{D/2} ]
HU_{G, \bar{v}} ( t )^{-D/2}   
e^{-  \frac {HV_{G, \bar{v}} ( t , x_e , p_{\bar v})}{HU_{G, \bar{v}} ( t )}},
\eea
where $K$ and $K'$ are some new inessential normalization constants
(which absorb in particular the factors 2 from $\sinh \alpha_l = 2 t_l  / (1-t_l^2)  $); $HU_G ( t )$ or $HU_{G, \bar{v}} ( t )$
are polynomials in the $t$ variables (there are no $c$'s because they are compensated
by the $t$'s coming from
$\sinh \alpha_l = 2 t_l  / (1-t_l^2)  $)
and $HV_G ( t , x_e )$ or $HV_{G, \bar{v}} ( t , x_e , p_{\bar v})$ are
quadratic forms in the external variables $x_e$ or  $(x_e , p_{\bar v})$ whose coefficients
are polynomials in the $t$ variables (again there are no $c$'s because they are compensated
by the $t$'s  which were included in the definition of $HU$, see below the difference between (\ref{HVgv1}) and (\ref{HVgv})).

There is a subtlety here. Overall approximate ``position conservation"
holds only for orientable graphs in the sense of \cite{GMRV}. 
Hyperbolic polynomials  for non orientable graphs  are
well defined through formulas  (\ref{hyperpcan})-(\ref{hyperpnoncan}) 
but they are significantly harder to compute than in the orientable case. 
Their amplitudes are also smaller in the ultraviolet,  and in
particular do not require any renormalization. Also many interesting 
non commutative theories such as the LSZ models \cite{Langmann:2003if}, the 
more general $(\bar\phi\phi)^2$ models of \cite{GMRV} and
the most natural Gross-Neveu models \cite{Propaga}, \cite{RenNCGN05} do not have any 
non orientable subgraphs. So for simplicity we shall restrict ourselves 
in this paper to examples 
of hyperbolic polynomials for orientable graphs; and
when identifying leading pieces 
under global scaling in the hyperbolic polynomials, something necessary for renormalization, we also limit ourselves to the orientable case.

We now proceed to the computation of these hyperbolic polynomials.

\subsection{Short and Long Variables}

This terminology was introduced in \cite{GMRV}.

We order each line $l$ joining corners $l=(i,j)$ 
(which in general do not belong to the same vertex), 
in an arbitrary way such that it exits $i$ and enters $j$. 
We define the incidence matrix 
between lines and corners $\epsilon_{l i}$ to be $1$ if $l$ enters in $v$, $-1$ if it exits at $i$
and 0 otherwise. 
Also we define $\eta_{l i}=|\epsilon_{l i}|$. We note the property:
\bea
\sum_l(\epsilon_{l i}\epsilon_{l j}+\eta_{l i}\eta_{l j})=2\delta_{i j} \ .
\eea
We now define the short variables $u$ and the long variables $v$ as
\be
v_l=\sum_i\frac{\eta_{l i}x_i}{\sqrt 2},~u_l=\sum_i\frac{\epsilon_{l i} x_i}{\sqrt 2}
;~x_i=\sum_l \frac{\eta_{l i}v_l+\epsilon_{l i}u_l}{\sqrt{2}}\ .
\ee
The Jacobian of this change of coordinates is $1$.
Moreover, in order to avoid unpleasant $\sqrt{2}$ factors we rescale the external positions to hold $\bar{x}_e=\sqrt{2}x_e$ and the internal hypermomenta $\bar{p}_v=p_v/\sqrt{2}$.
Note that if the graph is orientable we can choose $\epsilon_{l i}$ 
to be $(-1)^{i+1}$, so that the incidence matrix is consistent with the 
cyclic order at the vertices (halflines alternatively enter and go out).
The integral in the new variables is:
\bea
&&\int \prod_l \big[\frac{1-t_l^2}{t_l}\big]^{D/2} d\alpha_l  \int \prod_{i}d x_i  \prod_{v\neq\bar{v}} d p_v
\prod_l e^{-\frac{\Omega}{2}\coth(\frac{\alpha_l}{2})u_l^2-
\frac{\Omega}{2}\tanh(\frac{\alpha_l}{2})v_l^2}\nonumber\\
&&  \prod_v  e^{\frac{i}{4}\sum_{\substack{i < j;\\ i,j \in v}}(-1)^{i+j+1}(\eta_{l i}v_l+\epsilon_{l i}u_l)\theta^{-1}
(\eta_{l' j}v_{l'}+\epsilon_{l' j}u_{l'})}
\nonumber\\
&&\prod_v e^{ \bar{p}_v \sigma \sum_{i\in v}(-1)^{i+1}(\eta_{l i}v_l+\epsilon_{l i}u_l)}
\\
&&e^{\frac{i}{4}[\sum_{i \ne e}\omega(i,e)(\eta_{l i}v_l+\epsilon_{l i} u_l)
\theta^{-1}\bar{x}_e ]+\frac{i}{4}\sum_{e < e' } \bar{x}_e\theta^{-1}\bar{x}_{e'}+\sum_{e\in v}\bar{p}_v \sigma
(-1)^{e+1} \bar{x}_e}\ , \nonumber
\eea
where $\omega(i,e) = 1$ if $i<e$ and $\omega(i,e) = -1$ if $i>e$. When we write $i \in v$, it means that the corner $i$ belongs to $v$. 
>From now on we forget the bar over the rescaled variables.
We also concentrate on the computation of $HU_{G, {\bar v}}$ in (\ref{hyperpnoncan});
we indicate alongside the necessary modifications for $HU_{G}$ in (\ref{hyperpcan}).

We introduce the condensed notations:
\bea
{\cal A}_G  = \int \big[\frac{1-t^2}{t}\big]^{D/2} d\alpha \int d x d p e^{-\frac{\Omega}{2} X G X^t}
\eea
where 
\bea\label{defMPQ1}
X = \begin{pmatrix}
x_e & \bar{p} & u & v & p\\
\end{pmatrix} \ \ , \ \  G= \begin{pmatrix} M & P \\ P^{t} & Q \\
\end{pmatrix}\ .
\eea

Gaussian integration gives, up to inessential constants:
\bea\label{defMPQ2}
{\cal A}_G  = \int \big[\frac{1-t^2}{t}\big]^{D/2} d\alpha\frac{1}{\sqrt{Q}}
e^{- \frac{\Omega}{2}   
\begin{pmatrix} x_e & \bar{p} \\
\end{pmatrix} [M-P Q^{-1}P^{t}]
\begin{pmatrix} x_e \\ \bar{p} \\
\end{pmatrix}  }\ .
\eea

All we have to do now to get $HU$ and $HV$ is to compute the determinant and the minors of the matrix $Q$ for an arbitrary graph.
 
\section{The First Hyperbolic Polynomial $HU$}

We define $I_D$ to be the identity matrix in $D$ dimensions. We also put $d=2L+n-1$ so that
$Q$ can then be written as:
\bea
Q=A\otimes I_{D} +B\otimes \sigma
\eea
with $A$ is a $d$ by $d$ symmetric matrix (accounting for the contribution of the propagators in 
the Gaussian) and $B$ a antisymmetric matrix (accounting for the oscillation part in the Gaussian).

Note that the  symplectic pairs decouple completely so that 
$\det Q=[\det(A\otimes I_2+B\otimes\sigma_y)]^{D/2}$. 

\begin{lemma} For any two $n\times n$ matrices 
$A$ and $B$ let $R=A\otimes I_2+B\otimes\sigma_y$. Then:
\bea \label{rdeterminant}
\det R=(-1)^n\det(A+B) \det(A-B)
\eea
and:
\bea\label{rinverse}
R^{-1}=\frac{[(A+B)^{-1}+(A-B)^{-1}]}{2}\otimes
I_2+\frac{[(A+B)^{-1}-(A-B)^{-1}]}{2}\otimes\sigma_y .
\eea
\end{lemma}

\prf We express the determinant as a Grassmann-Berezin integral:
\bea
&&\Delta=\det(A\otimes I_2+B\otimes \sigma_y)\nonumber\\
&=&\int \prod_k d\bar{\psi}^1_k d\psi^1_k
d\bar{\psi}^2_k d^2\psi_k
e^{-\begin{pmatrix}
\bar{\psi}^1_i & \bar{\psi}^{2}_i\\
\end{pmatrix}
(a_{i j}\otimes I_2+b_{i j}\otimes\sigma)
\begin{pmatrix}
\psi^{1}_{j} \\ \psi^{2}_{j} \\
\end{pmatrix}
}\nonumber\\
&=&\int \prod_k d\bar{\psi}^1_k d\psi^1_k
d\bar{\psi}^2_k d^2\psi_k
e^{-[a_{i j}(\bar{\psi}^1_i\psi^1_j+
\bar{\psi}^2_i\psi^2_j)
+ib_{i j}(-\bar{\psi}^1_i\psi^2_j+\bar{\psi}^2_i\psi^1_j)]}\ .
\eea

We perform a change of variables of Jacobian $-1$ to:
\bea
\chi^1_i=\frac{\psi^1_i+i\psi^2_i}{\sqrt{2}};
\chi^2_i=\frac{\psi^1_i-i\psi^2_i}{\sqrt{2}}\ .
\eea
As:
\bea
\bar{\chi}^1_i\chi^1_j=
\frac{1}{2}(\bar{\psi}^1_i\psi^1_j-i\bar{\psi}^2_i\psi^1_j+
i\bar{\psi}^1_i\psi^2_j+\bar{\psi}^2_i\psi^2_j),
\eea
we see that:
\bea
\Delta=(-1)^n \int\prod_k d\bar{\chi}^1_k d\chi^1_k
d\bar{\chi}^2_k d\chi^2_k
e^{-[a_{i j}(\bar{\chi}^1_i\chi^1_j+\bar{\chi}^2_i\chi^2_j)-
b_{i j}(\bar{\chi}^1_i\chi^1_j-\bar{\chi}^2_i\chi^2_j)]} \ .
\eea
Separating the terms in $\bar{\chi}^1\chi^1$ and $\bar{\chi}^2\chi^2$ proves (\ref{rdeterminant}).

The inverse matrix is divided into $2\times 2$ blocs with indices $ij$, according 
to the values $a,b=1,2$.
\bea
(R^{-1})^{ab}_{i j}=
\frac{\int d\bar{\psi}^1{d\psi}^1 d\bar{\psi}^2{d\psi}^2 \psi^{a}_{i}\bar{\psi^{b}_{j}}
e^{-\bar{\psi}A\psi}}
{\int d\bar{\psi}^1{d\psi}^1 d\bar{\psi}^2{d\psi}^2 e^{-\bar{\psi}A\psi}}\ .
\eea

The four elements of the blocs are 
given by (taking into account that the integral decouples so that all the crossed terms are zero):
\bea
\psi^{1}_{i}\bar{\psi}^1_{j}&=&
\frac{(\chi^{1}_{i}+\chi^2_{i})(\bar{\chi}^1_j+\bar{\chi}^2_j)}{2}=
\frac{1}{2}(\chi^{1}_{i}\bar{\chi}^1_j+\chi^2_{i}\bar{\chi}^2_j),
\nonumber\\
\psi^1_{i}\bar{\psi}^2_j&=&
\frac{(\chi^{1}_{i}+\chi^2_{i})(\bar{\chi}^1_j-\bar{\chi}^2_j)}{2(-i)}=
\frac{i}{2}(\chi^{1}_{i}\bar{\chi}^1_j-\chi^2_{i}\bar{\chi}^2_j),
\nonumber\\
\psi^2_i\bar{\psi}^1_j&=&\frac{(\chi^{1}_{i}-\chi^2_{i})(\bar{\chi}^1_j+\bar{\chi}^2_j)}{2i}=
\frac{-i}{2}(\chi^{1}_{i}\bar{\chi}^1_j-\chi^2_{i}\bar{\chi}^2_j),
\nonumber\\
\psi^2_i\bar{\psi}^2_j&=&
\frac{(\chi^{1}_{i}-\chi^2_{i})(\bar{\chi}^1_j-\bar{\chi}^2_j)}{2(-i)i}=
\frac{1}{2}(\chi^{1}_{i}\bar{\chi}^1_j+\chi^2_{i}\bar{\chi}^2_j).
\eea
(\ref{rinverse}) follows then easily.
\qed

Returning to our initial problem we remark that the matrix $A$
is the symmetric part coming from the propagator, and the oscillating part,
when symmetrized, leads naturally to an antisymmetric matrix $B$ times
the antisymmetric $\sigma_y$, so that in our case 
\bea
\det Q &=& [\det (A+B)(A-B)]^{D/2}\nonumber\\ 
&=& [\det (A+B)(A^t + B^t)]^{D/2} =  
[\det(A+B)]^D .
\eea

The propagator part is:
\bea \label{defmatrixa}
 A=\begin{pmatrix} S & 0 & 0\\ 0  & T & 0 \\ 0&0&0\\
\end{pmatrix}
\eea 
where $S$ and $T$ are the two diagonal $L$ by $L$ matrices
with diagonal elements $c_l = \coth(\frac{\alpha_l}{2}) = 1/t_l $, 
and $t_l = \tanh(\frac{\alpha_l}{2}) $, 
and the last lines and columns of zeroes reflect the purely
oscillating nature of the hypermomenta integrals.

The hypermomenta oscillations are (in the case of (\ref{hyperpnoncan})):
\bea
C_{v l}=\begin{pmatrix}
\sum_{i\in v}(-1)^{i+1}\epsilon_{l i} \\
\sum_{i\in v}(-1)^{i+1}\eta_{l i} \\
\end{pmatrix}\ .
\eea
Remark that the elements of $C$ are integers which can take only values $0$ or $ \pm 1$.
It is easy to check that for a connected graph $G$ the rank of the matrix $C$ is maximal, 
namely $n-1$. Picking a tree of $G$ proves that this is even true for 
the $L$ by $n$ lower part of $C$, corresponding to the long variables $v$ only.

To generalize to (\ref{hyperpcan}), we simply need to add another column to $C$, the one
corresponding to $p_{\bar v}$. The rank of the extended $2L$ by $n$ matrix $\bar C$
is then $n$, but the rank of the restriction of  $\bar C$ to its lower part 
corresponding to
the long variables $v$ is either $n-1$ or $n$ depending on whether the graph is orientable or not \cite{GMRV}.
This has important consequences for power counting.

The determinant of the quadratic form is the square of the determinant of the matrix $A + B$,
where
\bea\label{defmatrixb}
B= \begin{pmatrix}\frac{1}{4\theta\Omega} E & C \\
-C^t & 0 \\
\end{pmatrix}\ .
\eea
We can explicitate the oscillation part between the $u,v$ variables as  the $2L$ by $2L$ matrix E. This matrix  $E=\begin{pmatrix}E^{uu} & E^{uv} \\ E^{vu} & E^{vv} \\
\end{pmatrix}$  represents the vertices oscillations. One can check
\bea
E^{vv}_{l,l'}&=&\sum_v
\sum_{i\ne j ; \; i, j \in v}  (-1)^{i+j+1} \omega(i,j)\eta_{l i}\eta_{l' j},
\nonumber\\
E^{uu}_{l,l'}&=&\sum_v
\sum_{i\ne j ; \; i, j \in v}  (-1)^{i+j+1} \omega(i,j)\epsilon_{l i}\epsilon_{l' j},
\nonumber\\
E^{uv}_{l,l'}&=&\sum_v
\sum_{i\ne j ; \; i, j \in v}  (-1)^{i+j+1} \omega(i,j)\epsilon_{l i}\eta_{l' j},
\eea 
where we recall that $\omega(i,j) = 1$ if $i < j$ and  $\omega(i,j) = -1$ if $i > j$;
moreover $E^{vu}_{l,l'} = - E^{uv}_{l',l}$. Remark that
the matrix elements of $E$ are integers and can in fact
only take values $0, \pm 1, \pm 2$. Moreover $E_{l,l'}$ is zero
if $l$ and $l'$ do not hook to any common vertex; it can take value $\pm 2$ 
only if the two lines hook to at most two vertices in total, which is not generic, at least
for large graphs.

\begin{lemma}
Let $A=(a_i\delta_{i j})_{i,j\in\{1,\dotsc,N\}}$ 
be diagonal and $B=(b_{i j})_{i,j\in\{1,\dotsc,N\}}$ 
be such that $b_{ii}=0$ (we need not require $B$ antisymmetric). We have:
\be      
\det(A+B)=\sum_{K\subset \{1,\dotsc ,N\}}\det(B_{\hat{K}}) \prod_{i\in K}a_i 
\ee
where $B_{\hat{K}}$ is the matrix obtained from $B$ by deleting the lines and 
columns with indices in $K$.
\end{lemma}

\prf The proof is straightforward. We have:
\bea
&&\det(A+B)=\sum_{\sigma\in {\mathfrak S}_N}\epsilon_{\sigma}\prod_{i\in\{1,\dotsc,N\}}
(a_{i\sigma(i)}+b_{i\sigma(i)}) \nonumber\\
&&=\sum_{K\subset\{1,\dotsc,N\}}\prod_{i\in K}a_{i}
\sum_{\substack{\sigma\in {\mathfrak S}_N \\ \sigma(i)=i~\forall i\in K}}
\prod_{k\in \{1,\dotsc,N\}\setminus K}\epsilon(\sigma)b_{k\sigma(k)}
\eea
and the lemma follows.
\qed

In our case the matrix $B=\begin{pmatrix} \frac{1}{4\theta\Omega}E & C \\
-C^t & 0 \\ \end{pmatrix} $ is antisymmetric. 
Remark that the matrix 
\bea B'=\begin{pmatrix} E & C \\
-C^t & 0 \\ \end{pmatrix} 
\eea 
has integer coefficients.
Moreover $A$ in  (\ref{defmatrixa})  has zero diagonal
in the $n-1$ by $n-1$ lower right corner 
corresponding to hypermomenta. Exploiting these facts
we can develop $\det (A+B) $  into Pffafians to get:
\begin{lemma}
With $A$ and $B$ given by (\ref{defmatrixa}) and (\ref{defmatrixb})
\bea\label{pffafianfor}
\det (A+B) = \sum_{\substack{I\subset \{1\dotsc L\},  J\subset \{L+1\dotsc 2L\},\\ n + |I|+|J|\; {\rm odd}}}
 (4\theta\Omega)^{|I|+|J| + n-1-2L}   n^2_{I J}
\prod_{l\in I}c_l\prod_{l'\in J}t_{l'}
\eea
with $n_{I J}=\mathrm{Pf}(B'_{\hat{I}\hat{J}})$, the Pffafian of the oscillation matrix 
with deleted lines and columns $I$ among the first $L$ indices 
(corresponding to short variables $u$) and $J$ among the next $L$ 
indices (corresponding to long variables $v$). 
\end{lemma}
\prf
Since $A$ has the form  (\ref{defmatrixa}), the part $K$ of the previous lemma
has to be the disjoint union of two sets $I$ and $J$ respectively corresponding to
short and long variables. Once these sets are deleted from the matrix $B$
we obtain a matrix $B_{\hat{I}\hat{J}}$ which has size $2L -|I|-|J| +  n-1$. 
This matrix is antisymmetric, so its determinant is the square of the corresponding triangular Pfaffian.
The Pfaffian of such a matrix is zero unless its size $2L -|I|-|J| +  n-1 = 2p$ is even, in which case it is a sum, 
with signs, over the pairings of the $2p$ lines into $p$ pairs of the products of the corresponding
matrix elements. Now from the particular form of matrix $B$ which has a lower right block $0$, 
we know that any pairing
of the $n-1$ hypermomentum variables must be with an $u$ or $v$ variable. Hence
any pairing contributing to the Pfaffian has necessarily $n-1$ terms of the $C$ type, hence 
$[(2L -|I|-|J|)-(n-1)] /2$ terms of the $E/4\theta\Omega$ type, hence 
\bea \mathrm{Pf}(B_{\hat{I}\hat{J}}) = \frac{1}{(4\theta\Omega)^{L-(n+|I|+|J|-1)/2}} 
\mathrm{Pf}(B'_{\hat{I}\hat{J}})\ .
\eea
Therefore
\bea \det (B_{\hat{I}\hat{J}}) = \frac{1}{(4\theta\Omega)^{2L-n-|I|-|J|+1}} 
\mathrm{Pf}^2(B'_{\hat{I}\hat{J}}),
\eea
hence the Lemma holds, with $n_{I,J}=\mathrm{Pf}(B'_{\hat{I}\hat{J}})$ which must be
an integer since any Pfaffian with integer entries is integer.
\qed

We have thus expressed the determinant of
$Q$ as a sums of positive terms. 

Recalling that $\det Q = (\det (A+B) )^{D}$,
the amplitude ${\cal A}_{G,{\bar v}} (0)$ with external arguments $x_e$ and 
$p_{\bar v}$ put to 0
is nothing but (up to an inessential normalization)
\bea
{\cal A}_{G,{\bar v}} (0) = \int_{0}^{\infty} \det (A+B) ]^{-D/2}\prod_l 
\big[\frac{1-t_l^2}{t_l}\big]^{D/2} d\alpha_l  .
\eea
Putting $s=(4\theta\Omega)^{-1}$, we use the relation $2-2g = n-L+F$
and define the integer $k_{I,J} = \vert I\vert+\vert J\vert - L - F +1$ to get:
\bea\label{hypnoncanpfaff}
HU_{G,{\bar v}} (t) &=&  \sum_{I,J}  s^{2g-k_{I,J}} \ n_{I,J}^2
\prod_{l \not\in I} t_l \prod_{l' \in J} t_{l'}\ .
\eea
This is our precise definition of the normalization of the polynomial 
$HU_{G,{\bar v}}$ in the variables $t_l$ introduced in (\ref{hyperpnoncan}).
This normalization is adapted so that the limit $s \to 0$ will give back the ordinary Symanzik polynomial at leading order as $t$'s go to 0 (the ultraviolet limit), as shown in the next section.

To get the polynomial $HU_{G}$ in (\ref{hyperpcan}), we proceed exactly in the same
way, replacing $C$ by $\bar C$, and obtain that it is also
a polynomial in the variables $t_l$ with positive coefficients, which (up
to the factors in $4\theta\Omega$) are squares of integers. 
But we will see that the leading terms studied 
in the next section will be quite different in this case.

\subsection{Leading terms in the First Polynomial $HU$}

By leading terms, we mean terms which have the smallest global degree 
in the $t$ variables, since we are interested in power counting in the ``ultraviolet"
regime where all $t$'s are scaled to 0.
Such terms are obtained by taking $\vert I\vert $ maximal and $\vert J\vert $ minimal  in (\ref{pffafianfor}).
We shall compute the leading terms corresponding to $I=[1,...,L]$ hence taking all 
the $c_l$ elements of the diagonal 
and $J$ minimal so that the remaining minor is non zero
\footnote{These are not the only globally leading terms; there are terms whose global scaling is equivalent,
for example the $t^2_3$ term in (\ref{hueye1})-(\ref{hueye2}). By the positivity of $HU$ they can certainly not deteriorate the power counting established in this section, but only improve it in certain particular ``Hepp sectors".}.
Below we prove that such terms have $\vert J \vert =F-1$, 
This explains the normalization in (\ref{hypnoncanpfaff}).

\begin{figure}
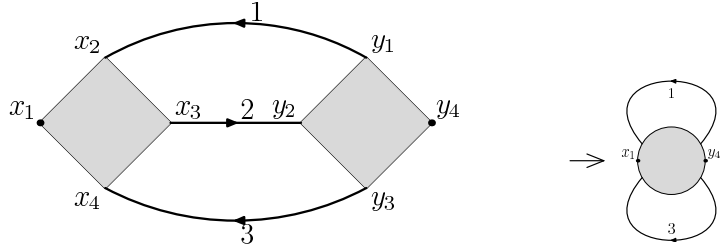

\centerline{\epsfig{figure=figraz-3.ps,width=6cm} \hfil \epsfig{figure=figraz-8.ps,width=2cm}}
\caption{The First Filk Move on the Sunshine Graph}\label{firstfilk}
\end{figure}

\begin{figure}
\centerline{\epsfig{figure=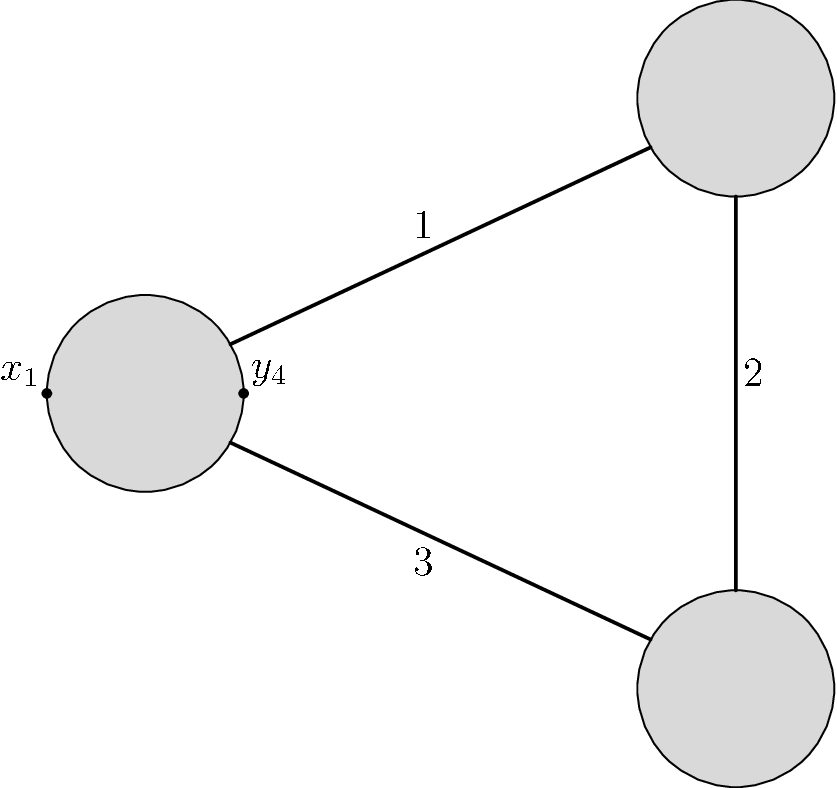,width=4cm} \hfil \epsfig{figure=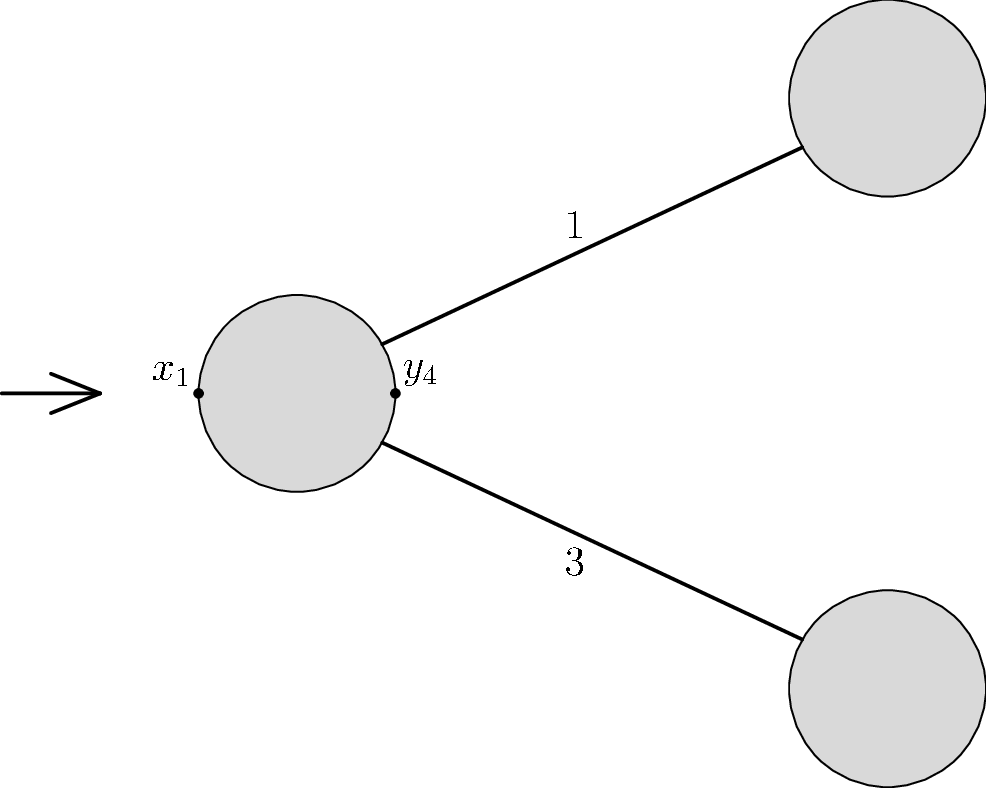,width=4cm}}
\caption{The First Filk Move on the dual of the Sunshine Graph}\label{dualsunshine}
\end{figure}

To analyze such leading terms we generalize the method of Filk's moves \cite{Filk:1996dm},
defining three distinct topological operations on a ribbon graph.
The first one is a regular ``first Filk move", namely reduction of a tree line of the graph. This amounts to glue 
the two end vertices of the line (of coordination $p$ and $q$) to get a "fatter" vertex of coordination $p+q-2$. The new graph thus obtained has one vertex less and one line less. 
Since $2-2g=n-L+F$, this operation conserves the genus.
On Figure \ref{firstfilk} the contraction of the central line of the Sunshine Graph 
(also pictured on Figure \ref{figex2}) is shown.
In the dual graph this operation deletes the direct tree line, as shown on Figure \ref{dualsunshine}.

\begin{figure}
\centerline{\epsfig{figure=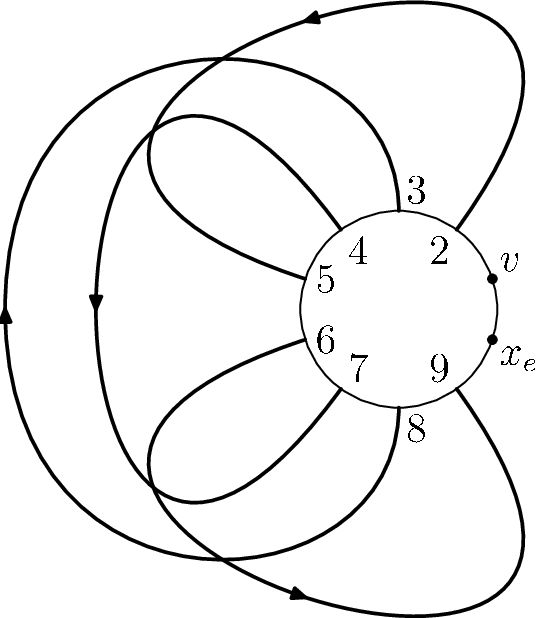,width=5cm}}
\caption{A Rooted Rosette}\label{rosette}
\end{figure}

Iterating this operation maximally we can always reduce a 
{\it spanning tree} in the direct graph,
with $n-1$ lines, obtaining a {\it rosette}. 
We recall that a rosette is simply a ribbon graph with a single vertex.
The rosettes we consider all have a root (i.e. an external line on $\bar v$), 
and a cyclic ordering to turn around, 
e.g. counterclockwise. We always draw the rooted rosette 
so that no line arches above the root. This defines uniquely 
a numbering of the halflines (see Figure \ref{rosette}, where the arrows represent the former line orientations\footnote{For an {\it orientable} graph these arrows
are compatible with the numbering, in the sense that halflines with even numbers
enter the rosette and halflines with odd numbers exit the rosette.}).

\begin{figure}
\centerline{\epsfig{figure=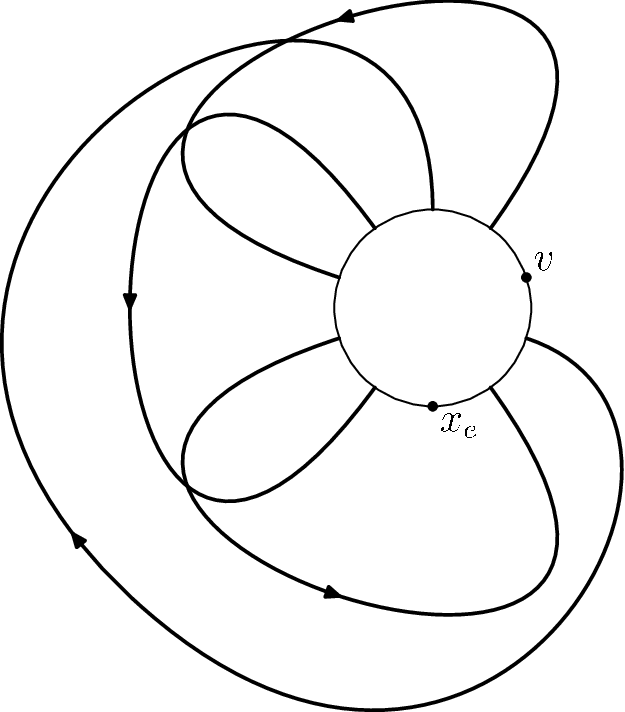,width=5cm}}
\caption{A SuperRosette}\label{superrosette}
\end{figure}

The second topological operation is the reduction of a tree line in the dual graph, exactly 
like the previous operation. Therefore it deletes this line in the direct graph. The resulting direct graph 
again keeps the same genus (remember that the genus of a graph is the same as the one of its dual). 
Iterating these two operations maximally we can always reduce completely a direct tree with $n-1$
lines {\it and} a dual tree with $f-1$ lines. We end up with a graph which we call a 
{\it superrosette}, which has only one vertex and one face 
(therefore its dual has one vertex and one face and is also a superrosette) (see Figure \ref{superrosette}).

The third operation is a genus reduction on a rosette. 
We define a {\it nice crossing} in a rosette to be a pair of lines such that the end point of the first is the successor in the rosette of the starting point of the other (in the natural cyclic order of the rosette). 
This ensures that the two lines have a common ``internal face".
When there are crossings in the rosette, it is easy to check that there exists at least one such nice crossing, for instance lines 2-5 and 4-8 in Figure \ref{rosette}.

\begin{figure}
\centerline{\epsfig{figure=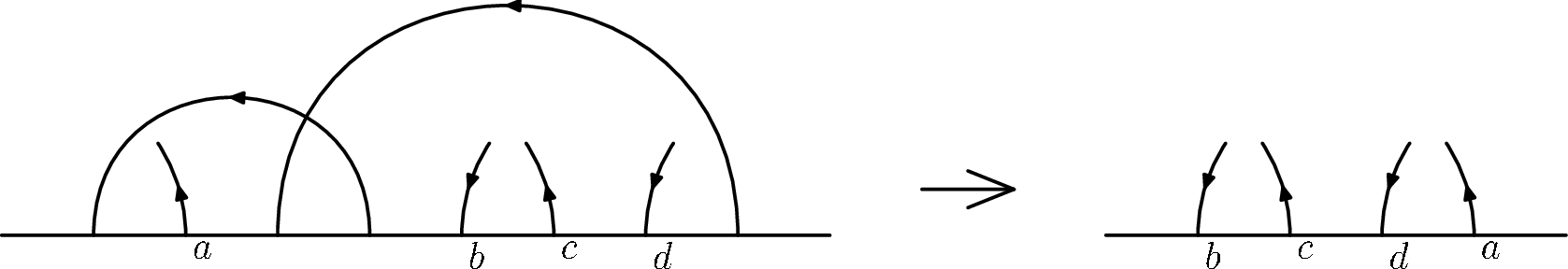,width=12cm,height=2cm}}
\caption{The Third Filk Move}\label{thirdfilk}
\end{figure}

The genus reduction consists in deleting the lines of a nice crossing
and interchanging all the halflines encompassed by the first line with those encompassed by the second line, see Figure \ref{thirdfilk}. 
This operation which we call the ``third Filk Move"\footnote{The second Filk move is 
the trivial simplification of non crossing lines in the rosette.} 
decreases the number of lines by two, glues again the faces in a coherent way, and decreases the genus by one.

We need then to compute the determinant of $B'$ matrices 
corresponding to reduced graphs of the type:
\bea
\begin{pmatrix}
\sum_{i\ne j} \omega(i,j) \eta_{l i}(-1)^{i+j+1}\eta_{l' j} 
& \sum_{i\in v} (-1)^{i+1}\eta_{l i} \\
\sum_{i\in v} (-1)^{i}\eta_{l i} & 0  \\
\end{pmatrix} \ .
\eea
As the graph is orientable and up to a possible overall sign we can cast the matrix into the form:
\bea
\begin{pmatrix}
\sum_{i\ne j}\omega(i,j)\epsilon_{l i}\epsilon_{l' j}
& \sum_{i\in v} \epsilon_{l i} \\
-\sum_{i\in v} \epsilon_{l i} & 0  \\
\end{pmatrix} \ .
\eea
We claim:
\begin{lemma}
\label{LemmaPfaff}
The above determinant is:
\begin{itemize}
\item  $0$ if the graph has more than one face,
\item $2^{2g}$ if the graph has exactly one face.
\end{itemize}
\end{lemma}

\prf  We reduce a tree. At each step we have a big vertex (the ``rosette in the making") $V$ and a small vertex $v$ bound to $V$  by a line $l=(i,j)$ which we contract. We then have at each step a Pfaffian 
$\int\prod_{d\chi_l d\psi_v}e^{-B'}$ where
\bea
B'&=&\sum_{l,l'}\chi_{l}\big(\sum_{i\ne j;i,j\in V}\omega(i,j)
\epsilon_{l i}\epsilon_{l' j} +
\sum_{i\ne j;i,j\in v}\omega(i,j)\epsilon_{l i}
\epsilon_{l' j}\big)\chi_{l'}\nonumber\\
&+&\sum_{l,v}\chi_l\epsilon_{l v}\psi_v\ .
\eea
At each step we use a permutation to put $l$ at the first place in the matrix. 
Note that this permutation has nothing to do with the ordering of the halflines. 
The terms containing $\chi_l$ or $\psi_v$ at each step are:
\bea
B'_l=&&\chi_l\epsilon_{l i}\sum_{l'}\big(\sum_{i\ne p;p  \in V}
\omega(i,p)\epsilon_{l' p} 
+\sum_{j\ne k;k  \in v}\omega(j,k)\epsilon_{l' k}\big)\chi_{l'}\nonumber\\
&&+\chi_l \epsilon_{l j}\psi_v-
\sum_{l''}\sum_{k\in v;k \neq j}\chi_{l''}\epsilon_{l'' k}\psi_v\ .
\eea

We perform the triangular change of variables:
\bea
\bar{\chi_l}&=&\chi_l-\epsilon_{l j}\sum_{l''}\sum_{k \in v;k\neq j}
\epsilon_{l'' k}\chi_{l''}\\
\bar{\psi_v}&=&\epsilon_{l j}\psi_v+\epsilon_{l i}\sum_{l'}
\big(\sum_{i\ne p;p \in V}\omega(i,p) \epsilon_{l' p}+
\sum_{i<k;k  \in v} \omega(i,k)\epsilon_{l' k}\big)\chi_{l'} 
\nonumber
\eea
Under this change of variable:
\bea
B'_l&=&\chi_l\bar{\psi_v}-\sum_{l''}\sum_{k\in v;k\neq j}\chi_{l''}
\epsilon_{l'' k} \psi_v\nonumber\\
&=&\bar{\chi_l}\bar{\psi_v}
-\epsilon_{l i}\epsilon_{l j}\sum_{l''}\sum_{l'}\chi_{l''}
\big(\sum_{k\in v;k \neq j}\sum_{p\in V;i\ne p}\omega(i,p)
\epsilon_{l'' k}\epsilon_{l' p}
\nonumber\\
&+&\sum_{k\in v;k\neq j}\sum_{r\in v;j\ne r}\omega(j,r)
\epsilon_{l'' k}\epsilon_{l' r}\big)\chi_{l'}
\eea
 
As $l$ is orientable $\epsilon_{l i}\epsilon_{l j}=-1$, 
so that the new term corresponds exactly to a new big vertex $\tilde{V}$ where the ordered halflines $k$ of the small vertex $v$ replace the halfline $i\in l$.   
We continue this procedure until we have reduced a complete tree in our graph. 
The remaining Pfaffian is of the form:
\bea
B'=\sum_{l<l';l\cap l'}\chi_l\big(\sum_{i\ne j;i,j\in V}\omega(i,j)\epsilon_{l i}\epsilon_{l'j}\big)\chi_{l'}
\eea
where $l<l'$ means that the starting halfline of $l$ precedes the end halfline of $l'$
in the final rosette vertex $R$.

When two lines in the rosette do not cross, in both sums over $i$ and $j$ above 
the two endpoints of any of the two lines add and give zero\footnote{This is the content of the ``second Filk move" of \cite{Filk:1996dm}.}.
Consider now two lines $l_1=(i_1,j_1)$ and $l_2=(i_2,j_2)$ which cross each other, i.e. such that $i_1<i_2<j_1<j_2$. 
We have:
\bea
\chi_{l_1}(\epsilon_{l_1 i_1}\epsilon_{l_2 i_2}+\epsilon_{l_1 i_1}
\epsilon_{l_2 j_2}+
\epsilon_{l_1 j_1}\epsilon_{l_2 j_2}-\epsilon_{l_1 j_1}\epsilon_{l_2 i_2})
\chi_{l_2}=2\chi_{l_1}\epsilon_{l_1 i_1}\epsilon_{l_2 i_2}\chi_{l_2}.
\eea

Changing the variables $\chi_{l}\rightarrow\epsilon_{l i}\chi_l$
and writing $l\cap l'$ if $l$ crosses $l'$ we have:
\bea\label{prepthird}
 \frac{B'}{2}=\sum_{l<l';l\cap l'}\chi_{l}\chi_{l'}\ .
\eea

We perform a new simplification trick which we call the ``third Filk move".
We can pick two lines $l_1$ and $l_2$ which form a ``nice crossing", i.e. 
the start of $l_2$ immediately precedes the end of $l_1$ in the rosette.
We define the variables:
\bea
\bar\chi_{l_1}=\chi_{l_1}+\sum_{l'<l_2;l'\cap l_2}\chi_{l'}-\sum_{l''>l_2;l''
\cap l_2}\chi_{l''},\nonumber\\
\bar\chi_{l_2}=\chi_{l_2}-\sum_{l'<l_1;l'\cap l_1}\chi_{l'}+\sum_{l''>l_1;l''
\cap l_1}\chi_{l''}\; ,
\eea
and get:
\bea
\bar\chi_{l_1}\bar\chi_{l_2}&=&\chi_{l_1}\chi_{l_2}-\chi_{l_1}\sum_{l'<l_1;l'
\cap l_1}\chi_{l'}+
\chi_{l_1}\sum_{l''>l_1;l''\cap l_1}\chi_{l''}\nonumber\\
&&+\sum_{l'<l_2;l'\cap l_2}\chi_{l'}\chi_{l_2}-\sum_{l''>l_2;l''
\cap l_2}\chi_{l''}\chi_{l_2}\\
&& +(\sum_{l'<l_2;l'\cap l_2}\chi_{l'}-\sum_{l''>l_2;l''\cap l_2}\chi_{l''})
(-\sum_{l'<l_1;l'\cap l_1}\chi_{l'}+\sum_{l''>l_1;l''\cap l_1}\chi_{l''})\ .
\nonumber
\eea
Denoting $B'_{l_1 l_2}$ all the terms which contain either $l_1$ or $l_2$ in $B'$ and consistently denoting $l_p$ the lines which cross $l_1$ and $l_q$ those which cross $l_2$ we have:
\bea
 B'_{l_1 l_2}&=&\bar{\chi_{l_1}}\bar{\chi_{l_2}}+
 \sum_{l_q<l_2;l_q\cap l_2}\chi_{l_q}\sum_{l_p<l_1;l_p\cap l_1}\chi_{l_p}
 -\sum_{l_q>l_2;l_q\cap l_2}\chi_{l_q}\sum_{l_p<l_1;l_p\cap l_1}\chi_{l_p}
\nonumber\\
 &-&\sum_{l_q<l_2;l_q\cap l_2}\chi_{l_q}\sum_{l_p>l_1;l_p\cap l_1}\chi_{l_p}+
 \sum_{l_q>l_2;l_q\cap l_2}\chi_{l_q}\sum_{l_p>l_1;l_p\cap l_1}\chi_{l_p}\ .
\label{eq:terms}
\eea

Suppose that $l_q$ is the first line crossing $l_2$ and $l_p$ is the last line crossing $l_1$. Suppose moreover that $l_q<l_2$, $l_p<l_1$. When changing to the new variables we must add to the rosette factor a term of the first type in the above expression $\chi_{l_q}\chi_{l_p}$. What is the effect of such a term? 
If $l_p<l_q$ we have a crossing $\chi_{l_p}\chi_{l_q}$ and adding the above term gives zero. If $l_p>l_q$ adding the extra term acts like a new crossing. In both cases this amounts to permute the endpoints of $l_p$ and $l_q$. We can check that this holds
in fact in all cases, and that by induction the extra terms in equation
(\ref{eq:terms}) permute all the legs of $l_p$ type with all the legs
of $l_q$ type. We conclude that:
\bea
B'=\bar{\chi_{l_1}}\bar{\chi_{l_2}}+B'_{l_p\leftrightarrow l_q} \ ,
\eea
which is the content of our ``third Filk move".

With this change of variables in the Pfaffian we see that 
contracting the lines of the tree through first Filk moves, and 
reducing the genus on the rosette through third Filk moves, we end up
with a final Pfaffian which is $\pm 2^g$, taking into account the factors $2$
in (\ref{prepthird}), hence a determinant which is $2^{2g}$.

Suppose that the initial graph has $L$ propagators, $n$ vertices and $F$ faces.
We have $2-2g=n-L+F$. We can reduce $n-1$ lines of a tree and $g$ 
nice crossings. The remaining rosette will 
have $L_{\mathrm{rest}}=F-1$ propagators, with no crossing. 
The only way for the remaining Pfaffian not to be zero is to have  $F=1$. This 
completes the proof of Lemma \ref{LemmaPfaff}.
\qed

Returning to the initial problem, we know that in order 
for the final graph to have a single face we need to reduce a tree $\tilde{T}$ 
in the dual graph ${\cal G}$. So when $I$ is maximal, $J$ must contain a
tree in ${\cal G}$. But $J$ cannot be too big either, because the complement of $J$
must contain a tree in $G$ otherwise we cannot pair all hypermomenta variables.
We say that $J$ is {\it admissible} if 
\begin{itemize}
\item it contains a tree $\tilde T$ in the dual graph
\item its complement contains a tree $T$ in the direct graph
\item The rosette obtained by removing the lines of $J$ and contracting the lines of $T$
is a superrosette, hence has a single face. 
\end{itemize}

In particular if $J$ is admissible,  we have $F-1 \le \vert J \vert \le F-1 +2g$, and $k_j = \vert J \vert - F +1$
is even and obeys $0\le k_J \le 2g$.

We have altogether proved that:
\begin{lemma}
Suppose $I$ is maximal, i.e. contains all lines.
The integer $n_{I,J}$ in (\ref{hypnoncanpfaff}) is non zero 
if and only if $J$ is admissible, in which case $n_{I,J}^2=2^{2g- k_J}$
\end{lemma}

For $I$ maximal and $J$ admissible we have $k_{I,J}= k_J $.
Hence using positivity to keep the terms with maximal $I$ that we have identified, we get
\bea\label{firstbound}
HU_{G,{\bar v}} \ge  \sum_{J \ {\rm admissible}} (2s)^{2g- k_J}
\prod_{l\in J}t_l   \ .
\eea
>From this formula power counting follows easily  by 
finding the leading terms under scaling of all $t's$ to 0, which are the ones with minimal $J$.
They correspond to $J$'s which are trees in ${\cal G}$, hence which have $k_J = 0$. 
Keeping only these terms in (\ref{firstbound}) we have the weaker bound: 
\bea\label{secondbound}
HU_{G,{\bar v}} \ge (2s)^{2g}\sum_{J \ {\rm tree}\ \in\cal{G}}\prod_{l\in J }t_l\, .
\eea
At $D=4$, and for a graph of genus $g$ with $N$ external lines, using $2-2g =n+F-L$ and $L=2n-N/2$, we get from (\ref{secondbound}) at least 
a power counting in $\lambda^{2g + (N-4)/2}d\lambda$, so that
we recover the correct power counting
as function of the genus\footnote{We recall this power 
counting is understood easily in the
matrix base representation \cite{GrWu03-1,GrWu04-3,Rivasseau2005bh}, but 
was not fully derived up to now in direct position
space \cite{GMRV} because the ``third Filk move"
was not performed there.}.

The commutative limit can be recovered easily
as the limit $s\to 0$, in which case only the terms with $k_J = 2g$ survive.
These $J$'s are exactly the complements of the trees in $G$ so that 
$HU_{G,{\bar v}}$ reduce to the usual commutative 
Symanzik polynomial $U_G$ in the limit $s  \to 0$. 

It is interesting to notice that since $s=1/4\theta\Omega$ this limit $s  \to 0$ seems 
to correspond to $\theta \to \infty$. But this is an artefact of our conventions and use of
the direct space representation. Indeed in the limit $\theta \to \infty$ the vertex in $x$-space
becomes the usual vertex in $p$-space of the commutative theory, whereas in the limit $\theta \to 0$ 
the vertex in $x$-space becomes the usual vertex in $x$-space of the commutative theory!
This explains this paradox (remark that the usual parametric representation 
(\ref{symanzik}) has $p$-space external variables).

Remark also that for planar graphs the complement of a tree in the dual graph is a tree
in the ordinary graph, hence (\ref{firstbound}) and (\ref{secondbound})
are identical in this case, as the sum over compatible $J$'s reduce
to a sum over trees of either $G$ or ${\cal G}$.

The canonical polynomial $HU_G$ can be analyzed in a similar way, but there we need
to take out one factor $c$, to pair to the additional hypermomentum in the Pfaffian,
so that $\vert I\vert $ is at most $L-1$; and the leading terms 
have one additional $t$ factor when compared to $HU_{G,{\bar v}}$
(see Section \ref{secexamp} for examples).

The power counting improvement in the number of faces broken by 
external fields is obtained after smearing external positions with
smooth test functions. For this improvement 
we analyze now the second hyperbolic polynomial.

\section{The Second Hyperbolic Polynomial}

We analyse only the $HV_{G,{\bar v}}$ polynomial, as the canonical $HV_{G}$ can be obtained easily afterwards.
The $P$ matrix in (\ref{defMPQ1})-(\ref{defMPQ2}) has elements:
\bea \label{formuleavecs}
P_{eu_l}&= -\sum_{i\neq e}\omega(i,e)\epsilon_{li}\frac{\sigma}{4\theta\Omega}\,,
\ 
P_{ev_l} &= -\sum_{i\neq e}\omega(i,e)\eta_{li}\frac{\sigma}{4\theta\Omega}\,,
\\
P_{ep_v}&=-\sum_{v;e\in v}(-1)^{e+1}\frac{\sigma}{\Omega}\,,
\
P_{\bar{p}u_l}&=\frac{\sigma}{\Omega}\,.
\eea

Note that all the elements of $P$ are multiples of $\sigma$. Upon transposition
of $P$ and multiplication we will recover a minus sign. Therefore the 
quadratic form in the external positions (and hypermomentum $\bar{p}$) in (\ref{defMPQ2}) is:
\bea
\begin{pmatrix}
x_e & \bar{p} \\
\end{pmatrix}
PQ^{-1}P^{t}
\begin{pmatrix} x_e \\ \bar{p}
\end{pmatrix}
&=&-x_{e_1}P_{e_1\tau}Q^{-1}_{\tau\tau'}P_{e_2\tau'}x_{e_2}-
x_{e_1}P_{e_1\tau}Q^{-1}_{\tau\tau'}P_{\bar{p}\tau'}\bar{p}\nonumber\\
&&-\bar{p}P_{\bar{p}\tau}Q^{-1}_{\tau\tau'}P_{e_2\tau'}x_{e_2}
-\bar{p}P_{\bar{p}\tau}Q^{-1}_{\tau\tau'}P_{\bar{p}\tau'}\bar{p}\,.
\eea

The inverse matrix of $Q$ being of the form
\bea\label{formulinvers}
 Q^{-1}_{\tau\tau'}=\frac{(A+B)^{-1}_{\tau\tau'}+(A-B)^{-1}_{\tau\tau'}}{2}\otimes I_D+
 \frac{(A+B)^{-1}-(A-B)^{-1}}{2}\otimes \sigma\, ,
\eea
we conclude that the quadratic form has a real part given by the $ I_D$ terms and an imaginary (oscillating) part 
given by the $\sigma$ terms:  the power counting we are looking for can be deduced solely from the former. 
To ease the writing we generically denote the set $x_e,\bar{p}$ by $x_e$. Also, let 
$\mathrm{Pf}(B_{\hat{K}\hat{\tau}})$ be the Pfaffian of the matrix obtained from $B$ by deleting 
the lines and columns in the set ${K,\tau}$, where again $K= I \cup J$
can be decomposed according to short and long variables. We have the analog of (\ref{pffafianfor}):
\bea
\frac{HV_{G,{\bar v}}}{HU_{G,{\bar v}}}(X_e)=
\frac{1}{\det (A+B)} \sum_K\prod_{i\in K}a_{ii}
\Big{[}\sum_{e_1}x_{e_1}\sum_{\tau\notin K}P_{e_1\tau}\epsilon_{K\tau}
\mathrm{Pf}(B_{\hat{K}\hat{\tau}})\Big{]}^2 \, .
\label{HVgv1}
\eea 

Multiplying by the product of $t$'s to compensate for the same
product in $HU_{G,{\bar v}}$ we get:
\begin{lemma}
\label{lemmapfaffsecond}
The real part $HV^R_{G,v}$ of the quadratic form in the external positions is:
\bea
\frac{HV^R_{G,{\bar v}}}{HU_{G,{\bar v}}}(x_e)=
\frac{1}{HU_{G,{\bar v}}} \sum_K\prod_{i\not\in K}t _{i}
\Big{[}\sum_{e_1}x_{e_1}\sum_{\tau\notin K}P_{e_1\tau}\epsilon_{K\tau}
\mathrm{Pf}(B_{\hat{K}\hat{\tau}})\Big{]}^2 \,.  
\label{HVgv}
\eea 
\end{lemma}

\prf:
We represent the matrix elements $Q^{-1}_{\tau\tau';\otimes K_d}$ by Grassmann integrals. As $(A+B)=(A-B)^t$ we have, for the first part of (\ref{formulinvers}):
\bea
Q^{-1}_{\tau\tau';\otimes I_D}&=&\frac{1}{2}[(A+B)^{-1}_{\tau\tau'}+(A-B)^{-1}_{\tau\tau'})\nonumber\\
&=&\frac{1}{2\det(A+B)}\int(d\bar{\psi}d\psi)[\psi_{\tau}\bar{\psi}_{\tau'}+
\psi_{\tau'}\bar{\psi}_{\tau}]e^{-\bar{\psi}(A+B)\psi} \,.
\eea

We perform the Pfaffian change of variables (of Jacobian $\imath$):
\bea
\psi_j=\frac{\chi_j+\imath \eta_j}{\sqrt{2}};~\bar{\psi}=\frac{\chi_j-\imath \eta_j}{\sqrt{2}}\,,
\eea
and we get (recalling that $d=2L+n-1$):
\bea
Q^{-1}_{\tau\tau';\otimes I_D}
&=&\frac{1}{2\det(A+B)}\int \prod_j(d\eta_j d\chi_j) \imath^{d}[\imath(\eta_{\tau}\chi_{\tau'}+\eta_{\tau'}
\chi_{\tau})]
\nonumber\\
&&e^{-\frac{1}{2}(\chi A \chi-\imath\eta A\chi+\chi B\chi-
\imath \eta B\chi+\imath\chi A\eta+\eta A \eta+\imath\chi B\eta+\eta B\eta)}\,.
\eea
As $B$ is antisymmetric the crossed terms in $B$ add to zero; as $A$ is diagonal the crossed terms 
are the only ones which survive. Reordering the measure and developping the exponential term 
in $A$ we get:
\bea
&&\frac{1}{2\det(A+B)}\int \prod_j(d\eta_j d\chi_j) \imath^{d}[-\imath(\chi_{\tau'}\eta_{\tau}+
\chi_{\tau}\eta_{\tau'})]
e^{-\frac{1}{2}(\chi B\chi+\eta B\eta)+\imath\eta A\chi}\nonumber\\
 &&=\frac{1}{2\det(A+B)}\int \prod d\eta \prod d\chi (-1)^{\frac{d^2}{2}}
(-\imath)(\chi_{\tau'}\eta_{\tau}+\chi_{\tau}\eta_{\tau'})\nonumber\\
&&\sum_{K}\prod_{i\in K}\imath^{|K|}a_{ii}\eta_i\chi_i
e^{-\frac{1}{2}(\chi B \chi+\eta B\eta)}\,.
\eea
Factorizing the sums over elements in $A$ and reordering the variables we finally get:
\bea
&&Q^{-1}_{\tau\tau;\otimes I_D}
=\frac{1}{2\det(A+B)}\sum_K\prod_{i\in K}a_{ii}
(-1)^{\frac{d^2}{2}}(-\imath)\imath^{|K|}(-1)^{\frac{|K|(|K|+1)}{2}}\nonumber\\
&&\int  d\eta \int d\chi \prod_{i\in K}\chi_i\prod_{i\in K}\eta_i
(\chi_{\tau'}\eta_{\tau}+\chi_{\tau}\eta_{\tau'})
 e^{-\frac{1}{2}(\chi B \chi+\eta B\eta)}\,.
\eea
Note that the last integrals are nonzero only if $d-|K|-1$ is even, which implies that the global sign in the above expression is always plus. The remaining Grassmann integrals can be expressed as:
\bea
\int \prod_{\alpha=1\dotsc d}d\chi_{\alpha}\prod_{i\in K}\chi_i\chi_{\tau}e^{-\frac{1}{2}\chi B\chi}=
\epsilon_{K\tau}\mathrm{Pf}(B_{\hat{K}\hat{\tau}})\,,
\eea
where $\epsilon_{K\tau}$ is the signature of the permutation 
\bea
1\dotsc d\rightarrow 1\dotsc\hat{i_1}\dotsc \hat{i_p}\dotsc \hat{i_{\tau}}\dotsc di_{\tau}i_p\dotsc i_1\,.
\eea

We have then:
\bea
 Q^{-1}_{\tau\tau';\otimes I_D}=\frac{1}{\det(A+B)}\sum_K\prod_{i\in K}a_{ii}\epsilon_{K\tau}
 \mathrm{Pf}(B_{\hat{K}\hat{\tau}})\epsilon_{K\tau'}\mathrm{Pf}(B_{\hat{K}\hat{\tau'}})\,,
\eea
and the real part of the quadratic form is:
\bea
&& \sum_{e_1, e_2, \tau , \tau'}  -x_{e_1}P_{e_1\tau} \frac{1}{\vert A+B \vert}\sum_K\prod_{i\in K}a_{ii}
\epsilon_{K\tau}\mathrm{Pf}(B_{\hat{K}\hat{\tau}})
 \epsilon_{K\tau'}\mathrm{Pf}(B_{\hat{K}\hat{\tau'}})
P_{e_2\tau'}x_{e_2}\nonumber\\
&&=-\frac{1}{\det(A+B)}\sum_K\prod_{i\in K}a_{ii}
\Big{[}\sum_{e_1}x_{e_1}\sum_{\tau\notin K}P_{e_1\tau}\epsilon_{K\tau}\mathrm{Pf}(B_{\hat{K}\hat{\tau}})
\Big{]}^2\,.
\eea
This ends the proof of (\ref{HVgv}). \qed

Using similar methods one can prove that the imaginary part of the inverse matrix elements is:
\bea
Q^{-1}_{\tau\tau';\otimes\sigma}=\frac{1}{\det(A+B)}\sum_K\prod_{i\in K}a_{ii}
\epsilon_{K\tau\tau'}\mathrm{Pf}(B_{\hat{K}\hat{\tau}\hat{\tau'}})\epsilon_K\mathrm{Pf}(B_{\hat{K}})\,,
\eea
and consequently the contribution to the quadratic form is:
\bea
&&-\frac{1}{\det(A+B)}\sum_K\prod_{i\in K}a_{ii}\epsilon_K\mathrm{Pf}(B_{\hat{K}})
\nonumber\\
&&\Big{[}\sum_{e_1,e_2} \Big{(}
\sum_{\tau\tau'}P_{e_1\tau}\epsilon_{K\tau\tau'}\mathrm{Pf}(B_{\hat{K}\hat{\tau}\hat{\tau'}})
P_{e_2\tau'}\Big{)}
x_{e_1}\sigma x_{e_2}\Big{]}\,.
\eea

\subsection{Leading terms}

The last step of our analysis is to find the leading terms in eq. (\ref{HVgv}). In order to do this one must find under which conditions Pfaffians like:
\bea
\epsilon_{K\tau}\mathrm{Pf}(B_{\hat{K}\hat{\tau}})=
\int \prod_{\alpha=1\dotsc d}d\chi_{\alpha}\prod_{i\in K}\chi_i\chi_{\tau}e^{-\frac{1}{2}\chi B\chi}\,,
\eea
are nonzero.

A priori one can exploit the $\delta$ functions associated with each external vertex to simplify the oscillating factor involving the external position. These manipulations do not have any effect on the form of $HV$ but can be used to set some $P_{ev_l}$ 
and $P_{eu_l}$ to zero. This is always the case if our graph does not have any vertex with two opposite external points.
We conclude that the power counting behavior of the second polynomial should entirely 
be given by $P_{ev_e}$ and 
consequently by terms like:
\bea
\epsilon_{Kv_e}\mathrm{Pf}(B_{\hat{K}\hat{v_e}})=
\int \prod_{\alpha=1\dotsc d}d\chi_{\alpha}\prod_{i\in K}\chi_i\psi_{v_e}e^{-\frac{1}{2}\chi B\chi}\,.
\eea

The reasoning is similar to that we used to find the leading contributions in the $HU_{G,v}$. We must find a nonzero Pfaffian multiplied by all the $c$'s and the smallest number of $t$'s possible, hence corresponding to
$K= I \cup J$ with $I=[1,...L]$ maximal and $J$ minimal.
One could find a change of variables similar to those of lemma \ref{LemmaPfaff}, but we will use here a slightly different approach.
 
We introduce a dummy Grassmann variable $\Psi$ in the Pfaffian integrals, to have:
\bea
\epsilon_{Kv_e}\mathrm{Pf}(B_{\hat{K}\hat{v_e}})=
\int \prod_{\alpha=1\dotsc d}d\chi_{\alpha}d\Psi\ \Psi \prod_{i\in K}\chi_i\psi_{v_e}e^{-\frac{1}{2}\chi B\chi}\,.
\eea

Next we exponentiate $\Psi\psi_{v_e}$ and pass to the Pfaffian of a modified matrix $B'$ (corresponding to a modified graph $G'$). The modified graph is obtained from $G$ by adding a line $l_0$ from $x_e$ to the root external line. Moreover, following the reasoning of lemma \ref{LemmaPfaff} we see that our Pfaffian is not modified if we impose that the dummy line is constrained to be a tree line in the direct graph $G'$, i.e. has to pair with
an hypermomentum variable.
We then have a one to one correspondence between the leading term in $HV_{G,{\bar v}}$ and the leading terms for the first polynomial of the modified graph $HU_{G',{\bar v}}$ 
in which the dummy line $l_0$ is chosen as a tree line.

In order to conclude we need only to compute the genus of the modified graph $G'$.  
Suppose the external point $x_e$ broke another face than the root external point. 
The dummy line we added will identify the two faces so that $G'$ has
$n'=n$, $L'=L+1$, $F'=F-1$ and
$n'-L'+F'=2-2g'$, so that $g'=g+1$.
If, on the other hand, $x_e$ broke the same face as the root halfline, the dummy line will part the latter in two different faces. 
We then have $F'=F+1$ and consequently $g'=g$.

We note that to any tree in $G'$ constrained to contain the dummy line $l_0$ there corresponds 
a two-tree $T_2$ in $G$, that is a tree minus a line (by removing this dummy line).

Therefore by Lemma III.4 the Pfaffian we are considering is non zero only if the new reduced rosette, 
where the dummy line is contracted as a tree line, has exactly one face and $2g'$ lines. 

We obtain therefore the real part of $V_{G,{\bar v}}$ as a sum of 
positive terms and exact 
analogs of bounds (\ref{firstbound}) and (\ref{secondbound}). 
We say that $J$ is 2-admissible in $G$ if $J$ is admissible in $G'$ and
the dummy line $l_0$ is in a tree of $G'$ contained in the complement of $J$.

Let $\tilde{G}$, be the graph obtained from $G$ by 
deleting the lines in $J$ and contracting the two-tree $T_2$. It has two faces, the one broken by the root
 and another one, $F_J$. This $F_J$ contains typically the external points belonging to several broken faces 
 in the initial graph.
The dummy line $l_0$ will link this two faces, but the topological structure of $G'$ is the same no matter 
which external points are chosen in $F_J$. We will therefore obtain a sum over all this possible choices.
We conclude that the bound analog to (\ref{firstbound}) is
\bea\label{firstboundext}
HV^R_{G,{\bar v}}(x_e) \ge \sum_{ J \ {\rm 2-admissible \ in} \ G} 
(2s)^{2g'-k_J}\prod_{l\in J}t_l [ \sum_{e \in F_J  } (-1)^e x_e ]^2 \; .
\eea
 
The analog of (\ref{secondbound}) is
\bea\label{secondboundext}
HV^R_{G,{\bar v}}(x_e) \ge \sum_{ J \ {\rm tree\  in} \ {\cal G}'} 
(2s)^{2g'}\prod_{l\in J} t_l \  [\sum_{e \in F_J  } (-1)^e x_e ]^2 \; .
\eea
This bound is the one useful to extract
the power counting in the broken faces.
Indeed when integrating the remaining external variables
against fixed test functions and scaling each $t$ variable by $\lambda \to 0$, 
we recover the full exact power counting of $G$ namely in dimension $D=4$ the scaling
$\lambda^{2g +(B-1) + (N-4)/2}d\lambda$.
Indeed each broken face leads to a term in $e^{-  [x_e+\dotsc]^2}$ for some external variable $x_e$ of the broken face, hence to an improved factor $\lambda$ when integrated against a fixed test function. 

It is also possible to recover the second Symanzik polynomial as $s \to 0$ in 
(\ref{firstboundext}); indeed for $k_J = 2g'$, we find that $J$ is the complement of a tree in $G'$
containing $l_0$, hence of a two-tree in $G$. Moreover, in this case the face $F_J$ will be the external face 
of the conected component $E(T_{2})\in T_2$ not containing the root, and we recover the known invariant 
$(\sum_{e\in E(T_2)}(-1)^{e+1}x_e)^2$.

It remains to discuss the case with special ``diagonal" vertices, that is with opposite 
external arguments as in Figure \ref{figex4}.
In this case bounds (\ref{firstboundext}) and (\ref{secondboundext}) still hold because
although there is a sum of two Paffians in Lemma \ref{lemmapfaffsecond}, 
they cannot add up to 0; they correspond to graphs $G'_1$ (of the same kind as before) and 
$G'_2$ (of a new type, with one line erased) with genuses
$g'_1$ and $g'_2 =g'_1 - 1$. These Pfafians have the same scaling in $s$ because 
there is an additional $s$ for $G'_2$ coming from formula (\ref{formuleavecs}), but they have not the same
power of 2 hence their sum cannot be zero again!

There is another modification: the alternate sum over a face no longer appear
in (\ref{firstboundext}) and  in  (\ref{secondboundext})  
if the root vertex itself is of this diagonal type which is the case
for the ``Broken Bubble" graph of Figure \ref{figex4}. These modifications 
do not affect the power counting and their verification is left to the reader.

\section{Examples}
\label{secexamp}

In this section we give the exact expressions for several of our polynomials. We recall that 
$s =(4\theta\Omega)^{-1}$.

\begin{figure}[ht]
\centerline{\epsfig{figure=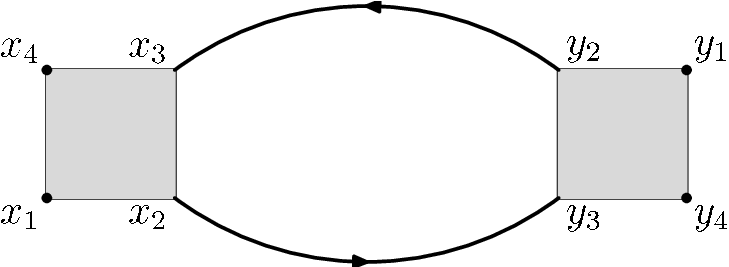,width=6cm}}
\caption{The Bubble graph}\label{figex1}
\end{figure}

We start by the bubble graph, Figure \ref{figex1}:

\bea
HU_{G,v}&=&(1+4s^2)(t_1+t_2+t_1^2t_2+t_1t_2^2)\,,\nonumber\\
HV_{G,v}&=&t_2^2\Big{[}p_2+2s(x_4-x_1)\Big{]}^2+t_1t_2\Big{[}2p_2^2+(1+16s^4)(x_1-x_4)^2
                \Big{]}\,,\nonumber\\
                &&+t_1^2\Big{[}p_2+2s(x_1-x_4)\Big{]}^2\nonumber\\
HU_{G}&=&4(t_1+t_2)^2\,,\nonumber\\
HV_{G}&=&(1+4s^2)\Big{[}(x_1-x_4+y_1-y_4)^2(t_1+t_2)\nonumber\\
                   && +(x_1-x_4-y_1+y_4)^2(t_1t_2^2+t_2t_1^2)\Big{]}\,.
\eea

The first two terms in $HU_{G,v}$ are the leading ones we previously exhibited. For the $HV_{G,v}$ we see that the scaling of the quadratic form will be in $O(1)$, which was expected as we do have only one broken face. Furthermore $HU_{G}$ scales in $t^2$, as expected, and the first term in the quadratic form $HU_{G}/HV_{G}$ is exactly the required one to reconstitute the $\delta$ function on the external legs in the $UV$ region.
The term $t_1t_2(x_1-x_4)^2$ in  $HV_{G,\bar{v}}$ is one of the leading terms previousely computed. It comes from the graphs $G'$ in which either $x_1$ or $x_4$ are linked to the root by the dummy line, and both lines $1$ and $2$ are chosen in the set $J$ in $\cal{G'}$.

Then comes the sunshine graph Fig. \ref{figex2}:
\begin{figure}[h]
\centerline{\epsfig{figure=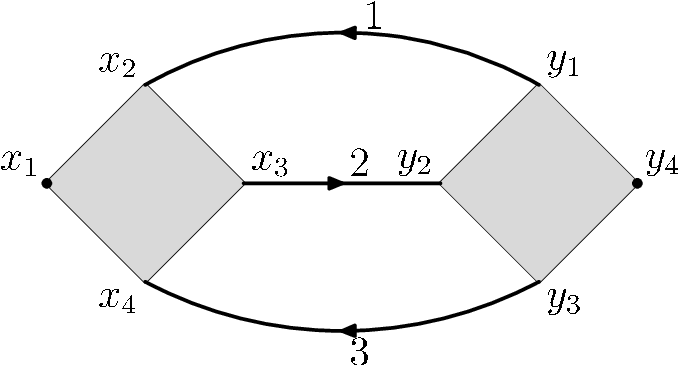,width=6cm}}
\caption{The Sunshine graph}\label{figex2}
\end{figure}

\bea
HU_{G,v}&=&\Big{[} t_1t_2+t_1t_3+t_2t_3+t_1^2t_2t_3+t_1t_2^2t_3+t_1t_2t_3^2\Big{]}
(1+8s^2+16s^4)\nonumber\\
&&+16s^2(t_2^2+t_1^2t_3^2)\, ,\nonumber\\
HU_{G}&=&t_1t_2t_3(1+64s^4)\nonumber\\
&&+\Big{[}t_1^2t_2+t_1t_2^2+t_1^2t_3+t_2^2t_3+t_1t_3^2+t_2t_3^2\Big{]}(4+16s^2)\,.
\eea

Here we identify also the leading contributions in $HU_{G,v}$, and the extra scaling factor in the $HU_{G}$.

For the nonplanar sunshine graph (see Fig. \ref{figex3}) we have:
\begin{figure}[ht]
\centerline{\epsfig{figure=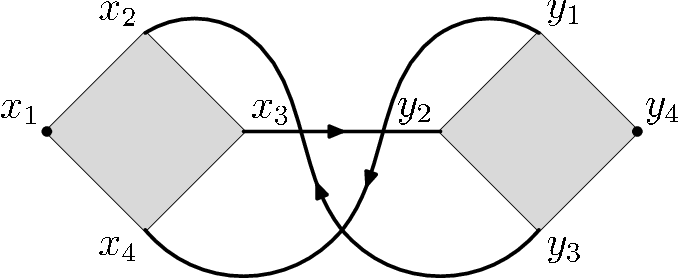,width=6cm}}
\caption{The Non-planar Sunshine graph}\label{figex3}
\end{figure}
\bea
HU_{G,v}&=&\Big{[} t_1t_2+t_1t_3+t_2t_3+t_1^2t_2t_3+t_1t_2^2t_3+t_1t_2t_3^2\Big{]}
(1+8s^2+16s^4)\nonumber\\
&&+4s^2\Big{[}1+t_1^2+t_2^2+t_1^2t_2^2+t_3^2+t_1^2t_3^2+t_2^2t_3^2+
t_1^2t_2^2t_3^2\Big{]}\,,\nonumber\\
HU_{G}&=&4\Big{[}t_1^2(t_2+t_3)+t_2t_3(t_2+t_3)+t_1(t_2^2+3t_2t_3+t_3^2)\Big{]}\nonumber\\
&&+16s^2\Big{[}t_1+t_2+t_3+t_1t_2^2t_3^2+t_1^2t_2t_3(t_2+t_3)\Big{]}
+64t_1t_2t_3s^4\,.
\eea

We note the improvement in the genus, as both $HU_{G,v}$ and $HU_{G,v}$ scale in $t^{-2}$ with respect to there planar counterparts.

For the broken bubble graph (see Fig. \ref{figex4}) we have:
\begin{figure}[ht]
\centerline{\epsfig{figure=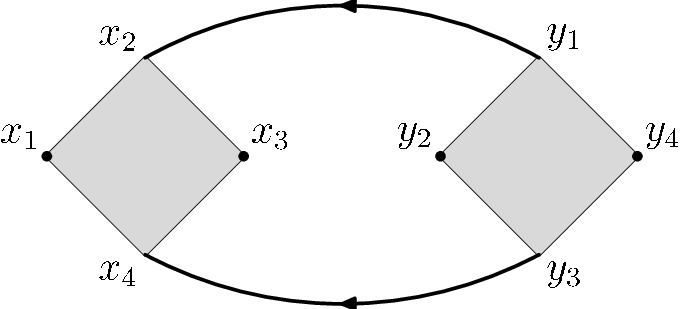,width=5cm}}
\caption{The Broken Bubble Graph}\label{figex4}
\end{figure}

\bea
HU_{G,v}&=&(1+4s^2)(t_1+t_2+t_1^2t_2+t_1t_2^2)\,,\nonumber\\
HV_{G,v}&=& t_2^2 \Big{[}4s^2(x_1+y_2)^2+(p_2-2s(x_3+y_4))^2\Big{]}+t_1^2\Big{[}p_2
+2s(x_3-y_4) \Big{]}^2\,,\nonumber\\
&&+t_1t_2\Big{[}8s^2y_2^2+2(p_2-2sy_4)^2+(x_1+x_3)^2+16s^4(x_1-x_3)^2\Big{]}\nonumber\\
&&+t_1^2t_2^24s^2(x_1-y_2)^2\,,\nonumber\\
HU_{G}&=&4(t_1+t_2)^2\,,\\
HV_{G}&=&(t_1+t_2)\Big{[}(y_2+y_4-x_1-x_3)^2+4s^2(y_2-y_4-x_1-3x_3)^2\Big{]}\nonumber\\
&&+(t_1^2t_2+t_1t_2^2)\Big{[}(y_2+y_4+x_1+x_3)^2+4s^2(y_2-y_4+x_1+3x_3)^2\Big{]}\,,\nonumber
\eea

Note that $HU_{G,v}$ and $HU_{G}$ are identical with those of the bubble with only one broken face. It is natural, as the amelioration in the broken faces for a given graph can be seen only in the $HV_{G,v}$ and $HV_{G}$.  Take $HV_{G}$. We see that we have two linear combinations which in the $UV$ region become an approximate $\delta$ function, whereas in the bubble graph with one broken face we had only one, therefore giving us the improvement in the broken faces. Similarely for the $HV_{G,v}$ we see that for the broken bubble we have three independent linear combinations which scale in $t^{-1}$ whereas for the bubble with only one broken face we only had two.
The term $t_1t_2(x_1+x_3)^2$ in $HV_{G,\bar{v}}$ is one of those computed in the previous section. Even if the two external points $x_1$ and $x_3$ do not belong to the same face they still appear summed, as we can chose the root halfline to be either
$y_2$ and $y_4$ and we must add the contributions for each choice. This is an example of the slight modifications generated by the presence of "diagonal" vertices.

\begin{figure}[ht]
\centerline{\epsfig{figure=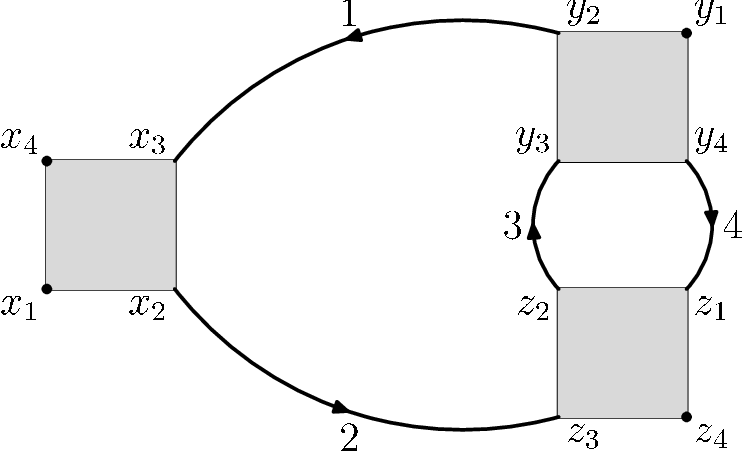,width=6cm}}
\caption{The Half-Eye Graph}\label{figeye}
\end{figure}
Finally, for the half-eye graph (see Fig. \ref{figeye}), we start by defining:
\bea
A_{24}=t_1t_3+t_1t_3t_2^2+t_1t_3t_4^2+t_1t_3t_2^2t_4^2\,.
\eea
The $HU_{G,v}$ polynomial with fixed hypermomentum corresponding to the vertex with two external legs is:
\bea\label{hueye1}
&&(A_{24}+A_{14}+A_{23}+A_{13}+A_{12})(1+8s^2+16s^4)\nonumber\\
&&+t_1t_2t_3t_4(8+16s^2+256s^4)+4t_1t_2t_3^2+4t_1t_2t_4^2\nonumber\\
&&+16s^2(t_3^2+t_2^2t_4^2+t_1^2t_4^2+t_1^2t_2^2t_3^2)
+64s^4(t_1t_2t_3^2+t_1t_2t_4^2)\,,\nonumber
\eea
whereas with another fixed hypermomentum we get:
\bea\label{hueye2}
HU_{G,v_2}&=&(A_{24}+A_{14}+A_{23}+A_{13}+A_{12})(1+8s^2+16s^4)\nonumber\\
&&+t_1t_2t_3t_4(4+32s^2+64s^4)+32s^2t_1t_2t_3^2+32s^2t_1t_2t_4^2\nonumber\\
&&+16s^2(t_3^2+t_1^2t_4^2+t_2^2t_4^2+t_1^2t_2^3t_3^2)\,.   
\eea

Note that the leading terms are identical, the choice of the root perturbing only the non-leading ones. Moreover note the presence of the $t_3^2$ term. Its presence can be understood by the fact that in the sector $t_1,t_2,t_4>t_3$ the subgraph formed by the lines $1,2,4$ has two broken faces. This is the sign of 
a power counting improvement due to the additional broken face in that sector. 
To exploit it, we have just to integrate over the variables of line $3$
in that sector, using the second polynomial $HV_{G',v}$ for the triangle subgraph $G'$ 
made of lines $1,2,4$.  

Finally the canonical $HU_{G}$ polynomial is:
\bea
HU_{G}&=&(4+16s^2)(t_1^2t_3+t_1t_3^2+t_2^2t_3+t_2t_3^2+t_1^2t_4+t_1t_4^2+
t_2^2t_4+t_2t_4^2\nonumber\\
&&+t_3^2t_4+t_3t_4^2
+t_1^2t_2t_3^2+t_1t_2^2t_3^2+t_1^2t_2t_4^2+t_1t_2^2t_4^2
+t_1^2t_2^2t_3^2t_4\nonumber\\
&&+t_1^2t_2^2t_3t_4^2)
+(8+32s^2)(t_1t_2t_3+t_1t_2t_4+t_1t_2t_3^2t_4+t_1t_2t_3t_4^2)\nonumber\\
&&+(12+64s^4)
(t_1t_3t_4+t_2t_3t_4+t_1^2t_2t_3t_4+t_1t_2^2t_3t_4)\,.
\eea

\subsubsection*{Acknowledgment}
We are indebted to A.~Abdesselam for his inspiring reference \cite{A}
and his help on the track to the Pfaffian analysis of Sections III-IV. 
We also thank  M.~Disertori, J.~Magnen and F.~Vignes-Tourneret 
for useful discussions during preparation of this work.

\chapter{Dimensional renormalization of NCQFT}
\label{chap:dimreg}

\begin{center} 
{\sf   R. Gurau, }{\sf A. Tanas\u a}
\end{center}

\medskip

\noindent
Using the recently introduced parametric representation of non-commutative quantum field theory, we implement here the dimensional regularization and renormalization of the vulcanized $\Phi^{\star 4}_4$ model on the Moyal space.

Keywords: non-commutative quantum field theory, dimensional regularization,
dimensional renormalization

\section{Introduction and motivation}
\setcounter{equation}{0}

Non-commutative geometry (see \cite{Connes}) is one of the
most appealing frameworks for the quantification of gravitation.
Quantum field theory (QFT) on these type of spaces, called non-commutative quantum
field theory (NCQFT) - for a general review see
\cite{DN,nc2}- is now
one of the most appealing candidates for new physics beyond the Standard
Model. Also, NCQFT arises as the effective limit of some string
theoretical models \cite{SeiWitt,a.connes98:noncom}.

Moreover, NCQFT is well suited to the description of the physics in background fields and with non-local interactions, like for example the
fractional quantum Hall effect \cite{Susk,Polych,Raam}.

However, naive NCQFT suffers from a new type of non renormalizable divergences, known as the ultraviolet (UV)/ infrared (IR) mixing. The simplest example of this kind of divergences is given by the nonplanar tadpole: it is UV convergent, but inserting it an arbitrary number of times in a loop gives rise to IR divergences.

Interest in NCQFT has been recently revived with the introduction
of the Grosse-Wulkenhaar scalar $\Phi^{\star 4}_4$ model, in which the UV/IR mixing is cured: the model is
renormalizable at all orders in perturbation theory \cite{GrWu03-1,GrWu04-3}. The idea of Grosse-Wulkenhaar was to modify the kinetic part of the action in order to satisfy the Langmann-Szabo duality \cite{LaSz} (which relates the infrared and
ultraviolet regions). We refer to this modified theory as the vulcanized 
$\Phi^{\star 4}_4$ model.

A general proof, using position space and multiscale analysis, has then been given in \cite{GMRV}, and the parametric representation of this model was computed in \cite{gurauhypersyman}. Furthermore, it was recently proved
that that the vulcanized $\Phi^{\star 4}_4$ is better behaved than
the commutative $\phi^4_4$ model: it does {\bf not} have a Landau ghost \cite{GrWu04-2,DR,landau3}.

In commutative QFT dimensional renormalization is the only scheme which respects the symmetries of gauge theories see \cite{reg, ren, nemtii}. It also is the appropriate setup for the Connes-Kreimer Hopf algebra approach to renormalization (see 
\cite{CK,kreimer} for the case of commutative QFT).
 
 A second class of renormalizable NCQFT exists. These models, called {\it covarinat}, are characterized
 by a propagator which decays in position space as $x-y$ tends to infinity (like the Grosse-Wulkenhaar propagator
of eq. (\ref{propa})) but it oscillates when $x+y$ goes to infinity, rather than decaying. In
this class of NCQFT models enters the
non-commutative Gross-Neveu model and the Langmann-Szabo-Zarembo model
\cite{Langmann:2003if}. The non-commutative orientable Gross-Neveu model was proven to be
renormalizable at any order in perturbation theory \cite{RenNCGN05}. The
parametric representation was extended to this class of models \cite{RivTan}.
For a general review of recent developments in the field of renormalizable NCQFT see \cite{sefu}.

The parametric representation introduced in \cite{gurauhypersyman} is the starting point for the dimensional regularization and renormalization performed in this paper. Our proof follows that of the commutative $\Phi^{\star 4}_4$ model, as presented in \cite{reg, ren}.

This paper is organized as follows. Section \ref{sec:NCPHI44} is a summary of the parametric representation
of the vulcanized $\Phi^{\star 4}_4$ model. The non-commutative equivalent $HU_G$ and $HV_G$ of the Symanzik polynomials $U_G$
and $V_G$ are recalled.
In section \ref{sec:furtherLeading} we prove the existence in the polynomial $HU_G$ of some further leading terms in the ultraviolet (UV) regime.  This is an improvement of the results of \cite{gurauhypersyman}, needed to correctly identify the meromorphic structure of the Feynman amplitudes. 
In section \ref{sec:factorisation} we prove the factorization properties of the Feynman amplitudes. These properties are needed in order to prove that the pole extraction is equivalent to adding counterterms of the form of the initial lagrangean. This factorization is essential for the definition of a coproduct $\Delta$ necessary for the implementation of a Hopf algebra structure in NCQFT \cite{progres}. 
Section \ref{sec:NCdimreg} uses the results of the previous sections to perform the dimensional regularization, prove the counterterm structure for NCQFT and complete the dimensional renormalization program. 
Section \ref{sec:conclusion} is devoted to some conclusion and perspectives. 

\section{The non-commutative model}
\label{sec:NCPHI44}
\setcounter{equation}{0}

In this section we give a brief overview of the Grosse-Wulkenhaar $\Phi^4$ model. Our notations and conventions as well as some notions of diagrammatics and the results of the parametric representation follow \cite{gurauhypersyman}.

To define the Moyal space of dimension $D$, we introduce the deformed Moyal product $\star$ on ${\mathbb R}^D$ so that
\beqa
\label{2D}
[x^\mu, x^\nu]=i \Theta^{\mu \nu},
\eeqa
\noi
where the the matrix $\Theta$ is
\begin{eqnarray}
\label{theta}
  \Theta= 
  \begin{pmatrix}
    \begin{matrix} 0 &\theta \\ 
      \hspace{-.5em} -\theta & 0
    \end{matrix}    &&     0
    \\ 
    &\ddots&\\
    0&&
    \begin{matrix}0&\theta\\
      \hspace{-.5em}-\theta & 0
    \end{matrix}
  \end{pmatrix}.
\end{eqnarray}
\noi
The associative Moyal product of two functions 
$f$ and $g$ on the Moyal space writes
\beqa
\label{moyal-product} 
 (f\star g)(x)&=&\int \frac{d^{D}k}{(2\pi)^{D}}d^{D}y\, f(x+{\textstyle\frac 12}\Theta\cdot
  k)g(x+y)e^{\imath k\cdot y}\nonumber\\
  &=&\frac{1}{\pi^{D}|\det\Theta|}\int d^{D}yd^{D}z\,f(x+y)
  g(x+z)e^{-2\imath y\Theta^{-1}z}\; .
\eeqa
The Euclidian action introduced in \cite{GrWu04-3} is 
\beqa
\label{lag-init}
S=\int d^4 x \left(\frac{1}{2} \partial_\mu \phi
\star \partial^\mu \phi +\frac{\Omega^2}{2} (\tilde{x}_\mu \phi )\star
(\tilde{x}^\mu \phi ) 
+ \frac{1}{2} m^2 \,\phi \star \phi
+  \phi^{\star 4}\right)\, ,
\eeqa
\noi
where 
\beqa
\label{tildex}
\tilde{x}_\mu = 2 (\Theta^{-1})_{\mu \nu} x^\nu \, .
\eeqa
\noi
The propagator of this model is the inverse of the operator
\beqa
\label{inverse-propa}
-\Delta+\Omega^{2}\tilde x^{2}.
\eeqa
\noi
The results we establish here hold for orientable models (in the
sense of subsection \ref{grafuri}). This corresponds to a Grosse-Wulkenhaar
model of a complex scalar field
\beqa
\label{lag}
S=\int d^4 x \left(\frac{1}{2} \partial_\mu \bar \phi
\star \partial^\mu \phi +\frac{\Omega^2}{2} (\tilde{x}_\mu \bar \phi )\star
(\tilde{x}^\mu \phi ) 
+  \bar \phi \star \phi \star \bar \phi \star \phi \right) \, .
\eeqa
\noi
Introducing $\tom=2\Omega/\theta$, the kernel of the propagator is (Lemma $3.1$ of \cite{Propaga}) 
\beqa
\label{propa}
C(x,y)=\int_0^\infty \frac{\tom d\alpha}{[2\pi\sinh(\alpha)]^{D/2}}
e^{-\frac{\tom}{4}\coth(\frac{\alpha}{2})(x-y)^2-
\frac{\tom}{4}\tanh(\frac{\alpha}{2})(x+y)^2}\; .
\eeqa

Using eq. (\ref{moyal-product}) the interaction term in eq. (\ref{lag})
leads to the following vertex contribution in position space (see \cite{GMRV}
\beqa
\label{v1}
\delta (x_1 - x_2 + x_3 - x_4)e^{2i\sum_{1\le
    i <j\le 4}(-1)^{i+j+1}x_i\Theta^{-1}x_j} \, .
\eeqa
\noi
with $x_1,\ldots, x_4$ the $4-$vectors of the positions of the $4$
fields incident to the vertex.

To any such vertex $V$ one associates 
a hypermomentum $p_V$ using the relation
\beqa
\label{GTpbar1}
\delta(x_1 -x_2+x_3-x_4 ) 
= \int  \frac{d p_V}{(2 \pi)^4}
e^{p_V \sigma (x_1-x_2+x_3-x_4)} \, .
\eeqa
\noi

\subsection{Some diagrammatics for NCQFT; orientability}
\label{grafuri}

In this subsection we introduce some useful conventions and  definitions, some of them used in \cite{GMRV} and  \cite{Propaga} but also some new ones.

Let a graph $G$ with $n(G)$ vertices, $L(G)$ internal lines and $F(G)$
faces. The Euler characteristic of the graph is
\beqa
\label{genus}
2-2g(G)=n(G)-L(G)+F(G),
\eeqa
\noi
where $g(G)\in{\mathbb N}$ is the {\it genus} of the graph.
Graphs divide in two categories, {\it planar graph} with $g(G)=0$, and 
{\it non-planar graphs} with $g(G)>0$.
Let also $B(G)$ denote the number of faces broken by external lines and
$N(G)$ be number of external points of the graph.

The ``orientable'' form eq .(\ref{v1}) of the
vertex contribution of our model allows us to associate ``+'' sign to a corner $\bar \phi$ and a  ``-'' sign to a corner $\phi$ of the vertex.
These signs alternate when turning around a vertex.
As the propagator allways relates a $\bar \phi$ to a $\phi$, the action in eq. (\ref{lag}) has orientable lines, that is any internal line joins a ``-'' corner to a ``+'' corner \footnote{The orientability of our theory allows us to simplify the proofs. It should however be possible, although tedious, to follow the same procedure for the non orientable model.}.

Consider a spanning tree $ {\cal T} $ in $G$. In has $n-1$ lines and 
the remaining $L-(n-1)$ lines form the set $\cal L$ of loop lines. 
Amongst the vertices $V$ one chooses a special one $\bar V_{G}$, the {\it
  root} of the tree. 
One associates to any vertex $V$ the unique tree line which hooks to $V$ and goes towards the root. 

We introduce now some topological operations on the graph which allow one to reexpress the oscillating factors coming from the vertices of the graph $G$.

\begin{figure}
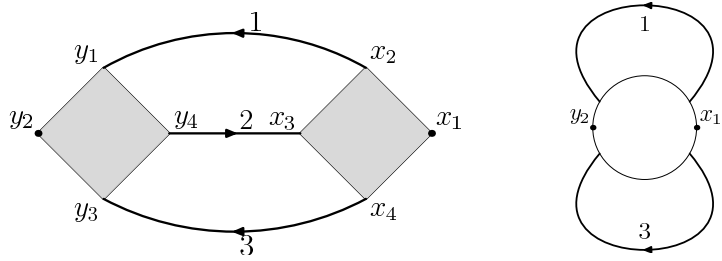

  \centerline{\epsfig{figure=sunshine-corectat.ps,width=6cm} \hfil \epsfig{figure=filk-corectat.ps,width=2cm}}
\caption{The first Filk move: the line $2$ is reduced and the $2$ vertices merge}\label{GTfirstfilk}
\end{figure}

Let a tree line in the graph $\ell=(i,j)$ and its endpoints $i$ and $j$. Suppose it connects to the root vertex $\bar V_G$ at $i$ and to another vertex $V$ at $j$.
In Fig.\ref{GTfirstfilk}, $\ell_2$ is the tree line, $y_4$ is $i$ and $x_1$ is $j$.
{\bf The first Filk move}, inspired by \cite{Filk:1996dm}, consists in removing such a line from the graph and gluing the two vertices together respecting the ordering. Thus the point $i$ on the root vertex is replaced by the neighbors of $j$ on $V$. 
This is represented in Fig. \ref{GTfirstfilk} where the new root vertex is
$y_2,y_3,x_4,x_1,x_2,y_1$.

Note that the number of faces or the genus of the graph do not change 
under this operation.

A technical point to be noted here is that one must chose the field $j$ on the vertex $V$ to be either the first (if the line $\ell$ enters $V$) or the last 
(if the line $\ell$ exits $V$) in the ordering of $V$. Of course this is allways possible by the use of the $\delta$ functions in the vertex contribution.

\begin{figure}[ht]
\centerline{\epsfig{figure=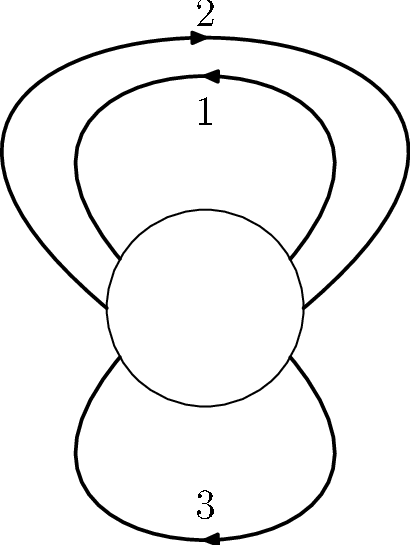,width=3cm}\hfil
  \epsfig{figure=rozeta-np.ps,width=3cm}}
\caption{Two rosettes obtained by contracting a tree {\it via} the first Filk}\label{rozeta}
\end{figure}

Iterating this operation for the $n-1$ tree lines, one obtains a single final
vertex with all the loop lines hooked to it - a {\it rosette} (see Fig. \ref{rozeta}).

The rosette contains all the topological information of the graph. If no two lines cross (on the left in Fig. \ref{rozeta}) the graph is planar. If on the contrary we have at least a crossing (on the right in Fig. \ref{rozeta}) the graph is non planar (for details see \cite{gurauhypersyman}).

For a nonplanar graph we define a {\it nice crossing} in a rosette as a pair of lines such that the end point of the first is the successor in the rosette of the
starting point of the other. A {\it genus line} of a graph is a loop line which is part of a nice crossing on the rosette (lines $2$ and $4$ on the right of Fig. \ref{rozeta}).

In the sequel we are interested in performing this operation in a way adapted to the scales introduced by the Hepp sectors: we  perform the first Filk move only for a subgraph $S$ (we iterate it only for a tree in $S$).
Thus, the subgraph $S$ will be shrunk to its
corresponding rosette inside the graph $G$. If $S$ is not primitively divergent we have a convergent sum over its associated Hepp parameter.

We will prove later that $S$ is primitively divergent if and only if $g(S)=0$, $B(S)=1$, $N(S)=2,4$.
For primitively divergent subgraphs the first Filk move above shrinks $S$ to a Moyal vertex inside the graph
$G$. 

\begin{figure}[ht]
\centerline{\epsfig{figure=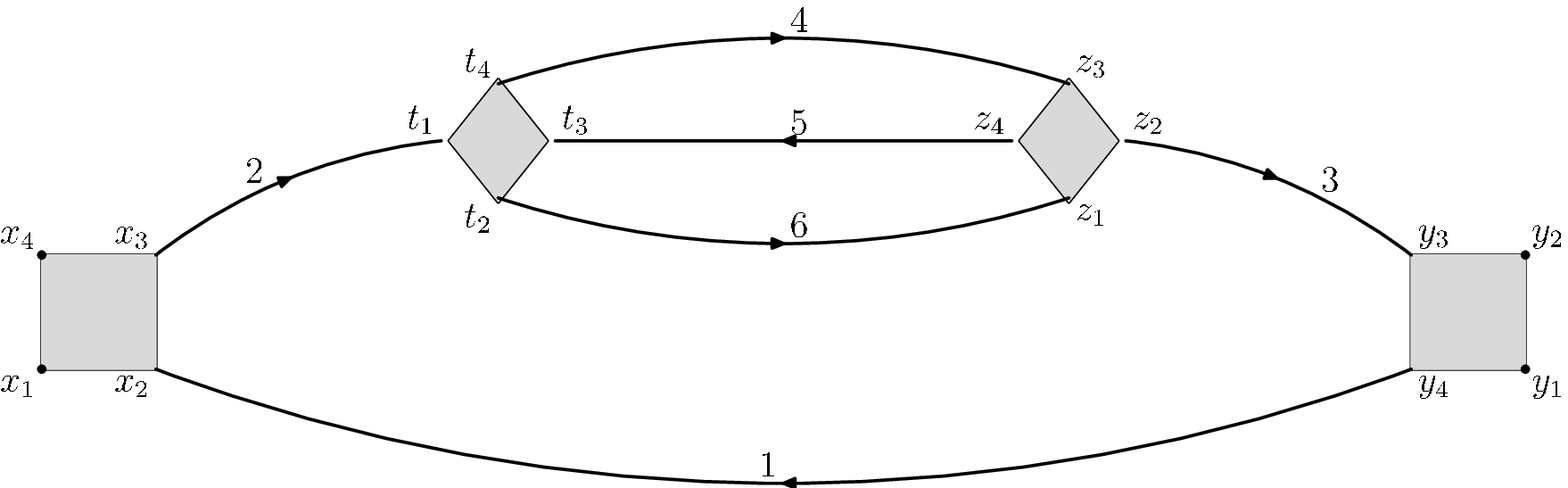,width=10cm}}
\caption{A graph containing a primitive divergent subgraph given by the lines  $\ell_4,\ell_5$ and $\ell_6$}\label{hyper}
\end{figure}

\begin{figure}[ht]
\centerline{\epsfig{figure=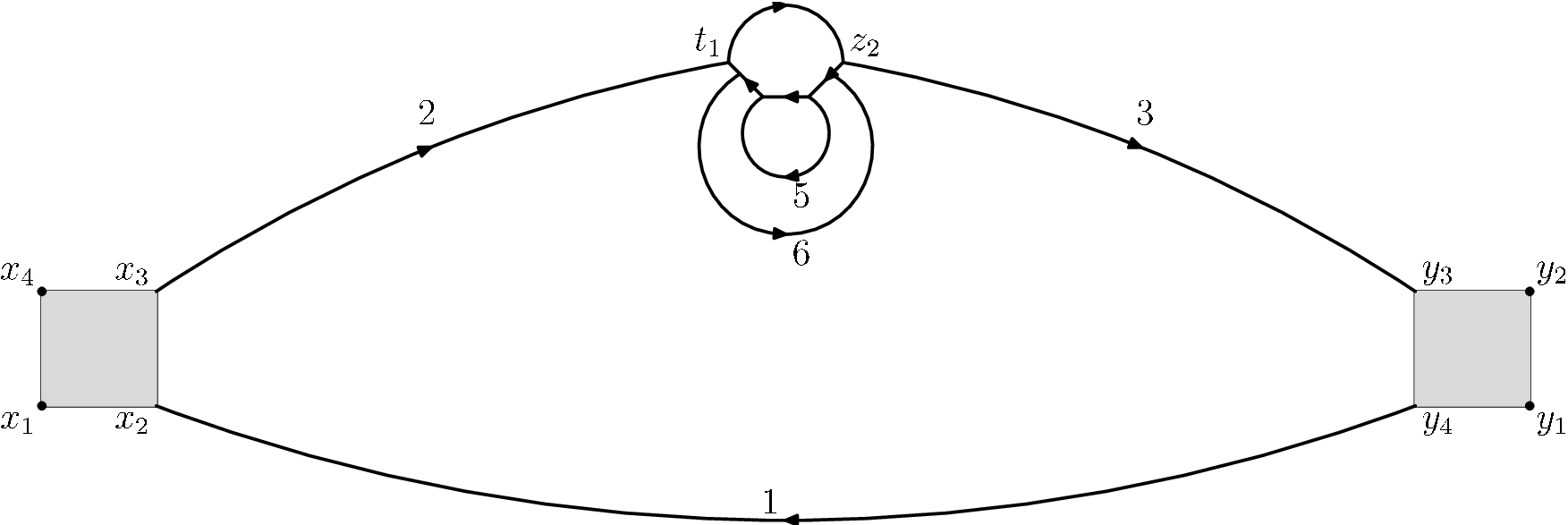,width=10cm}}
\caption{The graph $G$ with $S$ reduced to a rosette}
\label{fig:mese}
\end{figure}

\begin{figure}[ht]
\centerline{\epsfig{figure=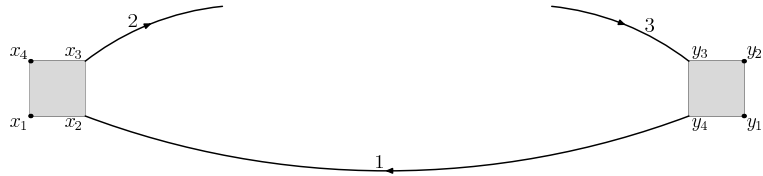,width=10cm}}
\caption{The graph $G-S$ obtained by erasing the lines
 and vertices of the primitive divergent subgraph $S$}
\label{G-S}
\end{figure}

\begin{figure}[ht]
\centerline{\epsfig{figure=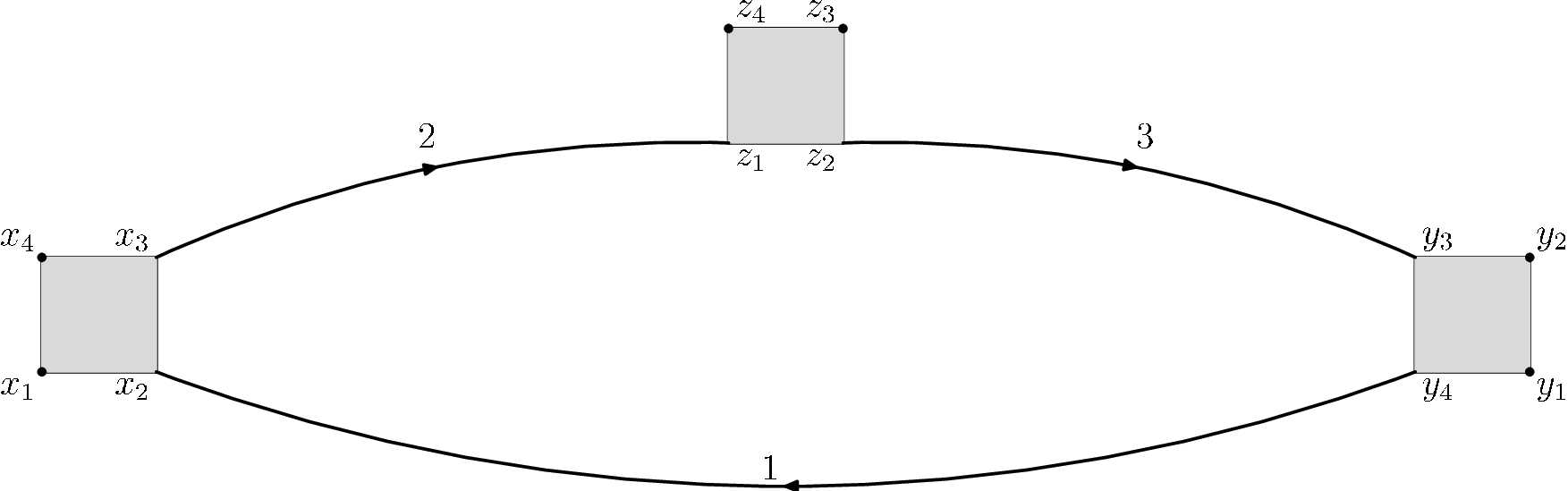,width=10cm}}
\caption{The graph $G/S$ obtained by shrinking to a Moyal vertex  the sunshine
  primitive divergent subgraph $S$}
\label{bula-deformata}
\end{figure}

For example, consider the graph $G$ of Fig. \ref{hyper} and its divergent
sunshine subgraph $S$ given by the set of lines $\ell_4,\ell_5$ and $\ell_6$. 
Under the first Filk move for the subgraph $S$, $G$  will have a rosette vertex insertion like in Fig. \ref{fig:mese},
Denote $G-S$ graph $G$ with its subgraph $S$ erased (see Fig.  \ref{G-S}). It becomes the graph $G/S$ with a Moyal vertex like in Fig. \ref{bula-deformata}.

In the commutative case, this operation corresponds to the shrinking of $S$ to a point: it represents the ''Moyality" (instead of locality) of the theory.

\subsection{Parametric representation for NCQFT}
\label{para-nc}

In this subsection we recall the definitions and results obtained in
\cite{gurauhypersyman} for the parametric
representations of the model defined by eq. (\ref{lag}).
First let us recall that, when considering the parametric
representation for commutative QFT, one has translation invariance in position
space. As a consequence of this invariance, the first polynomial vanishes when
integrating over all internal positions. Therefore, one has to integrate over
all internal positions (which correspond to vertices) save one, which is thus
marked. However, the polynomial is a still a canonical object, {\it
  i. e.} it {\it does not depend} of the choice of this
particular vertex.

As stated in \cite{gurauhypersyman}, in the non-commutative case translation
invariance is lost (because of non-locality). Therefore, one can integrate
over all internal positions and hypermomenta. However, in order to be able to recover the commutative limit, we
also mark a particular vertex $\bar V$; we do not integrate on its associate hypermomenta $p_{\bar V}$. This particular vertex is the root vertex. Because there is no translation invariance, the polynomial does depend on on the choice of the root; however the leading ultraviolet terms do not.

We define the $(L\times 4)$-dimensional incidence matrix $\e^V$ for each of the vertices $V$. Since the graph is orientable (in the sense defined in subsection
\ref{grafuri} above) we can choose
\beqa
\label{r1}
\e_{\ell i}^V= (-1)^{i+1}, \mbox { if the line $\ell$ hooks to the vertex $V$
  at corner $i$.}
\eeqa
\noi
Let also
\beqa
\label{r2}
\eta^V_{\ell i}=\vert \e^V_{\ell i}\vert, \mbox { } V=1,\ldots, n,\, 
\ell=1,\ldots, L \mbox{ and } i=1,\ldots, 4. 
\eeqa
From eq . (\ref{r1}) and (\ref{r2}) one has
\beqa
\label{r3}
\eta^V_{\ell i} = (-1)^{i+1}\e_{\ell i}.
\eeqa
We introduce withe the ''short" $u$ and ''long" $v$ variables by 
\beqa
v_\ell&=&\frac{1}{\sqrt{2}} \sum_V \sum_i \eta^V_{\ell i} x^V_i,\nonumber\\
u_\ell&=&\frac{1}{\sqrt{2}} \sum_V \sum_i \e^V_{\ell i} x^V_i.
\eeqa
Conversely, one has
\beqa
 x^V_i= \frac{1}{\sqrt{2}}\left(\eta^V_{\ell i}v_\ell+\e^V_{\ell i}u_\ell \right).
\eeqa

From the propagator \ref{propa} and vertices
contributions \ref{v1} one is able to write the amplitude ${\cal A}_{G,{\bar
    V}}$ of the graph $G$ (with the marked root $\bar V$) in terms of
the non-commutative polynomials $HU_{G, \bar{V}}$ and $HV_{G, \bar{V}}$ as (see
\cite{gurauhypersyman} for details)
\beqa
\label{HUGV}
{\cal A}_{G,{\bar V}}  (x_e,\;  p_{\bar V}) = \left(\frac{\tom}{2^{\frac D2
      -1}}\right)^L  \int_{0}^{\infty} \prod_{\ell=1}^L  [ d t_\ell
(1-t_\ell^2)^{\frac D2 -1} ]
\frac
{e^{-  \frac {HV_{G, \bar{V}} ( t_\ell , x_e , p_{\bar v})}
{HU_{G, \bar{V}} ( t )}}}
{HU_{G, \bar{V}} ( t )^{\frac D2}},
\eeqa
with $x_e$ the external positions of the graph and 
\beqa
\label{t}
t_\ell = {\rm tanh} \frac{\alpha_\ell}{2}, \ \ell=1,\ldots, L.
\eeqa
where  $\alpha_\ell$ are the parameters associated by eq. (\ref{propa}) to
the propagators of the graph. 
In \cite{gurauhypersyman} it was proved that  $HU$ and $HV$ are polynomials in the set of variables $t$. The first polynomial is given by (see again \cite{gurauhypersyman})
\beqa
\label{huqv12}
HU_{G, \bar{V}}=({\rm det} Q)^{\frac 1D} \prod_{\ell=1}^L t_\ell \, ,
\eeqa
where
\beqa
\label{Q}
Q= A\otimes 1_D - B \otimes \sigma \, ,
\eeqa
with $A$ a diagonal matrix and $B$ an antisymmetric matrix. The matrix $A$ writes 
\beqa 
\label{GTdefmatrixa}
 A=\begin{pmatrix} S & 0 & 0\\ 0  & T & 0 \\ 0&0&0\\
\end{pmatrix} \, ,
\eeqa 
where $S$ and resp. $T$ are the two diagonal $L$ by $L$ matrices
with diagonal elements $c_\ell = \coth(\frac{\alpha_\ell}{2}) = 1/t_\ell$, 
and resp. $t_\ell$.
The last $(n-1)$ lines and columns are have $0$ entries.

The antisymmetric part $B$ is
\beqa
\label{b}
B= \begin{pmatrix}{s} E & C \\
-C^t & 0 \\
\end{pmatrix}\, , 
\eeqa
with 
\beqa
s=\frac{2}{\theta\tom}=\frac{1}{\om}\, ,
\eeqa
and 
\beqa
\label{c}
C_{\ell V}=\begin{pmatrix}
\sum_{i=1}^4(-1)^{i+1}\epsilon^V_{\ell i} \\
\sum_{i=1}^4(-1)^{i+1}\eta^V_{\ell i} \\
\end{pmatrix}\ ,
\eeqa
\beqa
\label{e}
E=\begin{pmatrix}E^{uu} & E^{uv} \\ E^{vu} & E^{vv} \\
\end{pmatrix}.
\eeqa
The blocks of the matrix $E$ are
\beqa
\label{ee}
E^{vv}_{\ell,\ell'}&=&\sum_V
\sum_{i,j=1}^4  (-1)^{i+j+1} \omega(i,j)\eta_{\ell i}^V\eta_{\ell' j}^V,
\nonumber\\
E^{uu}_{\ell,\ell'}&=&\sum_V
\sum_{i,j=1}^4  (-1)^{i+j+1} \omega(i,j)\epsilon_{\ell i}^V\epsilon_{\ell' j}^V,
\nonumber\\
E^{uv}_{\ell,\ell'}&=&\sum_V
\sum_{i,j=1}^4  (-1)^{i+j+1} \omega(i,j)\epsilon_{\ell i}^V\eta_{\ell'
  j}^V. 
\eeqa
The symbol $\omega(i,j)$ takes the values $\omega(i,j)=1$ if $i<j$, 
$\omega(i,j)=-1$ is $j<i$ and $\omega(i,j)=0$ if $i=j$.
In eq. (\ref{c}) of the matrix $C$ we have rescaled by $s$ the hypermomenta $p_V$. 
For further reference we introduce the integer entries matrix:
\beqa
\label{b'}
B'= \begin{pmatrix} E & C \\
-C^t & 0 \\
\end{pmatrix}\ .
\eeqa

In \cite{gurauhypersyman} it was proven that
\beqa
\label{qm}
{\rm det} Q= ({\rm det} M)^D \, ,
\eeqa
where
\beqa
\label{M}
M= A+B \, .
\eeqa
Thus eq. (\ref{huqv12}) becomes:
\beqa
\label{hugvq2}
HU_{G, \bar{V}}={\rm det} M \prod_{\ell=1}^L t_\ell \, .
\eeqa

Let $I$ and resp. $J$ be two subsets of $\{1,\ldots,L\}$, of cardinal $\vert I \vert$ and $\vert J \vert$. Let
\beqa
\label{kij}
k_{I,J} = \vert I\vert+\vert J\vert - L - F +1 \, ,
\eeqa
and $n_{I J}=\mathrm{Pf}(B'_{\hat{I}\hat{J}})$, the Pffafian of the  matrix
$B'$ with deleted lines and columns $I$ among the first $L$ indices 
(corresponding to short variables $u$) and $J$ among the next $L$ 
indices (corresponding to long variables $v$).

The specific form \ref{GTdefmatrixa}
allows one to write the polynomial $HU$ as a sum of positive terms:
\beqa
\label{suma}
HU_{G,{\bar V}} (t) &=&  \sum_{I,J}  s^{2g-k_{I,J}} \ n_{I,J}^2
\prod_{\ell \not\in I} t_\ell \prod_{\ell' \in J} t_{\ell'}\ .
\eeqa

In \cite{gurauhypersyman}, non-zero {\it leading terms} ({\it i. e.} terms with
the smallest global degree in the $t$ variables) were identified. 
These terms are dominant in the UV regime. 
Some of them correspond to subsets $I=\{1,\ldots, L\}$ and $J$  {\it admissible}, that is 
\begin{itemize}
\item $J$ contains a tree $\tilde {\cal T}$ in the dual graph,
\item the complement of $J$ contains a tree $\cal T$ in the direct graph.
\end{itemize}
Associated to such $I$ and $J$ one has $n_{I,J}^2=2^{2g}$.

However, the list of these leading terms, as already remarked in
\cite{gurauhypersyman}, is not exhaustive. In the next section we complete this list
with further terms, necessary for the sequel \footnote{For the purposes of \cite{gurauhypersyman}, the existence of some non-zero leading
terms was sufficient.}.

\begin{figure}[ht]
\centerline{\epsfig{figure=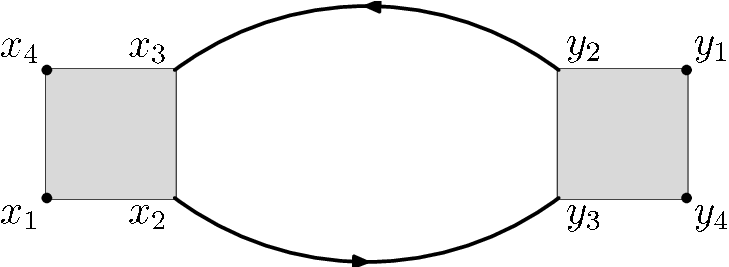,width=6cm}}
\caption{The bubble graph}\label{bubble}
\end{figure}

\begin{figure}[ht]
\centerline{\epsfig{figure=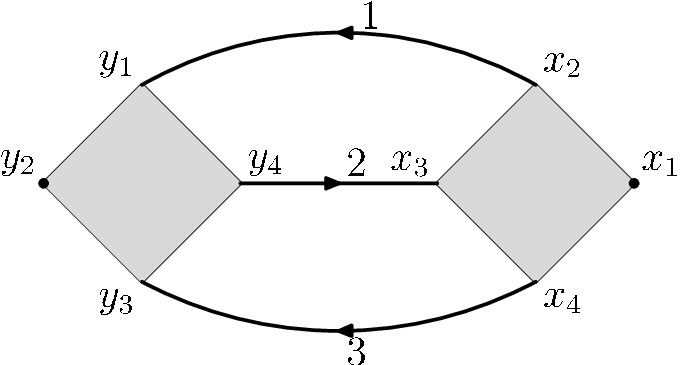,width=6cm}}
\caption{The sunshine graph}\label{sunshine}
\end{figure}

We end this section with the explicit example of the bubble and the sunshine graph (Fig. \ref{bubble} and Fig. \ref{sunshine}).

For the bubble graph one has
\beqa
\label{pol-bula}
HU_{G,\bar V}&=&(1+4s^2)(t_1+t_2+t_1^2t_2+t_1t_2^2).
\eeqa

For the sunshine graph one has
\beqa
\label{pol-sunshine}
HU_{G,\bar V}&=&\Big{[} t_1t_2+t_1t_3+t_2t_3+t_1^2t_2t_3+t_1t_2^2t_3+t_1t_2t_3^2\Big{]}
(1+4s^2)^2\nonumber\\
&&+16s^2(t_2^2+t_1^2t_3^2).
\eeqa

For further reference we also give the polynomial of the graph of Fig. \ref{bula-deformata}
\beqa
\label{bula-traficata}
HU_{G,\bar V}=(1+4s^2) (t_1 + t_2 + t_3 + t_1 t_2 t_3) (1+ t_2 t_3 + t_1 (t_2+ t_3)).
\eeqa

The polynomial $HV$ is more involved. One has
\beqa
\label{eq:HVGV}
\frac{HV_{G,\bar V}}{HU_{G,\bar V}}=\frac{\tom}{2}
\begin{pmatrix}x_e & p_{\bar V} \end{pmatrix}
PQ^{-1}P^t
\begin{pmatrix}x_e \\ p_{\bar V} \end{pmatrix},
\eeqa 
where $P$ is some matrix coupling the external positions $x_e$ and the root
hypermomenta $p_{\bar V}$ with the short $u$ and long $v$ variables and the
rest of the hypermomenta $p_V$ ($V\ne \bar V$). Explicit expressions can be found in \cite{gurauhypersyman}.

\section{Further leading terms in HU}
\label{sec:furtherLeading}
\setcounter{equation}{0}

To procede with the dimensional regularization one needs first to correctly isolate the divergent subgraphs in different Hepp sectors. Let a subgraph $S$ in the graph G. If $S$ is nonplanar the leading terms in $HU_G$ suffice to prove that $S$ is convergent in every Hepp sector as will be explained in section \ref{sec:NCdimreg}. 

This is not the case if $S$ is planar with more that one broken face. 
Let $\tilde \ell$ be the line of $G$ which breaks an internal face of $S$.

If  $\tilde \ell$ is a genus line in $G$ (see subsection \ref{grafuri}), one could still use only the leading terms in $HU_G$ to prove that $S$ is convergent.

But one still needs to prove that $S$ is convergent if the line $\tilde \ell$ is {\bf not} a genus line in $G$. This is for instance the case of the sunshine graph: in the Hepp sector $t_1< t_3 <t_2$ one must prove that the subgraph formed by the lines $l_1$ and $l_3$ is convergent. 
This is true due to the term $16 s^2 t_2^2$ in  eq. (\ref{pol-sunshine}).
Note that the variable $t_2$ is associated to the
line which breaks the internal face of the subgraph $\ell_1$, $\ell_3$.
One needs to prove that for arbitrary $G$ and $S$ we have such terms.

If we reduce $S$ to a rosette there exists a loop line $\ell_2\in S$ which 
 either  crosses $\tilde \ell$ or encompasses it. This line separates the two 
broken faces of $S$.

\begin{definition}
Let $J_0$ a subset of the internal lines of the graph $G$. $J_0$ is called
{\it pseudo-admissible} if: 
\begin{itemize}
\item its complement is the union of tree $\cal T$ in $G$ and $\ell_2$,
\item neither $\tilde \ell$ nor $\ell_2$ belong to $\cal T$,
\end{itemize}
\end{definition}

Let $I_0=\{\ell_1 \hdots \ell_L\} - \tilde \ell \equiv I- \tilde
\ell$. This implies $|I|=L(G)-1$ and $|J|=F(G)-2+2g(G)$. For the sunshine graph (see Fig. \ref{sunshine}) $I=\{ \ell_1, \ell_3 \}$ and $J=\{ \ell_2 \}$. 
One has the theorem

\begin{theorem}
\label{completare}
In the sum \ref{suma} the term associated to $I_0$ and $J_0$ above is
\beqa
n_{I_0, J_0}^2&=&4, \mbox{ if $\tilde \ell$ is a genus line, in $G$}\nonumber\\
&=& 16, \mbox{ if $\tilde \ell$ is not a genus line in $G$.}\nonumber
\eeqa
\end{theorem}
{\bf Proof:} 
The proof is similar to the one concerning the leading terms of
$HU$ given in Lemma $III.1$ of \cite{gurauhypersyman}, being however more involved. 

The matrix whose determinant we must compute is obtained from $B$ by deleting the
lines and columns corresponding to the subsets $I_0$ and $J_0$ (as explained
in the previous section).
 
The matrix $B'_{\hat I_0,\hat J_0}$ has
\begin{itemize}
\item a line and column corresponding to $u_{\tilde \ell}$, the short variable of $\tilde \ell$
\item $n$ lines and columns corresponding to the $v$ variables of the $n-1$ tree lines of $\cal T$ and the supplementary line $\ell_2$
\item $n-1$ lines and columns associated to the hypermomenta.
\end{itemize}

We represent the determinant of this matrix by the Grassmann integral:
\beqa
\label{d11}
\det B'_{\hat I_0,\hat J_0}=\int  d \bar \psi^u_{\tilde \ell} 
 d \psi^u_{\tilde \ell}
d \bar \psi^v d \psi^v
d\bar \psi^p d\psi^p
 e^{-\bar \psi B'\psi}
\eeqa
The quadratic form in the Grassman variables in the above integral is:
\beqa
\label{eq:formQuad}
&&-\sum_V \sum_{l;i,j} \bar \psi^u_{\tilde \ell} (-1)^{i+1}\omega(i,j) \e^V_{\tilde \ell i} \e^V_{\ell j} \psi^v_{\ell}
\nonumber\\
&&-\sum_V \sum_{l;i,j} \bar \psi^v_{\ell} \omega(i,j)\e^V_{\ell i} \e^V_{\tilde \ell j}(-1)^{j+1} \psi^u_{\tilde \ell}
\nonumber\\
&& +\sum_V \bar \psi^u_{\tilde \ell} (-1)^{i+1}\e^V_{\tilde \ell i} \psi^{p_V}
-\sum_V \bar \psi^{p_V} (-1)^{i+1}\e^V_{\tilde \ell i} \psi^u_{\tilde \ell}
\nonumber\\
&&-\sum_{V} \sum_{\ell,\ell'; i,j} \bar \psi^v_{\ell} \omega(i,j) \e^V_{\ell i} \e^V_{\ell' j} \psi^v_{\ell'}
\nonumber\\
&&+ \sum_V \sum_{\ell; i} \bar \psi^v_{\ell} \e^V_{\ell i} \psi^{p_V} - \sum_V \sum_{\ell; i} \bar \psi^{p_V} \e^V_{\ell i} \psi^v_{\ell} \, .
\eeqa

We implement the first Filk move as a Grasmann change of variales. At each step we reduce a tree line $\ell_1=(i,j)$ connecting the root vertex $\bar{V}_G$ to a normal vertex $V$ and gluing the two vertices. This is achieved by performing a change of variables for the hypermomenta and reinterpreting the quadratic form in the new variables as corresponding to a new vertex $\bar V'_G$: the quadratic form essentially reproduces itself under the change of variables!

Take $\ell_1=(i,j)$ a line connecting the ''root" vertex $\bar V_G$ to a vertex $V$. We make the change of variables
\beqa
\label{eq:magic1}
\psi^{p_V}&=&\chi^{p_V}+\sum_{\ell' \neq \ell_1}\sum_{k}
\left(
-\omega(i,k)\e^{\bar V_G}_{\ell' k}+\omega(j,k)\e^{V}_{\ell' k}
\right)\psi^v_{\ell'} 
\nonumber\\
&&+\sum_{k}\left(
-\omega(i,k)\e^{\bar V_G}_{\tilde \ell k}+\omega(j,k)\e^{V}_{\tilde \ell k}
\right) (-1)^{k+1}\psi^u_{\tilde \ell}
\nonumber\\
\bar \psi^{p_V}&=&\bar \chi^{p_V}+\sum_{\ell' \neq \ell_1}\sum_{k}
\left(
-\omega(i,k)\e^{\bar V_G}_{\ell' k}+\omega(j,k)\e^{V}_{\ell' k}
\right)\bar \psi^v_{\ell'}
\nonumber\\
&&+\sum_{k}\left(
-\omega(i,k)\e^{\bar V_G}_{\tilde \ell k}+\omega(j,k)\e^{V}_{\tilde \ell k}
\right) (-1)^{k+1} \bar \psi^u_{\tilde \ell} \,,
\eeqa
where the first sum is performed on the internal lines of $G$ (note that because of
the presence of the incidences matrices $\e$ this sum reduces to a sum on the
lines hooked to the two vertices $\bar{V}_G$ and $V$). 
The corners $i$ and resp. $j$ are the corners where the tree line $\ell_1$ hooks to the vertex $\bar V_G$ and resp. $V$.  

\vspace*{\stretch{1}}

At each step, let us consider the coupling between the variables associated to the line $\ell_1$ and to the hypermomentum $p_V$ and the rest of the variables.
Using eq. (\ref{eq:formQuad}) one  has
\beqa
\label{eq:magarie30}
&&-\bar \psi^u_{\tilde \ell}\sum_{p}(-1)^{p+1}
\left( \omega(p,i) \e^{V_G}_{\tilde \ell p} \e^{V_G}_{\ell_1 i} +
\omega(p,j)\e^{V}_{\tilde \ell p} \e^{V}_{\ell_1 j} \right)
\psi^v_{\ell_1}
\nonumber\\
&&-\bar \psi^v_{\ell_1} \sum_{p}
\left(\omega(i,p) \e^{V_G}_{\tilde \ell p} \e^{V_G}_{\ell_1 i} +
\omega(j,p)\e^{V}_{\tilde \ell p} \e^{V}_{\ell_1 j}
\right) \psi^u_{\tilde \ell}(-1)^{p+1}
\nonumber\\
&&+\sum_{V; p}\bar \psi^u_{\tilde \ell}(-1)^{p+1}\e^V_{\tilde \ell p} \psi^{p_V}
-\sum_{V; p}\bar \psi^{p_V} (-1)^{p+1}\e^V_{\tilde \ell p} \psi^u_{\tilde \ell}
\nonumber\\
&&-\bar \psi^v_{\ell_1}
\sum_{\ell'\neq \ell_1;k} 
\left(
\omega(i,k) \e^{\bar V_G}_{\ell_1 i} \e^{\bar V_G }_{\ell' k} 
+\omega(j,k) \e^{V}_{\ell_1 j} \e^{V}_{\ell' k}
\right)
\psi^v_{\ell'}
\nonumber\\
&&-\sum_{\ell'\neq \ell_1; k} \bar \psi^v_{\ell'} 
\left( \omega(k,i) \e^{\bar V_G}_{\ell' k} \e^{\bar V_G}_{\ell_1 i}
+ \omega(k,j) \e^{V}_{\ell' k} \e^{V}_{\ell_1 j}
\right) \psi^v_{\ell_1}
\nonumber\\
&&+\bar\psi^v_{\ell_1} \e^V_{\ell_1 j} \psi^{p_V}
+\sum_{\ell' \neq \ell_1; k} \bar \psi^v_{\ell'} \e^V_{\ell' k} \psi^{p_V}
\nonumber\\
&&-\bar\psi^{p_V} \e^V_{\ell_1 j} \psi^v_{\ell} -\sum_{\ell' \neq \ell_1; k} \bar \psi^{p_V} \e^V_{\ell' k} \psi^v_{\ell'} \, ,
\eeqa
which rewrites as
\beqa
\label{magarie2233}
&&-\Big{[}
\bar \psi^u_{\tilde \ell}\sum_{p}(-1)^{p+1}
\left( \omega(p,i) \e^{V_G}_{\tilde \ell p} \e^{V_G}_{\ell_1 i} +
\omega(p,j)\e^{V}_{\tilde \ell p} \e^{V}_{\ell_1 j} \right)
\nonumber\\
&&+\sum_{\ell'\neq \ell_1; k} \bar \psi^v_{\ell'} 
\left( \omega(k,i) \e^{\bar V_G}_{\ell' k} \e^{\bar V_G}_{\ell_1 i}
+ \omega(k,j) \e^{V}_{\ell' k} \e^{V}_{\ell_1 j}\right)
+\bar\psi^{p_V} \e^V_{\ell_1 j}
\Big{]}
\psi^v_{\ell_1}
\nonumber\\
&&-\bar \psi^v_{\ell_1} 
\Big{[}
\sum_{p}
\left(\omega(i,p) \e^{V_G}_{\tilde \ell p} \e^{V_G}_{\ell_1 i} +
\omega(j,p)\e^{V}_{\tilde \ell p} \e^{V}_{\ell_1 j}
\right) \psi^u_{\tilde \ell}(-1)^{p+1}
\nonumber\\
&&+\sum_{\ell'\neq \ell_1;k} 
\left(
\omega(i,k) \e^{\bar V_G}_{\ell_1 i} \e^{\bar V_G }_{\ell' k} 
+\omega(j,k) \e^{V}_{\ell_1 j} \e^{V}_{\ell' k}
\right)
\psi^v_{\ell'}
-\e^V_{\ell_1 j} \psi^{p_V}
\Big{]}
\nonumber\\
&&+\sum_{V; p}\bar \psi^u_{\tilde \ell}(-1)^{p+1}\e^V_{\tilde \ell p} \psi^{p_V}
-\sum_{V; p}\bar \psi^{p_V} (-1)^{p+1}\e^V_{\tilde \ell p} \psi^u_{\tilde \ell}
\nonumber\\
&&+\sum_{\ell' \neq \ell_1; k} \bar \psi^v_{\ell'} \e^V_{\ell' k} \psi^{p_V}
 -\sum_{\ell' \neq \ell_1; k} \bar \psi^{p_V} \e^V_{\ell' k} \psi^v_{\ell'} \, .
\eeqa
Performing now in \ref{eq:magarie30} the change of variable \ref{eq:magic1} 
for the hypermomentum $p_V$ of the vertex $V$ associated to the tree line $\ell_1$ of $G$ and taking into account that $\e_{\ell_1 i}=-\e_{\ell_1 j}$ the first two lines of \ref{magarie2233} are simply
\beqa
\label{eq:micamaga}
-\bar \chi^{p_V} \e^V_{\ell_1 j} \psi^v_{\ell_1}+
\bar \psi^v_{\ell_1} \e^V_{\ell_1 j} \chi^{p_V} \, .
\eeqa

\vspace*{\stretch{1}}

As $\psi^v_{\ell_1}$ and $\bar \psi^v_{\ell_1}$ do not appear anymore in the rest of the terms we are forced to pair them with $\bar \chi^{p_V}$ and $\chi^{p_V}$. The rest of the terms in the quadratic form are:
\beqa
\label{eq:magaroi}
&&+\sum_{V; p}\bar \psi^u_{\tilde \ell}(-1)^{p+1}\e^V_{\tilde \ell p} 
\Big{[}
\sum_{\ell' \neq \ell_1}\sum_{k}
\left(
-\omega(i,k)\e^{\bar V_G}_{\ell' k}+\omega(j,k)\e^{V}_{\ell' k}
\right)\psi^v_{\ell'} 
\nonumber\\
&&+\sum_{k}\left(
-\omega(i,k)\e^{\bar V_G}_{\tilde \ell k}+\omega(j,k)\e^{V}_{\tilde \ell k}
\right) (-1)^{k+1}\psi^u_{\tilde \ell}
\Big{]}
\nonumber\\
&&+\sum_{\ell' \neq \ell_1; p} \bar \psi^v_{\ell'} \e^V_{\ell' p} 
\Big{[}
\sum_{\ell' \neq \ell_1}\sum_{k}
\left(
-\omega(i,k)\e^{\bar V_G}_{\ell' k}+\omega(j,k)\e^{V}_{\ell' k}
\right)\psi^v_{\ell'} 
\nonumber\\
&&+\sum_{k}\left(
-\omega(i,k)\e^{\bar V_G}_{\tilde \ell k}+\omega(j,k)\e^{V}_{\tilde \ell k}
\right) (-1)^{k+1}\psi^u_{\tilde \ell}
\Big{]}
\nonumber\\
&&-\sum_{V; p}
\Big{[}
\sum_{\ell' \neq \ell_1}\sum_{k}
\left(
-\omega(i,k)\e^{\bar V_G}_{\ell' k}+\omega(j,k)\e^{V}_{\ell' k}
\right)\bar \psi^v_{\ell'}
\nonumber\\
&&+\sum_{k}\left(
-\omega(i,k)\e^{\bar V_G}_{\tilde \ell k}+\omega(j,k)\e^{V}_{\tilde \ell k}
\right) (-1)^{k+1} \bar \psi^u_{\tilde \ell}
\Big{]}
(-1)^{p+1}\e^V_{\tilde \ell p} \psi^u_{\tilde \ell}
\nonumber\\
&& -
\Big{[}
\sum_{\ell' \neq \ell_1}\sum_{k}
\left(
-\omega(i,k)\e^{\bar V_G}_{\ell' k}+\omega(j,k)\e^{V}_{\ell' k}
\right)\bar \psi^v_{\ell'}
\nonumber\\
&&+\sum_{k}\left(
-\omega(i,k)\e^{\bar V_G}_{\tilde \ell k}+\omega(j,k)\e^{V}_{\tilde \ell k}
\right) (-1)^{k+1} \bar \psi^u_{\tilde \ell}
\Big{]}
\sum_{\ell' \neq \ell_1; p}
\e^V_{\ell' p} \psi^v_{\ell'}\; .
\eeqa

We analyse the different terms in the above equation.
The term $\bar \psi^u_{\tilde \ell} \psi^u_{\tilde \ell}$ is:
\beqa
&&\sum_{p,k}\e^V_{\tilde\ell p} (-1)^{p+1}\Big{[}
- \omega(i,k) \e^{\bar V_G}_{\tilde \ell k} + \omega(j,k) \e^{V}_{\tilde \ell k} 
\Big{]}(-1)^{k+1}
\nonumber\\
&&-\sum_{p,k}\Big{[}
- \omega(i,k) \e^{\bar V_G}_{\tilde \ell k} + \omega(j,k) \e^{V}_{\tilde \ell k} 
\Big{]}(-1)^{k+1}\e^V_{\tilde\ell p} (-1)^{p+1}=0 \, .
\eeqa

The term in $\bar \psi^u_{\tilde \ell} \psi^v_{\ell'}$ is given by:
\beqa
&&+\sum_{p}\bar \psi^u_{\tilde \ell}(-1)^{p+1}\e^V_{\tilde \ell p} 
\sum_{\ell' \neq \ell_1}\sum_{k}
\left(
-\omega(i,k)\e^{\bar V_G}_{\ell' k}+\omega(j,k)\e^{V}_{\ell' k}
\right)\psi^v_{\ell'}
\nonumber\\ 
&& -
\sum_{p}\left(
-\omega(i,p)\e^{\bar V_G}_{\tilde \ell p}+\omega(j,p)\e^{V}_{\tilde \ell p}
\right) (-1)^{p+1} \bar \psi^u_{\tilde \ell}
\sum_{\ell' \neq \ell_1; k}
\e^V_{\ell' k} \psi^v_{\ell'} \, .
\eeqa
Setting $j$ to be either the first or the last halfline on the vertex $V$ we see that the last terms in the two lines above cancel eachother. The first two terms hold:
\beqa
\sum_{p} \sum_{\ell' \neq \ell_1} \sum_{k} \bar \psi^u_{\tilde \ell} \psi^v_{\ell'}
(-1)^{p+1}\Big{[}
-\omega(i,k)\e^{\bar V_G}_{\ell' k}\e^V_{\tilde \ell p}
+\omega(i,p)\e^{\bar V_G}_{\tilde \ell p}\e^V_{\ell' k}
\Big{]} \, .
\eeqa
This can be rewritten in the form:
\beqa
-\sum_{p} \sum_{\ell' \neq \ell_1} \sum_{k} \bar \psi^u_{\tilde \ell} \psi^v_{\ell'}
(-1)^{p+1} \omega(p,k) \e_{\tilde \ell p}^{\bar V_G'}\e_{\ell' k}^{\bar V_G'} \, ,
\eeqa
where $\bar V_G'$ is a new root vertex in which the vertex $V$ has been glued to the vertex $\bar V_G$ and the halflines on the vertex $V$ have been inserted on the new vertex at the place of the halfline $i$.

The coupling between $\psi^v$'s with $\psi^v$'is
\beqa
\label{eq:maga333}
&+&\sum_{\ell\neq\ell_1,k}\bar \psi^v_{\ell} \e^{V}_{\ell k} 
 \sum_{\ell'}\left(-\sum_p \e^{\bar{V}_G}_{\ell' p} \omega(i,p)
+\sum_{k'=1}^4 \e^{V}_{\ell' k'} \omega(j,k')\right) \psi^v_{\ell'}
\nonumber\\
&-&\sum_{\ell' \neq \ell_1} \left( - \sum_p \e^{\bar{V}_G}_{\ell' p} \omega(i,p) + \sum_{k=1}^4 \e^{V}_{\ell' k} \omega(j,k)\right) \bar{\psi}^v_{\ell'}
 \sum_{\ell\neq\ell_1;k'}\e^{V}_{\ell k'}\psi^v_{\ell} \ .
\eeqa

Again, as $j$ is either the first or the last halfline on the vertex $V$, the last two terms in the two lines of eq. (\ref{eq:maga333}) cancel each other. The rest of the terms give exactly the contacts between the $\bar\psi^v$ and $\psi^v$ on a new vertex $\bar V_G'$ obtained by gluing $V$ on $\bar{V}_G$
This is the first Filk move on the line $\ell_1$ and its associated vertex $V$. 
One iterates now this mechanism for the rest of the tree lines of $G$. 
Hence we reduce the graph $G$ to a rosette (see subsection \ref{grafuri}). 

The quadratic form writes finally as
\beqa
&&-\sum_{\ell'; p,k} \bar \psi^u_{\tilde \ell} \psi^v_{\ell'}
(-1)^{p+1} \omega(p,k) \e_{\tilde \ell p}^{\bar V_G}\e_{\ell' k}^{\bar V_G}
-\sum_{\ell'; p,k }
\bar\psi^v_{\ell'} \psi^u_{\tilde \ell}(-1)^{p+1}
\omega(k,p) \e_{\tilde \ell p}^{\bar V_G}\e_{\ell' k}^{\bar V_G}
\nonumber\\
&&-\sum_{\ell, \ell';j,k} \bar \psi^v_\ell \e^{\bar {V}_G}_{\ell j} \e^{\bar {V}_G} _{\ell' k} \omega (j,k) \psi^v_{\ell'} \, .
\eeqa
The sum concerns only the rosette vertex $\bar {V}_G$. Therefore the last line is $0$, as by the first Filk move we have exhausted all the tree lines of $\cal T$.

As $\tilde \ell$ breaks the face separated from the external face by $\ell_2$ we have  
$k_1< p < k_2$. By a direct inspection of the terms above, one obtains the
requested result.

\qed

\section{Factorization of the Feynman amplitudes}
\label{sec:factorisation}
\setcounter{equation}{0}

Take now $S$ to be a primitively divergent subgraph.
We now prove that the Feynman amplitude ${\cal A}_G$ factorizes into two
parts, one corresponding to the primitive divergent subgraph $S$ and the other to the graph $G/S$ (defined in section \ref{grafuri}). This is needed in order to prove that divergencies are cured by Moyal counterterms
\footnote{
 It is also the property needed for the definition of a coproduct $\Delta$ for a Connes-Kreimer Hopf algebra structure (see \cite{CK, kreimer}). Details of this construction are given elsewhere \cite{progres}.}.

In section \ref{sec:NCdimreg} we will prove that only the planar ($g=0$), one broken face $B=1$, $N=2$ or $N=4$ external legs subgraphs are primitively divergent. We now deal only with such subgraphs.

\subsection{Factorization of the polynomial $HU$}

We denote all the leading terms in the first polynomial associated to a graph $S$ by $HU^{l}_S$. If we rescale all the $t_{\ell}$ parameters corresponding to a subgraph $S$ by $\rho^2$, $HU_G$ becomes a polynomial in $\rho$. We denote the terms of minimal degree in $\rho$ in this polynomial by $HU_G^{l( \rho)}$. It is easy to see that for the subgraph $S$ we have 
$HU^{l(\rho)}_{S}=\rho^{2[L(S)-n(S)+1]}HU^{l}_{S}\vert_{\rho=1}$.
We have the following theorem
\begin{theorem}
\label{mare}
Under the rescaling
\beqa
\label{rescalare}
t_{\alpha}\mapsto \rho^2 t_{\alpha}
\eeqa
of the parameter corresponding to a divergent subgraph $S$ of any Feynman
graph $G$, the following factorization property holds
\beqa
HU^{l(\rho)}_{G,\bar V}=HU^{l(\rho)}_{S,\bar V_S} HU_{G/S,\bar V} \; .
\eeqa
\end{theorem}
{\bf Proof:} \\ 
In the matrix $M$ defined in eq. (\ref{M}) (corresponding to the graph $G$) we can rearrange the lines and columns so that we  place the matrix $M_S$ (corresponding to the subgraph $S$) into the upper left corner.
We place the line (and resp. the column) associated to the hypermomentum
of the root vertex of $S$ to be the last line (and resp. column) of $M$ 
(without loss of generality we consider that the root of the subgraph $S$ is not
the root of $G$). $M$ takes the form
\beqa
\label{M1}
\hskip -.2cm
\begin{pmatrix} E^{u^Su^S} & E^{u^Sv^S} & C^{u^Sp^S}  & E^{u^Su^{G-S}}&
  E^{u^Sv^{G-S}} & C^{u^Sp^{G-S}} \\
 E^{v^Su^S} & E^{v^Sv^S} & C^{v^Sp^S}  & E^{v^Su^{G-S}}&
  E^{v^Sv^{G-S}} & C^{v^Sp^{G-S}} \\
 C^{p^Su^S} & C^{p^Sv^S} & 0  & C^{p^Su^{G-S}} &
  C^{p^Sv^{G-S}} & 0\\
 E^{u^{G-S}u^S} & E^{u^{G-S}v^S} & C^{u^{G-S}p^S}  & E^{u^{G-S}u^{G-S}}&
  E^{u^{G-S}v^{G-S}} & C^{u^{G-S}p^{G-S}}\\
 E^{v^{G-S}u^S} & E^{v^{G-S}v^S} & C^{v^{G-S}p^S}  & E^{v^{G-S}u^{G-S}}&
  E^{v^{G-S}v^{G-S}} & C^{v^{G-S}p^{G-S}}\\
 C^{p^{G-S}u^S} & C^{p^{G-S}v^S} & 0  & C^{p^{G-S}u^{G-S}}&
  C^{p^{G-S}v^{G-S}} & 0
\end{pmatrix} \nonumber
\eeqa
where we have denoted by $E^{u^Su^S}$ a  coupling between two short
variables corresponding to internal lines of $S$ {\it etc.}.

{\bf I.} 
We first write the determinant of the matrix above under the form of a Grassmannian
integral 
\beqa
\label{d1}
\det M=\int  d \bar \psi^u d \psi^u  d \bar \psi^v d \psi^v d\bar \psi^p d\psi^p
 e^{- \bar \psi M \psi} \, .
\eeqa

Denote a generic line of the subgraph $S$ by $\ell_S$ and a generig line of the subgraph $G-S$ by $\ell_{G-S}$.

We perform a Grassmann change of variables of Jacobian $1$. The value of the integral (\ref{d1}) does not change under this change of variables. We will prove that the following properties hold for the different terms in the Grassmanian quadratic form
\beqa
\label{eq:scop}
&&E'^{v^Sv^S}=\mathrm{diag}(t_l), \quad E'^{v^Su^{G-S}}=0,
\quad E'^{v^Sv^{G-S}}=0\nonumber\\
&&E'^{u^{G-S}u^{G-S}}=E'^{u^{G/S}u^{G/S}}, \quad E'^{u^{G-S}v^{G-S}}=E'^{u^{G/S}v^{G/S}} \nonumber\\
&&E'^{v^{G-S}v^{G-S}}=E'^{v^{G/S}v^{G/S}}.
\eeqa
The new matrix of the quadratic form $M'$will now be 
\beqa
\label{M2}
\hskip -.2cm
\begin{pmatrix} E'^{u^Su^S} & E'^{u^Sv^S} & C^{u^Sp^S}  & E'^{u^Su^{G-S}}&
  E'^{u^Sv^{G-S}} & C^{u^Sp^{G-S}} \\
 E'^{v^Su^S} & E'^{v^Sv^S} & C^{v^Sp^S}  & 0 &
  0  & C^{v^Sp^{G-S}} \\
 C^{p^Su^S} & C^{p^Sv^S} & 0  & C^{p^Su^{G-S}} &
  C^{p^Sv^{G-S}} & 0\\
 E'^{u^{G-S}u^S} & 0 & C^{u^{G-S}p^S}  & E'^{u^{G/S}u^{G/S}}&
  E'^{u^{G/S}v^{G/S}} & C^{u^{G-S}p^{G-S}}\\
 E'^{v^{G-S}u^S} & 0  & C^{v^{G-S}p^S}  & E'^{v^{G/S}u^{G/S}}&
  E'^{v^{G/S}v^{G/S}} & C^{v^{G-S}p^{G-S}}\\
 C^{p^{G-S}u^S} & C^{p^{G-S}v^S} & 0  & C^{p^{G-S}u^{G-S}}&
  C^{p^{G-S}v^{G-S}} & 0
\end{pmatrix} \nonumber
\eeqa

The first part of the proof follows that of section \ref{sec:furtherLeading}, where we replace 
\beqa
\label{eq:improuvement}
 \psi^u_{\tilde \ell} \rightarrow \sum_{\ell'\in G-S} \psi^u_{\ell'}
 \qquad
  \bar \psi^{u}_{\tilde \ell} \rightarrow \sum_{\ell'\in G-S} \bar \psi^{u}_{\ell'} \, .
\eeqa
Thus the appropriate change of variables is now
\beqa
\label{eq:magic1'}
\psi^{p_V}&=&\chi^{p_V}+\sum_{\ell' \neq \ell_1; k}
\left(
-\omega(i,k)\e^{\bar V_S}_{\ell' k}+\omega(j,k)\e^{V}_{\ell'; k}
\right)\psi^v_{\ell'} 
\nonumber\\
&&+\sum_{\genfrac{}{}{0pt}{}{\ell'; k }{\ell'\in G-S}}\left(
-\omega(i,k)\e^{\bar V_S}_{\ell' k}+\omega(j,k)\e^{V}_{ \ell' k}
\right) (-1)^{k+1}\psi^u_{\ell'}
\nonumber\\
\bar \psi^{p_V}&=&\bar \chi^{p_V}+\sum_{\ell' \neq \ell_1; k}
\left(
-\omega(i,k)\e^{\bar V_S}_{\ell' k}+\omega(j,k)\e^{V}_{\ell' k}
\right)\bar \psi^v_{\ell'}
\nonumber\\
&&+\sum_{\genfrac{}{}{0pt}{}{\ell'; k }{\ell'\in G-S}}\left(
-\omega(i,k) \e^{\bar V_S}_{ \ell' k} + \omega(j,k)\e^{V}_{ \ell' k}
\right) (-1)^{k+1} \bar \psi^u_{ \ell'} \; .
\eeqa

We emphasize that this Grassmann change of variables can be viewed as forming appropriate linear combinations of lines and columns. As we only use lines and columns associated to hypermomenta $p_S$, this manipulations can not change the value of the determinant in the upper left corner: it will allways correspond precisely to the first polynomial of the subgraph $S$.

The relevant terms in the quadratic form are those of  eq. (\ref{eq:magarie30}) 
and eq. (\ref{magarie2233}), with the substitutions (\ref{eq:improuvement}).
Again, after the change of variables \ref{eq:magic1'} the only surviving contacts of $\psi^v_{\ell_1}$ and $\bar \psi^v_{\ell_1}$ are given by 
eq. (\ref{eq:micamaga}).
Finally, the remaining terms are given by  eq. (\ref{eq:magaroi}) with the substitutions (\ref{eq:improuvement}).

One needs again to analyse the different terms in this equation.
The quadratic term in $\bar\psi^u \psi^u$ is:
\beqa
&&\sum_{\ell',\ell''\in G-S}\bar \psi ^u_{\ell'}\psi ^u_{\ell''}(-1)^{k+p+1}
\Big{(}
\omega(i,k)\e^{V}_{\ell' p}\e^{\bar V_S}_{\ell'' k}-
\omega(j,k)\e^{V}_{\ell' p}\e^{\bar V}_{\ell'' k}
\nonumber\\
&&-\omega(i,k)\e^{\bar V_S}_{\ell' k}\e^{V}_{\ell'' p}
+\omega(j,k)\e^{V}_{\ell' k}\e^{V}_{\ell'' p}
\Big{)}\, .
\eeqa
As $j$ is either the first or the last halfline on the vertex $V$, we see that the last terms in the two lines above cancel. The remaining two terms give the contacts amongst $\bar\psi^u$ and $\psi^u$ on the rosette new vertex $ \bar V'_S$, obtained by gluing $\bar V_S$ and $V$ respecting the ordering.

The contacts between $\bar\psi^u$ and $\psi^v$ become
\beqa
&&+\sum_{\ell'}\sum_{p}\bar \psi^u_{\ell'}(-1)^{p+1}\e^V_{\ell' p} 
\sum_{\ell'' \neq \ell_1}\sum_{k}
\left(
-\omega(i,k)\e^{\bar V_G}_{\ell'' k}+\omega(j,k)\e^{V}_{\ell'' k}
\right)\psi^v_{\ell''}
\nonumber\\ 
&& -
\sum_{\ell'}\sum_{p}\left(
-\omega(i,p)\e^{\bar V_G}_{ \ell' p}+\omega(j,p)\e^{V}_{ \ell' p}
\right) (-1)^{p+1} \bar \psi^u_{\ell'}
\sum_{\ell'' \neq \ell_1; k}
\e^V_{\ell'' k} \psi^v_{\ell''} \, . \nonumber
\eeqa 
Again that the last terms in the two lines above cancel. Rearanging the rest as before we end recover again the terms corresponding to a rosette.

Finally for the $\bar\psi^v$ and $\psi^v$ contacts eq .(\ref{eq:maga333})  goes through.

We iterate the change of variables {\bf only} for a tree in the subgraph $S$, hence we reduce the subgraph $S$ to a rosette (see subsection \ref{grafuri}). As the quadratic form reproduced itself, the root vertex $\bar V_S$ is now a Moyal vertex, with either $2$ or $4$ external legs.

Let $r$ be an external half line of the subgraph $S$. As $S$ is planar one broken face any line $l_S=(p,q)$ will either have
$r<p,q$ or $p,q<r$. Thus
\beqa
E'^{v^Sv^{G-S}}=\sum_{l,p,r}\omega(p,r)\e^{\bar V_S}_{\ell_S p}\e^{\bar V_S}_{ \ell_{G-S} r}=0 \,.
\eeqa
The reader can check by similar computations that eq. (\ref{eq:scop}) holds. 

\medskip

\noi{\bf II.} To obtain in the  lower right corner the matrix corresponding
to the graph $G/S$, we just have to add the lines and
columns of the hypermomenta corresponding to the vertices of $S$ to the ones
corresponding to the root of $S$. Furthermore, by performing this operation,
the block $ C^{v^Sp^{G-S}}$ (which had only one non-trivial column, the column
corresponding to the hypermomentum $p_{\bar V_S}$) becomes identically $0$. Forgetting the primes, the matrix in the quadratic becomes:
\beqa
\label{M3}
\begin{pmatrix} E^{u^Su^S} & E^{u^Sv^S} & C^{u^Sp^S}  & E^{u^Su^{G-S}}&
  E^{u^Sv^{G-S}} & C^{u^Sp^{G-S}} \\
 E^{v^Su^S} & t_{\ell} \delta_{v^Sv^S} &  C^{v^Sp^S} & 0 &
  0  & 0\\
 C^{p^Su^S} & C^{p^Sv^S} & 0  & C^{p^Su^{G-S}} &
  C^{p^Sv^{G-S}} & 0\\
 E^{u^{G-S}u^S} & 0 & C^{u^{G-S}p^S}  & E^{u^{G/S}u^{G/S}}&
  E^{u^{G/S}v^{G/S}} & C^{u^{G/S}p^{G/S}}\\
 E^{v^{G-S}u^S} & 0  & C^{v^{G-S}p^S}  & E^{v^{G/S}u^{G/S}}&
  E^{v^{G/S}v^{G/S}} & C^{v^{G/S}p^{G/S}}\\
 C^{p^{G-S}u^S} & 0 & 0  & C^{p^{G/S}u^{G/S}}&
  C^{p^{G/S}v^{G/S}} & 0
\end{pmatrix}. \nonumber
\eeqa
This is equivalent to the Grassmannian change of variables:
\beqa
\chi'^{p^S}&=&\chi^{p^S}+\psi^{p_{\bar V_S}}\, ,\nonumber\\
\bar \chi'^{p^S}&=&\bar \chi^{p^S}+\bar \psi^{p_{\bar V_S}}\, .
\eeqa

\medskip

\noi
{\bf III.} We finally proceed with the rescaling 
 with $\rho^2$ of all the parameter $t_\alpha$ coresponding
to the divergent subgraph $S$ (see \ref{rescalare}). Recall that these parameters are present as
$\frac{1}{t_\alpha}$ on the diagonal of the block $E^{u^Su^S}$ and as
$t_\alpha$  on the diagonal of the block $E^{v^Sv^S}$.
 
We factorize $\frac{1}{\rho}$ on the first $L(S)$ lines of
columns of $M$ (corresponding to the $u^S$ variables) and $\rho$ on the next $L(S)$ lines and columns (coresponding to the $v^S$ variables). 
We also factorize $\frac{1}{\rho}$ on the $n(S)-1$ lines and columns
corresponding to the hypermomenta $p^S$. This is given by the Grassmann change of variables
\beqa
\label{ultima}
\psi'^u&=&\frac{1}{\rho}\psi^u\nonumber\\
\psi'^v&=&{\rho}\psi^v\nonumber\\
\chi'^p&=&\frac{1}{\rho}\chi^p\, .
\eeqa

In the new variables the matrix of the quadratic form is
\beqa
\label{M4}
\hskip -.2cm
\begin{pmatrix} c_l\delta_{uu'}+\rho^2 E'^{uu'} & E'^{uv} & \rho^2 C^{up}  & \rho E'^{uu}&
 \rho E'^{uv} & \rho C'^{up} \\
 E'^{vu} & t_l\delta_{vv'} & C^{vp}  & 0 & 0  & 0 \\
 \rho ^2 C^{pu} & C^{pv} & 0  & \rho C^{pu} & \rho C^{pv} & 0\\
 \rho E'^{uu} & 0 & \rho C^{up}  & E'^{u^{G/S}u^{G/S}}&
  E'^{u^{G/S}v^{G/S}} & C'^{u^{G/S}p^{G/S}}\\
 \rho E'^{vu} & 0  & \rho C^{vp}  & E'^{v^{G/S}u^{G/S}}&
  E'^{v^{G/S}v^{G/S}} & C'^{v^{G/S}p^{G/S}}\\
 \rho C'^{pu} & 0 & 0  & C'^{p^{G/S}u^{G/S}}& C'^{p^{G/S}v^{G/S}} & 0
\end{pmatrix} \nonumber
\eeqa

The determinant of the original matrix is obtained by multiplying the overall factor $\rho^{-2n(S)+2}$ (coming from the Jacobian of the change of variables) with the determinant of the matrix (\ref{M4}). To obtain $HU_G$, according to eq. (\ref{hugvq2}) we must multiply this determinant by a product over all lines of $t_{\ell}$.

The determinant upper left corner, coresponding to the subgraph $S$, multiplied by the appropriate product of $t_{\ell}$ as in eq. (\ref{hugvq2}) and by the Jacobian factor holds the complete polynomial $HU_S$. At leading order in $\rho$ it is $HU_S^{l(\rho)}$ 

The determinant of the lower right corner, multiplied by its corresponding product of $t_{\ell}$ holds the complete polynomial $HU_{G/S}$ and no factor $\rho$.

At the leading order in $\rho$ the off diagonal blocks become $0$. Therefore we have
\beqa
  HU^{l(\rho)}_G=HU^{l(\rho)}_S HU_{G/S} \, .
\eeqa

\qed

Let us illustrate all this with the example of the graph of Fig. \ref{hyper}, where the primitive divergent subgraph is taken to be the sunshine graph of lines
$\ell_4,\ell_5$ and $\ell_6$. A direct computation showed that, under the
rescaling
\beqa
t_4\to \rho^2 t_4,\ t_5\to \rho^2 t_5,\ t_6\to \rho^2 t_6,\
\eeqa
the leading terms in $\rho$ of the polynomial $HU_G$ factorize as
\beqa
\rho^{4}\left( (1+4s^2) t_4 (t_5+t_6) +t_5 (t_6+8s^2(2 t_5 +t_6+ 2 s^2
  t_6)\right)\nonumber\\
(1+4s^2) (t_1 + t_2 + t_3 + t_1 t_2 t_3) (1+ t_2 t_3 + t_1 (t_2+ t_3)).
\eeqa
The first line of this,formula corresponds to the leading terms under the
rescaling with $\rho$ of $HU_S$, while the second line is nothing but the
polynomial $HU_{G/S}$ of eq. (\ref{bula-traficata}).

\subsection{The exponential part of the Feynman amplitude}

In order to perform the appropriate subtractions we need to check the factorization also at the level of the second polynomial.
Throughout this section we suppose that $S$ is a completely internal subgraph, that is none of its external points is an external point of $G$. The general case is treated by the same methods with only slight modifications.
We have the following lemma
\begin{proposition}
   Under the rescaling $t_{\alpha}\mapsto \rho^2 t_{\alpha}$ of all the lines of the subgraph $S$ we have:
   \beqa
    \frac{HV_G}{HU_G}\Big{\vert}_{\rho =0}=\frac{HV_{G/S}}{HU_{G/S}}\, .
   \eeqa
\end{proposition}
{\bf Proof:} The ratio $\frac{HV_G}{HU_G}$ is given by the inverse matrix $Q^{-1}$ due to eq. (\ref{eq:HVGV}) which in turn is given by $M^{-1}$ (see \cite{gurauhypersyman} for the exact relation). Thus any property which holds for $M^{-1}$ will also hold for $Q^{-1}$.

We write the matrix elements of the inverse of $M$ with the help of Grassmann variables
\beqa
(M^{-1})_{ij}=\frac{\int d\bar \psi d\psi \psi_i \bar \psi_j e^{-\bar \psi M \psi}}{\int d\bar \psi d\psi  e^{-\bar \psi M \psi}}
\, .
\eeqa

As $S$ is a completely internal subgraph we only must analyse the inverse matrix entries
\beqa
(M^{-1})_{G-S G-S}=\frac{\int d\bar \psi d\psi \psi_{G-S} \bar \psi_{G-S} e^{-\bar \psi M \psi}}{\int d\bar \psi d\psi  e^{-\bar \psi M \psi}} \, ,
\eeqa
the only ones which intervene in the quadratic form due to the matrix $P$ in \ref{eq:HVGV}.
None of the changes of variables of the previous section involve any Grassmann variable associated with the $G-S$ sector. We conclude that
\beqa 
(M^{-1})_{G-S G-S}=(M'^{-1})_{G-S G-S} \, ,
\eeqa
with $M'$ in (\ref{M3}). After the rescaling with $\rho$, the matrix $M'$ becomes
(\ref{M4}). At leading order we set $\rho$ to zero so that $M'$ becomes block diagonal. Consequently
\beqa
M^{-1}_{G-S G-S}=M'^{-1}_{G/SG/S} \, .
\eeqa

\qed

\subsection{The two point function}
\label{sec:2point}

The results proven above must be refined further for the two point function. The reason is that, as explained in section \ref{sec:NCdimreg}, the two point functions have two singularities so that one needs also to analyse subleading behaviour.

In the sequel we replace $Q_G$ by $M_G$, $Q_{G/S}$ by $M_{G/S}$, etc., the difference between the $Q$'s and the $M$'s being inessential.

When integrating over the internal variables of $G$ we start by integrating over the variables associated to $S$ first. All the variables $u$, $v$ and $p$ appearing in the sequel belong then to $G/S$.
The amplitude of the graph $G$ will then write, after rescaling of the parameters of the subgraph $S$ and having perform the first Filk move
\beqa
{\cal A_G}=\int [ d t_\ell d\rho] \int [dudvdp]^{G/S}
\rho^{2L(S)-1}\frac
{e^{- \begin{pmatrix}
       u & v & p 
      \end{pmatrix}
(M'_{G/S}+\rho^2 \delta M')
       \begin{pmatrix}
         u \\ v \\ p
      \end{pmatrix}
          }}
{HU_{S}^{\frac D2}(\rho)},
\eeqa
where $[dt d\rho]$ is a short hand notation for the measure of integration on the Schwinger parameters, to be developped further in the next section.
$[dudvdp]^{G/S}$ is the measure of integration for the internal variables of $G/S$, and $M'$ is given in eq. (\ref{M4}). We explicitate $\delta M'$ as
 \beqa
\label{eq:deltaM'}
\delta M'=&& 
  \begin{pmatrix}
 E'^{u^{G-S}u^S} & 0 & C^{u^{G-S}p^S} \\
  E'^{v^{G-S}u^S} & 0  & C^{v^{G-S}p^S}\\
 C'^{p^{G-S}u^S} & 0 & 0
  \end{pmatrix}
  \begin{pmatrix}
c_l\delta_{u^Su'^S} & E'^{u^Sv^S} & 0 \\
 E'^{v^Su^S} & t_l\delta_{v^Sv'^S} & C^{v^Sp^S} \\
 0 & C^{p^Sv^S} & 0
  \end{pmatrix}^{-1}
\nonumber\\
&&
  \begin{pmatrix}
  E'^{u^Su^{G-S}}& E'^{u^Sv^{G-S}} & C'^{u^Sp^{G-S}} \\
   0 & 0  & 0 \\
   C^{p^Su^{G-S}} & C^{p^Sv^{G-S}} & 0\\
  \end{pmatrix} \, .
\eeqa

The Taylor development in $\rho$ of the exponential gives
\beqa
\label{eq:termsGT}
&&\int [ d t_\ell] \int [dudvdp]^{G/S}
{e^{- \begin{pmatrix}
       u & v & p 
      \end{pmatrix}
        M'_{G/S}
       \begin{pmatrix}
         u \\ v \\ p
      \end{pmatrix}
          }}\nonumber\\
&&\int d\rho \rho^{2L(S)-1}(-1)\frac
{1+ \rho^2\begin{pmatrix}
       u & v & p 
      \end{pmatrix}
       \delta M'
       \begin{pmatrix}
         u \\ v \\ p
      \end{pmatrix}
          }
{HU_{S}^{\frac D2}(\rho)} \, .
\eeqa

The first term in the integral over $\rho$ above corresponds to a (quadratic) mass divergence.

The second term (logarithmically divergent) coresponds to the insertion of some operator which we now to compute. 

The interaction is real. This means that we should symmetrize our amplitudes over complex conjugation of all 
vertices. For instance, at on loop one should allways symmetrize the left and right tadpoles \cite{GrWu04-2,DR,landau3}
\footnote{Following \cite{GMRV} one can prove using this argument that if terms like $x \partial$ do not appear in the initial 
lagrangean for the complex orientable model, they will not be generated by radiative corrections, which not proven there.}. 

Consequently, the inverse matrix in eq. (\ref{eq:deltaM'}) is actually a sum over the two possible choices of orientation of vertices. We must also sum over all possible choices of signs for the entries in the contact matrices in eq. (\ref{eq:deltaM'}) as a similar symmetrization must be performed for the hypermomenta.

As $\bar \psi^{u^S}$ couples only to the linear combination $\psi^{v^{G-S}}_{\ell}+\e^{V}_{\ell i}\psi^{u^{G-S}}_{\ell}$ in the initial matrix as well as in the change of variables, we have $E'^{u^Su^{G-S}}=\e^V_{l_{G-S}i} E'^{u^Sv^{G-S}}$.

Due to the sums over choices of signs, the only non zero entries in $\delta M'$ are $\delta M'_{u^S u^S}$, $\delta M'_{v^S v^S}$, $\delta M'_{v^S u^S}$, $\delta M'_{u^S v^S}$ and $\delta M'_{p_{\bar V_S} p_{\bar V_S}}$. 

We denote the two external lines of $S$ by $\ell_1$ and $\ell_2$. A tedious but straightforeward computation holds
\beqa
\begin{pmatrix}
       u & v & p 
      \end{pmatrix}
       \delta M'
       \begin{pmatrix}
         u \\ v \\ p
      \end{pmatrix}
&=&\Big{(}
A_1(\e^{\bar V_S}_{\ell_1 i}u_{\ell_1}+v_{\ell_1})^2+
A_2(\e^{V}_{\ell_2 i}u_{\ell_2}+v_{\ell_2})^2\nonumber\\
&&+B_1 p_{\bar V_S}^2
\Big{)} .
\eeqa

The first two terms are an insertion of the operator $\Omega x^2$ whereas the last is the insertion of an operator $-\Delta+x^2$, being of the form of the initial lagrangean.

Take the graph $G/S$ with the insertion of this operator at $S$, and with the adition of eventual mass subdivergencies (due to the first term). We denote it by an operator ${\cal O}_S$ action on the graph $G/S$. 

We can then sum up the results of this section in the formula:
\beqa
\label{eq:factfinal}
\frac{e^{-\frac{HV_G(\rho)}{HU_G(\rho)}}}{HU_G(\rho)^{D/2}}=
\frac{1}{[HU^{l(\rho)}_{S}]^{D/2}}(1+\rho^2{\cal O}_S) \frac{e^{-\frac{HV_{G/S}}{HU_{G/S}}}}{HU_{G/S}^{D/2}} \, .
\eeqa

\section{Renormalization of NCQFT}
\label{sec:NCdimreg}
\setcounter{equation}{0}

In this final section we proceed to the dimensional regularization and renormalization of NCQFT. We detail the meromorphic structure and give the form of the subtraction operator. Dimensional regularization and meromorphie of Feynam amplitudes for this model was also established in \cite{ncmellin}. However, for consistency reasons we will give here an independent proof of this results.

However, as the proof of convergence of the renormalized integral for this $\Phi^{\star 4}_4$ model is identical with that for the commutative $\Phi^4_4$ (up to substituting the commutative subtraction operator with our subtraction operator) we will not detail it here.

\subsection{Meromorphic structure of NCQFT}

In this subsection we prove the meromorphic structure of a Feynman amplitude ${\cal A}$. We follow here the approach of \cite{reg}. We express the amplitude by eq. (\ref{HUGV}) 
\beqa
\label{HUGV-t}
{\cal A}_{G,{\bar V}}  (x_e,\;  p_{\bar V}, D) = \left(\frac{\tom}{2^{\frac
    D2 -1}}\right)^L  \int_{0}^{1} \prod_{\ell=1}^L  
 dt_\ell (1-t_\ell^2)^{\frac D2 -1} 
\frac{
e^{-  \frac {HV_{G, \bar{V}} ( t_\ell , x_e , p_{\bar v})}{HU_{G, \bar{v}} ( t )}}
}
{HU_{G, \bar{V}} ( t )^{D/2}} \, .
\eeqa
We restrict our analysis to connected non-vacuum graphs. 
As in the commutative case we extend this expression to the entire complex plane. Take a Hepp sector $\sigma$ defined as 
\beqa
\label{hepp-nc}
0\le t_1 \le \ldots \le t_L \, ,
\eeqa
and perform the change of variables
\beqa
\label{change-nc}
t_\ell=\prod_{j=\ell}^L x_j^2,\ \ell =1,\ldots, L.
\eeqa
We denote by $G_i$ the subgraph composed by the lines $t_1$ to $t_i$. As before, we denote $L(G_i)=i$ the number of lines of $G_i$, $g(G_i)$ its genus, $F(G_i)$ its number of faces, etc..
The amplitude is
\beqa
{\cal A}_{G,\bar V}=&&\Big{(}\frac{\tilde \Omega}{2^{(D-4)/2}}\Big{)}^L
\int_{0}^{1}\prod_{i=1}^L 
\left(1-(\prod_{j=i}^L x_j^2)^2\right)^{\frac D2 -1}
dx_{i} 
\nonumber\\
&&\prod_{i=1}^{L}x_{i}^{2L(G_i)-1}
\frac{e^{-\frac{HV_{G,\bar V}(x^2)}{HU_{G,\bar V}(x^2)}}}{HU_{G,\bar V}(x^2)}\, .
\eeqa 
In the above equation we factor out in $HU_{G,\bar V}$the monomial with the smallest degree in each variable $x_i$
\beqa
\label{ampli-x}
{\cal A}_{G,{\bar V}}  (x_e,\;  p_{\bar v}) = &&\left(\frac{\tom}{2^\frac
    D2}\right)^L  \int_{0}^{1} \prod_{\ell=1}^L  
 dx_\ell \left(1-(\prod_{j=\ell}^L x_j^2)^2\right)^{\frac D2 -1} 
\nonumber\\ 
&& x_i^{2L(G_i)-1-D b'(G_i)}
 \frac{e^{-\frac {HV_{G, \bar V}}{HU_{G,\bar V}}}}{(a s^b+ F (x^2))^\frac D2}.
\eeqa
The last term in the above equation is always bounded by a constant. Divergences can arise only in the region $x_i$ close to zero (it is well known that this theory does not have an infrared problem, even at zero mass).

The integer $b'(G_i)$ is given by the topology of $G_i$. It is
\beqa
b'(G_i)=\begin{cases}
    {\displaystyle \le L(G_i)-[n(G_i)-1]-2g(G_i)} &\text{if } 
      g(G_i)>0
     \vspace{.3cm}\\
     {\displaystyle \le L(G_i)-n(G_i)} &\text{if }
      g(G_i)=0 \text{ and } B(G_i)>1
     \vspace{.3cm}\\
     {\displaystyle =L(G_i)-[n(G_i)-1]} &\text{if }
       g(G_i)=0 \text{ and } B(G_i)=1 \\
  \end{cases} .
\eeqa

To prove the first and the third line one must look at the scaling of a leading term with $I=\{1 \hdots L \}$ and $J$ admissible in $HU_G$. 
For the second line one must take $I=\{1 \hdots L \}-\tilde\ell$ 
and $J$ {\it pseudo-admissible}. We have proved that such terms exists in section 
\ref{sec:furtherLeading}.

We see that $b'(G_i)$ is at most $L(G_i)-n(G_i)+1$ and that the maximum is achieved if and only if $g(G_i)=0$ and $B(G_i)=1$.

The convergence in the UV regime ($x_i\to 0$) is ensured if
\beqa
 \Re [ 2L(G_i)-Db'(G_i) ]>0 ,\ i=1\ldots L \, .
\eeqa
As
\beqa
\Re[2L(G_i)-Db'(G_i)]>\Re\Big{(}2L(G_i)-D[L(G_i)-n(G_i)+1]\Big{)} \, ,
\eeqa
we always have convergence provided
\beqa
\Re D<2\le \frac{4n(G_i)-N(G_i)}{n(G_i)-N(G_i)/2+1} \le \frac{2L(G_i)}{L(G_i)-n(G_i)+1} \, .
\eeqa
where $N(G_i)$ is the number of external points of $G_i$
\footnote{We have used here the topological relation 
$4n(G_i)-N(G_i)=2L(G_i)$}.
Thus ${\cal A}_{G, \bar V}(D)$ is analytic in the strip
\beqa
\label{domain-nc}
{\cal D}^\sigma=\{ D\, | \, 0 < \Re\, D <  2 \}.
\eeqa

We extend now the ${\cal A}$ as a function of $D$ for $2 \le \Re D \le 4$. We claim that if
\begin{itemize}
\item $g(G_i)>0$
\item $g(G_i)=0$ and $B(G_i)>1$
\item $N(G_i)>4$ \, ,
\end{itemize}
the strip of analyticity can be immediately extended up to 
\beqa
{\cal D}^\sigma=\{ D\, | \, 0 < \Re\, D <  4+\e_G \}.
\eeqa
for some small positive number $\e_G$ depending on the graph.
Indeed, for the first two cases we have $b'(G_i)\le L(G_i)-n(G_i)$ so that the integral over $x_i$ converges for
\beqa
\Re D\le 4 < \frac{4n(G_i)-N(G_i)}{n(G_i)-N(G_i)/2}=\frac{2L(G_i)}{L(G_i)-n(G_i)}
\, ,
\eeqa
whereas in the third case, as $N(G_i)>4$ the integral over $x_i$ converges for
\beqa
\Re D\le 4 < \frac{4n(G_i)-N(G_i)}{n(G_i)-N(G_i)/2+1}=
\frac{2L(G_i)}{L(G_i)-n(G_i)+1} \, .
\eeqa

The only possible divergences in ${\cal A}_{G,\bar V}(D)$ are generated by planar one two or four external legs subgraphs with a single broken face. They are called primitively divergent subgraphs.

Let $S$ be a primitively divergent subgraph and call $\rho$ its associated Hepp parameter. Using eq. \ref{eq:factfinal} its contribution to the amplitude writes:
\beqa
\label{eq:polamp}
{\cal A}^{\rho}_{G,\bar V_G}\sim
\int_{0}d\rho \rho^{2L(S)-1-D[L(S)-n(S)+1]}
\frac{1+\rho^2 {\cal O}_S}{HU^l_{S,\bar V_{S}}|_{\rho=1}}
\frac{e^{-\frac{HV_{G/S}}{HU_{G/S}}}}{HU_{G/S}}\, .
\eeqa

The integral over $\rho$ is a meromorphic operator in $D$ with the divergent part given by
\beqa
\frac{r_1}{2L(S)-D[L(S)-n(S)+1]}+
     \frac{r_2}{2L(S)-D[L(S)-n(S)+1]+2}{\cal O}_S \, . \nonumber
\eeqa
We have a pole at $D=4$ if $S$ is a four point subgraph. If $S$ is a two point subgraph we have poles at $D=4-2/n(S)$ and $D=4$.
As all the singularities are of this type we conclude that ${\cal A}_G$ is a meromorphic function in the strip
\beqa
{\cal D}^\sigma=\{ D\, | \, 0 < \Re\, D < 4+\e_G \}.
\eeqa

\qed

\subsection{The subtraction operator}

The subtraction operator is similar to the usual one (see \cite{reg} \cite{ren}), with the notable difference that the set of primitively divergent subgraphs are different.
We give here a brief overview of its construction. For all functions $\rho^{\nu}g(\rho)$
with $g(0)\neq 0$, denote $E(\nu)$ the smallest integer such that $E(\nu)\ge \Re \nu$.
Let
\beqa
T_{\rho}^{q}=\sum_{k=0}^q\frac{1}{k!}g^{(k)}(0), \, q\ge 0;
\quad T_{\rho}^{q}=0, \, q<0\, ,
\eeqa
be the usual Taylor operator.
We define a generalized Taylor operator of order $n$ by
\beqa
\tau_{\rho}^n[\rho^{\nu}g(\rho)]=\rho^{\nu}
T_{\rho}^{n-E(\nu)}[g(\rho)] \, .
\eeqa

To each primitively divergent subgraph we associate a subtraction operator 
$\tau_S^{-2L(S)}$ acting on an integrand like
\beqa
\tau_S^{-2L(S)}\left( \frac{e^{-\frac{HV_G}{HU_G}}}{HU_G^{D/2}}
\right)=
\Big{[}
\tau_{\rho}^{-2L(S)}
\left(\frac{e^{-\frac{HV_G}{HU_G}}}{HU_G^{D/2}}\vert_{t_S\mapsto \rho^2 t_S}
\right)
\Big{]}_{\rho=1} \, .
\eeqa

Take the example of a bubble subgraph $S$. It is primitively divergent, and 
taking into account the factorization properties we have
\beqa
\tau_{S}^{-4}\left( \frac{e^{-\frac{HV_G}{HU_G}}}{HU_G^{D/2}}
\right)=
\left(\frac{e^{-\frac{HV_{G/S}}{HU_{G/S}}}}{HU_{G/S}^{D/2}}
\frac{1}{(HU^l_S)^{D/2}}
\right)
\Big{[}
\tau_{\rho}^{-4}(\frac{1}{\rho^D})
\Big{]}_{\rho=1} \, .
\eeqa
If $D<4$ $E(D)=-3$ and if $D\ge 4$ $E(D)=-4$. Consequently 
\beqa
\tau_S^{-4}=
\begin{cases}
     {\displaystyle \hspace{2cm}0} &\text{if } D<4,
     \vspace{.3cm}\\
     {\displaystyle \left(\frac{e^{-\frac{HV_{G/S}}{HU_{G/S}}}}{HU_{G/S}^{D/2}}
    \frac{1}{(HU^l_S)^{D/2}}\right)} &\text{if } D\ge 4.
     \vspace{.3cm}\\
\end{cases}
\eeqa

As expected the operator subtracts only for $D\ge 4$, and it exactly compensates the divergence in the expression (\ref{eq:polamp}).

We then define the complete subtraction operator as
\beqa
R=1+\sum_{{\cal F}}\prod_{S\in {\cal F}}(-\tau^{-2L(S)}_S) \, ,
\eeqa
where the sum runs over all forests of primitively divergent subgraphs.

From this point onward the classical proofs of (see \cite{reg}, \cite{ren}) go through. Theorems $1$, $2$ and $3$ of \cite{ren} so that we have the theorem
\begin{theorem}
The renormalized amplitude
\beqa
{\cal A}^r_{G}=R {\cal A}_{G}\, ,
\eeqa
is an analytic function of $D$ in the strip:
\beqa
{\cal D}^\sigma=\{ D\, | \, 0 < \Re\, D < 4+\e_G \},
\eeqa
for some small positive number $\e_G$.
\end{theorem}
\section{Conclusion and perspectives }
\label{sec:conclusion}

We have presented in this paper the dimensional regularization and
dimensional renormalization for the vulcanized $\Phi^{\star 4}_4$ model. 
The factorization results we have proven are the starting point
for the implementation of a Hopf algebra structure for NCQFT \cite{progres}.

The implementation of the dimensional
renormalization program for covariant NCQFT (e.g. non-commutative Gross-Neveu or the Langmann-Szabo-Zarembo model \cite{Langmann:2003if}, see section $1$) should follow the layout presented here.

One should try to extend the techniques presented here to 
non-commutative L-S dual gauge theories. Such models have recently been proposed, 
\cite{GrosseYM, WalletYM} but the reader should be aware that no proof of renormalizability of this models yet exists.

All the results mentioned here are obtained on a particular choice of non-commutative geometry, the Moyal space. One should try to extend this results to the NCQFT's on more involved geometries, like for example the non-commutative tori.

As mentioned in the introduction, NCQFT is a strong candidate for new
physics beyond the Standard Model. One could already study possible
phenomenological implications for Higgs physics of such non commutative renormalizable
$\Phi^{\star 4}_4$ models. The absence of Landau ghost makes this theory better behaved than its commutative counterpart. Also the Langmann-Szabo symmetry responsible for supressing the ghost could play a role similar to supersymmetry in taming UV divergencies.

\bigskip

{\bf Acknowledgment:} 
We thank Vincent Rivasseau for indicating us references \cite{reg} and \cite{ren}
and for fruitful discussions during the various stages of the preparation of this work.

\newpage
\thispagestyle{empty}

\chapter{CM Representation for NCQFT} 
\begin{center}
R. Gurau$^{(1)}$, A.P.C. Malbouisson$^{(2)}$\\ 
V. Rivasseau$^{(1)}$,
  A. Tanas{\u a}$^{(1)}$\\ 
1) Laboratoire de Physique Th\'eorique\\
CNRS UMR 8627, b\^at.\ 210\\ 
Universit\'e Paris XI,  F-91405 Orsay Cedex, France\\ 
2) Centro Brasileiro de Pesquisas F\'{\i}sicas,\\
Rua Dr.~Xavier Sigaud, 150, \\ 
22290-180 Rio de Janeiro, RJ, Brazil
\end{center}
 
\vskip 1cm 
We extend the complete Mellin (CM) representation of Feynman amplitudes 
to the non-commutative quantum field theories. This representation is a versatile tool. 
It provides a quick proof of meromorphy of Feynman amplitudes in parameters 
such as the dimension of space-time. In particular it 
paves the road for the dimensional renormalization of these theories. 
This complete Mellin representation also allows the study of asymptotic behavior 
under rescaling of arbitrary subsets of external invariants of any Feynman amplitude. 
 
\section{Introduction} 

Recently non-commutative quantum field theories such as the $\phi^{\star 4}$ and
the Gross-Neveu$_2$ models have  been shown renormalizable \cite{GrWu03-1,GWR2x2,GrWu04-3,RenNCGN05,Rivasseau2005bh,GMRV}
provided the propagator is modified to obey Langmann-Szabo duality \cite{LaSz}. These theories
are called ``vulcanized", and in \cite{gurauhypersyman,RivTan} the Feynman-Schwinger parametric 
representation was extended to them. For recent reviews, see \cite{sefu,RivTour}. 

In this paper we perform a further step, generalizing to these vulcanized NCQFT the 
complete Mellin (CM) representation.  This CM representation is derived from the 
parametric representation established in \cite{gurauhypersyman,RivTan} and provides a starting point 
for the study of dimensional renormalization and of the asymptotic behavior of Feynman amplitudes
under arbitrary rescaling of external invariants.

The study of asymptotic behaviors in 
conventional (commutative) field theories started in the 1970's with the papers \cite{BL,bergere3,bergere1} which use the BPHZ
renormalization scheme. 
Their approach is based on the Feynman-Schwinger parametric representation of
amplitudes. This representation involves ``Symanzik" or ``topological" polynomials
associated to the Feynman diagram. 
In commutative theories these ``topological polynomials" are written as sums 
over the spanning trees or 2-trees of the diagram \cite{nakanishi, itzykson}. 
The corresponding mathematical theory goes
back to the famous tree matrix theorem of Birkhoff (see \cite{A} and references therein).

In vulcanized NCQFTs, propagators are no longer based on the heat kernel
but on the Mehler kernel. The kernel being still quadratic in position space, explicit integration 
over all space variables is still possible. It leads to the 
NCQFT parametric representation. This representation no longer 
involves ordinary polynomials in the Schwinger parameters, 
but hyperbolic polynomials \cite{gurauhypersyman}. As
diagrams in NCQFT are ribbon graphs, these hyperbolic polynomials contain 
richer topological information than in the commutative case. In particular
they depend on the genus of the Riemann surface on which the graphs are defined.

Returning to the Mellin transform technique, it was
introduced in commutative field theory to prove theorems on the asymptotic
expansion of Feynman diagrams. The general idea 
was to prove the existence of an asymptotic series in powers of $\lambda $ and
powers of logarithms of $\lambda $, under rescaling by $\lambda$ of 
some (Euclidean) external invariants associated to the Feynman
amplitudes. The Mellin transform with respect to
the scaling parameter allows to obtain this result in some cases and to 
compute the series coefficients, even for renormalized diagrams \cite{bergerelam}.
But it did not work for\textit{ arbitrary} subsets of invariants.

In the subsequent years, the subject advanced further. 
The rescaling of internal squared masses (in order to study the infrared behavior 
of the amplitudes) was treated in \cite{malbouisson7}. There the concept of ``FINE" 
polynomials was introduced (that is, those being factorizable in each Hepp sector \cite{hepp}
of the variables\footnote{A Hepp sector is a complete ordering of the Schwinger parameters.}). It was then argued that the Mellin transform may be ``desingularized",
which means that the integrand of the inverse Mellin transform (which gives back the  
Feynman amplitude as a function of $\lambda $) has a meromorphic structure, so that the
residues of its various poles generate the asymptotic expansion in 
$\lambda $. However, this is not the case under arbitrary rescaling, 
because in many diagrams the FINE property simply does not occur. 

A first solution
to this problem was presented in \cite{malbouisson7} by introducing the
so-called ``multiple Mellin" representation, which consists in
splitting the Symanzik polynomials in a certain number of pieces, each one
of which having the FINE property. Then, after scaling by the parameter
 $\lambda $, an asymptotic expansion can be obtained as a sum over all Hepp
sectors. This is always possible if one adopts, as done in 
\cite{malbouisson, rivasseau, malbouisson5}, the extreme point of view to
split the Symanzik polynomials in all its monomials. Moreover, this apparent
complication is compensated by the fact that one can dispense with the use of 
Hepp sectors altogether. This is called the
``complete Mellin" (CM) representation.

The CM representation provides a general proof of the existence of an
asymptotic expansion in powers and powers of logarithms of the scaling
parameter in the most general case. Moreover the integrations over the 
Schwinger parameters can be explicitly performed, and we are left with the pure
geometrical study of convex polyhedra in the Mellin variables. The results of \cite{malbouisson7} are
obtained in a simpler way \cite{malbouisson}, and asymptotic expansions are computed in a more
compact form, without any division of the integral into Hepp sectors.

Moreover the CM representation allows a unified
treatment of the asymptotic behavior of both ultraviolet convergent and
divergent renormalized  amplitudes. Indeed, as shown in  \cite{malbouisson,rivasseau}, 
the renormalization procedure does not alter the algebraic
structure of integrands in the CM representation. It only changes the set of
relevant integration domains in the Mellin variables.
The method allows the study  of
dimensional regularization \cite{rivasseau,malbouisson5} and of the infrared behavior
of amplitudes relevant to critical phenomena \cite{malbouisson6}.

The CM representation is up to now the only tool which provides 
asymptotic expansions of Feynman amplitudes in powers and powers of logarithms in the most general
rescaling regimes. More precisely consider a Feynman amplitude $G(s_{k})$ written in terms
of its external invariants $s_{k}$ (including particle masses), and an asymptotic regime defined by 
\begin{equation}
s_{k}  \rightarrow \lambda ^{a_k} {s}_{k},  
\label{scaling1}
\end{equation}
where $a_k$ may be positive, negative, or zero.
Let $\lambda $ go to infinity  (in this way both ultraviolet and infrared behaviors 
are treated on the same footing). 
Then $G(s_{k})$ as a function of $\lambda $ has an 
asymptotic expansion of the form: 
\begin{equation}
G(\lambda;s_{k} )=\sum_{p=p_{\text{max}}}^{-\infty }\sum_{q=0}^{q_{\text{max}%
}(p)}G_{pq}(s_k)\lambda ^p\ln ^q\lambda ,
\label{expansao3}
\end{equation}
where $p$  runs over decreasing rational values, with $p_{max}$ as leading power, and $q$, for 
a given $p$, runs over a finite set of non-negative integer values. 

The CM representation proves this result for
any commutative field theory, including gauge theories in arbitrary gauges \cite{linhares}.

The above general result should hold {\it mutatis mutandis} for non-commutative 
field theories. In this paper we provide a starting point for this analysis by constructing 
the CM representation for $\phi^{*4}$ (Theorem \ref{cmrep} below). Moreover using this representation 
we check meromorphy of the Feynman amplitudes in the dimension of space-time (Theorem \ref{meromorph}). 

The main difference with the commutative case is that this integral representation (previously true 
in the sense of \textit{functions} of the external invariants)
now holds only in the sense of {\it distributions}.  Indeed
the distributional character of commutative amplitudes (in momentum space) 
reduces to a single overall $\delta$-function of momentum conservation. This is no longer true
for vulcanized NCQFT amplitudes, which must be interpreted as distributions to be smeared
against test functions of the external variables.

\section{Commutative CM representation} 
 
For simplicity we consider a scalar Feynman amplitude.
To get the CM representation \cite{malbouisson}, rewrite the Symanzik polynomials as  
\begin{equation} 
U(\alpha )=\sum_j\prod_{\ell=1}^L\alpha_\ell^{u_{\ell j}}\equiv \sum_jU_j,\;\;\qquad 
V(\alpha )=\sum_k s_k\left( \prod_{\ell=1}^L\alpha _\ell^{v_{\ell k}}\right) \equiv 
\sum_k V_k ,  \label{sym} 
\end{equation} 
where $j$ runs over the set of spanning trees and $k$ over the set of the 2-trees, 
\begin{equation} 
u_{\ell j}=\left\{  
\begin{array}{ll} 
0 & \text{if the line }\ell\text{ belongs to the 1-tree}\  j \\  
1 & \text{otherwise} 
\end{array} 
\right. 
\end{equation} 
and  
\begin{equation} 
v_{\ell k}=\left\{  
\begin{array}{ll} 
0 & \text{if the line }\ell\text{ belongs to the 2-tree}\  k\\  
1 & \text{otherwise.} 
\end{array} 
\right. 
\end{equation} 
 
The Mellin transform relies on the fact that for any function $f(u)$, piecewise smooth for $u>0$, if the integral  
\begin{equation} 
g(x)=\int_0^\infty du\,u^{-x-1}f(u) 
\end{equation} 
is absolutely convergent for $\alpha <$ Re $x<\beta $, then  for $\alpha <\sigma <\beta $ 
\begin{equation} 
f(u)=\frac 1{2\pi i}\int_{\sigma -i\infty }^{\sigma +i\infty }dx\,g(x)\,u^x .
\end{equation} 

Consider $D$ the space-time dimension to be for the moment real positive. 
Taking  $f(u) = e^{-u}$, and applying to $u=V_k/U$ one gets  
\begin{equation} 
e^{-V_k/U}=\int_{\tau _k}\Gamma (-y_k)\left( \frac{V_k}U\right) ^{y_k}, 
\label{gamayk} 
\end{equation} 
where $\int_{\tau _k}$ is a short notation for $\int_{-\infty }^{+\infty }%
\frac{d(\text{Im }y_k)}{2\pi }$, with Re $y_k$ fixed at $\tau _k<0$. We may 
now recall the identity 
\begin{equation} 
\Gamma (u)\left( A+B\right) ^{-u}=\int_{-\infty }^\infty \frac{d(\text{Im }x)%
}{2\pi }\Gamma (-x)A^x\Gamma (x+u)B^{-x-u} . \label{gamauamaisb} 
\end{equation} 
Taking $A\equiv U_1(x)$ and $B\equiv U_2+U_3+\cdots $ and using 
iteratively the identity above, leads, for $u=\sum_k y_k+D/2$, to 
\begin{equation} 
\Gamma \left( \sum_k y_k+\frac D2\right) U^{-\sum_k y_k-\frac D2}=\int_\sigma 
\prod_j\Gamma (-x_j)U_j^{x_j},  \label{gamayk2} 
\end{equation} 
with Re $x_j=\sigma _j<0$, Re $\left( \sum_k y_k +\frac D2\right) =\sum_k\tau 
_k+\frac D2>0$, and $\int_\sigma $ means $\int_{-\infty }^{+\infty }\prod_j
\frac{d(\text{Im }x_j)}{2\pi }$ with $\sum_jx_j+\sum_k y_k =-\frac D2$. Then, 
using (\ref{gamayk}) and (\ref{gamayk2}), the amplitude is written as  
\begin{equation} 
I_\cG (s_k ,m_l^2)=\int_\Delta \frac{\prod_j\Gamma (-x_j)}{\Gamma (-\sum_jx_j)}
\prod_k s_k ^{y_k }\Gamma (-y_k)\int_0^\infty \prod_l\,d\alpha _l\,\alpha 
_l^{\phi _l-1}\,e^{-\sum_l\alpha _lm_l^2},  \label{equacao} 
\end{equation} 
where  
\begin{equation} 
\phi _i\equiv \sum_ju_{ij}x_j+\sum_k v_{i k }y_k+1 . 
\end{equation} 
The symbol $\int_\Delta $ means integration over the independent 
variables $\frac{\text{Im }x_j}{2\pi }$, $\frac{\text{Im }y_k }{2\pi }$
in the convex domain $\Delta$ defined by ($\sigma $ and $\tau $ standing 
respectively for $\text{Re }x_j$ and $\text{Re }y_k$):  
\begin{equation} 
\Delta =\left\{ \sigma ,\tau \left|  
\begin{array}{l} 
\sigma _j<0;\;\tau _k<0;\;\sum_jx_j+\sum_k y_k =-\frac D2; \\  
\forall i,\;\text{Re }\phi _i\equiv \sum_ju_{ij}\sigma _j+\sum_k v_{ik }\tau 
_k +1>0 
\end{array} 
\right. \right\} \ . \label{domain} 
\end{equation} 
This domain $\Delta$ is non-empty as long as $D$ is positive and 
small enough so that every subgraph of $G$ has convergent power counting \cite{malbouisson},
hence in particular for the $\phi^4$ theory it is always non empty for any graph
for $0<D<2$. 
 
The $\alpha $ integrations may be performed, using the well-known 
representation for the gamma function, so that we have  
\begin{equation} 
\int_0^\infty d\alpha _l\,e^{-\alpha _lm_l^2}\alpha _l^{\phi _l-1}=\Gamma 
\left( \phi _l\right) \left( m_l^2\right) ^{-\phi _l}  \label{alphaint} 
\end{equation} 
and we finally get the CM representation of the amplitude in 
the scalar case:  
\begin{equation} 
I_G(s_k ,m_l^2)=\int_\Delta \frac{\prod_j\Gamma (-x_j)}{\Gamma (-\sum_jx_j)}%
\prod_k s_k ^{y_k }\Gamma (-y_k )\prod_l\left( m_l^2\right) ^{-\phi _l}\Gamma 
\left( \phi _l\right) .  \label{Mellin1} 
\end{equation} 
This representation can now be extended to complex values of $D$. For instance for a massive
$\phi^4$ graph, it is analytic in $D$ for $\Re D < 2$, and meromorphic in $D$ in the whole complex plane 
with singularities at rational values; furthermore its dimensional analytic continuation has
the same unchanged $CM$ integrand but translated integration contours \cite{rivasseau,malbouisson5}.

\section{Non-commutative parametric representation} 
\label{hyperbo} 
 
Let us summarize the results of \cite{gurauhypersyman}. 
Define the antisymmetric matrix $\sigma$ as 
\begin{align} 
\sigma=\begin{pmatrix} \sigma_2 & 0 \\ 0 & \sigma_2 \end{pmatrix} \mbox{ with}
\ \sigma_2=\begin{pmatrix} 0 & -i \\ i & 0 \end{pmatrix}\ . 
\end{align} 
The $\delta-$functions appearing in the vertices can be 
rewritten as an integral over some new variables $p_V$ called {\it hypermomenta},
{\it via} the relation 
\begin{align} 
\label{pbar1} 
\delta(x_1 -x_2+x_3-x_4 ) = \int  \frac{d p'}{(2 \pi)^4} 
e^{ip' (x_1 -x_2 +x_3 -x_4)}
=\int  \frac{d p}{(2 \pi)^4} 
e^{p \sigma (x_1-x_2+x_3-x_4)}. 
\end{align}

Let $\cG$ be a ribbon graph. Choosing a particular root vertex $\bar{V}$,
the parametric representation for the amplitude of $\cG$ is
expressed in terms of $t_\ell = \tanh \alpha_\ell /2 $ (where $\alpha_\ell$ is the former
Schwinger parameter) as
\begin{align} 
\label{a-condens} 
{\cal A}_{\cG} =& \int_0^1 \prod_{\ell} d t_\ell (1-t_{\ell}^2)^{\frac D2 -1}  
\int d x d p e^{-\frac {\Omega}{2} X G X^t} 
\end{align} 
where $D$ is the space-time dimension, $X$ summarizes all positions and hypermomenta 
and $G$ is a certain quadratic form. Calling
$x_e$ and $p_{\bar{V}}$ the external variables and $x_i, p_i$ 
the internal ones, we decompose $G$ into an internal  
quadratic form $Q$, an external one $M$ and a coupling part $P$ so that  
\begin{align} 
\label{defX} 
X =& \begin{pmatrix} 
x_e & p_{\bar{V}} & x_i & p_i\\ 
\end{pmatrix} , \ \  G= \begin{pmatrix} M & P \\ P^{t} & Q \\ 
\end{pmatrix}\ , 
\end{align} 
Performing the Gaussian integration over all internal variables one gets the non-commutative
parametric representation: 
\begin{align} 
\label{aQ} 
{\cal A}_{\cG}  =&  \int \prod_{\ell}d t_\ell (1-t_{\ell}^2)^{\frac D2 -1} 
\frac{1}{\sqrt{\det Q}} 
e^{-\frac{\tom}{2} 
\begin{pmatrix} x_e & p_{\bar V} \\ 
\end{pmatrix} \big[ M-P Q^{-1}P^{t}\big ] 
\begin{pmatrix} x_e \\  p_{\bar V}\\ 
\end{pmatrix}}\ . 
\end{align} 
This representation leads to new polynomials $HU_{\cG, \bar{V}}$ and $HV_{\cG, \bar{V}}$
 in the $t_\ell$ ($\ell =1,\ldots, L$) variables, analogs  
of the Symanzik polynomials $U$ and $V$ of the commutative case,
through
\begin{align} 
\label{polv} 
{\cal A}_{{\cG}}  (\{x_e\},\;  p_{\bar V}) =& K'  \int_{0}^{1} 
\prod_{\ell}d t_\ell (1-t_{\ell}^2)^{\frac D2 -1} 
\frac{e^{-  \frac {HV_{\cG, \bar{V}} ( t , x_e , p_{\bar v})}{HU_{\cG, \bar{V}} ( t )}}}
{HU_{\cG, \bar{V}} ( t )^{\frac D2}} .
\end{align} 
 
The main results of \cite{gurauhypersyman,RivTan} are 
 
\begin{itemize} 
 \item The polynomial $HU_{\cG, \bar{V}}$ and the real part of $HV_{\cG, \bar{V}}$  
have a \emph{positivity property}. They are sums of monomials 
with positive integer coefficients, which are squares of Pfaffians  
with integer entries.
 
\item {Leading terms} can be identified in a given ``Hepp sector'',  
at least for \textit{orientable graphs}. 
In $HU_{\cG, \bar{V}}$ they correspond to hyper-trees 
which are the disjoint union of a tree in the direct graph and an other tree in the dual graph. Any 
connected graph has such hypertrees. Similarly ``hyper-two-trees'' 
govern the leading behavior of $HV_{\cG, \bar{V}}$ in any Hepp sector. 
\end{itemize} 

Let us relabel the $2L$ internal positions of $\cG$ as $L$ short and
$L$ long variables in the terminology of \cite{GMRV}.
It has been shown in \cite{gurauhypersyman} that  for the Grosse-Wulkenhaar $\phi^{\star 4}$ model:
\beqa\label{hypnoncanpfaff111}
HU_{\cG,{\bar V}} (t) &=&   \sum_{{K_U}= I\cup J, \  n + |{K_U}|\; {\rm odd}}  s^{2g-k_{{K_U}}} \ n_{{K_U}}^2
\prod_{\ell \not\in I} t_\ell \prod_{\ell' \in J} t_{\ell'}\ 
\nonumber \\
&=&\sum_{{K_U}}a_{K_U} \prod_\ell t_\ell^{u_{\ell {K_U}}}\equiv 
\sum_{K_U} HU_{K_U} .
\eeqa
where  
\begin{itemize}

\item $I$ is a subset of the first $L$ indices (corresponding to the short variables) with $\vert I \vert $ elements,
and $J$ a subset of the next $L$ indices (corresponding to the long ones) with $\vert J\vert $ elements.
 
\item $B$ is the antisymmetric part of the quadratic form $Q$ restricted to these $2L$ short and long
variables (i.e. omitting hypermomenta)

\item $n_{{K_U}}=\mathrm{Pf}(B_{\widehat{{K_U}}})$ is the Pfaffian of the antisymmetric matrix 
obtained from $B$ by deleting  the lines and columns in the set ${K_U}= I \cup J$.

\item $k_{{K_U}}$ is $ \vert {K_U}\vert - L - F +1$, where $F$ is the number
of faces of the graph, $g$ is the genus of $\cG$ and $s$ is a constant.

\item $ a_{K_U} = s^{2g-k_{{K_U}}}n_{{K_U}}^2$, and
\begin{equation}\label{ul0}
u_{\ell {K_U}}=\left\{  
\begin{array}{ll} 
0 & \mbox{ if } \ell \in I \mbox{ and } \ell\notin J \\  
1 & \mbox{ if } (\ell \notin I \mbox{ and } \ell\notin J) \mbox{ or } (\ell \in I \mbox{ and } \ell\notin J)\\
2 &  \mbox{ if } \ell \notin I \mbox{ and } \ell\in J
\end{array} 
\right. .
\end{equation}

\end{itemize}

 The second polynomial $HV$ has both a real part $HV^\cR$ and an imaginary part
 $HV^\cI$, more difficult to write down. We need to introduce beyond $I$ and $J$
as above a particular line $\tau \notin I$ which is the analog of a two-tree cut. 
We define $\mathrm{Pf}(B_{\hat{{K_V}}\hat{\tau}})$ as the Pfaffian of the matrix obtained from $B$ by deleting 
the lines and columns in the sets $I$, $J$ and $\tau$. Moreover we
define $\epsilon_{I,\tau}$ to be the signature of the permutation obtained
from $(1,\ldots, d)$ by extracting the positions belonging to $I$ and
replacing them at the end in the order
\beqa
1,\dotsc ,d\rightarrow 1,\dotsc,\hat{i_1},\dotsc ,\hat{i_{\vert I\vert }},\dotsc
,\hat{i_{\tau}},\dotsc ,d, i_{\tau},i_{\vert I\vert }\dotsc, i_1\,.
\eeqa
where $d$ is the dimension of the matrix $Q$. Then:
\beqa
HV^\cR_{\cG,{\bar V}}&=&
 \sum_{{K_V}= I \cup J }\prod_{\ell \notin I} t_\ell
 \prod_{\ell' \in J} t_{\ell'}
\Big{[}\sum_{e_1}x_{e_1}\sum_{\tau\notin {K_V}}P_{e_1\tau}\epsilon_{{K_V}\tau}
\mathrm{Pf}(B_{\hat{{K_V}}\hat{\tau}})\Big{]}^2 \, .
\label{HVgv1Melin}\nonumber\\
&=&
\sum_{K_V} s^\cR_{K_V}\left( \prod_{\ell=1}^L t_\ell^{v_{\ell {K_V}}}\right) \equiv 
\sum_{K_V} HV_{K_V}^\cR
\eeqa
where
\beqa s_{K_V}^\cR= \left(
\sum_e x_e \sum_{\tau\notin {K_V}} P_{e\tau} \e_{{K_V} \tau} 
\mathrm{Pf}(B_{\hat{{K_V}}\hat{\tau}})
\right)^2 \eeqa
and $v_{\ell {K_V}}$ given by the same formula as $u_{\ell K_U}$.

The imaginary part involves {\it pairs} of lines $\tau, \tau'$ and corresponding
signatures (see \cite{gurauhypersyman,RivTan} for details):
\beqa \label{imaghv}
HV_{\cG,\bar V}^\cI&=&\sum_{{K_V}= I \cup J }\prod_{\ell \notin I} t_\ell
 \prod_{\ell' \in J} t_{\ell'}\nonumber \\
&&
\epsilon_{K_V}\mathrm{Pf}(B_{\hat{{K_V}}})
\Big{[}\sum_{e_1,e_2} \Big{(}
\sum_{\tau\tau'}P_{e_1\tau}\epsilon_{{K_V}\tau\tau'}\mathrm{Pf}(B_{\hat{{K_V}}\hat{\tau}\hat{\tau'}})
P_{e_2\tau'}\Big{)} x_{e_1}\sigma x_{e_2}\Big{]}\,.\nonumber
\\
&=&
\sum_{K_V} s^\cI_{K_V}\left( \prod_{\ell=1}^L t_\ell^{v_{\ell {K_V}}}\right) \equiv 
\sum_{K_V} HV_{K_V}^\cI
\eeqa
where
\beqa s_{K_V}^\cI=\epsilon_{K_V} \mathrm{Pf}(B_{\hat{{K_V}}})
\left(
\sum_{e, e'} (\sum_{\tau, \tau'}   P_{e \tau} \e_{{K_V} \tau \tau'} 
\mathrm{Pf}(B_{\hat{{K_V}}\hat{\tau}\hat{\tau'}}) P_{e'\tau'})x_e \sigma x_{e'}
\right) . \eeqa

\medskip
A similar development exists for the LSZ model (\cite{RivTan}), although more involved. 
The first polynomial in this case writes 
\beqa 
\label{pol-f} 
HU_{\cG, {\bar V}} (t) &=&   
\sum_{\substack{I\subset \{1\dotsc L\},\\ n + |I|\; {\rm odd}}}  s^{2g-k_{I}} \ n_{I}^2 
\prod_{l \in I} \frac{1+t_\ell^2}{2t_\ell} \prod_{l' \in \{1,\ldots,L\}} t_{l'}
\nonumber \\
&=&\sum_{{K_U}= I \cup J}a_{K_U} \prod_\ell t_\ell^{u_{\ell {K_U}}}\equiv 
\sum_{K_U} HU_{K_U} .
\eeqa 
Again $I$ runs over the subsets of $\{ 1, \ldots, L\}$. But $J$ now runs over subsets of 
$I$ (representing the expansion
of $\prod_{\ell \in I} [1 + t_\ell^2]$), and one has again
\begin{equation} 
\label{ul}
u_{\ell {K_U}}=\left\{  
\begin{array}{ll} 
0 & \mbox{ if } \ell\in I \mbox{ and } \ell \in J\\  
1 & \mbox{ if } \ell\notin I\\
2 &  \mbox{ if } \ell\in I \mbox{ and } \ell \notin J  
\end{array} 
\right. \ \ , \ \  
a_{K_U} = \frac{1}{2^{|I|}}s^{2g-k_{I}}n_I^2.
\end{equation} 
There are similar expressions for the second polynomial.

To summarize, the main differences of the NC parametric representation 
with respect to the commutative case are:
\begin{itemize}
\item the presence of the constants $a_j$ in the form
(\ref{hypnoncanpfaff111}) of $HU$,
\item the presence of an imaginary part $i\, HV^\cI$ in
$HV$,  see (\ref{imaghv}),
\item the fact that the parameters $u_{\ell j}$ and $v_{\ell k}$ in the formulas above
can have also the value $2$ (and not only $0$ or $1$).
\end{itemize}

\section{Non-commutative CM representation}

For the real part $HV^\cR$ of $HV$ one uses again the identity \eqref{gamayk} as
\begin{equation} 
\label{yR}
e^{-HV^\cR_{K_V}/U}=\int_{\tau_{K_V}^\cR}\Gamma (-y_{K_V}^\cR)\left( \frac{HV^\cR_{K_V}}U\right) ^{y_{K_V}^\cR}, 
\end{equation} 
which introduces the set of Mellin parameters $y_{K_V}^\cR$.

However for the imaginary part one cannot apply anymore the same identity. Nevertheless it remains
true {\bf in the sense of
distributions}. More precisely we have for $HV^\cI_{K_V}/U >0$ and $-1 < \tau_{K_V}^\cI <0$ 
\begin{equation} 
\label{yI}
e^{-i\, HV^\cI_{K_V}/U}=\int_{\tau_{K_V}^\cI}\Gamma (-y_{K_V}^\cI)\left( \frac{i\, HV^\cI_{K_V}}U\right) ^{y_{K_V}^\cI}.
\end{equation} 
The proof is given in Appendix B.

Note that this introduces another  set of Mellin parameters $y_{K_V}^\cI$. 
The distributional sense of formula \eqref{yI} is
a translation of the distributional form of a vertex contribution,
and the major difference with respect to the commutative case.

\medskip

For the polynomial $HU$ one can use again the formula \eqref{gamayk2} which rewrites here as
\begin{equation} 
\Gamma \left( \sum_{K_V}y_{K_V}+\frac D2\right) (HU)^{-\sum_{K_V} (y_{K_V}^\cR+ y_{K_V}^\cI) -\frac D2}=\int_\sigma 
\prod_{K_U}\Gamma (-x_{K_U})U_{K_U}^{x_{K_U}},  \label{x} 
\end{equation}
Note that the r\^ole of the Mellin parameters  $y_{K_V}$ of the commutative case is
now played by the sum $y_{K_V}^\cR + y_{K_V}^\cI$. 

\medskip

As in the commutative case, we now insert the distribution formulas
\eqref{yR}, \eqref{yI} and \eqref{x} in the general form of the Feynman
amplitude. This gives
\beqa 
\label{aproape}
{\cal A}_G &=&{\rm K'}\int_\Delta \frac{\prod_{K_U} a_{K_U}^{x_{K_U}} \Gamma (-x_{K_U})}{\Gamma (-\sum_{K_U}x_{K_U})}
\left( \prod_{K_V} (s_{K_V}^\cR)^{y_{K_V}^\cR}\Gamma (-y_{K_V}^\cR) \right)
\nonumber\\
&&\left( \prod_{K_V} (s_{K_V}^\cI)^{y_{K_V}^\cI}\Gamma (-y_{K_V}^\cI) \right)\int_0^1 \prod_{\ell =1}^L dt_\ell (1-t_\ell^2)^{\frac D2 -1}
t_\ell^{\phi_\ell -1}
\eeqa
where  
\begin{equation} 
\phi_\ell\equiv \sum_{K_U} u_{\ell {K_U}}x_{K_U}+\sum_{K_V}
(v^\cR_{\ell {K_V}}y_{K_V}^\cR+v^\cI_{\ell {K_V}}y_{K_V}^\cI) +1 .
\end{equation} 
Here $\int_\Delta $ means integration over the
variables $\frac{\text{Im }x_{K_U}}{2\pi i}$, $\frac{\text{Im }y_{K_V}^\cR}{2\pi i}$ and 
$\frac{\text{Im }y_{K_V}^\cI}{2\pi i}$, where
$\Delta $ is the convex domain:  
\begin{equation} 
\Delta =\left\{ \sigma ,\tau^\cR, \tau^\cI \left|  
\begin{array}{l} 
\sigma _{K_U}<0;\;\tau^\cR_{K_V}<0;\; -1 < \tau^\cI_{K_V}<0;\;
\\
\sum_{K_U}x_{K_U}+\sum_{K_V} (y_{K_V}^\cR+ y_{K_V}^\cI)=-\frac D2; \\  
\forall \ell,\;\text{Re }\phi_\ell\equiv \sum_{K_U}u_{\ell {K_U}}\sigma _{K_U}\\
+\sum_{K_V}
(v^\cR_{\ell {K_V}}\tau^\cR_{K_V}+v^\cI_{\ell {K_V}}\tau^\cI_{K_V})+1>0 
\end{array} 
\right. \right\}  \label{nc-domain} 
\end{equation} 
and $\sigma $, $\tau^\cR $ and $\tau^\cI $ stand 
for $\text{Re }x_{K_U}$, $\text{Re }y_{K_V}^\cR$   and $\text{Re }y_{K_V}^\cI$.

The $dt_\ell$ integrations in \eqref{aproape} may be performed using the  
representation for the beta function
\beqa
\label{beta}
\int_0^1 \prod_{\ell =1}^L dt_\ell (1-t_\ell^2)^{\frac D2 -1}
t_\ell^{\phi_\ell -1}=\frac 12 \beta (\frac{\phi_\ell}2, \frac D2).
\eeqa
Furthermore one has
$$  \beta (\frac{\phi_\ell}2, \frac D2) =\frac {\Gamma (\frac{\phi_\ell}2)
  \Gamma (\frac D2)}{\Gamma (\frac{\phi_\ell+D}2)}. $$
This representation is convergent for  $0<\Re D<2$ and we get:
  
\begin{theorem}\label{cmrep}
Any Feynman amplitude of a $\phi^{\star 4}$ graph
is analytic at least in the strip $0<\Re D<2$ where it
admits the following CM representation 
\beqa 
\label{final}
{\cal A}_\cG &=&{\rm K'}\int_\Delta \frac{\prod_{K_U} a_{K_U}^{x_{K_U}} \Gamma (-x_{K_U})}{\Gamma (-\sum_{K_U}x_{K_U})}
\left( \prod_{K_V} (s_{K_V}^\cR)^{y_{K_V}^\cR}\Gamma (-y_{K_V}^\cR) \right)
\nonumber\\
&& \left( \prod_{K_V} (s_{K_V}^\cI)^{y_{K_V}^\cI}\Gamma (-y_{K_V}^\cI) \right)
\left( \prod_{\ell=1}^L \frac {\Gamma (\frac{\phi_\ell}2)
  \Gamma (\frac D2)}{2\Gamma (\frac{\phi_\ell+D}2)} \right).\nonumber\\
\eeqa
which holds as tempered distribution of the external invariants.
\end{theorem}
\noindent{\bf Proof}
We have to check that for any $\phi^{\star 4}$ graph, the domain $\Delta$ 
is non empty for $0<\Re D<2$. It is obvious to check that $\Delta$ is non empty for $0<\Re D<1$;
indeed since the integers $u$ and $v$ are bounded by 2, we can put an arbitrary single $x_{K_U}$
close to $-D/2$ and the others with negative real parts close to 0 and the conditions 
$\phi_\ell >0$ will be satisfied. With a little extra care, one also gets easily that
$\Delta$ is non empty for $0<\Re D<2$. Indeed one can put again one single $x_{K_U}$
close to $-D/2$, provided for any $\ell$ we have $u_{\ell,{K_U}} \ne 2$. For the ordinary 
Grosse-Wulkenhaar model one can always find ${K_U}={I\cup J}$
where $I=\{1,...,L\}$ is the full set of lines \cite{gurauhypersyman}, hence  $u_{\ell,{K_U}} \ne 2$ for any $\ell$
by (\ref{ul0}).
In the ``covariant" case treated in \cite{RivTan}, one can simply take ${K_U}$ with $J=I$.
Considering (\ref{ul}) ensures again obviously that  $u_{\ell,{K_U}} \ne 2$ for any $\ell$.

The analyticity statement is part of a more 
precise meromorphy statement proved in the next section.
\qed

\medskip
We have thus obtained the {\bf complete Mellin representation of Feynman
amplitudes for non-commutative QFT}. Let us
point out first that the distributional aspect does not affect
the last product, the one involving the dimensionality of space-time
$D$. Therefore, one can still prove the meromorphy of the Feynman amplitude
${\cal A}_\cG (D) $ in the space-time dimension $D$ 
essentially as in the commutative case. This is done in the next section.

Another difference is that the $dt_\ell$ integration \eqref{beta}
brought $\beta$ functions, hence more complicated ratios of $\Gamma$ 
functions than in the commutative case. In particular 
singularities appear for negative $D$ which come from the integration near
$t_\ell =1$, hence the ``infrared side". They have absolutely no analog
in the commutative case and are due to the hyperbolic nature
of the Mehler kernel. At zero mass these singularities
occur for $D$ real negative even integer and in the massive case
they occur for $D$ real but sufficiently negative depending on the mass. 

\section{Meromorphy in the space-time dimension $D$} 
 
The complete Mellin representation \eqref{final} for non-commutative
theories  allows a quick proof of the meromorphy of the Feynman
amplitude ${\cal A}_G $ in the variable $D$, since $D$ appears only as argument
of $\Gamma$ functions which are meromorphic. 

\begin{theorem}\label{meromorph}
Any amplitude ${\cal A}_\cG $ is a tempered meromorphic distribution in $D$, that is 
 ${\cal A}_\cG $ smeared against any fixed Schwarz-class test function
of the external invariants yields a meromorphic function in $D$
in the entire complex plane, with singularities located among a discrete rational 
set $S_G$ which depends only on the graph $G$ not on the test function. 
In the $\phi^4$ case, no such singularity can occur in the strip $0<\Re D<2$, 
a region which is therefore a germ of analyticity common to the whole theory.
\end{theorem}

\noindent{\bf Proof}
 $\Gamma$ functions and their inverse
are meromorphic in the entire complex plane.
Formula (\ref{final}) is a tempered distribution 
which is the Fourier transform of a slow-growth function, see Appendix A.
It must be smeared against a Schwarz-class test function and by definition the result  
can be computed in Fourier space, as an ordinary integral of the product of
the slow-growth function by the rapid decaying Fourier transform of the test function.
At this stage the $D$-dependence occurs in fact in the integrand of an ordinary integral.

We can now copy the analysis of \cite{rivasseau,malbouisson5}. We introduce
general notations $z_\nu$ for the variables $x$ and $y$ and the lattices of polar varieties 
$\phi_{r}= -n_r$ where the different $\Gamma$ functions have their singularities. 
This defines convex cells, and singularities in $D$ can occur only when the hyperplane
$P(D)= \{\sum_\nu z_\nu = -D/2\}$ crosses the vertices of that lattice.
Because the coefficients $u$ and $v$ can only take integer values (in fact 0, 1 and 2)
this give a discrete set of rational values $S_G$ which depend on the graph, but not on the test-function. 
Out of these values,
to check that this integral is holomorphic around any dimension which does not cross the lattice vertices
is a simple consequence of commuting the Cauchy-Riemann $\partial_{\overline{D}}$ 
operator with the integral through Lebesgue's dominated convergence theorem.

Finally to check meromorphy, fix a particular $D_0 \in S_G$. We can write the Laurent series expansion of the 
integrand near that $D_0$ which starts as $a_{n_0} (D-D_0)^{-n_0}$. Multiplying
by $(D-D_0)^{n_0}$ we get a finite limit as $D \to D_0$, hence the isolated singularity at $D_0$
cannot be essential. Repeating the argument at each point of $S_G$ we obtain
meromorphy in the space time dimension $D$ in the whole complex plane $\mathbb{C}$,
apart from some discrete rational set where poles of finite order can occur.
\qed

Recall that this theorem is the starting point for the 
dimensional renormalization of NCQFTs 
which will be published elsewhere \cite{GurTan}.

\medskip
\noindent{\bf Appendix A: Proof of the formula \eqref{yI}}
\medskip

Let us return to formula (\ref{yI}). It follows from 
\begin{lemma}
For $-1 < s <0$, as tempered distribution in $u$
\begin{equation} 
\label{yIsimp}
e^{-iu}\chi (u)=\chi(u)
\int_{-\infty}^{+ \infty}\Gamma (-s-it)\left( iu \right) ^{s+it} \frac{d t}{2\pi }.
\end{equation}  
where $\chi$ is the Heaviside step function ensuring $u >0$.
\end{lemma}

\noindent{\bf Proof}
The left hand side is a locally integrable function bounded 
by one (together with all its derivatives) times the Heaviside function, so it is a tempered 
distribution. The right hand side is in fact a Fourier transform in $t$ but taken at a value $\log u$. 
To check that this is the same tempered distribution as $e^{-iu}\chi (u)$, let us apply 
it to a Schwartz test function $f$. Writing the usual representation for
the $\Gamma$ function we have to compute
\begin{equation} 
\label{ysimp2} I(f)=\frac{1}{2\pi }
\int_{0}^{\infty}  f(u) du 
 \int_{-\infty}^{+\infty} (iu)^{s+it} dt \int_{0}^{\infty} x^{-s-it-1} e^{-x}dx.
\end{equation} 
But the integrand is analytic in $x$ in the open upper right quarter of the complex plane. Let us
rotate the contour from $x$ to $ix$. The contribution of the quarter circle with radius sent
to infinity tends to zero because of the condition $-1 < s <0$, and we find 
\begin{eqnarray} 
\label{ysimp3} I(f)&=&\frac{1}{2\pi }
\int_{0}^{\infty}  f(u) du 
\int_{-\infty}^{+\infty} (iu)^{s+it} dt \lim_{R \to \infty}
\int_{0}^{R} (ix)^{-s-it-1} e^{-ix} i dx \nonumber
\\&=&\frac{1}{2\pi } \int_{-\infty}^{+\infty} dt  
\int_{-\infty}^{\infty}  f(e^v) e^{v(s+1)}  e^{itv} dv   \lim_{R\to \infty}\int_{0}^{R} x^{-s-it-1}e^{-ix} dx
\nonumber
\\&=&\frac{1}{2\pi } \int_{-\infty}^{+\infty} dt  \hat G (t)  \lim_{R\to \infty}\int_{0}^{R} 
  x^{-s-it-1}e^{-ix} dx
  \nonumber
\\&=& \lim_{R\to \infty}\int_{0}^{R}dx  e^{-ix}   x^{-s-1}  
\frac{1}{2\pi }\int_{-\infty}^{+\infty} dt  \hat G (t) e^{-it\log x}
\nonumber   
\\&=& \int_{0}^{\infty} f(x)  e^{-ix} dx\ .
\end{eqnarray} 
where $G(v) = f(e^v) e^{v (s+1)}$ is still rapid decay and
in the second line we change variables from $u$ to $v=\log u$ and use the 
definition of Fourier transform for distributions.
\qed

\medskip
\noindent{\bf Appendix B: The massive case}
\medskip

In the massive case we have to modify equation \eqref{beta} to include the additional
$e^{- \alpha_\ell m^2}$ per line, but unfortunately 
we have $e^{- \alpha_\ell} = \frac{1-t_\ell}{1+t_\ell}$ so that 
\eqref{beta} becomes
\beqa
\label{beta1}
\int_0^1 \prod_{\ell =1}^L dt_\ell   (1-t_\ell^2)^{\frac D2 -1}  \biggl(\frac{1-t_\ell}{1+t_\ell}\biggr)^{m^2}
t_\ell^{\phi_\ell -1}
=\frac 12 \beta_{m} (\frac{\phi_\ell}2, \frac D2).
\eeqa
where the modified ``$\beta_m$"  function is no longer an explicit quotient of $\Gamma$
functions, and depends of the mass $m$. 

It is rather easy to check that the analyticity and meromorphy properties
of this modified beta function are similar to those of the ordinary beta function,
at least for $\Re D >0$. Indeed
\beqa
\beta_{m} (\frac{\phi_\ell}2, \frac D2)-\beta (\frac{\phi_\ell}2, \frac D2)
=2\int_0^1 \prod_{\ell =1}^L dt_\ell   (1-t_\ell^2)^{\frac D2 -1}  \Big{[}
\biggl(\frac{1-t_\ell}{1+t_\ell}\biggr)^{m^2}-1
 \Big{]}
t_\ell^{\phi_\ell -1}.
\eeqa

The integral above is convergent for $t_\ell$ close to $1$ if $\Re D>0$ and for $t_\ell$ close to $0$ if $\Re \phi_l>-1$ so that it is an analytic function of $D$ and $\phi_l$ in this domain, hence $\beta$ and $\beta_m$
have the same singularity structure.
 
Nevertheless this CM representation becomes less explicit and therefore less attractive in this massive case.
Remark however that for $\Re D>0$, masses are not essential to the analysis of 
vulcanized NC field theories  which have no infrared divergencies and only ``half-a-direction" for 
their renormalization group anyway. As remarked earlier, masses can only push further away
the infrared singularities at negative $D$ that come form the hyperbolic nature of the Mehler kernel.
But it is not clear whether such infrared singularities at negative non commutative dimensions
have any physical interpretation.

\newpage
\thispagestyle{empty}

\newpage

\end{document}